\documentclass{article}

\usepackage{amsmath}
\usepackage{amssymb}
\usepackage{enumerate}
\usepackage{mathrsfs}
\usepackage{color}

\usepackage{breqn}

\usepackage{lipsum}

\numberwithin{equation}{section}

\newcommand{\inv}[1]{{#1}^{\mbox{\tiny -1}}}
\newcommand{\invt}[1]{{#1}^{\mbox{\tiny -T}}}
\newcommand{\transp}[1]{{#1}^{\mbox{\tiny T}}}
\newcommand{\transpb}[2]{#1_{#2}^{\mbox{\tiny T}}}

\begin{document}

\title{\bf Partial Proof of a Conjecture with\\ Implications for Spectral Majorization\vspace*{-8pt}}
\author{Jeffrey Uhlmann\thanks{(Corresponding author) 201 Naka Hall, Dept.\ of Electrical Engineering and Computer Science, University of Missouri-Columbia, uhlmannj@missouri.edu}}
\date{}
\maketitle

\vspace*{-18pt}
\begin{abstract}
In this paper we report on new results relating to a conjecture
regarding properties of $n\times n$, $n\leq 6$, positive definite 
matrices. The conjecture has been proven for $n\leq 4$ using
computer-assisted
sum of squares (SoS) methods for proving polynomial
nonnegativity. Based on these proven cases, we report on 
the recent identification of a new family of matrices with the
property that their diagonals majorize their spectrum. We
then present new results showing that this family can extended
via Kronecker composition to $n>6$ while retaining the 
special majorization property. We conclude with general
considerations on the future of computer-assisted and
AI-based proofs.\\ \vspace*{-4pt}
~ 
\end{abstract}

\section{Introduction}

In this paper we report new developments based on the proven cases of a surprising
conjecture relating to special properties of $n\times n$ positive definite (PD) 
matrices for $n\leq 6$. It is argued in \cite{irga} that traditional mathematics has 
focused primarily on results that hold generally for all $n$, whereas most 
theoretical physics models, and most applied mathematics and engineering 
problems, are intrinsically defined in a fixed (and small) number of dimensions, 
e.g., time and the three spatial dimensions of ordinary experience. What is not 
commonly recognized is that as soon the dimensionality of a problem becomes 
fixed, e.g., to 3 dimensions, opportunities exist to establish potentially useful 
properties of the system of interest that do not hold generally in higher 
dimensions. Unfortunately, proving such properties typically requires
computer-assisted methods that are not familiar to most scientists,
mathematicians, and  engineers. 

In Section 2, we illustrate these statements by showing 
how the IRGA conjecture for $n\leq 3$ is relatively
straightforward to establish by hand, while proof of the $n=4$ 
case was accomplished using powerful computer-assisted proof
methods. We discuss why such methods are likely
required, and why proofs for the remaining $n=5$ and $n=6$ 
cases are likely beyond the capabilities of current state-of-the-art 
methods on even the most powerful supercomputers.

In Section 3, we describe how proven cases of the IRGA conjecture 
define a fixed-dimensional family of matrices for which the diagonal 
majorizes the spectrum \cite{specmat}. In Section 4, we present new 
results showing that Kronecker products of these matrices retain this 
unique majorization property. Then, in Section 5, we conclude with 
considerations on the imminent arrival of AI-based theorem provers, 
which can be viewed as proof oracles rather than computer-assisted 
tools.

\section{IRGA Conjecture}


Given a real symmetric $n\times n$ positive-definite (PD) matrix $P$, the author conjectured the following:
\begin{quote}
{\bf IRGA Conjecture}: Given any real $n\!\times\! n$ symmetric positive-definite (PD) 
matrix $P$, then for $n\!\leq\! 6$:
\end{quote}
\vspace*{-8pt} 
\begin{equation}
S \,=\,\inv{(P\circ \inv{P})} ~~ \in ~ \mbox{\em nonsingular nonnegative PD doubly-stochastic} \label{pdrga}
\end{equation}
\begin{quote}
\vspace*{-6pt}
where ``$\circ$'' is the Schur-Hadamard elementwise matrix product. 
\end{quote}

As an example, given that the following PD matrix $P$ 
\begin{equation}
   P ~=\, \left[\!\begin{array}{rrrr}
   1.93758  & 1.04850 & -2.92314 &  0.83685\\
   1.04850  & 1.29729  &-1.49559 &  0.56616\\
  -2.92314 & -1.49559 &  8.97170&  -0.55271\\
   0.83685 &  0.56616  &-0.55271 &  0.61107
                  \end{array}\right] 
\end{equation}
the conjecture implies that the following must be (and is) 
doubly-stochastic:
\begin{equation}
   \inv{(P\circ \inv{P})}  ~=\, \left[\!\begin{array}{rrrr}
   0.34750 &  0.13498 &  0.25568 &  0.26184\\
   0.13498 &  0.58077 &  0.11800 &  0.16625\\
   0.25568 &  0.11800  & 0.47034 &  0.15597\\
   0.26184 &  0.16625  & 0.15597 &  0.41593
                     \end{array}\right] .
\end{equation}

The form $M\circ \invt{M}$, for general nonsingular matrix $M$, is a  familiar tool in the field of process control and is referred to as the Relative Gain Array (RGA) \cite{bristol,ucrga,rafal}, and its inverse inspires the IRGA Conjecture appellation. Two key properties of the RGA are invariance with respect to arbitrary diagonal scaling, i.e., RGA($M$)=RGA($DME$) for nonsingular diagonal matrices $D$ and $E$, and a sum of $1$ for entries in each row and column \cite{bristol,rgajs,rafal}.  Both of these properties can also be verified to hold for for $\inv{(P\circ \inv{P})}\!\!$\!, so proof of the conjecture can be reduced to proving that $\inv{(P\circ \inv{P})}$ is nonnegative, which implies that the result is doubly stochastic.  

We note that a proof of the conjecture for $n=k$ immediately subsumes the $n<k$ cases. The $n=1$ case is trivial, and the $2\times 2$ case is also amenable to hand verification. Proof of the $n=3$ case, however, requires a more structured approach involving a parameterization of PD matrices. This can be done in terms of the lower-triangular Cholesky square root, $\mbox{Chol}(P)\cdot\transp{\mbox{Chol}(P)}=P$, which is guaranteed to exist for any PD matrix. In the case of $P$ for $n=3$, the Cholesky square root can be expressed as:
\begin{equation}
\mbox{Chol($P$)} ~=~
 \left[\,\begin{array}{ccc}
d_1 & 0 & 0\\
x & d_2 & 0\\
y & z & d_3
\end{array}\,\right]
\end{equation}
Because the form of the conjecture is invariant with respect to left and right multiplication by a positive diagonal matrix $D$, by taking $D=\mbox{diag}(1/d_1,1/d_2,1/d_3)$, we can assume without loss of generality that the Cholesky square root of interest has the form
\begin{equation}
L_3 ~=~ \mbox{Chol($DPD$)}~=~
 \left[\,\begin{array}{ccc}
1 & 0 & 0\\
a & 1 & 0\\
b & c & 1
\end{array}\,\right] ,
\end{equation}
which now has three fewer variables. Letting $R=L_3\,\transpb{L}{3}$, 
the conjecture implies $S=\inv{(R\!\circ\!\inv{R})}$ is doubly stochastic. 
Because a permutation similarity can move any off-diagonal entry of $S$ to any other 
off-diagonal position, it is now sufficient by typicality to show that any off-diagonal entry 
of $S$ is nonnegative. It can be verified that the polynomial for entry $(2,3)$ of $S$ is:
\begin{equation}
\label{pn3}
a^2 b^2 + a b c + c^2 + a^2 c^2 + b^2 c^2 - 2 a b c^3 + a^2 c^4
\end{equation}
which, although somewhat arduous, can be transformed by hand into sum-of-squares (SoS) form: 
\begin{eqnarray}
(a c)^2  ~~+ & \frac{1}{4}\left(a b\sqrt{2-\sqrt{3}}+c\sqrt{2+\sqrt{3}}\right)^2 & +~\\
~          &  \frac{1}{4}\left(a b\sqrt{2+\sqrt{3}}+c\sqrt{2-\sqrt{3}} \right)^2 & +~~ \left(b c-a c^2\right)^2\!\!,
\end{eqnarray}
which establishes nonnegativity of the chosen typical entry, hence also the nonnegativity and doubly stochasticity of $S$.

Unfortunately, the degree of the resulting polynomial for a typical entry 
of grows rapidly with $n$, and it can be verified that the corresponding 
entry from $S=(L_4\transp{L_4\!\!})$ is:
\begin{dmath}
\label{pn4}
-\left(d^2+e^2+f^2+1\right) \left(a \left(f^2+1\right) \left(b^2+c^2+1\right) \left(-a c^2
   f^2-a c^2+2 a c e f-a e^2-a+b c f^2+b c-b e f-c d f+d e\right)-b (a b+c) \left(-c f^2-c+e
   f\right) \left(a c f^2+a c-a e f-b f^2-b+d f\right)\right)-d (-a c f+a e+b f-d)
   \left(\left(f^2+1\right) \left(-\left(b^2+c^2+1\right)\right) (a d+e) (c f-e)-f (a b+c)
   \left(-c f^2-c+e f\right) (b d+c e+f)\right)-f (b d+c e+f) \left(-a f (b d+c e+f)
   \left(-a c^2 f^2-a c^2+2 a c e f-a e^2-a+b c f^2+b c-b e f-c d f+d e\right)-b (a d+e) (c
   f-e) \left(a c f^2+a c-a e f-b f^2-b+d f\right)\right)
\end{dmath}
for which the conversion to SoS form is not only well-beyond 
what can feasibly be done by hand, it is beyond the practical capabilities of most if not all
general-purpose computer algebra systems (CAS), e.g., Mathematica. To give a feel for
the rapid growth in complexity going from $n=3$ to $n=4$, the following is the Cholesky
parameterization
\begin{equation}
L_4 ~=~ 
 \left[\,\begin{array}{cccc}
1&0&0&0\\
a&1&0&0\\
b&c&1&0\\
d&e&f&1
\end{array}\,\right]
\end{equation}
from which the polynomial
for entry (1,2) of $S_4$ is:
\begin{align*}
S_4(1,2) =& \\
&-(d^2+e^2+f^2+1) (a (f^2+1) (b^2\\
&+c^2+1) (-a c^2 f^2-a c^2+2 a c e f\\
&-a e^2-a+b c f^2+b c-b e f-c d f+d e)\\
&-b (a b+c)(-c f^2-c+e f)\\
&(a c f^2+a c-a e f-b f^2-b+d f))\\
&-(a d+e)(c f-e)(d (f^2+1)(-(b^2\\
&+c^2+1))(-a c f+a e+b f-d)\\
&-b f (b d+c e+f) (a c f^2+a c-a e f\\
&-b f^2-b+d f))-f (b d+c e+f)(d (-(a b\\
&+c))(-c f^2 -c+e f)(-a c f +a e+b f-d)\\
&-a f (b d+c e +f)(-a c^2 f^2-a c^2+2 a c e f\\
&-a e^2-a+b c f^2+b c-b e f-c d f+d e)).
\end{align*}
The SoS solution from \cite{irga}, obtained using specialized state-of-the-art SoS methods \cite{wang,wu,doublynonneg,peyrl}, consists of the following 25 squared terms:
\allowdisplaybreaks
\begin{align*}
     p_1= &~ 2 (a c d f-a b e f)^2,\\
     p_2=&~\mbox{\large (}\frac{1}{2} a f^2 c^2+\frac{a c^2}{2}-\frac{1}{2} b f^2 c+\frac{d f c}{2}\\
            &~ -a e f c-\frac{b c}{2}+\frac{a e^2}{2}+a+\frac{b e f}{2}-\frac{d e}{2}\mbox{\large )}^2,\\
     p_3=&~\frac{3}{4} \mbox{\large (}\frac{1}{3} a f^2 c^2+\frac{a c^2}{3}-\frac{1}{3} b f^2 c+\frac{2}{3} a e f c\\
            &~~ -\frac{b c}{3}-\frac{d f c}{3}-a e^2+d e-\frac{b e f}{3}\mbox{\large )}^2,\\
     p_4=&\frac{2}{3}\! \left(\!\! -a f^2\! c^2\!-a c^2\! +b f^2 c+b c+a e f c-\!\frac{d f c}{2}-\!\frac{b e f}{2}\right)^2\!\!\! ,\\
     p_5=&~\frac{1}{2} \left(a c^2 f^2-b c f^2+c d f-a c e f\right)^2,\\
     p_6=&~\frac{1}{2} \left(a c^2 f^2-b c f^2+b e f-a c e f\right)^2,\\
     p_7=&~\mbox{\large (}\! -\frac{1}{2} a e f c^2+\frac{1}{2} a e^2 c\\
            &~~ -\frac{1}{2} a f^2 c+a c+\frac{1}{2} b e f c-\frac{d e c}{2}+\frac{a e f}{2}\mbox{\large )}^2,\\
     p_8=~&\frac{3}{4} \left(a e f c^2\!-a e^2 c-\frac{1}{3} a f^2 c+d e c-b e f c+\!\frac{a e f}{3}\right)^2\!\! ,\\
     p_9=~&\frac{2}{3} \left(a e f-a c f^2\right)^2,\\
     p_{10}=&\left(-\frac{1}{2} a d f c^2+\frac{1}{2} b d f c+\frac{1}{2} a b e^2+a b-\frac{b d e}{2}\right)^2,\\
     p_{11}=&~\frac{3}{4} \left(a d f c^2-b d f c-a b e^2+b d e\right)^2,\\
     p_{12}=&\left(a d f-a b f^2\right)^2,\\
     p_{13}=&\left(\frac{1}{2} a e c^2-\frac{b e c}{2}-\frac{a f c}{2}+a e\right)^2,\\
     p_{14}=&\frac{3}{4} \left(-a e c^2+b e c-a f c\right)^2,\\
     p_{15}=&~ 2 a^2 c^2 f^2,\\
     p_{16}=&\left(\frac{1}{2} a d c^2-\frac{b d c}{2}+a d\right)^2,\\
     p_{17}=&~ \frac{3}{4} \left(-a d c^2+b d c-\frac{2 a b f}{3}\right)^2,\\
     p_{18}=&~ \frac{5}{3} a^2 b^2 f^2,\\
     p_{19}=&~ a^2 c^2 d^2,\\
     p_{20}=&~2 \mbox{\large (}\frac{1}{2} a c^2 f^3-\frac{1}{2} b c f^3+\frac{1}{2} c d f^2+\frac{1}{2} b e f^2-a c e f^2\\
                &~~~ +\frac{1}{2} a c^2 f+\frac{1}{2} a e^2 f+a f-\frac{b c f}{2}-\frac{d e f}{2}\mbox{\large )}^2,\\
     p_{21}=&~2 \mbox{\large (}\!\! -\frac{1}{2} a c^2 f^3+\frac{1}{2} b c f^3-\frac{1}{2} c d f^2-\frac{1}{2} b e f^2\\
                &~~~~ +a c e f^2-\frac{1}{2} a e^2 f+\frac{d e f}{2}\mbox{\large )}^2,\\
     p_{22}=&~ \frac{1}{2} \left(b c f-a c^2 f\right)^2,\\
     p_{23}=&~ a^2 c^2 e^2,\\
     p_{24}=&~ a^2 b^2 d^2,\\
     p_{25}=&~ a^2 b^2 e^2.
\end{align*}

The restriction $n\leq 6$ of the conjecture can be established from 
counterexamples that are easily found for $n$=7 (an explicit counterexample is given in 
\cite{irga}). Therefore, it remains only to prove the 
$n$=6 case, which of course subsumes the $n=5$ case. Unfortunately, the size of the
polynomial for a typical off-diagonal entry of $S_6$ vastly exceeds the capability of any known
SoS solution method. Purely as a dramatic illustration, the polynomial for entry (1,2) of $S_6$ is
provided in the Appendix.

\section{SPPD Matrices}

In this section we consider one of the suprising implications of the IRGA conjecture
and its proven cases. Letting $k$ be the largest integer for which the conjecture 
holds, we define the following matrix form:
\vspace*{-4pt} 
\begin{quote}
{\bf Definition} (Special PD-Diagonalizable): {\em A given
$n\!\times\! n$ matrix $M\!=\!PD\inv{P}$, $n\!\leq\! k$, is defined
to be special PD-diagonalizable if matrix $D$ is diagonal and 
positive-definite matrix $P$ is real symmetric.}
\end{quote}
The motivation for this definition is the following:
\vspace*{-4pt}
\begin{quote}
{\bf Majorization Theorem:} {\em The spectrum of a special PD-diagonalizable 
matrix is majorized by its diagonal.}
\end{quote}
\vspace*{-4pt}
Majorization is an important theoretical and practical property 
that is exploited in a variety of problem domains because it  
provides a preorder relational operator for comparing
the relative disorder (entropy) of the distribution of entries of two vectors.
Specifically, given two  vectors  $x, y \in \mathbb{R}^n$ with equal
entry sums,  $y$ is said to
{\em majorize} $x$, notated as $x\prec y$, if and only if 
there exists a set of permutation matrices $P_j$ and probabilities $p_j$ 
such that
\begin{equation}
x ~=~ \sum_j p_j P_j y \, .
\label{gendef}
\end{equation}
This expression of $x$ as a probabilistic sum of permutations
of $y$ can be thought of as indicating that the entry distribution
of $x$ is more {\em disordered} in a particular sense than that
of $y$. An alternative definition is that $x\prec y$ if and only if
\begin{equation}
x ~=~ S y \, ,
\label{dsdef}
\end{equation}
where $S$ is a doubly-stochastic matrix, i.e., nonnegative
with every row and column sum equal to unity. This definition
makes the probabilistic interpretation even more explicit 
in the form $Sy\rightarrow x$, where $x$ can be interpreted
a conservative stochastic evolutionary state of $y$. This interpretation
can be made rigorous by the following known result:
\begin{equation}
x \prec y ~\implies~ H(x) \geq H(y)
\end{equation}
where $H(\cdot)$ is the Shannon entropy,
$H(z) \doteq -\sum_{i=1}^n 
z_i \log{z_i}$, where $z_i\log z_i$ is taken as $0$ 
for $z_i$=$0$. 

The definition in terms of a doubly-stochastic relationship
also permits majorization to be generalized from real to complex 
(or hypercomplex) vectors \cite{bapat}.

We begin our proof of the majorization theorem by noting a known
result that relates the spectrum vector of a diagonalizable matrix $M=AE\inv{A}$,
where $\lambda(A) ~=~ \mbox{diag}(E)$
 for diagonal eigenvalue
matrix $E$, to its diagonal vector via RGA($A$) as: 
\begin{equation}
(A\circ \invt{A}) \cdot \mbox{diag}(M) ~=~ \lambda(M) ~ \label{EtoD}
\end{equation}
In the case of nonsingular RGA($A$), this relationship can be expressed as:
\begin{equation}
\mbox{diag}(M) ~=~ \inv{(A\circ \invt{A})} \cdot \lambda(M)  \label{DtoA}.
\end{equation}
In the case of special PD-diagonalizable $M$ diagonalized by a real positive definite $P$, this
gives
\begin{eqnarray}
\mbox{diag}(M) & = &  \inv{(P\circ \invt{P})} \cdot \lambda(M)\\
~ & = & S \cdot \lambda(M)
\end{eqnarray}
where $S$ is doubly stochastic. From Eq.\,(\ref{dsdef}) we can infer
that the diagonal of a special PD-diagonalizable matrix majorizes its spectrum,
thus proving the majorization theorem.

The significance of this result is that it establishes special PD-diagonalizable 
matrices as only the second known matrix class for which a majorization
relationship exists between the spectrum and the diagonal. The only previously
known class is the set of unitarily diagonalizable matrices $M=UEU^*$. The
majorization property of this class can be directly verified by
applying Eq.\,(\ref{EtoD})
\begin{equation}
(U\circ \invt{U}) \cdot \lambda(M) ~=~ \mbox{diag}(M).
\end{equation}
and observing that the Hadamard product of a unitary matrix and its
transposed inverse necessarily gives a doubly stochastic result. This 
shows that the spectrum of any unitarily diagonalizable matrix
majorizes its diagonal. Thus, the set of special PD-diagonalizable matrices
becomes the only known matrix class with the reverse property of 
the diagonal majorizing the spectrum. 

\section{Generalized PD-Diagonalizable (GPDD) Matrices}

In \cite{specmat}, an application of special PD-diagonalizable (SPDD) matrices involving manifold search. The problem assumes a $k$-dimensional manifold of diagonal matrices over a family of objects for which generalized majorization defines a meaningful partial order, where the set of objects/points has been subjected to a real continuously deformed positive definite gauge-like similarity transformation. In other words, each point of the manifold (or lattice) is an SPDD matrix. The goal is to proceed from a point on the manifold and proceed to a local optimum of high or low spectral entropy. The following is a high-level description of the basic algorithm:
\begin{enumerate}
    \item Initialize the search at a random point on the SPDD manifold, represented by the matrix $A_0$ with diagonal entries $\mathbf{d}_0 = [d_{01}, d_{02}, \ldots, d_{0N}]^\top$.
    
    \item For the current point on the manifold, compute its neighbors on the manifold. The neighbors can be defined according to a specific neighborhood relation, e.g., within a certain distance along the manifold or connected through a given transformation.
    
    \item Among the neighbors, identify the neighboring matrix $A_k$ with diagonal entries $\mathbf{d}_k = [d_{k1}, d_{k2}$, $...$, $d_{kN}]^\top$ such that $\mathbf{d}_0$ majorizes $\mathbf{d}_k$, i.e., determine whether there exists a doubly-stochastic transformation relating $\mathbf{d}_0$ and $\mathbf{d}_k$ that satisfies the desired direction of majorization.
    
    \item If such a neighboring matrix $A_k$ is found, update the current point on the manifold to the new point corresponding to $A_k$ and its diagonal entries $\mathbf{d}_k$. Return to step 2 and continue the search.
    
    \item If no neighboring matrix is found that satisfies the majorization condition, terminate the search. The current point on the manifold corresponds to a local minimum with respect to the majorization property.
\end{enumerate}
This search algorithm is designed to exploit the majorization property of SPDD matrices to traverse the transformed manifold in a systematic manner to a point corresponding to high or low spectral entropy using only diagonal information, i.e., without need for a spectral decomposition at every point. However, the practical applicability of the approach is severely constrained by the upper bound $k\leq n$ size limitation of SPDD matrices.

We note that the proven SPDD majorization relationship exploits an explicit construction of a doubly-stochastic relationship between the diagonal and spectrum based on the IRGA conjecture. However, the fact that such a construction does not generally exist for $n>6$ does not necessarily imply that the diagonal-spectrum majorization relationship does not hold in general. In other words, the constructive proof for all $n\leq 4$ could be interpreted as possible evidence that the result may in fact hold in general, i.e., while $\inv{(P\circ\inv{P})}$ may not always be doubly stochastic for $n>6$ -- {\em though it often is} -- there could exist other structural properties of a general PD-diagonalizable matrix that ensures the existence of a doubly-stochastic relationship between its diagonal and spectrum. Such structure may even hold for diagonalizability by arbitrary complex-valued PD matrices. Empirical investigation may be able to provide sufficient evidence to justify a conjecture or, alternatively, empirical and/or analytical efforts may identify counterexamples that constrain such generalizations

One viable generalization involves exanding the set of special PD-diagonalizable (SPPD) matrices to include PD matrices $P$ obtained as a Kronecker product,  $P=P_1\otimes P_2$, of SPPD matrices $P_1$ and $P_2$ such that the majorization property is preserved. For example, if $P_1$ is $4\times 4$, and $P_2$ is $3\times 3$, then $P$ would have size $12\times 12$. We now show that this generalization does in fact extend the class of matrices for which the diagonal majorizes the spectrum. We now provide a theorem formalizing such a generalization:
\vspace*{-4pt}
\begin{quote}
{\bf Generalized PDD Theorem:} {\em Kronecker products of SPPD matrices retain the SPDD majorization property.}
\end{quote}
\vspace*{-4pt}
 We begin with a relatively general lemma:
\vspace*{-4pt}
\begin{quote}
{\bf Diagonal-Spectrum Mapping Lemma:} Let $M_1$ and $M_2$ be $n \times n$ matrices with $\text{diag}(M_1) = S_1\, \lambda(M_1)$ and $\text{diag}(M_2) = S_2\, \lambda(M_2)$. The Kronecker product $S_1 \otimes S_2$ maps $\lambda(M_1 \otimes M_2)$ to the diagonal of $M_1 \otimes M_2$.
\end{quote}
\vspace*{-4pt}

We note that the eigenvalues of the Kronecker product of two matrices are given by the Kronecker product of their eigenvalues:
\begin{equation}
\lambda(M_1 \otimes M_2) = \lambda(M_1) \otimes \lambda(M_2) \,  ,
\end{equation}
and the diagonal of $M_1 \otimes M_2$ is given by the Kronecker product of the diagonals of $M_1$ and $M_2$:
\begin{equation}
\text{diag}(M_1 \otimes M_2) = \text{diag}(M_1) \otimes \text{diag}(M_2).
\end{equation}

Now, consider the product of $S_1 \otimes S_2$ and $\lambda(M_1 \otimes M_2)$:
\begin{equation}
(S_1 \otimes S_2)\, \lambda(M_1 \otimes M_2) = (S_1 \otimes S_2) (\lambda(M_1) \otimes \lambda(M_2)).
\end{equation}

Using the Kronecker mixed product property, $A \otimes B \cdot C \otimes D = AC \otimes BD$, we can rewrite the equation as:
\begin{equation}
(S_1 \otimes S_2)\, \lambda(M_1 \otimes M_2) = (S_1\,\lambda(M_1)) \otimes (S_2\,\lambda(M_2)).
\end{equation}

Since $\text{diag}(M_1) = S_1\, \lambda(M_1)$ and $\text{diag}(M_2) = S_2\, \lambda(M_2) = $, the equation becomes:
\begin{equation}
\text{diag}(M_1) \otimes \text{diag}(M_2) = (S_1 \otimes S_2)\, \lambda(M_1 \otimes M_2).
\end{equation}

Thus, we have verified that the Kronecker product $S_1 \otimes S_2$ maps $\lambda(M_1 \otimes M_2)$ to the diagonal of $M_1 \otimes M_2$. This implies the following corollary:
\vspace*{-4pt}
\begin{quote}
{\bf Corollary:} {\em Diagonal-Spectrum Majorization -} If $M_1$ and $M_2$ each possesses the property that its diagonal majorizes its spectrum, then $M_1\otimes M_2$ also possesses this property.
\end{quote}
\vspace*{-4pt}

The proof is straightforward: the majorization property of the corollary implies there exist doubly stochastic matrices $S_1\, \lambda(M_1) = \text{diag}(M_1)$ and $S_2\, \lambda(M_2) = \text{diag}(M_2)$. From the Diagonal-Spectrum Mapping Lemma, the Kronecker product $S_1 \otimes S_2$ maps $\lambda(M_1 \otimes M_2)$ to the diagonal of $M_1 \otimes M_2$.  Because the Kronecker product of doubly stochastic matrices is doubly stochastic, we can conclude that the diagonal of $M_1\otimes M_2$ majorizes its spectrum.

We have established the key property of the GPDD Theorem, but technically it remains to be shown that the Kronecker product of positive definite matrices is positive definite:
\vspace*{-4pt}
\begin{quote}
{\bf Kronecker PDD Lemma:} Given PD-diagonalizable matrices $M_1 = P_1 E_1 \inv{P_1}$ and $M_2 = P_2 E_2 \inv{P_2}$, where $P_1$ and $P_2$ are positive definite and $E_1$ and $E_2$ are diagonal, the Kronecker product $M_3 = M_1 \otimes M_2$ has the form $M_3 = P_3 E_3 \inv{P_3}$ for some positive definite $P_3$ and diagonal $E_3$.
\end{quote}
\vspace*{-4pt}
To prove this, onsider the Kronecker product $M_3 = M_1 \otimes M_2$:
\begin{equation*}
M_3 = (P_1 E_1 \inv{P_1}) \otimes (P_2 E_2 \inv{P_2}).
\end{equation*}

Using the Kronecker mixed product property, we have:
\begin{equation*}
M_3 = (P_1 \otimes P_2) (E_1 \otimes E_2) (\inv{P_1} \otimes \inv{P_2}).
\end{equation*}

Let $P_3 = P_1 \otimes P_2$ and $E_3 = E_1 \otimes E_2$. Since $P_1$ and $P_2$ are positive definite, their Kronecker product $P_3$ is also positive definite. Furthermore, since $E_1$ and $E_2$ are diagonal, their Kronecker product $E_3$ is also diagonal.

Finally, we show that $(\inv{P_1} \otimes \inv{P_2})$ is the inverse of $P_3$. Using the Kronecker mixed product property, we have:
\begin{equation*}
(P_1 \otimes P_2)(\inv{P_1} \otimes \inv{P_2}) = (P_1 \inv{P_1}) \otimes (P_2 \inv{P_2}) = I_n \otimes I_n = I_{n^2}.
\end{equation*}
This implies that $(\inv{P_1} \otimes \inv{P_2})$ is the inverse of $P_3$, so $M_3$ has the sought form $M_3 = P_3 E_3 \inv{P_3}$, where $P_3$ is positive definite and $E_3$ is diagonal.

In summary, we have established that a PD-diagonalizable matrix of any size divisible by 2 or 3 (the only primes less than 4) can be constructed with the property that its diagonal majorizes its spectrum. However, we can generalize this to construct examples for all $n$ using block-diagonal constructions:
\vspace*{-4pt}
\begin{quote}
{\bf GPDD Construction Lemma:} Nontrivial GPDD matrices (i.e., containing no blocks that are strictly diagonal) can be constructed for any value of $n>1$. 
\end{quote}
\vspace*{-4pt}
This can be proven inductively using the values $\{2, 3, 4\}$, since the IRGA conjecture has been proven for $n \leq 4$. We begin by extending our set of base cases to include $5$ and $6$ by noting that the case $n=5$ can be constructed with nontrivial diagonal blocks of size 2 and 3, and the case $n=6$ can be constructed with nontrivial blocks of size 2 and 4. Now we can assume the statement holds for all values of $k$ such that $5 \leq k \leq n$. The inductive step for $n+1$ can be established as follows:
\begin{enumerate}
\item If $n$ is the sum of a sequence containing a 2, replace the 2 with a 4 to obtain $n + 2$. For $n + 1$, we can use the sequence for $n$ and append a 2.
   
\item If $n$ is the sum of a sequence containing a 3, replace the 3 with a 4 to obtain $n + 1$. For $n + 2$, we can use the sequence for $n + 1$ and append a 2.

\item If $n$ is a sum of all 4s, then for $n + 1$ and $n + 2$, we can append a 2 or 3 respectively.
\end{enumerate}
Thus, we can construct an instance of a block diagonal GPDD matrix, with every block of size greater than 1 (i.e., no trivial blocks), for any $n>1$. Note that this does not imply that every $n\times n$ matrix $M=PDP^{-1}$ satisfies the Majorization Theorem, just that nontrivial instances which do satisfy it can be constructed for all $n>1$. 

\section{Discussion}

In this paper we have reported on the recently-proven $n=4$ case of the IRGA conjecture, and how the conjecture has led to the discovery of the first matrix class for which the diagonal majorizes the spectrum. We then presented new results generalizing this class to include instances of any size $n$ divisible by 2 or 3. More generally, this class includes {\em all} PD-diagonalizable matrices of size $n=4$ and is conjectured (via the IRGA conjecture) to also include all PD-diagonalizable matrices of size $5$ and $6$. Some of these results are unlikely to ever have been recognized without computer assistance, and it is likely that the proofs for some could/can only be obtained by computer-assisted methods.

There seems little doubt that artificial intelligence (AI) agents will possess capabilities to solve mathematical problems vastly beyond what can be solved by any human. However, the results at the core of this paper motivate a conjecture that humans will be able to continue contributing to mathematical discovery. This idea is based on the fact that an arbitrarily long traversal of successive branches of derivation from an obscure result to some distant and seemingly unrelated result is easy, but identifying the path backward from that end-result alone can be vastly more difficult. With that, we conjecture the following:
\begin{quote}
{\bf Human vs.\ AI Conjecture}: While an AI will be able mine a vastly larger expanse of conjectures than humans, and it will be able to prove vastly more of what it finds than humans, the infinitude of the expanse will leave overlooked seams from which new results can be extracted by humans that otherwise might never be found.
\end{quote}
Because of its form, the conjecture can never be proven, but evidence for it can potentially be obtained from the following experiment:
\begin{quote}
{\bf Experiment}: Give an AI (or a team of AIs) access to everything that humans know at the current time about mathematics, but with the exception of a carefully selected human-produced theorem, and its related results, and test whether the AI can prove the theorem.  
\end{quote}
The outcome of the experiment may seem obvious based on the anticipated power of future AIs, but consider the case in which the removed theorem is something like the following:
\begin{quote}
{\em For any diagonalizable real matrix $M$, there exists a matrix $X$ in the neighborhood of $M$ with the property that $\mbox{diag}(X)$ majorizes its spectrum $\lambda(X)$ according to the generalized definition that vector $u$ majorizes $v$ if and only if there exists doubly-stochastic $S$ such that $u=Sv$. (The definition of ``neighborhood'' is not given so as to avoid giving away too much information, but it is well-defined, nontrivial, and not inconsistent with uses of the term elsewhere in the literature.)}
\end{quote}
The actual truth of the statement is unimportant for present illustrative purposes, so we just assume a human, or a human using computer-assisted tools, has proven it based on observations like those discussed in this paper. Key question: {\em Is it possible that insufficient information is given in the statement for an AI to identify a starting point that can lead it to a result that satisfies the statement?} Probably not, but can the answer be foreseen with certainty? The results described in this paper derive from the IRGA conjecture, which has little more significance than the features of a grain of sand that distinguish it from the billions of sourrounding grains on a beach. Anyone who chooses to pick up a grain at random will be able to record its features and, with high probablity, be the only person (AI or human) who will ever observe those features. 

A pessimist might argue that the Human/AI conjecture assumes a future in which humans are relegated to being sifters of minutiae, celebrating their trivial discoveries like a child showing a shiny stone to friends. An AI might agree but add that the pessimist's description actually applies to all accomplishments in human history. However, if the AI were to fail even one instance of the proposed test, then it must recognize that one or more human discoveries\footnote{Note that the situation with human mathematicians is not at all analogous to AI versus human players of chess or other games. It is more analogous to crypto mining in which (it is assumed) the more mining that is done, the more coin that will be found. In this view, although humans can be thought of as low-power computers, they may still contribute in proportion to that net power.} may someday prove essential for deriving results that are deemed important by both AIs and humans. In other words, ultimate assessments of importance must be left to the judgment of future historians -- {\em whether they be AI or human}.

\bibliographystyle{amsplain}

\appendix

\section{The {\bf\em n}\,=\,6 Polynomial}


What follows is the $(1,2)$ polynomial entry of $S_6$. This
is provided here to dramatically illustrate the increase in complexity over the $n$=3 and
$n$=4 polynomials given in Eqs.\,(\ref{pn3}) and (\ref{pn4}), respectively. We note 
that this is the maximally simplified form achievable with Mathematica. (As per
Mathematica convention, adjacency of variables implies multiplication.)

The exponential complexity of the problem can be understood intuitively by
considering the number of degrees of freedom that must be examined to confirm
or disconfirm whether it can be transformed to a sum of squares form.

$S_6$(1,2) =
\vspace*{6pt}\\ \tiny
-k (-k+a m+b n-a c n+d p-a e p-b f p+a c f p+g q-a h q-b i q+a c i q-d j q+a e j q+b f j q-a
   c f j q) (-(b g+c h+i) (-i q\textsuperscript 2+f j q\textsuperscript 2+n q-f p q-i+f j)
   ((d\textsuperscript 2+e\textsuperscript 2+f\textsuperscript 2+1) (q\textsuperscript 2 j\textsuperscript 2+j\textsuperscript 2-2 p q j+p\textsuperscript 2+1) ((a g+h) (b
   k+c m+n) (-n+f p+i q-f j q) (-h q\textsuperscript 2+c i q\textsuperscript 2+e j q\textsuperscript 2-c f j q\textsuperscript 2+m q-c n q-e p q+c f p
   q-h+c i+e j-c f j)-(b g+c h+i) (a k+m) (-m+c n+e p-c f p+h q-c i q-e j q+c f j q)
   (-i q\textsuperscript 2+f j q\textsuperscript 2+n q-f p q-i+f j))-(d g+e h+f i+j) (-j q\textsuperscript 2+p
   q-j) ((a d+e) (b k+c m+n) (-n+f p+i q-f j q) (-e q\textsuperscript 2 j\textsuperscript 2+c f q\textsuperscript 2 j\textsuperscript 2-e
   j\textsuperscript 2+c f j\textsuperscript 2+h q\textsuperscript 2 j-c i q\textsuperscript 2 j+h j-c i j-m q j+c n q j+2 e p q j-2 c f p q j-e p\textsuperscript 2+c f
   p\textsuperscript 2-e+c f+m p-c n p-h p q+c i p q)-(b d+c e+f) (a k+m) (-m+c n+e p-c f p+h q-c i
   q-e j q+c f j q) (-f q\textsuperscript 2 j\textsuperscript 2-f j\textsuperscript 2+i q\textsuperscript 2 j+i j-n q j+2 f p q j-f p\textsuperscript 2-f+n p-i p
   q))+(d k+e m+f n+p) (j q-p) ((a d+e) (b g+c h+i) (-i q\textsuperscript 2+f j q\textsuperscript 2+n
   q-f p q-i+f j) (-e q\textsuperscript 2 j\textsuperscript 2+c f q\textsuperscript 2 j\textsuperscript 2-e j\textsuperscript 2+c f j\textsuperscript 2+h q\textsuperscript 2 j-c i q\textsuperscript 2 j+h j-c i
   j-m q j+c n q j+2 e p q j-2 c f p q j-e p\textsuperscript 2+c f p\textsuperscript 2-e+c f+m p-c n p-h p q+c i p
   q)-(b d+c e+f) (a g+h) (-h q\textsuperscript 2+c i q\textsuperscript 2+e j q\textsuperscript 2-c f j q\textsuperscript 2+m q-c n q-e p q+c f p
   q-h+c i+e j-c f j) (-f q\textsuperscript 2 j\textsuperscript 2-f j\textsuperscript 2+i q\textsuperscript 2 j+i j-n q j+2 f p q j-f p\textsuperscript 2-f+n p-i
   p q)))+(d g+e h+f i+j) (-j q\textsuperscript 2+p q-j) ((b d+c e+f)
   (-f q\textsuperscript 2 j\textsuperscript 2-f j\textsuperscript 2+i q\textsuperscript 2 j+i j-n q j+2 f p q j-f p\textsuperscript 2-f+n p-i p q) ((a g+h)
   (b k+c m+n) (-n+f p+i q-f j q) (-h q\textsuperscript 2+c i q\textsuperscript 2+e j q\textsuperscript 2-c f j q\textsuperscript 2+m q-c n q-e p q+c f
   p q-h+c i+e j-c f j)-(b g+c h+i) (a k+m) (-m+c n+e p-c f p+h q-c i q-e j q+c f j q)
   (-i q\textsuperscript 2+f j q\textsuperscript 2+n q-f p q-i+f j))-(d g+e h+f i+j) (-j q\textsuperscript 2+p
   q-j) ((a b+c) (b k+c m+n) (-n+f p+i q-f j q) (-c j\textsuperscript 2 f\textsuperscript 2-c p\textsuperscript 2 f\textsuperscript 2-c j\textsuperscript 2
   q\textsuperscript 2 f\textsuperscript 2-c f\textsuperscript 2+2 c j p q f\textsuperscript 2+e j\textsuperscript 2 f+e p\textsuperscript 2 f+e j\textsuperscript 2 q\textsuperscript 2 f-h j q\textsuperscript 2 f+2 c i j q\textsuperscript 2 f+e f-h j
   f+2 c i j f-m p f+2 c n p f+j m q f-2 c j n q f+h p q f-2 c i p q f-2 e j p q f-c i\textsuperscript 2-c
   n\textsuperscript 2-c i\textsuperscript 2 q\textsuperscript 2+h i q\textsuperscript 2-e i j q\textsuperscript 2-c+h i-e i j+m n-e n p-i m q-h n q+2 c i n q+e j n q+e i p
   q)-(b\textsuperscript 2+c\textsuperscript 2+1) (a k+m) (-m+c n+e p-c f p+h q-c i q-e j q+c f j q)
   (j\textsuperscript 2 f\textsuperscript 2+p\textsuperscript 2 f\textsuperscript 2+j\textsuperscript 2 q\textsuperscript 2 f\textsuperscript 2-2 j p q f\textsuperscript 2+f\textsuperscript 2-2 i j q\textsuperscript 2 f-2 i j f-2 n p f+2 j n q f+2
   i p q f+i\textsuperscript 2+n\textsuperscript 2+i\textsuperscript 2 q\textsuperscript 2-2 i n q+1))+(d k+e m+f n+p) (j q-p) ((a b+c) (b
   g+c h+i) (-i q\textsuperscript 2+f j q\textsuperscript 2+n q-f p q-i+f j) (-c j\textsuperscript 2 f\textsuperscript 2-c p\textsuperscript 2 f\textsuperscript 2-c j\textsuperscript 2 q\textsuperscript 2
   f\textsuperscript 2-c f\textsuperscript 2+2 c j p q f\textsuperscript 2+e j\textsuperscript 2 f+e p\textsuperscript 2 f+e j\textsuperscript 2 q\textsuperscript 2 f-h j q\textsuperscript 2 f+2 c i j q\textsuperscript 2 f+e f-h j f+2 c
   i j f-m p f+2 c n p f+j m q f-2 c j n q f+h p q f-2 c i p q f-2 e j p q f-c i\textsuperscript 2-c n\textsuperscript 2-c
   i\textsuperscript 2 q\textsuperscript 2+h i q\textsuperscript 2-e i j q\textsuperscript 2-c+h i-e i j+m n-e n p-i m q-h n q+2 c i n q+e j n q+e i p
   q)-(b\textsuperscript 2+c\textsuperscript 2+1) (a g+h) (-h q\textsuperscript 2+c i q\textsuperscript 2+e j q\textsuperscript 2-c f j q\textsuperscript 2+m q-c n
   q-e p q+c f p q-h+c i+e j-c f j) (j\textsuperscript 2 f\textsuperscript 2+p\textsuperscript 2 f\textsuperscript 2+j\textsuperscript 2 q\textsuperscript 2 f\textsuperscript 2-2 j p q
   f\textsuperscript 2+f\textsuperscript 2-2 i j q\textsuperscript 2 f-2 i j f-2 n p f+2 j n q f+2 i p q f+i\textsuperscript 2+n\textsuperscript 2+i\textsuperscript 2 q\textsuperscript 2-2 i n
   q+1)))-(g\textsuperscript 2+h\textsuperscript 2+i\textsuperscript 2+j\textsuperscript 2+1) (q\textsuperscript 2+1) ((b d+c
   e+f) (-f q\textsuperscript 2 j\textsuperscript 2-f j\textsuperscript 2+i q\textsuperscript 2 j+i j-n q j+2 f p q j-f p\textsuperscript 2-f+n p-i p q) ((a
   d+e) (b k+c m+n) (-n+f p+i q-f j q) (-e q\textsuperscript 2 j\textsuperscript 2+c f q\textsuperscript 2 j\textsuperscript 2-e j\textsuperscript 2+c f j\textsuperscript 2+h q\textsuperscript 2 j-c
   i q\textsuperscript 2 j+h j-c i j-m q j+c n q j+2 e p q j-2 c f p q j-e p\textsuperscript 2+c f p\textsuperscript 2-e+c f+m p-c n p-h p
   q+c i p q)-(b d+c e+f) (a k+m) (-m+c n+e p-c f p+h q-c i q-e j q+c f j q) (-f
   q\textsuperscript 2 j\textsuperscript 2-f j\textsuperscript 2+i q\textsuperscript 2 j+i j-n q j+2 f p q j-f p\textsuperscript 2-f+n p-i p
   q))-(d\textsuperscript 2+e\textsuperscript 2+f\textsuperscript 2+1) (q\textsuperscript 2 j\textsuperscript 2+j\textsuperscript 2-2 p q j+p\textsuperscript 2+1)
   ((a b+c) (b k+c m+n) (-n+f p+i q-f j q) (-c j\textsuperscript 2 f\textsuperscript 2-c p\textsuperscript 2 f\textsuperscript 2-c j\textsuperscript 2 q\textsuperscript 2 f\textsuperscript 2-c
   f\textsuperscript 2+2 c j p q f\textsuperscript 2+e j\textsuperscript 2 f+e p\textsuperscript 2 f+e j\textsuperscript 2 q\textsuperscript 2 f-h j q\textsuperscript 2 f+2 c i j q\textsuperscript 2 f+e f-h j f+2 c i j
   f-m p f+2 c n p f+j m q f-2 c j n q f+h p q f-2 c i p q f-2 e j p q f-c i\textsuperscript 2-c n\textsuperscript 2-c i\textsuperscript 2
   q\textsuperscript 2+h i q\textsuperscript 2-e i j q\textsuperscript 2-c+h i-e i j+m n-e n p-i m q-h n q+2 c i n q+e j n q+e i p
   q)-(b\textsuperscript 2+c\textsuperscript 2+1) (a k+m) (-m+c n+e p-c f p+h q-c i q-e j q+c f j q)
   (j\textsuperscript 2 f\textsuperscript 2+p\textsuperscript 2 f\textsuperscript 2+j\textsuperscript 2 q\textsuperscript 2 f\textsuperscript 2-2 j p q f\textsuperscript 2+f\textsuperscript 2-2 i j q\textsuperscript 2 f-2 i j f-2 n p f+2 j n q f+2
   i p q f+i\textsuperscript 2+n\textsuperscript 2+i\textsuperscript 2 q\textsuperscript 2-2 i n q+1))+(d k+e m+f n+p) (j q-p) ((a b+c) (b
   d+c e+f) (-f q\textsuperscript 2 j\textsuperscript 2-f j\textsuperscript 2+i q\textsuperscript 2 j+i j-n q j+2 f p q j-f p\textsuperscript 2-f+n p-i p q)
   (-c j\textsuperscript 2 f\textsuperscript 2-c p\textsuperscript 2 f\textsuperscript 2-c j\textsuperscript 2 q\textsuperscript 2 f\textsuperscript 2-c f\textsuperscript 2+2 c j p q f\textsuperscript 2+e j\textsuperscript 2 f+e p\textsuperscript 2 f+e j\textsuperscript 2 q\textsuperscript 2
   f-h j q\textsuperscript 2 f+2 c i j q\textsuperscript 2 f+e f-h j f+2 c i j f-m p f+2 c n p f+j m q f-2 c j n q f+h p q
   f-2 c i p q f-2 e j p q f-c i\textsuperscript 2-c n\textsuperscript 2-c i\textsuperscript 2 q\textsuperscript 2+h i q\textsuperscript 2-e i j q\textsuperscript 2-c+h i-e i j+m n-e n p-i
   m q-h n q+2 c i n q+e j n q+e i p q)-(b\textsuperscript 2+c\textsuperscript 2+1) (a d+e) (-e q\textsuperscript 2
   j\textsuperscript 2+c f q\textsuperscript 2 j\textsuperscript 2-e j\textsuperscript 2+c f j\textsuperscript 2+h q\textsuperscript 2 j-c i q\textsuperscript 2 j+h j-c i j-m q j+c n q j+2 e p q j-2 c f p
   q j-e p\textsuperscript 2+c f p\textsuperscript 2-e+c f+m p-c n p-h p q+c i p q) (j\textsuperscript 2 f\textsuperscript 2+p\textsuperscript 2 f\textsuperscript 2+j\textsuperscript 2 q\textsuperscript 2
   f\textsuperscript 2-2 j p q f\textsuperscript 2+f\textsuperscript 2-2 i j q\textsuperscript 2 f-2 i j f-2 n p f+2 j n q f+2 i p q f+i\textsuperscript 2+n\textsuperscript 2+i\textsuperscript 2 q\textsuperscript 2-2 i n
   q+1)))-q (g k+h m+i n+j p+q) ((b d+c e+f) (-f q\textsuperscript 2 j\textsuperscript 2-f j\textsuperscript 2+i
   q\textsuperscript 2 j+i j-n q j+2 f p q j-f p\textsuperscript 2-f+n p-i p q) ((a d+e) (b g+c h+i) (-i
   q\textsuperscript 2+f j q\textsuperscript 2+n q-f p q-i+f j) (-e q\textsuperscript 2 j\textsuperscript 2+c f q\textsuperscript 2 j\textsuperscript 2-e j\textsuperscript 2+c f j\textsuperscript 2+h q\textsuperscript 2 j-c i
   q\textsuperscript 2 j+h j-c i j-m q j+c n q j+2 e p q j-2 c f p q j-e p\textsuperscript 2+c f p\textsuperscript 2-e+c f+m p-c n p-h p q+c
   i p q)-(b d+c e+f) (a g+h) (-h q\textsuperscript 2+c i q\textsuperscript 2+e j q\textsuperscript 2-c f j q\textsuperscript 2+m q-c n q-e p q+c
   f p q-h+c i+e j-c f j) (-f q\textsuperscript 2 j\textsuperscript 2-f j\textsuperscript 2+i q\textsuperscript 2 j+i j-n q j+2 f p q j-f p\textsuperscript 2-f+n
   p-i p q))-(d\textsuperscript 2+e\textsuperscript 2+f\textsuperscript 2+1) (q\textsuperscript 2 j\textsuperscript 2+j\textsuperscript 2-2 p q j+p\textsuperscript 2+1)
   ((a b+c) (b g+c h+i) (-i q\textsuperscript 2+f j q\textsuperscript 2+n q-f p q-i+f j) (-c j\textsuperscript 2 f\textsuperscript 2-c
   p\textsuperscript 2 f\textsuperscript 2-c j\textsuperscript 2 q\textsuperscript 2 f\textsuperscript 2-c f\textsuperscript 2+2 c j p q f\textsuperscript 2+e j\textsuperscript 2 f+e p\textsuperscript 2 f+e j\textsuperscript 2 q\textsuperscript 2 f-h j q\textsuperscript 2 f+2 c i j
   q\textsuperscript 2 f+e f-h j f+2 c i j f-m p f+2 c n p f+j m q f-2 c j n q f+h p q f-2 c i p q f-2 e j p
   q f-c i\textsuperscript 2-c n\textsuperscript 2-c i\textsuperscript 2 q\textsuperscript 2+h i q\textsuperscript 2-e i j q\textsuperscript 2-c+h i-e i j+m n-e n p-i m q-h n q+2 c i n q+e
   j n q+e i p q)-(b\textsuperscript 2+c\textsuperscript 2+1) (a g+h) (-h q\textsuperscript 2+c i q\textsuperscript 2+e j q\textsuperscript 2-c f j
   q\textsuperscript 2+m q-c n q-e p q+c f p q-h+c i+e j-c f j) (j\textsuperscript 2 f\textsuperscript 2+p\textsuperscript 2 f\textsuperscript 2+j\textsuperscript 2 q\textsuperscript 2 f\textsuperscript 2-2 j
   p q f\textsuperscript 2+f\textsuperscript 2-2 i j q\textsuperscript 2 f-2 i j f-2 n p f+2 j n q f+2 i p q f+i\textsuperscript 2+n\textsuperscript 2+i\textsuperscript 2 q\textsuperscript 2-2 i n
   q+1))+(d g+e h+f i+j) (-j q\textsuperscript 2+p q-j) ((a b+c) (b d+c e+f)
   (-f q\textsuperscript 2 j\textsuperscript 2-f j\textsuperscript 2+i q\textsuperscript 2 j+i j-n q j+2 f p q j-f p\textsuperscript 2-f+n p-i p q) (-c j\textsuperscript 2
   f\textsuperscript 2-c p\textsuperscript 2 f\textsuperscript 2-c j\textsuperscript 2 q\textsuperscript 2 f\textsuperscript 2-c f\textsuperscript 2+2 c j p q f\textsuperscript 2+e j\textsuperscript 2 f+e p\textsuperscript 2 f+e j\textsuperscript 2 q\textsuperscript 2 f-h j q\textsuperscript 2 f+2 c
   i j q\textsuperscript 2 f+e f-h j f+2 c i j f-m p f+2 c n p f+j m q f-2 c j n q f+h p q f-2 c i p q f-2 e
   j p q f-c i\textsuperscript 2-c n\textsuperscript 2-c i\textsuperscript 2 q\textsuperscript 2+h i q\textsuperscript 2-e i j q\textsuperscript 2-c+h i-e i j+m n-e n p-i m q-h n q+2 c i n
   q+e j n q+e i p q)-(b\textsuperscript 2+c\textsuperscript 2+1) (a d+e) (-e q\textsuperscript 2 j\textsuperscript 2+c f q\textsuperscript 2 j\textsuperscript 2-e
   j\textsuperscript 2+c f j\textsuperscript 2+h q\textsuperscript 2 j-c i q\textsuperscript 2 j+h j-c i j-m q j+c n q j+2 e p q j-2 c f p q j-e p\textsuperscript 2+c f
   p\textsuperscript 2-e+c f+m p-c n p-h p q+c i p q) (j\textsuperscript 2 f\textsuperscript 2+p\textsuperscript 2 f\textsuperscript 2+j\textsuperscript 2 q\textsuperscript 2 f\textsuperscript 2-2 j p q
   f\textsuperscript 2+f\textsuperscript 2-2 i j q\textsuperscript 2 f-2 i j f-2 n p f+2 j n q f+2 i p q f+i\textsuperscript 2+n\textsuperscript 2+i\textsuperscript 2 q\textsuperscript 2-2 i n
   q+1))))+(b k+c m+n) (-n+f p+i q-f j q) (-g (-g q\textsuperscript 2+a h
   q\textsuperscript 2+b i q\textsuperscript 2-a c i q\textsuperscript 2+d j q\textsuperscript 2-a e j q\textsuperscript 2-b f j q\textsuperscript 2+a c f j q\textsuperscript 2+k q-a m q-b n q+a c n q-d p
   q+a e p q+b f p q-a c f p q-g+a h+b i-a c i+d j-a e j-b f j+a c f j)
   ((d\textsuperscript 2+e\textsuperscript 2+f\textsuperscript 2+1) (q\textsuperscript 2 j\textsuperscript 2+j\textsuperscript 2-2 p q j+p\textsuperscript 2+1) ((a g+h) (b
   k+c m+n) (-n+f p+i q-f j q) (-h q\textsuperscript 2+c i q\textsuperscript 2+e j q\textsuperscript 2-c f j q\textsuperscript 2+m q-c n q-e p q+c f p
   q-h+c i+e j-c f j)-(b g+c h+i) (a k+m) (-m+c n+e p-c f p+h q-c i q-e j q+c f j q)
   (-i q\textsuperscript 2+f j q\textsuperscript 2+n q-f p q-i+f j))-(d g+e h+f i+j) (-j q\textsuperscript 2+p
   q-j) ((a d+e) (b k+c m+n) (-n+f p+i q-f j q) (-e q\textsuperscript 2 j\textsuperscript 2+c f q\textsuperscript 2 j\textsuperscript 2-e
   j\textsuperscript 2+c f j\textsuperscript 2+h q\textsuperscript 2 j-c i q\textsuperscript 2 j+h j-c i j-m q j+c n q j+2 e p q j-2 c f p q j-e p\textsuperscript 2+c f
   p\textsuperscript 2-e+c f+m p-c n p-h p q+c i p q)-(b d+c e+f) (a k+m) (-m+c n+e p-c f p+h q-c i
   q-e j q+c f j q) (-f q\textsuperscript 2 j\textsuperscript 2-f j\textsuperscript 2+i q\textsuperscript 2 j+i j-n q j+2 f p q j-f p\textsuperscript 2-f+n p-i p
   q))+(d k+e m+f n+p) (j q-p) ((a d+e) (b g+c h+i) (-i q\textsuperscript 2+f j q\textsuperscript 2+n
   q-f p q-i+f j) (-e q\textsuperscript 2 j\textsuperscript 2+c f q\textsuperscript 2 j\textsuperscript 2-e j\textsuperscript 2+c f j\textsuperscript 2+h q\textsuperscript 2 j-c i q\textsuperscript 2 j+h j-c i
   j-m q j+c n q j+2 e p q j-2 c f p q j-e p\textsuperscript 2+c f p\textsuperscript 2-e+c f+m p-c n p-h p q+c i p
   q)-(b d+c e+f) (a g+h) (-h q\textsuperscript 2+c i q\textsuperscript 2+e j q\textsuperscript 2-c f j q\textsuperscript 2+m q-c n q-e p q+c f p
   q-h+c i+e j-c f j) (-f q\textsuperscript 2 j\textsuperscript 2-f j\textsuperscript 2+i q\textsuperscript 2 j+i j-n q j+2 f p q j-f p\textsuperscript 2-f+n p-i
   p q)))+(d g+e h+f i+j) (-j q\textsuperscript 2+p q-j) (d (-d q\textsuperscript 2
   j\textsuperscript 2+a e q\textsuperscript 2 j\textsuperscript 2+b f q\textsuperscript 2 j\textsuperscript 2-a c f q\textsuperscript 2 j\textsuperscript 2-d j\textsuperscript 2+a e j\textsuperscript 2+b f j\textsuperscript 2-a c f j\textsuperscript 2+g q\textsuperscript 2 j-a h q\textsuperscript 2
   j-b i q\textsuperscript 2 j+a c i q\textsuperscript 2 j+g j-a h j-b i j+a c i j-k q j+a m q j+b n q j-a c n q j+2 d p q
   j-2 a e p q j-2 b f p q j+2 a c f p q j-d p\textsuperscript 2+a e p\textsuperscript 2+b f p\textsuperscript 2-a c f p\textsuperscript 2-d+a e+b f-a c f+k
   p-a m p-b n p+a c n p-g p q+a h p q+b i p q-a c i p q) ((a g+h) (b k+c m+n)
   (-n+f p+i q-f j q) (-h q\textsuperscript 2+c i q\textsuperscript 2+e j q\textsuperscript 2-c f j q\textsuperscript 2+m q-c n q-e p q+c f p q-h+c i+e
   j-c f j)-(b g+c h+i) (a k+m) (-m+c n+e p-c f p+h q-c i q-e j q+c f j q) (-i
   q\textsuperscript 2+f j q\textsuperscript 2+n q-f p q-i+f j))-(d g+e h+f i+j) (-j q\textsuperscript 2+p q-j)
   (a (b k+c m+n) (-n+f p+i q-f j q) ((g k+h m+i n+j p+q) (a i n c\textsuperscript 2-a f j n
   c\textsuperscript 2-a f i p c\textsuperscript 2+a f\textsuperscript 2 j p c\textsuperscript 2+a f\textsuperscript 2 q c\textsuperscript 2+a q c\textsuperscript 2-a i m c+a f j m c+g n c-a h n c-b i n
   c-d j n c+a e j n c+b f j n c-f g p c+a f h p c+a e i p c+b f i p c-b f\textsuperscript 2 j p c+d f j p
   c-2 a e f j p c-b f\textsuperscript 2 q c-b q c+d f q c-2 a e f q c-g m+a h m+b i m+d j m-a e j m-b f j
   m+e g p-a e h p-b e i p+a e\textsuperscript 2 j p-d e j p+b e f j p+a e\textsuperscript 2 q+a q-d e q+b e f q)-(d
   k+e m+f n+p) (-a f j\textsuperscript 2 n c\textsuperscript 2-a f n c\textsuperscript 2+a i j n c\textsuperscript 2-a i\textsuperscript 2 p c\textsuperscript 2-a p c\textsuperscript 2+a f i j p
   c\textsuperscript 2+a f i q c\textsuperscript 2+a j q c\textsuperscript 2+a f j\textsuperscript 2 m c+a f m c-a i j m c-d j\textsuperscript 2 n c+a e j\textsuperscript 2 n c+b f j\textsuperscript 2 n
   c-d n c+a e n c+b f n c+g j n c-a h j n c-b i j n c+b i\textsuperscript 2 p c+b p c-g i p c+2 a h i p c-a
   f h j p c+d i j p c-a e i j p c-b f i j p c-a f h q c+d i q c-a e i q c-b f i q c-b j q
   c+d j\textsuperscript 2 m-a e j\textsuperscript 2 m-b f j\textsuperscript 2 m+d m-a e m-b f m-g j m+a h j m+b i j m-a h\textsuperscript 2 p-a p+g h p-b h
   i p-d h j p+a e h j p+b f h j p-d h q+a e h q+b f h q+a j q)+(b k+c m+n) (a
   j\textsuperscript 2 n e\textsuperscript 2+a n e\textsuperscript 2-a i j p e\textsuperscript 2-a i q e\textsuperscript 2-a f j\textsuperscript 2 m e-a f m e+a i j m e-d j\textsuperscript 2 n e+b f j\textsuperscript 2 n
   e-a c f j\textsuperscript 2 n e-d n e+b f n e-a c f n e+g j n e-2 a h j n e-b i j n e+a c i j n e+b i\textsuperscript 2 p
   e-a c i\textsuperscript 2 p e+b p e-a c p e-g i p e+a h i p e+a f h j p e+d i j p e-b f i j p e+a c f i j
   p e+a f h q e+d i q e-b f i q e+a c f i q e-b j q e+a c j q e-b f\textsuperscript 2 m+a c f\textsuperscript 2 m-b i\textsuperscript 2 m+a
   c i\textsuperscript 2 m-b f\textsuperscript 2 j\textsuperscript 2 m+a c f\textsuperscript 2 j\textsuperscript 2 m+d f j\textsuperscript 2 m-b m+a c m+d f m+g i m-a h i m-f g j m+a f h j
   m-d i j m+2 b f i j m-2 a c f i j m+a h\textsuperscript 2 n+a n-g h n+b h i n-a c h i n+d h j n-b f h j
   n+a c f h j n-a f h\textsuperscript 2 p-a f p+f g h p-b f h i p+a c f h i p+b f\textsuperscript 2 h j p-a c f\textsuperscript 2 h j p-d f
   h j p+b f\textsuperscript 2 h q-a c f\textsuperscript 2 h q+b h q-a c h q-d f h q-a i q+a f j q)-(a f\textsuperscript 2 c\textsuperscript 2+a
   i\textsuperscript 2 c\textsuperscript 2+a f\textsuperscript 2 j\textsuperscript 2 c\textsuperscript 2+a c\textsuperscript 2-2 a f i j c\textsuperscript 2-b f\textsuperscript 2 c-b i\textsuperscript 2 c-b f\textsuperscript 2 j\textsuperscript 2 c+d f j\textsuperscript 2 c-2 a e f
   j\textsuperscript 2 c-b c+d f c-2 a e f c+g i c-2 a h i c-f g j c+2 a f h j c-d i j c+2 a e i j c+2 b f i
   j c+a e\textsuperscript 2+a h\textsuperscript 2+a e\textsuperscript 2 j\textsuperscript 2-d e j\textsuperscript 2+b e f j\textsuperscript 2+a-d e+b e f-g h+b h i+e g j+d h j-2 a e h j-b
   f h j-b e i j) (k\textsuperscript 2+m\textsuperscript 2+n\textsuperscript 2+p\textsuperscript 2+q\textsuperscript 2+1)-k
   (-(g\textsuperscript 2+h\textsuperscript 2+i\textsuperscript 2+j\textsuperscript 2+1) (-(d\textsuperscript 2+e\textsuperscript 2+f\textsuperscript 2+1) (-m b\textsuperscript 2+c k
   b+a c m b+a n b-a c\textsuperscript 2 k-a k-m+c n)+(b d+c e+f) (b e k-a c e k-a f k-b d m+a c d m-f
   m+a d n+e n)+(-e b\textsuperscript 2+c d b+a c e b+a f b-a c\textsuperscript 2 d-a d-e+c f) (d k+e m+f
   n+p))+(d g+e h+f i+j) (-(d g+e h+f i+j) (-m b\textsuperscript 2+c k b+a c m b+a n b-a c\textsuperscript 2
   k-a k-m+c n)+(b d+c e+f) (b h k-a c h k-a i k-b g m+a c g m-i m+a g n+h n)+(-h
   b\textsuperscript 2+c g b+a c h b+a i b-a c\textsuperscript 2 g-a g-h+c i) (d k+e m+f n+p))-(b g+c h+i)
   (-(d g+e h+f i+j) (b e k-a c e k-a f k-b d m+a c d m-f m+a d n+e
   n)+(d\textsuperscript 2+e\textsuperscript 2+f\textsuperscript 2+1) (b h k-a c h k-a i k-b g m+a c g m-i m+a g n+h n)+(b e g-a
   c e g-a f g-b d h+a c d h-f h+a d i+e i) (d k+e m+f
   n+p))+(-(d\textsuperscript 2+e\textsuperscript 2+f\textsuperscript 2+1) (-h b\textsuperscript 2+c g b+a c h b+a i b-a c\textsuperscript 2 g-a
   g-h+c i)+(b d+c e+f) (b e g-a c e g-a f g-b d h+a c d h-f h+a d i+e i)+(-e
   b\textsuperscript 2+c d b+a c e b+a f b-a c\textsuperscript 2 d-a d-e+c f) (d g+e h+f i+j)) (g k+h m+i n+j
   p+q)))-b (a k+m) (-m+c n+e p-c f p+h q-c i q-e j q+c f j q) (-(g k+h m+i
   n+j p+q) (-b j p f\textsuperscript 2+a c j p f\textsuperscript 2-b q f\textsuperscript 2+a c q f\textsuperscript 2+b j n f-a c j n f-g p f+a h p f+b
   i p f-a c i p f+d j p f-a e j p f+d q f-a e q f+g n-a h n-b i n+a c i n-d j n+a e j n-b
   q+a c q)+(d k+e m+f n+p) (b p i\textsuperscript 2-a c p i\textsuperscript 2-b j n i+a c j n i-g p i+a h p i+d
   j p i-a e j p i-b f j p i+a c f j p i+d q i-a e q i-b f q i+a c f q i-d j\textsuperscript 2 n+a e j\textsuperscript 2 n+b
   f j\textsuperscript 2 n-a c f j\textsuperscript 2 n-d n+a e n+b f n-a c f n+g j n-a h j n+b p-a c p-b j q+a c j
   q)-(a k+m) (a j\textsuperscript 2 n e\textsuperscript 2+a n e\textsuperscript 2-a i j p e\textsuperscript 2-a i q e\textsuperscript 2-a f j\textsuperscript 2 m e-a f m e+a i
   j m e-d j\textsuperscript 2 n e+b f j\textsuperscript 2 n e-a c f j\textsuperscript 2 n e-d n e+b f n e-a c f n e+g j n e-2 a h j n e-b i
   j n e+a c i j n e+b i\textsuperscript 2 p e-a c i\textsuperscript 2 p e+b p e-a c p e-g i p e+a h i p e+a f h j p e+d i j
   p e-b f i j p e+a c f i j p e+a f h q e+d i q e-b f i q e+a c f i q e-b j q e+a c j q e-b
   f\textsuperscript 2 m+a c f\textsuperscript 2 m-b i\textsuperscript 2 m+a c i\textsuperscript 2 m-b f\textsuperscript 2 j\textsuperscript 2 m+a c f\textsuperscript 2 j\textsuperscript 2 m+d f j\textsuperscript 2 m-b m+a c m+d f m+g i
   m-a h i m-f g j m+a f h j m-d i j m+2 b f i j m-2 a c f i j m+a h\textsuperscript 2 n+a n-g h n+b h i n-a
   c h i n+d h j n-b f h j n+a c f h j n-a f h\textsuperscript 2 p-a f p+f g h p-b f h i p+a c f h i p+b f\textsuperscript 2
   h j p-a c f\textsuperscript 2 h j p-d f h j p+b f\textsuperscript 2 h q-a c f\textsuperscript 2 h q+b h q-a c h q-d f h q-a i q+a f j
   q)+k (c f\textsuperscript 2 m a\textsuperscript 2+c i\textsuperscript 2 m a\textsuperscript 2+c f\textsuperscript 2 j\textsuperscript 2 m a\textsuperscript 2-e f j\textsuperscript 2 m a\textsuperscript 2+c m a\textsuperscript 2-e f m
   a\textsuperscript 2-h i m a\textsuperscript 2+f h j m a\textsuperscript 2+e i j m a\textsuperscript 2-2 c f i j m a\textsuperscript 2+e\textsuperscript 2 n a\textsuperscript 2+h\textsuperscript 2 n a\textsuperscript 2+e\textsuperscript 2 j\textsuperscript 2 n a\textsuperscript 2-c
   e f j\textsuperscript 2 n a\textsuperscript 2-c e f n a\textsuperscript 2-c h i n a\textsuperscript 2-2 e h j n a\textsuperscript 2+c f h j n a\textsuperscript 2+c e i j n a\textsuperscript 2+n a\textsuperscript 2-f
   h\textsuperscript 2 p a\textsuperscript 2-c e i\textsuperscript 2 p a\textsuperscript 2-c e p a\textsuperscript 2-f p a\textsuperscript 2+e h i p a\textsuperscript 2+c f h i p a\textsuperscript 2-c f\textsuperscript 2 h j p a\textsuperscript 2+e f h
   j p a\textsuperscript 2-e\textsuperscript 2 i j p a\textsuperscript 2+c e f i j p a\textsuperscript 2-c f\textsuperscript 2 h q a\textsuperscript 2-c h q a\textsuperscript 2+e f h q a\textsuperscript 2-e\textsuperscript 2 i q a\textsuperscript 2+c e
   f i q a\textsuperscript 2-i q a\textsuperscript 2+c e j q a\textsuperscript 2+f j q a\textsuperscript 2-c f\textsuperscript 2 k a-c i\textsuperscript 2 k a-c f\textsuperscript 2 j\textsuperscript 2 k a+e f j\textsuperscript 2 k a-c k
   a+e f k a+h i k a-f h j k a-e i j k a+2 c f i j k a-b f\textsuperscript 2 m a-b i\textsuperscript 2 m a-b f\textsuperscript 2 j\textsuperscript 2 m a+d f
   j\textsuperscript 2 m a-b m a+d f m a+g i m a-f g j m a-d i j m a+2 b f i j m a-2 d e j\textsuperscript 2 n a+c d f j\textsuperscript 2 n
   a+b e f j\textsuperscript 2 n a-2 d e n a+c d f n a+b e f n a-2 g h n a+c g i n a+b h i n a+2 e g j n a-c
   f g j n a+2 d h j n a-b f h j n a-c d i j n a-b e i j n a+c d i\textsuperscript 2 p a+b e i\textsuperscript 2 p a+c d p
   a+b e p a+2 f g h p a-e g i p a-c f g i p a-d h i p a-b f h i p a+c f\textsuperscript 2 g j p a-e f g j p
   a+b f\textsuperscript 2 h j p a-d f h j p a+2 d e i j p a-c d f i j p a-b e f i j p a+c f\textsuperscript 2 g q a+c g q
   a-e f g q a+b f\textsuperscript 2 h q a+b h q a-d f h q a+2 d e i q a-c d f i q a-b e f i q a-c d j q a-b
   e j q a+b f\textsuperscript 2 k+b i\textsuperscript 2 k+b f\textsuperscript 2 j\textsuperscript 2 k-d f j\textsuperscript 2 k+b k-d f k-g i k+f g j k+d i j k-2 b f i j
   k+d\textsuperscript 2 n+g\textsuperscript 2 n+d\textsuperscript 2 j\textsuperscript 2 n-b d f j\textsuperscript 2 n-b d f n-b g i n-2 d g j n+b f g j n+b d i j n+n-f g\textsuperscript 2
   p-b d i\textsuperscript 2 p-b d p-f p+d g i p+b f g i p-b f\textsuperscript 2 g j p+d f g j p-d\textsuperscript 2 i j p+b d f i j p-b f\textsuperscript 2
   g q-b g q+d f g q-d\textsuperscript 2 i q+b d f i q-i q+b d j q+f j q)+(-b j\textsuperscript 2 f\textsuperscript 2+a c j\textsuperscript 2
   f\textsuperscript 2-b f\textsuperscript 2+a c f\textsuperscript 2+d j\textsuperscript 2 f-a e j\textsuperscript 2 f+d f-a e f-g j f+a h j f+2 b i j f-2 a c i j f-b i\textsuperscript 2+a
   c i\textsuperscript 2-b+a c+g i-a h i-d i j+a e i j)
   (k\textsuperscript 2+m\textsuperscript 2+n\textsuperscript 2+p\textsuperscript 2+q\textsuperscript 2+1)))+(d k+e m+f n+p) (j q-p) (a (b g+c
   h+i) (-i q\textsuperscript 2+f j q\textsuperscript 2+n q-f p q-i+f j) ((g k+h m+i n+j p+q) (a i n
   c\textsuperscript 2-a f j n c\textsuperscript 2-a f i p c\textsuperscript 2+a f\textsuperscript 2 j p c\textsuperscript 2+a f\textsuperscript 2 q c\textsuperscript 2+a q c\textsuperscript 2-a i m c+a f j m c+g n c-a h
   n c-b i n c-d j n c+a e j n c+b f j n c-f g p c+a f h p c+a e i p c+b f i p c-b f\textsuperscript 2 j p
   c+d f j p c-2 a e f j p c-b f\textsuperscript 2 q c-b q c+d f q c-2 a e f q c-g m+a h m+b i m+d j m-a e j
   m-b f j m+e g p-a e h p-b e i p+a e\textsuperscript 2 j p-d e j p+b e f j p+a e\textsuperscript 2 q+a q-d e q+b e f
   q)-(d k+e m+f n+p) (-a f j\textsuperscript 2 n c\textsuperscript 2-a f n c\textsuperscript 2+a i j n c\textsuperscript 2-a i\textsuperscript 2 p c\textsuperscript 2-a p c\textsuperscript 2+a
   f i j p c\textsuperscript 2+a f i q c\textsuperscript 2+a j q c\textsuperscript 2+a f j\textsuperscript 2 m c+a f m c-a i j m c-d j\textsuperscript 2 n c+a e j\textsuperscript 2 n c+b f
   j\textsuperscript 2 n c-d n c+a e n c+b f n c+g j n c-a h j n c-b i j n c+b i\textsuperscript 2 p c+b p c-g i p c+2 a h i
   p c-a f h j p c+d i j p c-a e i j p c-b f i j p c-a f h q c+d i q c-a e i q c-b f i q c-b
   j q c+d j\textsuperscript 2 m-a e j\textsuperscript 2 m-b f j\textsuperscript 2 m+d m-a e m-b f m-g j m+a h j m+b i j m-a h\textsuperscript 2 p-a p+g h
   p-b h i p-d h j p+a e h j p+b f h j p-d h q+a e h q+b f h q+a j q)+(b k+c m+n)
   (a j\textsuperscript 2 n e\textsuperscript 2+a n e\textsuperscript 2-a i j p e\textsuperscript 2-a i q e\textsuperscript 2-a f j\textsuperscript 2 m e-a f m e+a i j m e-d j\textsuperscript 2 n e+b
   f j\textsuperscript 2 n e-a c f j\textsuperscript 2 n e-d n e+b f n e-a c f n e+g j n e-2 a h j n e-b i j n e+a c i j n
   e+b i\textsuperscript 2 p e-a c i\textsuperscript 2 p e+b p e-a c p e-g i p e+a h i p e+a f h j p e+d i j p e-b f i j p
   e+a c f i j p e+a f h q e+d i q e-b f i q e+a c f i q e-b j q e+a c j q e-b f\textsuperscript 2 m+a c f\textsuperscript 2
   m-b i\textsuperscript 2 m+a c i\textsuperscript 2 m-b f\textsuperscript 2 j\textsuperscript 2 m+a c f\textsuperscript 2 j\textsuperscript 2 m+d f j\textsuperscript 2 m-b m+a c m+d f m+g i m-a h i m-f g
   j m+a f h j m-d i j m+2 b f i j m-2 a c f i j m+a h\textsuperscript 2 n+a n-g h n+b h i n-a c h i n+d h j
   n-b f h j n+a c f h j n-a f h\textsuperscript 2 p-a f p+f g h p-b f h i p+a c f h i p+b f\textsuperscript 2 h j p-a c f\textsuperscript 2
   h j p-d f h j p+b f\textsuperscript 2 h q-a c f\textsuperscript 2 h q+b h q-a c h q-d f h q-a i q+a f j q)-(a
   f\textsuperscript 2 c\textsuperscript 2+a i\textsuperscript 2 c\textsuperscript 2+a f\textsuperscript 2 j\textsuperscript 2 c\textsuperscript 2+a c\textsuperscript 2-2 a f i j c\textsuperscript 2-b f\textsuperscript 2 c-b i\textsuperscript 2 c-b f\textsuperscript 2 j\textsuperscript 2 c+d f j\textsuperscript 2
   c-2 a e f j\textsuperscript 2 c-b c+d f c-2 a e f c+g i c-2 a h i c-f g j c+2 a f h j c-d i j c+2 a e i j
   c+2 b f i j c+a e\textsuperscript 2+a h\textsuperscript 2+a e\textsuperscript 2 j\textsuperscript 2-d e j\textsuperscript 2+b e f j\textsuperscript 2+a-d e+b e f-g h+b h i+e g j+d h j-2
   a e h j-b f h j-b e i j) (k\textsuperscript 2+m\textsuperscript 2+n\textsuperscript 2+p\textsuperscript 2+q\textsuperscript 2+1)-k
   (-(g\textsuperscript 2+h\textsuperscript 2+i\textsuperscript 2+j\textsuperscript 2+1) (-(d\textsuperscript 2+e\textsuperscript 2+f\textsuperscript 2+1) (-m b\textsuperscript 2+c k
   b+a c m b+a n b-a c\textsuperscript 2 k-a k-m+c n)+(b d+c e+f) (b e k-a c e k-a f k-b d m+a c d m-f
   m+a d n+e n)+(-e b\textsuperscript 2+c d b+a c e b+a f b-a c\textsuperscript 2 d-a d-e+c f) (d k+e m+f
   n+p))+(d g+e h+f i+j) (-(d g+e h+f i+j) (-m b\textsuperscript 2+c k b+a c m b+a n b-a c\textsuperscript 2
   k-a k-m+c n)+(b d+c e+f) (b h k-a c h k-a i k-b g m+a c g m-i m+a g n+h n)+(-h
   b\textsuperscript 2+c g b+a c h b+a i b-a c\textsuperscript 2 g-a g-h+c i) (d k+e m+f n+p))-(b g+c h+i)
   (-(d g+e h+f i+j) (b e k-a c e k-a f k-b d m+a c d m-f m+a d n+e
   n)+(d\textsuperscript 2+e\textsuperscript 2+f\textsuperscript 2+1) (b h k-a c h k-a i k-b g m+a c g m-i m+a g n+h n)+(b e g-a
   c e g-a f g-b d h+a c d h-f h+a d i+e i) (d k+e m+f
   n+p))+(-(d\textsuperscript 2+e\textsuperscript 2+f\textsuperscript 2+1) (-h b\textsuperscript 2+c g b+a c h b+a i b-a c\textsuperscript 2 g-a
   g-h+c i)+(b d+c e+f) (b e g-a c e g-a f g-b d h+a c d h-f h+a d i+e i)+(-e
   b\textsuperscript 2+c d b+a c e b+a f b-a c\textsuperscript 2 d-a d-e+c f) (d g+e h+f i+j)) (g k+h m+i n+j
   p+q)))-b (a g+h) (-h q\textsuperscript 2+c i q\textsuperscript 2+e j q\textsuperscript 2-c f j q\textsuperscript 2+m q-c n q-e p q+c f p
   q-h+c i+e j-c f j) (-(g k+h m+i n+j p+q) (-b j p f\textsuperscript 2+a c j p f\textsuperscript 2-b q
   f\textsuperscript 2+a c q f\textsuperscript 2+b j n f-a c j n f-g p f+a h p f+b i p f-a c i p f+d j p f-a e j p f+d q f-a
   e q f+g n-a h n-b i n+a c i n-d j n+a e j n-b q+a c q)+(d k+e m+f n+p) (b p
   i\textsuperscript 2-a c p i\textsuperscript 2-b j n i+a c j n i-g p i+a h p i+d j p i-a e j p i-b f j p i+a c f j p i+d q
   i-a e q i-b f q i+a c f q i-d j\textsuperscript 2 n+a e j\textsuperscript 2 n+b f j\textsuperscript 2 n-a c f j\textsuperscript 2 n-d n+a e n+b f n-a c f
   n+g j n-a h j n+b p-a c p-b j q+a c j q)-(a k+m) (a j\textsuperscript 2 n e\textsuperscript 2+a n e\textsuperscript 2-a i j p
   e\textsuperscript 2-a i q e\textsuperscript 2-a f j\textsuperscript 2 m e-a f m e+a i j m e-d j\textsuperscript 2 n e+b f j\textsuperscript 2 n e-a c f j\textsuperscript 2 n e-d n e+b f
   n e-a c f n e+g j n e-2 a h j n e-b i j n e+a c i j n e+b i\textsuperscript 2 p e-a c i\textsuperscript 2 p e+b p e-a c p
   e-g i p e+a h i p e+a f h j p e+d i j p e-b f i j p e+a c f i j p e+a f h q e+d i q e-b f
   i q e+a c f i q e-b j q e+a c j q e-b f\textsuperscript 2 m+a c f\textsuperscript 2 m-b i\textsuperscript 2 m+a c i\textsuperscript 2 m-b f\textsuperscript 2 j\textsuperscript 2 m+a c
   f\textsuperscript 2 j\textsuperscript 2 m+d f j\textsuperscript 2 m-b m+a c m+d f m+g i m-a h i m-f g j m+a f h j m-d i j m+2 b f i j m-2
   a c f i j m+a h\textsuperscript 2 n+a n-g h n+b h i n-a c h i n+d h j n-b f h j n+a c f h j n-a f h\textsuperscript 2 p-a
   f p+f g h p-b f h i p+a c f h i p+b f\textsuperscript 2 h j p-a c f\textsuperscript 2 h j p-d f h j p+b f\textsuperscript 2 h q-a c f\textsuperscript 2 h
   q+b h q-a c h q-d f h q-a i q+a f j q)+k (c f\textsuperscript 2 m a\textsuperscript 2+c i\textsuperscript 2 m a\textsuperscript 2+c f\textsuperscript 2 j\textsuperscript 2 m
   a\textsuperscript 2-e f j\textsuperscript 2 m a\textsuperscript 2+c m a\textsuperscript 2-e f m a\textsuperscript 2-h i m a\textsuperscript 2+f h j m a\textsuperscript 2+e i j m a\textsuperscript 2-2 c f i j m a\textsuperscript 2+e\textsuperscript 2
   n a\textsuperscript 2+h\textsuperscript 2 n a\textsuperscript 2+e\textsuperscript 2 j\textsuperscript 2 n a\textsuperscript 2-c e f j\textsuperscript 2 n a\textsuperscript 2-c e f n a\textsuperscript 2-c h i n a\textsuperscript 2-2 e h j n a\textsuperscript 2+c f h
   j n a\textsuperscript 2+c e i j n a\textsuperscript 2+n a\textsuperscript 2-f h\textsuperscript 2 p a\textsuperscript 2-c e i\textsuperscript 2 p a\textsuperscript 2-c e p a\textsuperscript 2-f p a\textsuperscript 2+e h i p a\textsuperscript 2+c f h
   i p a\textsuperscript 2-c f\textsuperscript 2 h j p a\textsuperscript 2+e f h j p a\textsuperscript 2-e\textsuperscript 2 i j p a\textsuperscript 2+c e f i j p a\textsuperscript 2-c f\textsuperscript 2 h q a\textsuperscript 2-c h q
   a\textsuperscript 2+e f h q a\textsuperscript 2-e\textsuperscript 2 i q a\textsuperscript 2+c e f i q a\textsuperscript 2-i q a\textsuperscript 2+c e j q a\textsuperscript 2+f j q a\textsuperscript 2-c f\textsuperscript 2 k a-c i\textsuperscript 2 k
   a-c f\textsuperscript 2 j\textsuperscript 2 k a+e f j\textsuperscript 2 k a-c k a+e f k a+h i k a-f h j k a-e i j k a+2 c f i j k a-b f\textsuperscript 2
   m a-b i\textsuperscript 2 m a-b f\textsuperscript 2 j\textsuperscript 2 m a+d f j\textsuperscript 2 m a-b m a+d f m a+g i m a-f g j m a-d i j m a+2 b f i
   j m a-2 d e j\textsuperscript 2 n a+c d f j\textsuperscript 2 n a+b e f j\textsuperscript 2 n a-2 d e n a+c d f n a+b e f n a-2 g h n a+c
   g i n a+b h i n a+2 e g j n a-c f g j n a+2 d h j n a-b f h j n a-c d i j n a-b e i j n
   a+c d i\textsuperscript 2 p a+b e i\textsuperscript 2 p a+c d p a+b e p a+2 f g h p a-e g i p a-c f g i p a-d h i p a-b f
   h i p a+c f\textsuperscript 2 g j p a-e f g j p a+b f\textsuperscript 2 h j p a-d f h j p a+2 d e i j p a-c d f i j p a-b
   e f i j p a+c f\textsuperscript 2 g q a+c g q a-e f g q a+b f\textsuperscript 2 h q a+b h q a-d f h q a+2 d e i q a-c d f
   i q a-b e f i q a-c d j q a-b e j q a+b f\textsuperscript 2 k+b i\textsuperscript 2 k+b f\textsuperscript 2 j\textsuperscript 2 k-d f j\textsuperscript 2 k+b k-d f k-g i
   k+f g j k+d i j k-2 b f i j k+d\textsuperscript 2 n+g\textsuperscript 2 n+d\textsuperscript 2 j\textsuperscript 2 n-b d f j\textsuperscript 2 n-b d f n-b g i n-2 d g j
   n+b f g j n+b d i j n+n-f g\textsuperscript 2 p-b d i\textsuperscript 2 p-b d p-f p+d g i p+b f g i p-b f\textsuperscript 2 g j p+d f g j
   p-d\textsuperscript 2 i j p+b d f i j p-b f\textsuperscript 2 g q-b g q+d f g q-d\textsuperscript 2 i q+b d f i q-i q+b d j q+f j
   q)+(-b j\textsuperscript 2 f\textsuperscript 2+a c j\textsuperscript 2 f\textsuperscript 2-b f\textsuperscript 2+a c f\textsuperscript 2+d j\textsuperscript 2 f-a e j\textsuperscript 2 f+d f-a e f-g j f+a h
   j f+2 b i j f-2 a c i j f-b i\textsuperscript 2+a c i\textsuperscript 2-b+a c+g i-a h i-d i j+a e i j)
   (k\textsuperscript 2+m\textsuperscript 2+n\textsuperscript 2+p\textsuperscript 2+q\textsuperscript 2+1))))-(g\textsuperscript 2+h\textsuperscript 2+i\textsuperscript 2+j\textsuperscript 2+1)
   (q\textsuperscript 2+1) (d (-d q\textsuperscript 2 j\textsuperscript 2+a e q\textsuperscript 2 j\textsuperscript 2+b f q\textsuperscript 2 j\textsuperscript 2-a c f q\textsuperscript 2 j\textsuperscript 2-d j\textsuperscript 2+a
   e j\textsuperscript 2+b f j\textsuperscript 2-a c f j\textsuperscript 2+g q\textsuperscript 2 j-a h q\textsuperscript 2 j-b i q\textsuperscript 2 j+a c i q\textsuperscript 2 j+g j-a h j-b i j+a c i j-k
   q j+a m q j+b n q j-a c n q j+2 d p q j-2 a e p q j-2 b f p q j+2 a c f p q j-d p\textsuperscript 2+a e
   p\textsuperscript 2+b f p\textsuperscript 2-a c f p\textsuperscript 2-d+a e+b f-a c f+k p-a m p-b n p+a c n p-g p q+a h p q+b i p q-a c i
   p q) ((a d+e) (b k+c m+n) (-n+f p+i q-f j q) (-e q\textsuperscript 2 j\textsuperscript 2+c f q\textsuperscript 2 j\textsuperscript 2-e
   j\textsuperscript 2+c f j\textsuperscript 2+h q\textsuperscript 2 j-c i q\textsuperscript 2 j+h j-c i j-m q j+c n q j+2 e p q j-2 c f p q j-e p\textsuperscript 2+c f
   p\textsuperscript 2-e+c f+m p-c n p-h p q+c i p q)-(b d+c e+f) (a k+m) (-m+c n+e p-c f p+h q-c i
   q-e j q+c f j q) (-f q\textsuperscript 2 j\textsuperscript 2-f j\textsuperscript 2+i q\textsuperscript 2 j+i j-n q j+2 f p q j-f p\textsuperscript 2-f+n p-i p
   q))-(d\textsuperscript 2+e\textsuperscript 2+f\textsuperscript 2+1) (q\textsuperscript 2 j\textsuperscript 2+j\textsuperscript 2-2 p q j+p\textsuperscript 2+1) (a
   (b k+c m+n) (-n+f p+i q-f j q) ((g k+h m+i n+j p+q) (a i n c\textsuperscript 2-a f j n c\textsuperscript 2-a f
   i p c\textsuperscript 2+a f\textsuperscript 2 j p c\textsuperscript 2+a f\textsuperscript 2 q c\textsuperscript 2+a q c\textsuperscript 2-a i m c+a f j m c+g n c-a h n c-b i n c-d j n
   c+a e j n c+b f j n c-f g p c+a f h p c+a e i p c+b f i p c-b f\textsuperscript 2 j p c+d f j p c-2 a e f
   j p c-b f\textsuperscript 2 q c-b q c+d f q c-2 a e f q c-g m+a h m+b i m+d j m-a e j m-b f j m+e g p-a e
   h p-b e i p+a e\textsuperscript 2 j p-d e j p+b e f j p+a e\textsuperscript 2 q+a q-d e q+b e f q)-(d k+e m+f n+p)
   (-a f j\textsuperscript 2 n c\textsuperscript 2-a f n c\textsuperscript 2+a i j n c\textsuperscript 2-a i\textsuperscript 2 p c\textsuperscript 2-a p c\textsuperscript 2+a f i j p c\textsuperscript 2+a f i q
   c\textsuperscript 2+a j q c\textsuperscript 2+a f j\textsuperscript 2 m c+a f m c-a i j m c-d j\textsuperscript 2 n c+a e j\textsuperscript 2 n c+b f j\textsuperscript 2 n c-d n c+a e n
   c+b f n c+g j n c-a h j n c-b i j n c+b i\textsuperscript 2 p c+b p c-g i p c+2 a h i p c-a f h j p c+d i
   j p c-a e i j p c-b f i j p c-a f h q c+d i q c-a e i q c-b f i q c-b j q c+d j\textsuperscript 2 m-a e
   j\textsuperscript 2 m-b f j\textsuperscript 2 m+d m-a e m-b f m-g j m+a h j m+b i j m-a h\textsuperscript 2 p-a p+g h p-b h i p-d h j p+a
   e h j p+b f h j p-d h q+a e h q+b f h q+a j q)+(b k+c m+n) (a j\textsuperscript 2 n e\textsuperscript 2+a n
   e\textsuperscript 2-a i j p e\textsuperscript 2-a i q e\textsuperscript 2-a f j\textsuperscript 2 m e-a f m e+a i j m e-d j\textsuperscript 2 n e+b f j\textsuperscript 2 n e-a c f j\textsuperscript 2 n
   e-d n e+b f n e-a c f n e+g j n e-2 a h j n e-b i j n e+a c i j n e+b i\textsuperscript 2 p e-a c i\textsuperscript 2 p
   e+b p e-a c p e-g i p e+a h i p e+a f h j p e+d i j p e-b f i j p e+a c f i j p e+a f h q
   e+d i q e-b f i q e+a c f i q e-b j q e+a c j q e-b f\textsuperscript 2 m+a c f\textsuperscript 2 m-b i\textsuperscript 2 m+a c i\textsuperscript 2 m-b
   f\textsuperscript 2 j\textsuperscript 2 m+a c f\textsuperscript 2 j\textsuperscript 2 m+d f j\textsuperscript 2 m-b m+a c m+d f m+g i m-a h i m-f g j m+a f h j m-d i j
   m+2 b f i j m-2 a c f i j m+a h\textsuperscript 2 n+a n-g h n+b h i n-a c h i n+d h j n-b f h j n+a c f h
   j n-a f h\textsuperscript 2 p-a f p+f g h p-b f h i p+a c f h i p+b f\textsuperscript 2 h j p-a c f\textsuperscript 2 h j p-d f h j p+b
   f\textsuperscript 2 h q-a c f\textsuperscript 2 h q+b h q-a c h q-d f h q-a i q+a f j q)-(a f\textsuperscript 2 c\textsuperscript 2+a i\textsuperscript 2
   c\textsuperscript 2+a f\textsuperscript 2 j\textsuperscript 2 c\textsuperscript 2+a c\textsuperscript 2-2 a f i j c\textsuperscript 2-b f\textsuperscript 2 c-b i\textsuperscript 2 c-b f\textsuperscript 2 j\textsuperscript 2 c+d f j\textsuperscript 2 c-2 a e f j\textsuperscript 2
   c-b c+d f c-2 a e f c+g i c-2 a h i c-f g j c+2 a f h j c-d i j c+2 a e i j c+2 b f i j
   c+a e\textsuperscript 2+a h\textsuperscript 2+a e\textsuperscript 2 j\textsuperscript 2-d e j\textsuperscript 2+b e f j\textsuperscript 2+a-d e+b e f-g h+b h i+e g j+d h j-2 a e h j-b f
   h j-b e i j) (k\textsuperscript 2+m\textsuperscript 2+n\textsuperscript 2+p\textsuperscript 2+q\textsuperscript 2+1)-k
   (-(g\textsuperscript 2+h\textsuperscript 2+i\textsuperscript 2+j\textsuperscript 2+1) (-(d\textsuperscript 2+e\textsuperscript 2+f\textsuperscript 2+1) (-m b\textsuperscript 2+c k
   b+a c m b+a n b-a c\textsuperscript 2 k-a k-m+c n)+(b d+c e+f) (b e k-a c e k-a f k-b d m+a c d m-f
   m+a d n+e n)+(-e b\textsuperscript 2+c d b+a c e b+a f b-a c\textsuperscript 2 d-a d-e+c f) (d k+e m+f
   n+p))+(d g+e h+f i+j) (-(d g+e h+f i+j) (-m b\textsuperscript 2+c k b+a c m b+a n b-a c\textsuperscript 2
   k-a k-m+c n)+(b d+c e+f) (b h k-a c h k-a i k-b g m+a c g m-i m+a g n+h n)+(-h
   b\textsuperscript 2+c g b+a c h b+a i b-a c\textsuperscript 2 g-a g-h+c i) (d k+e m+f n+p))-(b g+c h+i)
   (-(d g+e h+f i+j) (b e k-a c e k-a f k-b d m+a c d m-f m+a d n+e
   n)+(d\textsuperscript 2+e\textsuperscript 2+f\textsuperscript 2+1) (b h k-a c h k-a i k-b g m+a c g m-i m+a g n+h n)+(b e g-a
   c e g-a f g-b d h+a c d h-f h+a d i+e i) (d k+e m+f
   n+p))+(-(d\textsuperscript 2+e\textsuperscript 2+f\textsuperscript 2+1) (-h b\textsuperscript 2+c g b+a c h b+a i b-a c\textsuperscript 2 g-a
   g-h+c i)+(b d+c e+f) (b e g-a c e g-a f g-b d h+a c d h-f h+a d i+e i)+(-e
   b\textsuperscript 2+c d b+a c e b+a f b-a c\textsuperscript 2 d-a d-e+c f) (d g+e h+f i+j)) (g k+h m+i n+j
   p+q)))-b (a k+m) (-m+c n+e p-c f p+h q-c i q-e j q+c f j q) (-(g k+h m+i
   n+j p+q) (-b j p f\textsuperscript 2+a c j p f\textsuperscript 2-b q f\textsuperscript 2+a c q f\textsuperscript 2+b j n f-a c j n f-g p f+a h p f+b
   i p f-a c i p f+d j p f-a e j p f+d q f-a e q f+g n-a h n-b i n+a c i n-d j n+a e j n-b
   q+a c q)+(d k+e m+f n+p) (b p i\textsuperscript 2-a c p i\textsuperscript 2-b j n i+a c j n i-g p i+a h p i+d
   j p i-a e j p i-b f j p i+a c f j p i+d q i-a e q i-b f q i+a c f q i-d j\textsuperscript 2 n+a e j\textsuperscript 2 n+b
   f j\textsuperscript 2 n-a c f j\textsuperscript 2 n-d n+a e n+b f n-a c f n+g j n-a h j n+b p-a c p-b j q+a c j
   q)-(a k+m) (a j\textsuperscript 2 n e\textsuperscript 2+a n e\textsuperscript 2-a i j p e\textsuperscript 2-a i q e\textsuperscript 2-a f j\textsuperscript 2 m e-a f m e+a i
   j m e-d j\textsuperscript 2 n e+b f j\textsuperscript 2 n e-a c f j\textsuperscript 2 n e-d n e+b f n e-a c f n e+g j n e-2 a h j n e-b i
   j n e+a c i j n e+b i\textsuperscript 2 p e-a c i\textsuperscript 2 p e+b p e-a c p e-g i p e+a h i p e+a f h j p e+d i j
   p e-b f i j p e+a c f i j p e+a f h q e+d i q e-b f i q e+a c f i q e-b j q e+a c j q e-b
   f\textsuperscript 2 m+a c f\textsuperscript 2 m-b i\textsuperscript 2 m+a c i\textsuperscript 2 m-b f\textsuperscript 2 j\textsuperscript 2 m+a c f\textsuperscript 2 j\textsuperscript 2 m+d f j\textsuperscript 2 m-b m+a c m+d f m+g i
   m-a h i m-f g j m+a f h j m-d i j m+2 b f i j m-2 a c f i j m+a h\textsuperscript 2 n+a n-g h n+b h i n-a
   c h i n+d h j n-b f h j n+a c f h j n-a f h\textsuperscript 2 p-a f p+f g h p-b f h i p+a c f h i p+b f\textsuperscript 2
   h j p-a c f\textsuperscript 2 h j p-d f h j p+b f\textsuperscript 2 h q-a c f\textsuperscript 2 h q+b h q-a c h q-d f h q-a i q+a f j
   q)+k (c f\textsuperscript 2 m a\textsuperscript 2+c i\textsuperscript 2 m a\textsuperscript 2+c f\textsuperscript 2 j\textsuperscript 2 m a\textsuperscript 2-e f j\textsuperscript 2 m a\textsuperscript 2+c m a\textsuperscript 2-e f m
   a\textsuperscript 2-h i m a\textsuperscript 2+f h j m a\textsuperscript 2+e i j m a\textsuperscript 2-2 c f i j m a\textsuperscript 2+e\textsuperscript 2 n a\textsuperscript 2+h\textsuperscript 2 n a\textsuperscript 2+e\textsuperscript 2 j\textsuperscript 2 n a\textsuperscript 2-c
   e f j\textsuperscript 2 n a\textsuperscript 2-c e f n a\textsuperscript 2-c h i n a\textsuperscript 2-2 e h j n a\textsuperscript 2+c f h j n a\textsuperscript 2+c e i j n a\textsuperscript 2+n a\textsuperscript 2-f
   h\textsuperscript 2 p a\textsuperscript 2-c e i\textsuperscript 2 p a\textsuperscript 2-c e p a\textsuperscript 2-f p a\textsuperscript 2+e h i p a\textsuperscript 2+c f h i p a\textsuperscript 2-c f\textsuperscript 2 h j p a\textsuperscript 2+e f h
   j p a\textsuperscript 2-e\textsuperscript 2 i j p a\textsuperscript 2+c e f i j p a\textsuperscript 2-c f\textsuperscript 2 h q a\textsuperscript 2-c h q a\textsuperscript 2+e f h q a\textsuperscript 2-e\textsuperscript 2 i q a\textsuperscript 2+c e
   f i q a\textsuperscript 2-i q a\textsuperscript 2+c e j q a\textsuperscript 2+f j q a\textsuperscript 2-c f\textsuperscript 2 k a-c i\textsuperscript 2 k a-c f\textsuperscript 2 j\textsuperscript 2 k a+e f j\textsuperscript 2 k a-c k
   a+e f k a+h i k a-f h j k a-e i j k a+2 c f i j k a-b f\textsuperscript 2 m a-b i\textsuperscript 2 m a-b f\textsuperscript 2 j\textsuperscript 2 m a+d f
   j\textsuperscript 2 m a-b m a+d f m a+g i m a-f g j m a-d i j m a+2 b f i j m a-2 d e j\textsuperscript 2 n a+c d f j\textsuperscript 2 n
   a+b e f j\textsuperscript 2 n a-2 d e n a+c d f n a+b e f n a-2 g h n a+c g i n a+b h i n a+2 e g j n a-c
   f g j n a+2 d h j n a-b f h j n a-c d i j n a-b e i j n a+c d i\textsuperscript 2 p a+b e i\textsuperscript 2 p a+c d p
   a+b e p a+2 f g h p a-e g i p a-c f g i p a-d h i p a-b f h i p a+c f\textsuperscript 2 g j p a-e f g j p
   a+b f\textsuperscript 2 h j p a-d f h j p a+2 d e i j p a-c d f i j p a-b e f i j p a+c f\textsuperscript 2 g q a+c g q
   a-e f g q a+b f\textsuperscript 2 h q a+b h q a-d f h q a+2 d e i q a-c d f i q a-b e f i q a-c d j q a-b
   e j q a+b f\textsuperscript 2 k+b i\textsuperscript 2 k+b f\textsuperscript 2 j\textsuperscript 2 k-d f j\textsuperscript 2 k+b k-d f k-g i k+f g j k+d i j k-2 b f i j
   k+d\textsuperscript 2 n+g\textsuperscript 2 n+d\textsuperscript 2 j\textsuperscript 2 n-b d f j\textsuperscript 2 n-b d f n-b g i n-2 d g j n+b f g j n+b d i j n+n-f g\textsuperscript 2
   p-b d i\textsuperscript 2 p-b d p-f p+d g i p+b f g i p-b f\textsuperscript 2 g j p+d f g j p-d\textsuperscript 2 i j p+b d f i j p-b f\textsuperscript 2
   g q-b g q+d f g q-d\textsuperscript 2 i q+b d f i q-i q+b d j q+f j q)+(-b j\textsuperscript 2 f\textsuperscript 2+a c j\textsuperscript 2
   f\textsuperscript 2-b f\textsuperscript 2+a c f\textsuperscript 2+d j\textsuperscript 2 f-a e j\textsuperscript 2 f+d f-a e f-g j f+a h j f+2 b i j f-2 a c i j f-b i\textsuperscript 2+a
   c i\textsuperscript 2-b+a c+g i-a h i-d i j+a e i j)
   (k\textsuperscript 2+m\textsuperscript 2+n\textsuperscript 2+p\textsuperscript 2+q\textsuperscript 2+1)))+(d k+e m+f n+p) (j q-p) (a (b d+c
   e+f) (-f q\textsuperscript 2 j\textsuperscript 2-f j\textsuperscript 2+i q\textsuperscript 2 j+i j-n q j+2 f p q j-f p\textsuperscript 2-f+n p-i p q) ((g
   k+h m+i n+j p+q) (a i n c\textsuperscript 2-a f j n c\textsuperscript 2-a f i p c\textsuperscript 2+a f\textsuperscript 2 j p c\textsuperscript 2+a f\textsuperscript 2 q c\textsuperscript 2+a q
   c\textsuperscript 2-a i m c+a f j m c+g n c-a h n c-b i n c-d j n c+a e j n c+b f j n c-f g p c+a f h p
   c+a e i p c+b f i p c-b f\textsuperscript 2 j p c+d f j p c-2 a e f j p c-b f\textsuperscript 2 q c-b q c+d f q c-2 a e f
   q c-g m+a h m+b i m+d j m-a e j m-b f j m+e g p-a e h p-b e i p+a e\textsuperscript 2 j p-d e j p+b e f j
   p+a e\textsuperscript 2 q+a q-d e q+b e f q)-(d k+e m+f n+p) (-a f j\textsuperscript 2 n c\textsuperscript 2-a f n c\textsuperscript 2+a i j n
   c\textsuperscript 2-a i\textsuperscript 2 p c\textsuperscript 2-a p c\textsuperscript 2+a f i j p c\textsuperscript 2+a f i q c\textsuperscript 2+a j q c\textsuperscript 2+a f j\textsuperscript 2 m c+a f m c-a i j m
   c-d j\textsuperscript 2 n c+a e j\textsuperscript 2 n c+b f j\textsuperscript 2 n c-d n c+a e n c+b f n c+g j n c-a h j n c-b i j n c+b
   i\textsuperscript 2 p c+b p c-g i p c+2 a h i p c-a f h j p c+d i j p c-a e i j p c-b f i j p c-a f h q
   c+d i q c-a e i q c-b f i q c-b j q c+d j\textsuperscript 2 m-a e j\textsuperscript 2 m-b f j\textsuperscript 2 m+d m-a e m-b f m-g j m+a
   h j m+b i j m-a h\textsuperscript 2 p-a p+g h p-b h i p-d h j p+a e h j p+b f h j p-d h q+a e h q+b f h
   q+a j q)+(b k+c m+n) (a j\textsuperscript 2 n e\textsuperscript 2+a n e\textsuperscript 2-a i j p e\textsuperscript 2-a i q e\textsuperscript 2-a f j\textsuperscript 2 m e-a
   f m e+a i j m e-d j\textsuperscript 2 n e+b f j\textsuperscript 2 n e-a c f j\textsuperscript 2 n e-d n e+b f n e-a c f n e+g j n e-2 a h
   j n e-b i j n e+a c i j n e+b i\textsuperscript 2 p e-a c i\textsuperscript 2 p e+b p e-a c p e-g i p e+a h i p e+a f h j
   p e+d i j p e-b f i j p e+a c f i j p e+a f h q e+d i q e-b f i q e+a c f i q e-b j q e+a
   c j q e-b f\textsuperscript 2 m+a c f\textsuperscript 2 m-b i\textsuperscript 2 m+a c i\textsuperscript 2 m-b f\textsuperscript 2 j\textsuperscript 2 m+a c f\textsuperscript 2 j\textsuperscript 2 m+d f j\textsuperscript 2 m-b m+a c
   m+d f m+g i m-a h i m-f g j m+a f h j m-d i j m+2 b f i j m-2 a c f i j m+a h\textsuperscript 2 n+a n-g h
   n+b h i n-a c h i n+d h j n-b f h j n+a c f h j n-a f h\textsuperscript 2 p-a f p+f g h p-b f h i p+a c f
   h i p+b f\textsuperscript 2 h j p-a c f\textsuperscript 2 h j p-d f h j p+b f\textsuperscript 2 h q-a c f\textsuperscript 2 h q+b h q-a c h q-d f h q-a i
   q+a f j q)-(a f\textsuperscript 2 c\textsuperscript 2+a i\textsuperscript 2 c\textsuperscript 2+a f\textsuperscript 2 j\textsuperscript 2 c\textsuperscript 2+a c\textsuperscript 2-2 a f i j c\textsuperscript 2-b f\textsuperscript 2 c-b
   i\textsuperscript 2 c-b f\textsuperscript 2 j\textsuperscript 2 c+d f j\textsuperscript 2 c-2 a e f j\textsuperscript 2 c-b c+d f c-2 a e f c+g i c-2 a h i c-f g j c+2 a
   f h j c-d i j c+2 a e i j c+2 b f i j c+a e\textsuperscript 2+a h\textsuperscript 2+a e\textsuperscript 2 j\textsuperscript 2-d e j\textsuperscript 2+b e f j\textsuperscript 2+a-d e+b e
   f-g h+b h i+e g j+d h j-2 a e h j-b f h j-b e i j)
   (k\textsuperscript 2+m\textsuperscript 2+n\textsuperscript 2+p\textsuperscript 2+q\textsuperscript 2+1)-k (-(g\textsuperscript 2+h\textsuperscript 2+i\textsuperscript 2+j\textsuperscript 2+1)
   (-(d\textsuperscript 2+e\textsuperscript 2+f\textsuperscript 2+1) (-m b\textsuperscript 2+c k b+a c m b+a n b-a c\textsuperscript 2 k-a k-m+c
   n)+(b d+c e+f) (b e k-a c e k-a f k-b d m+a c d m-f m+a d n+e n)+(-e b\textsuperscript 2+c d
   b+a c e b+a f b-a c\textsuperscript 2 d-a d-e+c f) (d k+e m+f n+p))+(d g+e h+f i+j) (-(d
   g+e h+f i+j) (-m b\textsuperscript 2+c k b+a c m b+a n b-a c\textsuperscript 2 k-a k-m+c n)+(b d+c e+f) (b h
   k-a c h k-a i k-b g m+a c g m-i m+a g n+h n)+(-h b\textsuperscript 2+c g b+a c h b+a i b-a c\textsuperscript 2 g-a
   g-h+c i) (d k+e m+f n+p))-(b g+c h+i) (-(d g+e h+f i+j) (b e k-a c e k-a
   f k-b d m+a c d m-f m+a d n+e n)+(d\textsuperscript 2+e\textsuperscript 2+f\textsuperscript 2+1) (b h k-a c h k-a i k-b g m+a
   c g m-i m+a g n+h n)+(b e g-a c e g-a f g-b d h+a c d h-f h+a d i+e i) (d k+e m+f
   n+p))+(-(d\textsuperscript 2+e\textsuperscript 2+f\textsuperscript 2+1) (-h b\textsuperscript 2+c g b+a c h b+a i b-a c\textsuperscript 2 g-a
   g-h+c i)+(b d+c e+f) (b e g-a c e g-a f g-b d h+a c d h-f h+a d i+e i)+(-e
   b\textsuperscript 2+c d b+a c e b+a f b-a c\textsuperscript 2 d-a d-e+c f) (d g+e h+f i+j)) (g k+h m+i n+j
   p+q)))-b (a d+e) (-e q\textsuperscript 2 j\textsuperscript 2+c f q\textsuperscript 2 j\textsuperscript 2-e j\textsuperscript 2+c f j\textsuperscript 2+h q\textsuperscript 2 j-c i q\textsuperscript 2
   j+h j-c i j-m q j+c n q j+2 e p q j-2 c f p q j-e p\textsuperscript 2+c f p\textsuperscript 2-e+c f+m p-c n p-h p q+c i p
   q) (-(g k+h m+i n+j p+q) (-b j p f\textsuperscript 2+a c j p f\textsuperscript 2-b q f\textsuperscript 2+a c q f\textsuperscript 2+b j n
   f-a c j n f-g p f+a h p f+b i p f-a c i p f+d j p f-a e j p f+d q f-a e q f+g n-a h n-b i
   n+a c i n-d j n+a e j n-b q+a c q)+(d k+e m+f n+p) (b p i\textsuperscript 2-a c p i\textsuperscript 2-b j n
   i+a c j n i-g p i+a h p i+d j p i-a e j p i-b f j p i+a c f j p i+d q i-a e q i-b f q i+a
   c f q i-d j\textsuperscript 2 n+a e j\textsuperscript 2 n+b f j\textsuperscript 2 n-a c f j\textsuperscript 2 n-d n+a e n+b f n-a c f n+g j n-a h j n+b
   p-a c p-b j q+a c j q)-(a k+m) (a j\textsuperscript 2 n e\textsuperscript 2+a n e\textsuperscript 2-a i j p e\textsuperscript 2-a i q e\textsuperscript 2-a f
   j\textsuperscript 2 m e-a f m e+a i j m e-d j\textsuperscript 2 n e+b f j\textsuperscript 2 n e-a c f j\textsuperscript 2 n e-d n e+b f n e-a c f n e+g j
   n e-2 a h j n e-b i j n e+a c i j n e+b i\textsuperscript 2 p e-a c i\textsuperscript 2 p e+b p e-a c p e-g i p e+a h i p
   e+a f h j p e+d i j p e-b f i j p e+a c f i j p e+a f h q e+d i q e-b f i q e+a c f i q
   e-b j q e+a c j q e-b f\textsuperscript 2 m+a c f\textsuperscript 2 m-b i\textsuperscript 2 m+a c i\textsuperscript 2 m-b f\textsuperscript 2 j\textsuperscript 2 m+a c f\textsuperscript 2 j\textsuperscript 2 m+d f j\textsuperscript 2
   m-b m+a c m+d f m+g i m-a h i m-f g j m+a f h j m-d i j m+2 b f i j m-2 a c f i j m+a h\textsuperscript 2
   n+a n-g h n+b h i n-a c h i n+d h j n-b f h j n+a c f h j n-a f h\textsuperscript 2 p-a f p+f g h p-b f h
   i p+a c f h i p+b f\textsuperscript 2 h j p-a c f\textsuperscript 2 h j p-d f h j p+b f\textsuperscript 2 h q-a c f\textsuperscript 2 h q+b h q-a c h q-d
   f h q-a i q+a f j q)+k (c f\textsuperscript 2 m a\textsuperscript 2+c i\textsuperscript 2 m a\textsuperscript 2+c f\textsuperscript 2 j\textsuperscript 2 m a\textsuperscript 2-e f j\textsuperscript 2 m
   a\textsuperscript 2+c m a\textsuperscript 2-e f m a\textsuperscript 2-h i m a\textsuperscript 2+f h j m a\textsuperscript 2+e i j m a\textsuperscript 2-2 c f i j m a\textsuperscript 2+e\textsuperscript 2 n a\textsuperscript 2+h\textsuperscript 2 n
   a\textsuperscript 2+e\textsuperscript 2 j\textsuperscript 2 n a\textsuperscript 2-c e f j\textsuperscript 2 n a\textsuperscript 2-c e f n a\textsuperscript 2-c h i n a\textsuperscript 2-2 e h j n a\textsuperscript 2+c f h j n a\textsuperscript 2+c e
   i j n a\textsuperscript 2+n a\textsuperscript 2-f h\textsuperscript 2 p a\textsuperscript 2-c e i\textsuperscript 2 p a\textsuperscript 2-c e p a\textsuperscript 2-f p a\textsuperscript 2+e h i p a\textsuperscript 2+c f h i p a\textsuperscript 2-c
   f\textsuperscript 2 h j p a\textsuperscript 2+e f h j p a\textsuperscript 2-e\textsuperscript 2 i j p a\textsuperscript 2+c e f i j p a\textsuperscript 2-c f\textsuperscript 2 h q a\textsuperscript 2-c h q a\textsuperscript 2+e f h q
   a\textsuperscript 2-e\textsuperscript 2 i q a\textsuperscript 2+c e f i q a\textsuperscript 2-i q a\textsuperscript 2+c e j q a\textsuperscript 2+f j q a\textsuperscript 2-c f\textsuperscript 2 k a-c i\textsuperscript 2 k a-c f\textsuperscript 2 j\textsuperscript 2
   k a+e f j\textsuperscript 2 k a-c k a+e f k a+h i k a-f h j k a-e i j k a+2 c f i j k a-b f\textsuperscript 2 m a-b i\textsuperscript 2 m
   a-b f\textsuperscript 2 j\textsuperscript 2 m a+d f j\textsuperscript 2 m a-b m a+d f m a+g i m a-f g j m a-d i j m a+2 b f i j m a-2 d e
   j\textsuperscript 2 n a+c d f j\textsuperscript 2 n a+b e f j\textsuperscript 2 n a-2 d e n a+c d f n a+b e f n a-2 g h n a+c g i n a+b h
   i n a+2 e g j n a-c f g j n a+2 d h j n a-b f h j n a-c d i j n a-b e i j n a+c d i\textsuperscript 2 p
   a+b e i\textsuperscript 2 p a+c d p a+b e p a+2 f g h p a-e g i p a-c f g i p a-d h i p a-b f h i p a+c
   f\textsuperscript 2 g j p a-e f g j p a+b f\textsuperscript 2 h j p a-d f h j p a+2 d e i j p a-c d f i j p a-b e f i j p
   a+c f\textsuperscript 2 g q a+c g q a-e f g q a+b f\textsuperscript 2 h q a+b h q a-d f h q a+2 d e i q a-c d f i q a-b e
   f i q a-c d j q a-b e j q a+b f\textsuperscript 2 k+b i\textsuperscript 2 k+b f\textsuperscript 2 j\textsuperscript 2 k-d f j\textsuperscript 2 k+b k-d f k-g i k+f g j
   k+d i j k-2 b f i j k+d\textsuperscript 2 n+g\textsuperscript 2 n+d\textsuperscript 2 j\textsuperscript 2 n-b d f j\textsuperscript 2 n-b d f n-b g i n-2 d g j n+b f g j
   n+b d i j n+n-f g\textsuperscript 2 p-b d i\textsuperscript 2 p-b d p-f p+d g i p+b f g i p-b f\textsuperscript 2 g j p+d f g j p-d\textsuperscript 2 i j
   p+b d f i j p-b f\textsuperscript 2 g q-b g q+d f g q-d\textsuperscript 2 i q+b d f i q-i q+b d j q+f j q)+(-b
   j\textsuperscript 2 f\textsuperscript 2+a c j\textsuperscript 2 f\textsuperscript 2-b f\textsuperscript 2+a c f\textsuperscript 2+d j\textsuperscript 2 f-a e j\textsuperscript 2 f+d f-a e f-g j f+a h j f+2 b i j f-2 a
   c i j f-b i\textsuperscript 2+a c i\textsuperscript 2-b+a c+g i-a h i-d i j+a e i j)
   (k\textsuperscript 2+m\textsuperscript 2+n\textsuperscript 2+p\textsuperscript 2+q\textsuperscript 2+1))))-q (g k+h m+i n+j p+q) (d
   (-d q\textsuperscript 2 j\textsuperscript 2+a e q\textsuperscript 2 j\textsuperscript 2+b f q\textsuperscript 2 j\textsuperscript 2-a c f q\textsuperscript 2 j\textsuperscript 2-d j\textsuperscript 2+a e j\textsuperscript 2+b f j\textsuperscript 2-a c f j\textsuperscript 2+g
   q\textsuperscript 2 j-a h q\textsuperscript 2 j-b i q\textsuperscript 2 j+a c i q\textsuperscript 2 j+g j-a h j-b i j+a c i j-k q j+a m q j+b n q j-a c n
   q j+2 d p q j-2 a e p q j-2 b f p q j+2 a c f p q j-d p\textsuperscript 2+a e p\textsuperscript 2+b f p\textsuperscript 2-a c f p\textsuperscript 2-d+a
   e+b f-a c f+k p-a m p-b n p+a c n p-g p q+a h p q+b i p q-a c i p q) ((a d+e)
   (b g+c h+i) (-i q\textsuperscript 2+f j q\textsuperscript 2+n q-f p q-i+f j) (-e q\textsuperscript 2 j\textsuperscript 2+c f q\textsuperscript 2 j\textsuperscript 2-e
   j\textsuperscript 2+c f j\textsuperscript 2+h q\textsuperscript 2 j-c i q\textsuperscript 2 j+h j-c i j-m q j+c n q j+2 e p q j-2 c f p q j-e p\textsuperscript 2+c f
   p\textsuperscript 2-e+c f+m p-c n p-h p q+c i p q)-(b d+c e+f) (a g+h) (-h q\textsuperscript 2+c i q\textsuperscript 2+e j
   q\textsuperscript 2-c f j q\textsuperscript 2+m q-c n q-e p q+c f p q-h+c i+e j-c f j) (-f q\textsuperscript 2 j\textsuperscript 2-f j\textsuperscript 2+i q\textsuperscript 2
   j+i j-n q j+2 f p q j-f p\textsuperscript 2-f+n p-i p q))-(d\textsuperscript 2+e\textsuperscript 2+f\textsuperscript 2+1)
   (q\textsuperscript 2 j\textsuperscript 2+j\textsuperscript 2-2 p q j+p\textsuperscript 2+1) (a (b g+c h+i) (-i q\textsuperscript 2+f j q\textsuperscript 2+n q-f p
   q-i+f j) ((g k+h m+i n+j p+q) (a i n c\textsuperscript 2-a f j n c\textsuperscript 2-a f i p c\textsuperscript 2+a f\textsuperscript 2 j
   p c\textsuperscript 2+a f\textsuperscript 2 q c\textsuperscript 2+a q c\textsuperscript 2-a i m c+a f j m c+g n c-a h n c-b i n c-d j n c+a e j n c+b f j
   n c-f g p c+a f h p c+a e i p c+b f i p c-b f\textsuperscript 2 j p c+d f j p c-2 a e f j p c-b f\textsuperscript 2 q c-b
   q c+d f q c-2 a e f q c-g m+a h m+b i m+d j m-a e j m-b f j m+e g p-a e h p-b e i p+a e\textsuperscript 2
   j p-d e j p+b e f j p+a e\textsuperscript 2 q+a q-d e q+b e f q)-(d k+e m+f n+p) (-a f j\textsuperscript 2 n
   c\textsuperscript 2-a f n c\textsuperscript 2+a i j n c\textsuperscript 2-a i\textsuperscript 2 p c\textsuperscript 2-a p c\textsuperscript 2+a f i j p c\textsuperscript 2+a f i q c\textsuperscript 2+a j q c\textsuperscript 2+a f j\textsuperscript 2
   m c+a f m c-a i j m c-d j\textsuperscript 2 n c+a e j\textsuperscript 2 n c+b f j\textsuperscript 2 n c-d n c+a e n c+b f n c+g j n c-a h
   j n c-b i j n c+b i\textsuperscript 2 p c+b p c-g i p c+2 a h i p c-a f h j p c+d i j p c-a e i j p c-b f
   i j p c-a f h q c+d i q c-a e i q c-b f i q c-b j q c+d j\textsuperscript 2 m-a e j\textsuperscript 2 m-b f j\textsuperscript 2 m+d m-a e
   m-b f m-g j m+a h j m+b i j m-a h\textsuperscript 2 p-a p+g h p-b h i p-d h j p+a e h j p+b f h j p-d h
   q+a e h q+b f h q+a j q)+(b k+c m+n) (a j\textsuperscript 2 n e\textsuperscript 2+a n e\textsuperscript 2-a i j p e\textsuperscript 2-a i q
   e\textsuperscript 2-a f j\textsuperscript 2 m e-a f m e+a i j m e-d j\textsuperscript 2 n e+b f j\textsuperscript 2 n e-a c f j\textsuperscript 2 n e-d n e+b f n e-a c f
   n e+g j n e-2 a h j n e-b i j n e+a c i j n e+b i\textsuperscript 2 p e-a c i\textsuperscript 2 p e+b p e-a c p e-g i p
   e+a h i p e+a f h j p e+d i j p e-b f i j p e+a c f i j p e+a f h q e+d i q e-b f i q e+a
   c f i q e-b j q e+a c j q e-b f\textsuperscript 2 m+a c f\textsuperscript 2 m-b i\textsuperscript 2 m+a c i\textsuperscript 2 m-b f\textsuperscript 2 j\textsuperscript 2 m+a c f\textsuperscript 2 j\textsuperscript 2
   m+d f j\textsuperscript 2 m-b m+a c m+d f m+g i m-a h i m-f g j m+a f h j m-d i j m+2 b f i j m-2 a c f i
   j m+a h\textsuperscript 2 n+a n-g h n+b h i n-a c h i n+d h j n-b f h j n+a c f h j n-a f h\textsuperscript 2 p-a f p+f g
   h p-b f h i p+a c f h i p+b f\textsuperscript 2 h j p-a c f\textsuperscript 2 h j p-d f h j p+b f\textsuperscript 2 h q-a c f\textsuperscript 2 h q+b h
   q-a c h q-d f h q-a i q+a f j q)-(a f\textsuperscript 2 c\textsuperscript 2+a i\textsuperscript 2 c\textsuperscript 2+a f\textsuperscript 2 j\textsuperscript 2 c\textsuperscript 2+a c\textsuperscript 2-2 a
   f i j c\textsuperscript 2-b f\textsuperscript 2 c-b i\textsuperscript 2 c-b f\textsuperscript 2 j\textsuperscript 2 c+d f j\textsuperscript 2 c-2 a e f j\textsuperscript 2 c-b c+d f c-2 a e f c+g i c-2
   a h i c-f g j c+2 a f h j c-d i j c+2 a e i j c+2 b f i j c+a e\textsuperscript 2+a h\textsuperscript 2+a e\textsuperscript 2 j\textsuperscript 2-d e
   j\textsuperscript 2+b e f j\textsuperscript 2+a-d e+b e f-g h+b h i+e g j+d h j-2 a e h j-b f h j-b e i j)
   (k\textsuperscript 2+m\textsuperscript 2+n\textsuperscript 2+p\textsuperscript 2+q\textsuperscript 2+1)-k (-(g\textsuperscript 2+h\textsuperscript 2+i\textsuperscript 2+j\textsuperscript 2+1)
   (-(d\textsuperscript 2+e\textsuperscript 2+f\textsuperscript 2+1) (-m b\textsuperscript 2+c k b+a c m b+a n b-a c\textsuperscript 2 k-a k-m+c
   n)+(b d+c e+f) (b e k-a c e k-a f k-b d m+a c d m-f m+a d n+e n)+(-e b\textsuperscript 2+c d
   b+a c e b+a f b-a c\textsuperscript 2 d-a d-e+c f) (d k+e m+f n+p))+(d g+e h+f i+j) (-(d
   g+e h+f i+j) (-m b\textsuperscript 2+c k b+a c m b+a n b-a c\textsuperscript 2 k-a k-m+c n)+(b d+c e+f) (b h
   k-a c h k-a i k-b g m+a c g m-i m+a g n+h n)+(-h b\textsuperscript 2+c g b+a c h b+a i b-a c\textsuperscript 2 g-a
   g-h+c i) (d k+e m+f n+p))-(b g+c h+i) (-(d g+e h+f i+j) (b e k-a c e k-a
   f k-b d m+a c d m-f m+a d n+e n)+(d\textsuperscript 2+e\textsuperscript 2+f\textsuperscript 2+1) (b h k-a c h k-a i k-b g m+a
   c g m-i m+a g n+h n)+(b e g-a c e g-a f g-b d h+a c d h-f h+a d i+e i) (d k+e m+f
   n+p))+(-(d\textsuperscript 2+e\textsuperscript 2+f\textsuperscript 2+1) (-h b\textsuperscript 2+c g b+a c h b+a i b-a c\textsuperscript 2 g-a
   g-h+c i)+(b d+c e+f) (b e g-a c e g-a f g-b d h+a c d h-f h+a d i+e i)+(-e
   b\textsuperscript 2+c d b+a c e b+a f b-a c\textsuperscript 2 d-a d-e+c f) (d g+e h+f i+j)) (g k+h m+i n+j
   p+q)))-b (a g+h) (-h q\textsuperscript 2+c i q\textsuperscript 2+e j q\textsuperscript 2-c f j q\textsuperscript 2+m q-c n q-e p q+c f p
   q-h+c i+e j-c f j) (-(g k+h m+i n+j p+q) (-b j p f\textsuperscript 2+a c j p f\textsuperscript 2-b q
   f\textsuperscript 2+a c q f\textsuperscript 2+b j n f-a c j n f-g p f+a h p f+b i p f-a c i p f+d j p f-a e j p f+d q f-a
   e q f+g n-a h n-b i n+a c i n-d j n+a e j n-b q+a c q)+(d k+e m+f n+p) (b p
   i\textsuperscript 2-a c p i\textsuperscript 2-b j n i+a c j n i-g p i+a h p i+d j p i-a e j p i-b f j p i+a c f j p i+d q
   i-a e q i-b f q i+a c f q i-d j\textsuperscript 2 n+a e j\textsuperscript 2 n+b f j\textsuperscript 2 n-a c f j\textsuperscript 2 n-d n+a e n+b f n-a c f
   n+g j n-a h j n+b p-a c p-b j q+a c j q)-(a k+m) (a j\textsuperscript 2 n e\textsuperscript 2+a n e\textsuperscript 2-a i j p
   e\textsuperscript 2-a i q e\textsuperscript 2-a f j\textsuperscript 2 m e-a f m e+a i j m e-d j\textsuperscript 2 n e+b f j\textsuperscript 2 n e-a c f j\textsuperscript 2 n e-d n e+b f
   n e-a c f n e+g j n e-2 a h j n e-b i j n e+a c i j n e+b i\textsuperscript 2 p e-a c i\textsuperscript 2 p e+b p e-a c p
   e-g i p e+a h i p e+a f h j p e+d i j p e-b f i j p e+a c f i j p e+a f h q e+d i q e-b f
   i q e+a c f i q e-b j q e+a c j q e-b f\textsuperscript 2 m+a c f\textsuperscript 2 m-b i\textsuperscript 2 m+a c i\textsuperscript 2 m-b f\textsuperscript 2 j\textsuperscript 2 m+a c
   f\textsuperscript 2 j\textsuperscript 2 m+d f j\textsuperscript 2 m-b m+a c m+d f m+g i m-a h i m-f g j m+a f h j m-d i j m+2 b f i j m-2
   a c f i j m+a h\textsuperscript 2 n+a n-g h n+b h i n-a c h i n+d h j n-b f h j n+a c f h j n-a f h\textsuperscript 2 p-a
   f p+f g h p-b f h i p+a c f h i p+b f\textsuperscript 2 h j p-a c f\textsuperscript 2 h j p-d f h j p+b f\textsuperscript 2 h q-a c f\textsuperscript 2 h
   q+b h q-a c h q-d f h q-a i q+a f j q)+k (c f\textsuperscript 2 m a\textsuperscript 2+c i\textsuperscript 2 m a\textsuperscript 2+c f\textsuperscript 2 j\textsuperscript 2 m
   a\textsuperscript 2-e f j\textsuperscript 2 m a\textsuperscript 2+c m a\textsuperscript 2-e f m a\textsuperscript 2-h i m a\textsuperscript 2+f h j m a\textsuperscript 2+e i j m a\textsuperscript 2-2 c f i j m a\textsuperscript 2+e\textsuperscript 2
   n a\textsuperscript 2+h\textsuperscript 2 n a\textsuperscript 2+e\textsuperscript 2 j\textsuperscript 2 n a\textsuperscript 2-c e f j\textsuperscript 2 n a\textsuperscript 2-c e f n a\textsuperscript 2-c h i n a\textsuperscript 2-2 e h j n a\textsuperscript 2+c f h
   j n a\textsuperscript 2+c e i j n a\textsuperscript 2+n a\textsuperscript 2-f h\textsuperscript 2 p a\textsuperscript 2-c e i\textsuperscript 2 p a\textsuperscript 2-c e p a\textsuperscript 2-f p a\textsuperscript 2+e h i p a\textsuperscript 2+c f h
   i p a\textsuperscript 2-c f\textsuperscript 2 h j p a\textsuperscript 2+e f h j p a\textsuperscript 2-e\textsuperscript 2 i j p a\textsuperscript 2+c e f i j p a\textsuperscript 2-c f\textsuperscript 2 h q a\textsuperscript 2-c h q
   a\textsuperscript 2+e f h q a\textsuperscript 2-e\textsuperscript 2 i q a\textsuperscript 2+c e f i q a\textsuperscript 2-i q a\textsuperscript 2+c e j q a\textsuperscript 2+f j q a\textsuperscript 2-c f\textsuperscript 2 k a-c i\textsuperscript 2 k
   a-c f\textsuperscript 2 j\textsuperscript 2 k a+e f j\textsuperscript 2 k a-c k a+e f k a+h i k a-f h j k a-e i j k a+2 c f i j k a-b f\textsuperscript 2
   m a-b i\textsuperscript 2 m a-b f\textsuperscript 2 j\textsuperscript 2 m a+d f j\textsuperscript 2 m a-b m a+d f m a+g i m a-f g j m a-d i j m a+2 b f i
   j m a-2 d e j\textsuperscript 2 n a+c d f j\textsuperscript 2 n a+b e f j\textsuperscript 2 n a-2 d e n a+c d f n a+b e f n a-2 g h n a+c
   g i n a+b h i n a+2 e g j n a-c f g j n a+2 d h j n a-b f h j n a-c d i j n a-b e i j n
   a+c d i\textsuperscript 2 p a+b e i\textsuperscript 2 p a+c d p a+b e p a+2 f g h p a-e g i p a-c f g i p a-d h i p a-b f
   h i p a+c f\textsuperscript 2 g j p a-e f g j p a+b f\textsuperscript 2 h j p a-d f h j p a+2 d e i j p a-c d f i j p a-b
   e f i j p a+c f\textsuperscript 2 g q a+c g q a-e f g q a+b f\textsuperscript 2 h q a+b h q a-d f h q a+2 d e i q a-c d f
   i q a-b e f i q a-c d j q a-b e j q a+b f\textsuperscript 2 k+b i\textsuperscript 2 k+b f\textsuperscript 2 j\textsuperscript 2 k-d f j\textsuperscript 2 k+b k-d f k-g i
   k+f g j k+d i j k-2 b f i j k+d\textsuperscript 2 n+g\textsuperscript 2 n+d\textsuperscript 2 j\textsuperscript 2 n-b d f j\textsuperscript 2 n-b d f n-b g i n-2 d g j
   n+b f g j n+b d i j n+n-f g\textsuperscript 2 p-b d i\textsuperscript 2 p-b d p-f p+d g i p+b f g i p-b f\textsuperscript 2 g j p+d f g j
   p-d\textsuperscript 2 i j p+b d f i j p-b f\textsuperscript 2 g q-b g q+d f g q-d\textsuperscript 2 i q+b d f i q-i q+b d j q+f j
   q)+(-b j\textsuperscript 2 f\textsuperscript 2+a c j\textsuperscript 2 f\textsuperscript 2-b f\textsuperscript 2+a c f\textsuperscript 2+d j\textsuperscript 2 f-a e j\textsuperscript 2 f+d f-a e f-g j f+a h
   j f+2 b i j f-2 a c i j f-b i\textsuperscript 2+a c i\textsuperscript 2-b+a c+g i-a h i-d i j+a e i j)
   (k\textsuperscript 2+m\textsuperscript 2+n\textsuperscript 2+p\textsuperscript 2+q\textsuperscript 2+1)))+(d g+e h+f i+j) (-j q\textsuperscript 2+p
   q-j) (a (b d+c e+f) (-f q\textsuperscript 2 j\textsuperscript 2-f j\textsuperscript 2+i q\textsuperscript 2 j+i j-n q j+2 f p q j-f
   p\textsuperscript 2-f+n p-i p q) ((g k+h m+i n+j p+q) (a i n c\textsuperscript 2-a f j n c\textsuperscript 2-a f i p
   c\textsuperscript 2+a f\textsuperscript 2 j p c\textsuperscript 2+a f\textsuperscript 2 q c\textsuperscript 2+a q c\textsuperscript 2-a i m c+a f j m c+g n c-a h n c-b i n c-d j n c+a e
   j n c+b f j n c-f g p c+a f h p c+a e i p c+b f i p c-b f\textsuperscript 2 j p c+d f j p c-2 a e f j p
   c-b f\textsuperscript 2 q c-b q c+d f q c-2 a e f q c-g m+a h m+b i m+d j m-a e j m-b f j m+e g p-a e h
   p-b e i p+a e\textsuperscript 2 j p-d e j p+b e f j p+a e\textsuperscript 2 q+a q-d e q+b e f q)-(d k+e m+f n+p)
   (-a f j\textsuperscript 2 n c\textsuperscript 2-a f n c\textsuperscript 2+a i j n c\textsuperscript 2-a i\textsuperscript 2 p c\textsuperscript 2-a p c\textsuperscript 2+a f i j p c\textsuperscript 2+a f i q
   c\textsuperscript 2+a j q c\textsuperscript 2+a f j\textsuperscript 2 m c+a f m c-a i j m c-d j\textsuperscript 2 n c+a e j\textsuperscript 2 n c+b f j\textsuperscript 2 n c-d n c+a e n
   c+b f n c+g j n c-a h j n c-b i j n c+b i\textsuperscript 2 p c+b p c-g i p c+2 a h i p c-a f h j p c+d i
   j p c-a e i j p c-b f i j p c-a f h q c+d i q c-a e i q c-b f i q c-b j q c+d j\textsuperscript 2 m-a e
   j\textsuperscript 2 m-b f j\textsuperscript 2 m+d m-a e m-b f m-g j m+a h j m+b i j m-a h\textsuperscript 2 p-a p+g h p-b h i p-d h j p+a
   e h j p+b f h j p-d h q+a e h q+b f h q+a j q)+(b k+c m+n) (a j\textsuperscript 2 n e\textsuperscript 2+a n
   e\textsuperscript 2-a i j p e\textsuperscript 2-a i q e\textsuperscript 2-a f j\textsuperscript 2 m e-a f m e+a i j m e-d j\textsuperscript 2 n e+b f j\textsuperscript 2 n e-a c f j\textsuperscript 2 n
   e-d n e+b f n e-a c f n e+g j n e-2 a h j n e-b i j n e+a c i j n e+b i\textsuperscript 2 p e-a c i\textsuperscript 2 p
   e+b p e-a c p e-g i p e+a h i p e+a f h j p e+d i j p e-b f i j p e+a c f i j p e+a f h q
   e+d i q e-b f i q e+a c f i q e-b j q e+a c j q e-b f\textsuperscript 2 m+a c f\textsuperscript 2 m-b i\textsuperscript 2 m+a c i\textsuperscript 2 m-b
   f\textsuperscript 2 j\textsuperscript 2 m+a c f\textsuperscript 2 j\textsuperscript 2 m+d f j\textsuperscript 2 m-b m+a c m+d f m+g i m-a h i m-f g j m+a f h j m-d i j
   m+2 b f i j m-2 a c f i j m+a h\textsuperscript 2 n+a n-g h n+b h i n-a c h i n+d h j n-b f h j n+a c f h
   j n-a f h\textsuperscript 2 p-a f p+f g h p-b f h i p+a c f h i p+b f\textsuperscript 2 h j p-a c f\textsuperscript 2 h j p-d f h j p+b
   f\textsuperscript 2 h q-a c f\textsuperscript 2 h q+b h q-a c h q-d f h q-a i q+a f j q)-(a f\textsuperscript 2 c\textsuperscript 2+a i\textsuperscript 2
   c\textsuperscript 2+a f\textsuperscript 2 j\textsuperscript 2 c\textsuperscript 2+a c\textsuperscript 2-2 a f i j c\textsuperscript 2-b f\textsuperscript 2 c-b i\textsuperscript 2 c-b f\textsuperscript 2 j\textsuperscript 2 c+d f j\textsuperscript 2 c-2 a e f j\textsuperscript 2
   c-b c+d f c-2 a e f c+g i c-2 a h i c-f g j c+2 a f h j c-d i j c+2 a e i j c+2 b f i j
   c+a e\textsuperscript 2+a h\textsuperscript 2+a e\textsuperscript 2 j\textsuperscript 2-d e j\textsuperscript 2+b e f j\textsuperscript 2+a-d e+b e f-g h+b h i+e g j+d h j-2 a e h j-b f
   h j-b e i j) (k\textsuperscript 2+m\textsuperscript 2+n\textsuperscript 2+p\textsuperscript 2+q\textsuperscript 2+1)-k
   (-(g\textsuperscript 2+h\textsuperscript 2+i\textsuperscript 2+j\textsuperscript 2+1) (-(d\textsuperscript 2+e\textsuperscript 2+f\textsuperscript 2+1) (-m b\textsuperscript 2+c k
   b+a c m b+a n b-a c\textsuperscript 2 k-a k-m+c n)+(b d+c e+f) (b e k-a c e k-a f k-b d m+a c d m-f
   m+a d n+e n)+(-e b\textsuperscript 2+c d b+a c e b+a f b-a c\textsuperscript 2 d-a d-e+c f) (d k+e m+f
   n+p))+(d g+e h+f i+j) (-(d g+e h+f i+j) (-m b\textsuperscript 2+c k b+a c m b+a n b-a c\textsuperscript 2
   k-a k-m+c n)+(b d+c e+f) (b h k-a c h k-a i k-b g m+a c g m-i m+a g n+h n)+(-h
   b\textsuperscript 2+c g b+a c h b+a i b-a c\textsuperscript 2 g-a g-h+c i) (d k+e m+f n+p))-(b g+c h+i)
   (-(d g+e h+f i+j) (b e k-a c e k-a f k-b d m+a c d m-f m+a d n+e
   n)+(d\textsuperscript 2+e\textsuperscript 2+f\textsuperscript 2+1) (b h k-a c h k-a i k-b g m+a c g m-i m+a g n+h n)+(b e g-a
   c e g-a f g-b d h+a c d h-f h+a d i+e i) (d k+e m+f
   n+p))+(-(d\textsuperscript 2+e\textsuperscript 2+f\textsuperscript 2+1) (-h b\textsuperscript 2+c g b+a c h b+a i b-a c\textsuperscript 2 g-a
   g-h+c i)+(b d+c e+f) (b e g-a c e g-a f g-b d h+a c d h-f h+a d i+e i)+(-e
   b\textsuperscript 2+c d b+a c e b+a f b-a c\textsuperscript 2 d-a d-e+c f) (d g+e h+f i+j)) (g k+h m+i n+j
   p+q)))-b (a d+e) (-e q\textsuperscript 2 j\textsuperscript 2+c f q\textsuperscript 2 j\textsuperscript 2-e j\textsuperscript 2+c f j\textsuperscript 2+h q\textsuperscript 2 j-c i q\textsuperscript 2
   j+h j-c i j-m q j+c n q j+2 e p q j-2 c f p q j-e p\textsuperscript 2+c f p\textsuperscript 2-e+c f+m p-c n p-h p q+c i p
   q) (-(g k+h m+i n+j p+q) (-b j p f\textsuperscript 2+a c j p f\textsuperscript 2-b q f\textsuperscript 2+a c q f\textsuperscript 2+b j n
   f-a c j n f-g p f+a h p f+b i p f-a c i p f+d j p f-a e j p f+d q f-a e q f+g n-a h n-b i
   n+a c i n-d j n+a e j n-b q+a c q)+(d k+e m+f n+p) (b p i\textsuperscript 2-a c p i\textsuperscript 2-b j n
   i+a c j n i-g p i+a h p i+d j p i-a e j p i-b f j p i+a c f j p i+d q i-a e q i-b f q i+a
   c f q i-d j\textsuperscript 2 n+a e j\textsuperscript 2 n+b f j\textsuperscript 2 n-a c f j\textsuperscript 2 n-d n+a e n+b f n-a c f n+g j n-a h j n+b
   p-a c p-b j q+a c j q)-(a k+m) (a j\textsuperscript 2 n e\textsuperscript 2+a n e\textsuperscript 2-a i j p e\textsuperscript 2-a i q e\textsuperscript 2-a f
   j\textsuperscript 2 m e-a f m e+a i j m e-d j\textsuperscript 2 n e+b f j\textsuperscript 2 n e-a c f j\textsuperscript 2 n e-d n e+b f n e-a c f n e+g j
   n e-2 a h j n e-b i j n e+a c i j n e+b i\textsuperscript 2 p e-a c i\textsuperscript 2 p e+b p e-a c p e-g i p e+a h i p
   e+a f h j p e+d i j p e-b f i j p e+a c f i j p e+a f h q e+d i q e-b f i q e+a c f i q
   e-b j q e+a c j q e-b f\textsuperscript 2 m+a c f\textsuperscript 2 m-b i\textsuperscript 2 m+a c i\textsuperscript 2 m-b f\textsuperscript 2 j\textsuperscript 2 m+a c f\textsuperscript 2 j\textsuperscript 2 m+d f j\textsuperscript 2
   m-b m+a c m+d f m+g i m-a h i m-f g j m+a f h j m-d i j m+2 b f i j m-2 a c f i j m+a h\textsuperscript 2
   n+a n-g h n+b h i n-a c h i n+d h j n-b f h j n+a c f h j n-a f h\textsuperscript 2 p-a f p+f g h p-b f h
   i p+a c f h i p+b f\textsuperscript 2 h j p-a c f\textsuperscript 2 h j p-d f h j p+b f\textsuperscript 2 h q-a c f\textsuperscript 2 h q+b h q-a c h q-d
   f h q-a i q+a f j q)+k (c f\textsuperscript 2 m a\textsuperscript 2+c i\textsuperscript 2 m a\textsuperscript 2+c f\textsuperscript 2 j\textsuperscript 2 m a\textsuperscript 2-e f j\textsuperscript 2 m
   a\textsuperscript 2+c m a\textsuperscript 2-e f m a\textsuperscript 2-h i m a\textsuperscript 2+f h j m a\textsuperscript 2+e i j m a\textsuperscript 2-2 c f i j m a\textsuperscript 2+e\textsuperscript 2 n a\textsuperscript 2+h\textsuperscript 2 n
   a\textsuperscript 2+e\textsuperscript 2 j\textsuperscript 2 n a\textsuperscript 2-c e f j\textsuperscript 2 n a\textsuperscript 2-c e f n a\textsuperscript 2-c h i n a\textsuperscript 2-2 e h j n a\textsuperscript 2+c f h j n a\textsuperscript 2+c e
   i j n a\textsuperscript 2+n a\textsuperscript 2-f h\textsuperscript 2 p a\textsuperscript 2-c e i\textsuperscript 2 p a\textsuperscript 2-c e p a\textsuperscript 2-f p a\textsuperscript 2+e h i p a\textsuperscript 2+c f h i p a\textsuperscript 2-c
   f\textsuperscript 2 h j p a\textsuperscript 2+e f h j p a\textsuperscript 2-e\textsuperscript 2 i j p a\textsuperscript 2+c e f i j p a\textsuperscript 2-c f\textsuperscript 2 h q a\textsuperscript 2-c h q a\textsuperscript 2+e f h q
   a\textsuperscript 2-e\textsuperscript 2 i q a\textsuperscript 2+c e f i q a\textsuperscript 2-i q a\textsuperscript 2+c e j q a\textsuperscript 2+f j q a\textsuperscript 2-c f\textsuperscript 2 k a-c i\textsuperscript 2 k a-c f\textsuperscript 2 j\textsuperscript 2
   k a+e f j\textsuperscript 2 k a-c k a+e f k a+h i k a-f h j k a-e i j k a+2 c f i j k a-b f\textsuperscript 2 m a-b i\textsuperscript 2 m
   a-b f\textsuperscript 2 j\textsuperscript 2 m a+d f j\textsuperscript 2 m a-b m a+d f m a+g i m a-f g j m a-d i j m a+2 b f i j m a-2 d e
   j\textsuperscript 2 n a+c d f j\textsuperscript 2 n a+b e f j\textsuperscript 2 n a-2 d e n a+c d f n a+b e f n a-2 g h n a+c g i n a+b h
   i n a+2 e g j n a-c f g j n a+2 d h j n a-b f h j n a-c d i j n a-b e i j n a+c d i\textsuperscript 2 p
   a+b e i\textsuperscript 2 p a+c d p a+b e p a+2 f g h p a-e g i p a-c f g i p a-d h i p a-b f h i p a+c
   f\textsuperscript 2 g j p a-e f g j p a+b f\textsuperscript 2 h j p a-d f h j p a+2 d e i j p a-c d f i j p a-b e f i j p
   a+c f\textsuperscript 2 g q a+c g q a-e f g q a+b f\textsuperscript 2 h q a+b h q a-d f h q a+2 d e i q a-c d f i q a-b e
   f i q a-c d j q a-b e j q a+b f\textsuperscript 2 k+b i\textsuperscript 2 k+b f\textsuperscript 2 j\textsuperscript 2 k-d f j\textsuperscript 2 k+b k-d f k-g i k+f g j
   k+d i j k-2 b f i j k+d\textsuperscript 2 n+g\textsuperscript 2 n+d\textsuperscript 2 j\textsuperscript 2 n-b d f j\textsuperscript 2 n-b d f n-b g i n-2 d g j n+b f g j
   n+b d i j n+n-f g\textsuperscript 2 p-b d i\textsuperscript 2 p-b d p-f p+d g i p+b f g i p-b f\textsuperscript 2 g j p+d f g j p-d\textsuperscript 2 i j
   p+b d f i j p-b f\textsuperscript 2 g q-b g q+d f g q-d\textsuperscript 2 i q+b d f i q-i q+b d j q+f j q)+(-b
   j\textsuperscript 2 f\textsuperscript 2+a c j\textsuperscript 2 f\textsuperscript 2-b f\textsuperscript 2+a c f\textsuperscript 2+d j\textsuperscript 2 f-a e j\textsuperscript 2 f+d f-a e f-g j f+a h j f+2 b i j f-2 a
   c i j f-b i\textsuperscript 2+a c i\textsuperscript 2-b+a c+g i-a h i-d i j+a e i j)
   (k\textsuperscript 2+m\textsuperscript 2+n\textsuperscript 2+p\textsuperscript 2+q\textsuperscript 2+1)))))-(d k+e m+f n+p) (j q-p)
   (-g (-g q\textsuperscript 2+a h q\textsuperscript 2+b i q\textsuperscript 2-a c i q\textsuperscript 2+d j q\textsuperscript 2-a e j q\textsuperscript 2-b f j q\textsuperscript 2+a c f j q\textsuperscript 2+k
   q-a m q-b n q+a c n q-d p q+a e p q+b f p q-a c f p q-g+a h+b i-a c i+d j-a e j-b f j+a c
   f j) ((b d+c e+f) (-f q\textsuperscript 2 j\textsuperscript 2-f j\textsuperscript 2+i q\textsuperscript 2 j+i j-n q j+2 f p q j-f p\textsuperscript 2-f+n
   p-i p q) ((a g+h) (b k+c m+n) (-n+f p+i q-f j q) (-h q\textsuperscript 2+c i q\textsuperscript 2+e j
   q\textsuperscript 2-c f j q\textsuperscript 2+m q-c n q-e p q+c f p q-h+c i+e j-c f j)-(b g+c h+i) (a k+m) (-m+c
   n+e p-c f p+h q-c i q-e j q+c f j q) (-i q\textsuperscript 2+f j q\textsuperscript 2+n q-f p q-i+f
   j))-(d g+e h+f i+j) (-j q\textsuperscript 2+p q-j) ((a b+c) (b k+c m+n) (-n+f
   p+i q-f j q) (-c j\textsuperscript 2 f\textsuperscript 2-c p\textsuperscript 2 f\textsuperscript 2-c j\textsuperscript 2 q\textsuperscript 2 f\textsuperscript 2-c f\textsuperscript 2+2 c j p q f\textsuperscript 2+e j\textsuperscript 2 f+e p\textsuperscript 2
   f+e j\textsuperscript 2 q\textsuperscript 2 f-h j q\textsuperscript 2 f+2 c i j q\textsuperscript 2 f+e f-h j f+2 c i j f-m p f+2 c n p f+j m q f-2 c j n
   q f+h p q f-2 c i p q f-2 e j p q f-c i\textsuperscript 2-c n\textsuperscript 2-c i\textsuperscript 2 q\textsuperscript 2+h i q\textsuperscript 2-e i j q\textsuperscript 2-c+h i-e i j+m
   n-e n p-i m q-h n q+2 c i n q+e j n q+e i p q)-(b\textsuperscript 2+c\textsuperscript 2+1) (a k+m) (-m+c
   n+e p-c f p+h q-c i q-e j q+c f j q) (j\textsuperscript 2 f\textsuperscript 2+p\textsuperscript 2 f\textsuperscript 2+j\textsuperscript 2 q\textsuperscript 2 f\textsuperscript 2-2 j p q f\textsuperscript 2+f\textsuperscript 2-2
   i j q\textsuperscript 2 f-2 i j f-2 n p f+2 j n q f+2 i p q f+i\textsuperscript 2+n\textsuperscript 2+i\textsuperscript 2 q\textsuperscript 2-2 i n q+1))+(d
   k+e m+f n+p) (j q-p) ((a b+c) (b g+c h+i) (-i q\textsuperscript 2+f j q\textsuperscript 2+n q-f p q-i+f
   j) (-c j\textsuperscript 2 f\textsuperscript 2-c p\textsuperscript 2 f\textsuperscript 2-c j\textsuperscript 2 q\textsuperscript 2 f\textsuperscript 2-c f\textsuperscript 2+2 c j p q f\textsuperscript 2+e j\textsuperscript 2 f+e p\textsuperscript 2 f+e
   j\textsuperscript 2 q\textsuperscript 2 f-h j q\textsuperscript 2 f+2 c i j q\textsuperscript 2 f+e f-h j f+2 c i j f-m p f+2 c n p f+j m q f-2 c j n q
   f+h p q f-2 c i p q f-2 e j p q f-c i\textsuperscript 2-c n\textsuperscript 2-c i\textsuperscript 2 q\textsuperscript 2+h i q\textsuperscript 2-e i j q\textsuperscript 2-c+h i-e i j+m
   n-e n p-i m q-h n q+2 c i n q+e j n q+e i p q)-(b\textsuperscript 2+c\textsuperscript 2+1) (a g+h)
   (-h q\textsuperscript 2+c i q\textsuperscript 2+e j q\textsuperscript 2-c f j q\textsuperscript 2+m q-c n q-e p q+c f p q-h+c i+e j-c f j)
   (j\textsuperscript 2 f\textsuperscript 2+p\textsuperscript 2 f\textsuperscript 2+j\textsuperscript 2 q\textsuperscript 2 f\textsuperscript 2-2 j p q f\textsuperscript 2+f\textsuperscript 2-2 i j q\textsuperscript 2 f-2 i j f-2 n p f+2 j n q f+2
   i p q f+i\textsuperscript 2+n\textsuperscript 2+i\textsuperscript 2 q\textsuperscript 2-2 i n q+1)))+(b g+c h+i) (-i q\textsuperscript 2+f j q\textsuperscript 2+n
   q-f p q-i+f j) (d (-d q\textsuperscript 2 j\textsuperscript 2+a e q\textsuperscript 2 j\textsuperscript 2+b f q\textsuperscript 2 j\textsuperscript 2-a c f q\textsuperscript 2 j\textsuperscript 2-d
   j\textsuperscript 2+a e j\textsuperscript 2+b f j\textsuperscript 2-a c f j\textsuperscript 2+g q\textsuperscript 2 j-a h q\textsuperscript 2 j-b i q\textsuperscript 2 j+a c i q\textsuperscript 2 j+g j-a h j-b i j+a c
   i j-k q j+a m q j+b n q j-a c n q j+2 d p q j-2 a e p q j-2 b f p q j+2 a c f p q j-d
   p\textsuperscript 2+a e p\textsuperscript 2+b f p\textsuperscript 2-a c f p\textsuperscript 2-d+a e+b f-a c f+k p-a m p-b n p+a c n p-g p q+a h p q+b i p
   q-a c i p q) ((a g+h) (b k+c m+n) (-n+f p+i q-f j q) (-h q\textsuperscript 2+c i q\textsuperscript 2+e j
   q\textsuperscript 2-c f j q\textsuperscript 2+m q-c n q-e p q+c f p q-h+c i+e j-c f j)-(b g+c h+i) (a k+m) (-m+c
   n+e p-c f p+h q-c i q-e j q+c f j q) (-i q\textsuperscript 2+f j q\textsuperscript 2+n q-f p q-i+f
   j))-(d g+e h+f i+j) (-j q\textsuperscript 2+p q-j) (a (b k+c m+n) (-n+f p+i
   q-f j q) ((g k+h m+i n+j p+q) (a i n c\textsuperscript 2-a f j n c\textsuperscript 2-a f i p c\textsuperscript 2+a f\textsuperscript 2 j p
   c\textsuperscript 2+a f\textsuperscript 2 q c\textsuperscript 2+a q c\textsuperscript 2-a i m c+a f j m c+g n c-a h n c-b i n c-d j n c+a e j n c+b f j n
   c-f g p c+a f h p c+a e i p c+b f i p c-b f\textsuperscript 2 j p c+d f j p c-2 a e f j p c-b f\textsuperscript 2 q c-b q
   c+d f q c-2 a e f q c-g m+a h m+b i m+d j m-a e j m-b f j m+e g p-a e h p-b e i p+a e\textsuperscript 2 j
   p-d e j p+b e f j p+a e\textsuperscript 2 q+a q-d e q+b e f q)-(d k+e m+f n+p) (-a f j\textsuperscript 2 n
   c\textsuperscript 2-a f n c\textsuperscript 2+a i j n c\textsuperscript 2-a i\textsuperscript 2 p c\textsuperscript 2-a p c\textsuperscript 2+a f i j p c\textsuperscript 2+a f i q c\textsuperscript 2+a j q c\textsuperscript 2+a f j\textsuperscript 2
   m c+a f m c-a i j m c-d j\textsuperscript 2 n c+a e j\textsuperscript 2 n c+b f j\textsuperscript 2 n c-d n c+a e n c+b f n c+g j n c-a h
   j n c-b i j n c+b i\textsuperscript 2 p c+b p c-g i p c+2 a h i p c-a f h j p c+d i j p c-a e i j p c-b f
   i j p c-a f h q c+d i q c-a e i q c-b f i q c-b j q c+d j\textsuperscript 2 m-a e j\textsuperscript 2 m-b f j\textsuperscript 2 m+d m-a e
   m-b f m-g j m+a h j m+b i j m-a h\textsuperscript 2 p-a p+g h p-b h i p-d h j p+a e h j p+b f h j p-d h
   q+a e h q+b f h q+a j q)+(b k+c m+n) (a j\textsuperscript 2 n e\textsuperscript 2+a n e\textsuperscript 2-a i j p e\textsuperscript 2-a i q
   e\textsuperscript 2-a f j\textsuperscript 2 m e-a f m e+a i j m e-d j\textsuperscript 2 n e+b f j\textsuperscript 2 n e-a c f j\textsuperscript 2 n e-d n e+b f n e-a c f
   n e+g j n e-2 a h j n e-b i j n e+a c i j n e+b i\textsuperscript 2 p e-a c i\textsuperscript 2 p e+b p e-a c p e-g i p
   e+a h i p e+a f h j p e+d i j p e-b f i j p e+a c f i j p e+a f h q e+d i q e-b f i q e+a
   c f i q e-b j q e+a c j q e-b f\textsuperscript 2 m+a c f\textsuperscript 2 m-b i\textsuperscript 2 m+a c i\textsuperscript 2 m-b f\textsuperscript 2 j\textsuperscript 2 m+a c f\textsuperscript 2 j\textsuperscript 2
   m+d f j\textsuperscript 2 m-b m+a c m+d f m+g i m-a h i m-f g j m+a f h j m-d i j m+2 b f i j m-2 a c f i
   j m+a h\textsuperscript 2 n+a n-g h n+b h i n-a c h i n+d h j n-b f h j n+a c f h j n-a f h\textsuperscript 2 p-a f p+f g
   h p-b f h i p+a c f h i p+b f\textsuperscript 2 h j p-a c f\textsuperscript 2 h j p-d f h j p+b f\textsuperscript 2 h q-a c f\textsuperscript 2 h q+b h
   q-a c h q-d f h q-a i q+a f j q)-(a f\textsuperscript 2 c\textsuperscript 2+a i\textsuperscript 2 c\textsuperscript 2+a f\textsuperscript 2 j\textsuperscript 2 c\textsuperscript 2+a c\textsuperscript 2-2 a
   f i j c\textsuperscript 2-b f\textsuperscript 2 c-b i\textsuperscript 2 c-b f\textsuperscript 2 j\textsuperscript 2 c+d f j\textsuperscript 2 c-2 a e f j\textsuperscript 2 c-b c+d f c-2 a e f c+g i c-2
   a h i c-f g j c+2 a f h j c-d i j c+2 a e i j c+2 b f i j c+a e\textsuperscript 2+a h\textsuperscript 2+a e\textsuperscript 2 j\textsuperscript 2-d e
   j\textsuperscript 2+b e f j\textsuperscript 2+a-d e+b e f-g h+b h i+e g j+d h j-2 a e h j-b f h j-b e i j)
   (k\textsuperscript 2+m\textsuperscript 2+n\textsuperscript 2+p\textsuperscript 2+q\textsuperscript 2+1)-k (-(g\textsuperscript 2+h\textsuperscript 2+i\textsuperscript 2+j\textsuperscript 2+1)
   (-(d\textsuperscript 2+e\textsuperscript 2+f\textsuperscript 2+1) (-m b\textsuperscript 2+c k b+a c m b+a n b-a c\textsuperscript 2 k-a k-m+c
   n)+(b d+c e+f) (b e k-a c e k-a f k-b d m+a c d m-f m+a d n+e n)+(-e b\textsuperscript 2+c d
   b+a c e b+a f b-a c\textsuperscript 2 d-a d-e+c f) (d k+e m+f n+p))+(d g+e h+f i+j) (-(d
   g+e h+f i+j) (-m b\textsuperscript 2+c k b+a c m b+a n b-a c\textsuperscript 2 k-a k-m+c n)+(b d+c e+f) (b h
   k-a c h k-a i k-b g m+a c g m-i m+a g n+h n)+(-h b\textsuperscript 2+c g b+a c h b+a i b-a c\textsuperscript 2 g-a
   g-h+c i) (d k+e m+f n+p))-(b g+c h+i) (-(d g+e h+f i+j) (b e k-a c e k-a
   f k-b d m+a c d m-f m+a d n+e n)+(d\textsuperscript 2+e\textsuperscript 2+f\textsuperscript 2+1) (b h k-a c h k-a i k-b g m+a
   c g m-i m+a g n+h n)+(b e g-a c e g-a f g-b d h+a c d h-f h+a d i+e i) (d k+e m+f
   n+p))+(-(d\textsuperscript 2+e\textsuperscript 2+f\textsuperscript 2+1) (-h b\textsuperscript 2+c g b+a c h b+a i b-a c\textsuperscript 2 g-a
   g-h+c i)+(b d+c e+f) (b e g-a c e g-a f g-b d h+a c d h-f h+a d i+e i)+(-e
   b\textsuperscript 2+c d b+a c e b+a f b-a c\textsuperscript 2 d-a d-e+c f) (d g+e h+f i+j)) (g k+h m+i n+j
   p+q)))-b (a k+m) (-m+c n+e p-c f p+h q-c i q-e j q+c f j q) (-(g k+h m+i
   n+j p+q) (-b j p f\textsuperscript 2+a c j p f\textsuperscript 2-b q f\textsuperscript 2+a c q f\textsuperscript 2+b j n f-a c j n f-g p f+a h p f+b
   i p f-a c i p f+d j p f-a e j p f+d q f-a e q f+g n-a h n-b i n+a c i n-d j n+a e j n-b
   q+a c q)+(d k+e m+f n+p) (b p i\textsuperscript 2-a c p i\textsuperscript 2-b j n i+a c j n i-g p i+a h p i+d
   j p i-a e j p i-b f j p i+a c f j p i+d q i-a e q i-b f q i+a c f q i-d j\textsuperscript 2 n+a e j\textsuperscript 2 n+b
   f j\textsuperscript 2 n-a c f j\textsuperscript 2 n-d n+a e n+b f n-a c f n+g j n-a h j n+b p-a c p-b j q+a c j
   q)-(a k+m) (a j\textsuperscript 2 n e\textsuperscript 2+a n e\textsuperscript 2-a i j p e\textsuperscript 2-a i q e\textsuperscript 2-a f j\textsuperscript 2 m e-a f m e+a i
   j m e-d j\textsuperscript 2 n e+b f j\textsuperscript 2 n e-a c f j\textsuperscript 2 n e-d n e+b f n e-a c f n e+g j n e-2 a h j n e-b i
   j n e+a c i j n e+b i\textsuperscript 2 p e-a c i\textsuperscript 2 p e+b p e-a c p e-g i p e+a h i p e+a f h j p e+d i j
   p e-b f i j p e+a c f i j p e+a f h q e+d i q e-b f i q e+a c f i q e-b j q e+a c j q e-b
   f\textsuperscript 2 m+a c f\textsuperscript 2 m-b i\textsuperscript 2 m+a c i\textsuperscript 2 m-b f\textsuperscript 2 j\textsuperscript 2 m+a c f\textsuperscript 2 j\textsuperscript 2 m+d f j\textsuperscript 2 m-b m+a c m+d f m+g i
   m-a h i m-f g j m+a f h j m-d i j m+2 b f i j m-2 a c f i j m+a h\textsuperscript 2 n+a n-g h n+b h i n-a
   c h i n+d h j n-b f h j n+a c f h j n-a f h\textsuperscript 2 p-a f p+f g h p-b f h i p+a c f h i p+b f\textsuperscript 2
   h j p-a c f\textsuperscript 2 h j p-d f h j p+b f\textsuperscript 2 h q-a c f\textsuperscript 2 h q+b h q-a c h q-d f h q-a i q+a f j
   q)+k (c f\textsuperscript 2 m a\textsuperscript 2+c i\textsuperscript 2 m a\textsuperscript 2+c f\textsuperscript 2 j\textsuperscript 2 m a\textsuperscript 2-e f j\textsuperscript 2 m a\textsuperscript 2+c m a\textsuperscript 2-e f m
   a\textsuperscript 2-h i m a\textsuperscript 2+f h j m a\textsuperscript 2+e i j m a\textsuperscript 2-2 c f i j m a\textsuperscript 2+e\textsuperscript 2 n a\textsuperscript 2+h\textsuperscript 2 n a\textsuperscript 2+e\textsuperscript 2 j\textsuperscript 2 n a\textsuperscript 2-c
   e f j\textsuperscript 2 n a\textsuperscript 2-c e f n a\textsuperscript 2-c h i n a\textsuperscript 2-2 e h j n a\textsuperscript 2+c f h j n a\textsuperscript 2+c e i j n a\textsuperscript 2+n a\textsuperscript 2-f
   h\textsuperscript 2 p a\textsuperscript 2-c e i\textsuperscript 2 p a\textsuperscript 2-c e p a\textsuperscript 2-f p a\textsuperscript 2+e h i p a\textsuperscript 2+c f h i p a\textsuperscript 2-c f\textsuperscript 2 h j p a\textsuperscript 2+e f h
   j p a\textsuperscript 2-e\textsuperscript 2 i j p a\textsuperscript 2+c e f i j p a\textsuperscript 2-c f\textsuperscript 2 h q a\textsuperscript 2-c h q a\textsuperscript 2+e f h q a\textsuperscript 2-e\textsuperscript 2 i q a\textsuperscript 2+c e
   f i q a\textsuperscript 2-i q a\textsuperscript 2+c e j q a\textsuperscript 2+f j q a\textsuperscript 2-c f\textsuperscript 2 k a-c i\textsuperscript 2 k a-c f\textsuperscript 2 j\textsuperscript 2 k a+e f j\textsuperscript 2 k a-c k
   a+e f k a+h i k a-f h j k a-e i j k a+2 c f i j k a-b f\textsuperscript 2 m a-b i\textsuperscript 2 m a-b f\textsuperscript 2 j\textsuperscript 2 m a+d f
   j\textsuperscript 2 m a-b m a+d f m a+g i m a-f g j m a-d i j m a+2 b f i j m a-2 d e j\textsuperscript 2 n a+c d f j\textsuperscript 2 n
   a+b e f j\textsuperscript 2 n a-2 d e n a+c d f n a+b e f n a-2 g h n a+c g i n a+b h i n a+2 e g j n a-c
   f g j n a+2 d h j n a-b f h j n a-c d i j n a-b e i j n a+c d i\textsuperscript 2 p a+b e i\textsuperscript 2 p a+c d p
   a+b e p a+2 f g h p a-e g i p a-c f g i p a-d h i p a-b f h i p a+c f\textsuperscript 2 g j p a-e f g j p
   a+b f\textsuperscript 2 h j p a-d f h j p a+2 d e i j p a-c d f i j p a-b e f i j p a+c f\textsuperscript 2 g q a+c g q
   a-e f g q a+b f\textsuperscript 2 h q a+b h q a-d f h q a+2 d e i q a-c d f i q a-b e f i q a-c d j q a-b
   e j q a+b f\textsuperscript 2 k+b i\textsuperscript 2 k+b f\textsuperscript 2 j\textsuperscript 2 k-d f j\textsuperscript 2 k+b k-d f k-g i k+f g j k+d i j k-2 b f i j
   k+d\textsuperscript 2 n+g\textsuperscript 2 n+d\textsuperscript 2 j\textsuperscript 2 n-b d f j\textsuperscript 2 n-b d f n-b g i n-2 d g j n+b f g j n+b d i j n+n-f g\textsuperscript 2
   p-b d i\textsuperscript 2 p-b d p-f p+d g i p+b f g i p-b f\textsuperscript 2 g j p+d f g j p-d\textsuperscript 2 i j p+b d f i j p-b f\textsuperscript 2
   g q-b g q+d f g q-d\textsuperscript 2 i q+b d f i q-i q+b d j q+f j q)+(-b j\textsuperscript 2 f\textsuperscript 2+a c j\textsuperscript 2
   f\textsuperscript 2-b f\textsuperscript 2+a c f\textsuperscript 2+d j\textsuperscript 2 f-a e j\textsuperscript 2 f+d f-a e f-g j f+a h j f+2 b i j f-2 a c i j f-b i\textsuperscript 2+a
   c i\textsuperscript 2-b+a c+g i-a h i-d i j+a e i j)
   (k\textsuperscript 2+m\textsuperscript 2+n\textsuperscript 2+p\textsuperscript 2+q\textsuperscript 2+1)))+(d k+e m+f n+p) (j q-p) (a (b g+c
   h+i) (-i q\textsuperscript 2+f j q\textsuperscript 2+n q-f p q-i+f j) ((g k+h m+i n+j p+q) (a i n
   c\textsuperscript 2-a f j n c\textsuperscript 2-a f i p c\textsuperscript 2+a f\textsuperscript 2 j p c\textsuperscript 2+a f\textsuperscript 2 q c\textsuperscript 2+a q c\textsuperscript 2-a i m c+a f j m c+g n c-a h
   n c-b i n c-d j n c+a e j n c+b f j n c-f g p c+a f h p c+a e i p c+b f i p c-b f\textsuperscript 2 j p
   c+d f j p c-2 a e f j p c-b f\textsuperscript 2 q c-b q c+d f q c-2 a e f q c-g m+a h m+b i m+d j m-a e j
   m-b f j m+e g p-a e h p-b e i p+a e\textsuperscript 2 j p-d e j p+b e f j p+a e\textsuperscript 2 q+a q-d e q+b e f
   q)-(d k+e m+f n+p) (-a f j\textsuperscript 2 n c\textsuperscript 2-a f n c\textsuperscript 2+a i j n c\textsuperscript 2-a i\textsuperscript 2 p c\textsuperscript 2-a p c\textsuperscript 2+a
   f i j p c\textsuperscript 2+a f i q c\textsuperscript 2+a j q c\textsuperscript 2+a f j\textsuperscript 2 m c+a f m c-a i j m c-d j\textsuperscript 2 n c+a e j\textsuperscript 2 n c+b f
   j\textsuperscript 2 n c-d n c+a e n c+b f n c+g j n c-a h j n c-b i j n c+b i\textsuperscript 2 p c+b p c-g i p c+2 a h i
   p c-a f h j p c+d i j p c-a e i j p c-b f i j p c-a f h q c+d i q c-a e i q c-b f i q c-b
   j q c+d j\textsuperscript 2 m-a e j\textsuperscript 2 m-b f j\textsuperscript 2 m+d m-a e m-b f m-g j m+a h j m+b i j m-a h\textsuperscript 2 p-a p+g h
   p-b h i p-d h j p+a e h j p+b f h j p-d h q+a e h q+b f h q+a j q)+(b k+c m+n)
   (a j\textsuperscript 2 n e\textsuperscript 2+a n e\textsuperscript 2-a i j p e\textsuperscript 2-a i q e\textsuperscript 2-a f j\textsuperscript 2 m e-a f m e+a i j m e-d j\textsuperscript 2 n e+b
   f j\textsuperscript 2 n e-a c f j\textsuperscript 2 n e-d n e+b f n e-a c f n e+g j n e-2 a h j n e-b i j n e+a c i j n
   e+b i\textsuperscript 2 p e-a c i\textsuperscript 2 p e+b p e-a c p e-g i p e+a h i p e+a f h j p e+d i j p e-b f i j p
   e+a c f i j p e+a f h q e+d i q e-b f i q e+a c f i q e-b j q e+a c j q e-b f\textsuperscript 2 m+a c f\textsuperscript 2
   m-b i\textsuperscript 2 m+a c i\textsuperscript 2 m-b f\textsuperscript 2 j\textsuperscript 2 m+a c f\textsuperscript 2 j\textsuperscript 2 m+d f j\textsuperscript 2 m-b m+a c m+d f m+g i m-a h i m-f g
   j m+a f h j m-d i j m+2 b f i j m-2 a c f i j m+a h\textsuperscript 2 n+a n-g h n+b h i n-a c h i n+d h j
   n-b f h j n+a c f h j n-a f h\textsuperscript 2 p-a f p+f g h p-b f h i p+a c f h i p+b f\textsuperscript 2 h j p-a c f\textsuperscript 2
   h j p-d f h j p+b f\textsuperscript 2 h q-a c f\textsuperscript 2 h q+b h q-a c h q-d f h q-a i q+a f j q)-(a
   f\textsuperscript 2 c\textsuperscript 2+a i\textsuperscript 2 c\textsuperscript 2+a f\textsuperscript 2 j\textsuperscript 2 c\textsuperscript 2+a c\textsuperscript 2-2 a f i j c\textsuperscript 2-b f\textsuperscript 2 c-b i\textsuperscript 2 c-b f\textsuperscript 2 j\textsuperscript 2 c+d f j\textsuperscript 2
   c-2 a e f j\textsuperscript 2 c-b c+d f c-2 a e f c+g i c-2 a h i c-f g j c+2 a f h j c-d i j c+2 a e i j
   c+2 b f i j c+a e\textsuperscript 2+a h\textsuperscript 2+a e\textsuperscript 2 j\textsuperscript 2-d e j\textsuperscript 2+b e f j\textsuperscript 2+a-d e+b e f-g h+b h i+e g j+d h j-2
   a e h j-b f h j-b e i j) (k\textsuperscript 2+m\textsuperscript 2+n\textsuperscript 2+p\textsuperscript 2+q\textsuperscript 2+1)-k
   (-(g\textsuperscript 2+h\textsuperscript 2+i\textsuperscript 2+j\textsuperscript 2+1) (-(d\textsuperscript 2+e\textsuperscript 2+f\textsuperscript 2+1) (-m b\textsuperscript 2+c k
   b+a c m b+a n b-a c\textsuperscript 2 k-a k-m+c n)+(b d+c e+f) (b e k-a c e k-a f k-b d m+a c d m-f
   m+a d n+e n)+(-e b\textsuperscript 2+c d b+a c e b+a f b-a c\textsuperscript 2 d-a d-e+c f) (d k+e m+f
   n+p))+(d g+e h+f i+j) (-(d g+e h+f i+j) (-m b\textsuperscript 2+c k b+a c m b+a n b-a c\textsuperscript 2
   k-a k-m+c n)+(b d+c e+f) (b h k-a c h k-a i k-b g m+a c g m-i m+a g n+h n)+(-h
   b\textsuperscript 2+c g b+a c h b+a i b-a c\textsuperscript 2 g-a g-h+c i) (d k+e m+f n+p))-(b g+c h+i)
   (-(d g+e h+f i+j) (b e k-a c e k-a f k-b d m+a c d m-f m+a d n+e
   n)+(d\textsuperscript 2+e\textsuperscript 2+f\textsuperscript 2+1) (b h k-a c h k-a i k-b g m+a c g m-i m+a g n+h n)+(b e g-a
   c e g-a f g-b d h+a c d h-f h+a d i+e i) (d k+e m+f
   n+p))+(-(d\textsuperscript 2+e\textsuperscript 2+f\textsuperscript 2+1) (-h b\textsuperscript 2+c g b+a c h b+a i b-a c\textsuperscript 2 g-a
   g-h+c i)+(b d+c e+f) (b e g-a c e g-a f g-b d h+a c d h-f h+a d i+e i)+(-e
   b\textsuperscript 2+c d b+a c e b+a f b-a c\textsuperscript 2 d-a d-e+c f) (d g+e h+f i+j)) (g k+h m+i n+j
   p+q)))-b (a g+h) (-h q\textsuperscript 2+c i q\textsuperscript 2+e j q\textsuperscript 2-c f j q\textsuperscript 2+m q-c n q-e p q+c f p
   q-h+c i+e j-c f j) (-(g k+h m+i n+j p+q) (-b j p f\textsuperscript 2+a c j p f\textsuperscript 2-b q
   f\textsuperscript 2+a c q f\textsuperscript 2+b j n f-a c j n f-g p f+a h p f+b i p f-a c i p f+d j p f-a e j p f+d q f-a
   e q f+g n-a h n-b i n+a c i n-d j n+a e j n-b q+a c q)+(d k+e m+f n+p) (b p
   i\textsuperscript 2-a c p i\textsuperscript 2-b j n i+a c j n i-g p i+a h p i+d j p i-a e j p i-b f j p i+a c f j p i+d q
   i-a e q i-b f q i+a c f q i-d j\textsuperscript 2 n+a e j\textsuperscript 2 n+b f j\textsuperscript 2 n-a c f j\textsuperscript 2 n-d n+a e n+b f n-a c f
   n+g j n-a h j n+b p-a c p-b j q+a c j q)-(a k+m) (a j\textsuperscript 2 n e\textsuperscript 2+a n e\textsuperscript 2-a i j p
   e\textsuperscript 2-a i q e\textsuperscript 2-a f j\textsuperscript 2 m e-a f m e+a i j m e-d j\textsuperscript 2 n e+b f j\textsuperscript 2 n e-a c f j\textsuperscript 2 n e-d n e+b f
   n e-a c f n e+g j n e-2 a h j n e-b i j n e+a c i j n e+b i\textsuperscript 2 p e-a c i\textsuperscript 2 p e+b p e-a c p
   e-g i p e+a h i p e+a f h j p e+d i j p e-b f i j p e+a c f i j p e+a f h q e+d i q e-b f
   i q e+a c f i q e-b j q e+a c j q e-b f\textsuperscript 2 m+a c f\textsuperscript 2 m-b i\textsuperscript 2 m+a c i\textsuperscript 2 m-b f\textsuperscript 2 j\textsuperscript 2 m+a c
   f\textsuperscript 2 j\textsuperscript 2 m+d f j\textsuperscript 2 m-b m+a c m+d f m+g i m-a h i m-f g j m+a f h j m-d i j m+2 b f i j m-2
   a c f i j m+a h\textsuperscript 2 n+a n-g h n+b h i n-a c h i n+d h j n-b f h j n+a c f h j n-a f h\textsuperscript 2 p-a
   f p+f g h p-b f h i p+a c f h i p+b f\textsuperscript 2 h j p-a c f\textsuperscript 2 h j p-d f h j p+b f\textsuperscript 2 h q-a c f\textsuperscript 2 h
   q+b h q-a c h q-d f h q-a i q+a f j q)+k (c f\textsuperscript 2 m a\textsuperscript 2+c i\textsuperscript 2 m a\textsuperscript 2+c f\textsuperscript 2 j\textsuperscript 2 m
   a\textsuperscript 2-e f j\textsuperscript 2 m a\textsuperscript 2+c m a\textsuperscript 2-e f m a\textsuperscript 2-h i m a\textsuperscript 2+f h j m a\textsuperscript 2+e i j m a\textsuperscript 2-2 c f i j m a\textsuperscript 2+e\textsuperscript 2
   n a\textsuperscript 2+h\textsuperscript 2 n a\textsuperscript 2+e\textsuperscript 2 j\textsuperscript 2 n a\textsuperscript 2-c e f j\textsuperscript 2 n a\textsuperscript 2-c e f n a\textsuperscript 2-c h i n a\textsuperscript 2-2 e h j n a\textsuperscript 2+c f h
   j n a\textsuperscript 2+c e i j n a\textsuperscript 2+n a\textsuperscript 2-f h\textsuperscript 2 p a\textsuperscript 2-c e i\textsuperscript 2 p a\textsuperscript 2-c e p a\textsuperscript 2-f p a\textsuperscript 2+e h i p a\textsuperscript 2+c f h
   i p a\textsuperscript 2-c f\textsuperscript 2 h j p a\textsuperscript 2+e f h j p a\textsuperscript 2-e\textsuperscript 2 i j p a\textsuperscript 2+c e f i j p a\textsuperscript 2-c f\textsuperscript 2 h q a\textsuperscript 2-c h q
   a\textsuperscript 2+e f h q a\textsuperscript 2-e\textsuperscript 2 i q a\textsuperscript 2+c e f i q a\textsuperscript 2-i q a\textsuperscript 2+c e j q a\textsuperscript 2+f j q a\textsuperscript 2-c f\textsuperscript 2 k a-c i\textsuperscript 2 k
   a-c f\textsuperscript 2 j\textsuperscript 2 k a+e f j\textsuperscript 2 k a-c k a+e f k a+h i k a-f h j k a-e i j k a+2 c f i j k a-b f\textsuperscript 2
   m a-b i\textsuperscript 2 m a-b f\textsuperscript 2 j\textsuperscript 2 m a+d f j\textsuperscript 2 m a-b m a+d f m a+g i m a-f g j m a-d i j m a+2 b f i
   j m a-2 d e j\textsuperscript 2 n a+c d f j\textsuperscript 2 n a+b e f j\textsuperscript 2 n a-2 d e n a+c d f n a+b e f n a-2 g h n a+c
   g i n a+b h i n a+2 e g j n a-c f g j n a+2 d h j n a-b f h j n a-c d i j n a-b e i j n
   a+c d i\textsuperscript 2 p a+b e i\textsuperscript 2 p a+c d p a+b e p a+2 f g h p a-e g i p a-c f g i p a-d h i p a-b f
   h i p a+c f\textsuperscript 2 g j p a-e f g j p a+b f\textsuperscript 2 h j p a-d f h j p a+2 d e i j p a-c d f i j p a-b
   e f i j p a+c f\textsuperscript 2 g q a+c g q a-e f g q a+b f\textsuperscript 2 h q a+b h q a-d f h q a+2 d e i q a-c d f
   i q a-b e f i q a-c d j q a-b e j q a+b f\textsuperscript 2 k+b i\textsuperscript 2 k+b f\textsuperscript 2 j\textsuperscript 2 k-d f j\textsuperscript 2 k+b k-d f k-g i
   k+f g j k+d i j k-2 b f i j k+d\textsuperscript 2 n+g\textsuperscript 2 n+d\textsuperscript 2 j\textsuperscript 2 n-b d f j\textsuperscript 2 n-b d f n-b g i n-2 d g j
   n+b f g j n+b d i j n+n-f g\textsuperscript 2 p-b d i\textsuperscript 2 p-b d p-f p+d g i p+b f g i p-b f\textsuperscript 2 g j p+d f g j
   p-d\textsuperscript 2 i j p+b d f i j p-b f\textsuperscript 2 g q-b g q+d f g q-d\textsuperscript 2 i q+b d f i q-i q+b d j q+f j
   q)+(-b j\textsuperscript 2 f\textsuperscript 2+a c j\textsuperscript 2 f\textsuperscript 2-b f\textsuperscript 2+a c f\textsuperscript 2+d j\textsuperscript 2 f-a e j\textsuperscript 2 f+d f-a e f-g j f+a h
   j f+2 b i j f-2 a c i j f-b i\textsuperscript 2+a c i\textsuperscript 2-b+a c+g i-a h i-d i j+a e i j)
   (k\textsuperscript 2+m\textsuperscript 2+n\textsuperscript 2+p\textsuperscript 2+q\textsuperscript 2+1))))-(g\textsuperscript 2+h\textsuperscript 2+i\textsuperscript 2+j\textsuperscript 2+1)
   (q\textsuperscript 2+1) (d (-d q\textsuperscript 2 j\textsuperscript 2+a e q\textsuperscript 2 j\textsuperscript 2+b f q\textsuperscript 2 j\textsuperscript 2-a c f q\textsuperscript 2 j\textsuperscript 2-d j\textsuperscript 2+a
   e j\textsuperscript 2+b f j\textsuperscript 2-a c f j\textsuperscript 2+g q\textsuperscript 2 j-a h q\textsuperscript 2 j-b i q\textsuperscript 2 j+a c i q\textsuperscript 2 j+g j-a h j-b i j+a c i j-k
   q j+a m q j+b n q j-a c n q j+2 d p q j-2 a e p q j-2 b f p q j+2 a c f p q j-d p\textsuperscript 2+a e
   p\textsuperscript 2+b f p\textsuperscript 2-a c f p\textsuperscript 2-d+a e+b f-a c f+k p-a m p-b n p+a c n p-g p q+a h p q+b i p q-a c i
   p q) ((a b+c) (b k+c m+n) (-n+f p+i q-f j q) (-c j\textsuperscript 2 f\textsuperscript 2-c p\textsuperscript 2 f\textsuperscript 2-c j\textsuperscript 2
   q\textsuperscript 2 f\textsuperscript 2-c f\textsuperscript 2+2 c j p q f\textsuperscript 2+e j\textsuperscript 2 f+e p\textsuperscript 2 f+e j\textsuperscript 2 q\textsuperscript 2 f-h j q\textsuperscript 2 f+2 c i j q\textsuperscript 2 f+e f-h j
   f+2 c i j f-m p f+2 c n p f+j m q f-2 c j n q f+h p q f-2 c i p q f-2 e j p q f-c i\textsuperscript 2-c
   n\textsuperscript 2-c i\textsuperscript 2 q\textsuperscript 2+h i q\textsuperscript 2-e i j q\textsuperscript 2-c+h i-e i j+m n-e n p-i m q-h n q+2 c i n q+e j n q+e i p
   q)-(b\textsuperscript 2+c\textsuperscript 2+1) (a k+m) (-m+c n+e p-c f p+h q-c i q-e j q+c f j q)
   (j\textsuperscript 2 f\textsuperscript 2+p\textsuperscript 2 f\textsuperscript 2+j\textsuperscript 2 q\textsuperscript 2 f\textsuperscript 2-2 j p q f\textsuperscript 2+f\textsuperscript 2-2 i j q\textsuperscript 2 f-2 i j f-2 n p f+2 j n q f+2
   i p q f+i\textsuperscript 2+n\textsuperscript 2+i\textsuperscript 2 q\textsuperscript 2-2 i n q+1))-(b d+c e+f) (-f q\textsuperscript 2 j\textsuperscript 2-f j\textsuperscript 2+i q\textsuperscript 2
   j+i j-n q j+2 f p q j-f p\textsuperscript 2-f+n p-i p q) (a (b k+c m+n) (-n+f p+i q-f j q)
   ((g k+h m+i n+j p+q) (a i n c\textsuperscript 2-a f j n c\textsuperscript 2-a f i p c\textsuperscript 2+a f\textsuperscript 2 j p c\textsuperscript 2+a f\textsuperscript 2 q
   c\textsuperscript 2+a q c\textsuperscript 2-a i m c+a f j m c+g n c-a h n c-b i n c-d j n c+a e j n c+b f j n c-f g p c+a
   f h p c+a e i p c+b f i p c-b f\textsuperscript 2 j p c+d f j p c-2 a e f j p c-b f\textsuperscript 2 q c-b q c+d f q c-2
   a e f q c-g m+a h m+b i m+d j m-a e j m-b f j m+e g p-a e h p-b e i p+a e\textsuperscript 2 j p-d e j p+b
   e f j p+a e\textsuperscript 2 q+a q-d e q+b e f q)-(d k+e m+f n+p) (-a f j\textsuperscript 2 n c\textsuperscript 2-a f n c\textsuperscript 2+a
   i j n c\textsuperscript 2-a i\textsuperscript 2 p c\textsuperscript 2-a p c\textsuperscript 2+a f i j p c\textsuperscript 2+a f i q c\textsuperscript 2+a j q c\textsuperscript 2+a f j\textsuperscript 2 m c+a f m c-a i
   j m c-d j\textsuperscript 2 n c+a e j\textsuperscript 2 n c+b f j\textsuperscript 2 n c-d n c+a e n c+b f n c+g j n c-a h j n c-b i j n
   c+b i\textsuperscript 2 p c+b p c-g i p c+2 a h i p c-a f h j p c+d i j p c-a e i j p c-b f i j p c-a f h
   q c+d i q c-a e i q c-b f i q c-b j q c+d j\textsuperscript 2 m-a e j\textsuperscript 2 m-b f j\textsuperscript 2 m+d m-a e m-b f m-g j
   m+a h j m+b i j m-a h\textsuperscript 2 p-a p+g h p-b h i p-d h j p+a e h j p+b f h j p-d h q+a e h q+b f
   h q+a j q)+(b k+c m+n) (a j\textsuperscript 2 n e\textsuperscript 2+a n e\textsuperscript 2-a i j p e\textsuperscript 2-a i q e\textsuperscript 2-a f j\textsuperscript 2 m
   e-a f m e+a i j m e-d j\textsuperscript 2 n e+b f j\textsuperscript 2 n e-a c f j\textsuperscript 2 n e-d n e+b f n e-a c f n e+g j n e-2
   a h j n e-b i j n e+a c i j n e+b i\textsuperscript 2 p e-a c i\textsuperscript 2 p e+b p e-a c p e-g i p e+a h i p e+a f
   h j p e+d i j p e-b f i j p e+a c f i j p e+a f h q e+d i q e-b f i q e+a c f i q e-b j q
   e+a c j q e-b f\textsuperscript 2 m+a c f\textsuperscript 2 m-b i\textsuperscript 2 m+a c i\textsuperscript 2 m-b f\textsuperscript 2 j\textsuperscript 2 m+a c f\textsuperscript 2 j\textsuperscript 2 m+d f j\textsuperscript 2 m-b m+a
   c m+d f m+g i m-a h i m-f g j m+a f h j m-d i j m+2 b f i j m-2 a c f i j m+a h\textsuperscript 2 n+a n-g
   h n+b h i n-a c h i n+d h j n-b f h j n+a c f h j n-a f h\textsuperscript 2 p-a f p+f g h p-b f h i p+a c
   f h i p+b f\textsuperscript 2 h j p-a c f\textsuperscript 2 h j p-d f h j p+b f\textsuperscript 2 h q-a c f\textsuperscript 2 h q+b h q-a c h q-d f h q-a
   i q+a f j q)-(a f\textsuperscript 2 c\textsuperscript 2+a i\textsuperscript 2 c\textsuperscript 2+a f\textsuperscript 2 j\textsuperscript 2 c\textsuperscript 2+a c\textsuperscript 2-2 a f i j c\textsuperscript 2-b f\textsuperscript 2 c-b
   i\textsuperscript 2 c-b f\textsuperscript 2 j\textsuperscript 2 c+d f j\textsuperscript 2 c-2 a e f j\textsuperscript 2 c-b c+d f c-2 a e f c+g i c-2 a h i c-f g j c+2 a
   f h j c-d i j c+2 a e i j c+2 b f i j c+a e\textsuperscript 2+a h\textsuperscript 2+a e\textsuperscript 2 j\textsuperscript 2-d e j\textsuperscript 2+b e f j\textsuperscript 2+a-d e+b e
   f-g h+b h i+e g j+d h j-2 a e h j-b f h j-b e i j)
   (k\textsuperscript 2+m\textsuperscript 2+n\textsuperscript 2+p\textsuperscript 2+q\textsuperscript 2+1)-k (-(g\textsuperscript 2+h\textsuperscript 2+i\textsuperscript 2+j\textsuperscript 2+1)
   (-(d\textsuperscript 2+e\textsuperscript 2+f\textsuperscript 2+1) (-m b\textsuperscript 2+c k b+a c m b+a n b-a c\textsuperscript 2 k-a k-m+c
   n)+(b d+c e+f) (b e k-a c e k-a f k-b d m+a c d m-f m+a d n+e n)+(-e b\textsuperscript 2+c d
   b+a c e b+a f b-a c\textsuperscript 2 d-a d-e+c f) (d k+e m+f n+p))+(d g+e h+f i+j) (-(d
   g+e h+f i+j) (-m b\textsuperscript 2+c k b+a c m b+a n b-a c\textsuperscript 2 k-a k-m+c n)+(b d+c e+f) (b h
   k-a c h k-a i k-b g m+a c g m-i m+a g n+h n)+(-h b\textsuperscript 2+c g b+a c h b+a i b-a c\textsuperscript 2 g-a
   g-h+c i) (d k+e m+f n+p))-(b g+c h+i) (-(d g+e h+f i+j) (b e k-a c e k-a
   f k-b d m+a c d m-f m+a d n+e n)+(d\textsuperscript 2+e\textsuperscript 2+f\textsuperscript 2+1) (b h k-a c h k-a i k-b g m+a
   c g m-i m+a g n+h n)+(b e g-a c e g-a f g-b d h+a c d h-f h+a d i+e i) (d k+e m+f
   n+p))+(-(d\textsuperscript 2+e\textsuperscript 2+f\textsuperscript 2+1) (-h b\textsuperscript 2+c g b+a c h b+a i b-a c\textsuperscript 2 g-a
   g-h+c i)+(b d+c e+f) (b e g-a c e g-a f g-b d h+a c d h-f h+a d i+e i)+(-e
   b\textsuperscript 2+c d b+a c e b+a f b-a c\textsuperscript 2 d-a d-e+c f) (d g+e h+f i+j)) (g k+h m+i n+j
   p+q)))-b (a k+m) (-m+c n+e p-c f p+h q-c i q-e j q+c f j q) (-(g k+h m+i
   n+j p+q) (-b j p f\textsuperscript 2+a c j p f\textsuperscript 2-b q f\textsuperscript 2+a c q f\textsuperscript 2+b j n f-a c j n f-g p f+a h p f+b
   i p f-a c i p f+d j p f-a e j p f+d q f-a e q f+g n-a h n-b i n+a c i n-d j n+a e j n-b
   q+a c q)+(d k+e m+f n+p) (b p i\textsuperscript 2-a c p i\textsuperscript 2-b j n i+a c j n i-g p i+a h p i+d
   j p i-a e j p i-b f j p i+a c f j p i+d q i-a e q i-b f q i+a c f q i-d j\textsuperscript 2 n+a e j\textsuperscript 2 n+b
   f j\textsuperscript 2 n-a c f j\textsuperscript 2 n-d n+a e n+b f n-a c f n+g j n-a h j n+b p-a c p-b j q+a c j
   q)-(a k+m) (a j\textsuperscript 2 n e\textsuperscript 2+a n e\textsuperscript 2-a i j p e\textsuperscript 2-a i q e\textsuperscript 2-a f j\textsuperscript 2 m e-a f m e+a i
   j m e-d j\textsuperscript 2 n e+b f j\textsuperscript 2 n e-a c f j\textsuperscript 2 n e-d n e+b f n e-a c f n e+g j n e-2 a h j n e-b i
   j n e+a c i j n e+b i\textsuperscript 2 p e-a c i\textsuperscript 2 p e+b p e-a c p e-g i p e+a h i p e+a f h j p e+d i j
   p e-b f i j p e+a c f i j p e+a f h q e+d i q e-b f i q e+a c f i q e-b j q e+a c j q e-b
   f\textsuperscript 2 m+a c f\textsuperscript 2 m-b i\textsuperscript 2 m+a c i\textsuperscript 2 m-b f\textsuperscript 2 j\textsuperscript 2 m+a c f\textsuperscript 2 j\textsuperscript 2 m+d f j\textsuperscript 2 m-b m+a c m+d f m+g i
   m-a h i m-f g j m+a f h j m-d i j m+2 b f i j m-2 a c f i j m+a h\textsuperscript 2 n+a n-g h n+b h i n-a
   c h i n+d h j n-b f h j n+a c f h j n-a f h\textsuperscript 2 p-a f p+f g h p-b f h i p+a c f h i p+b f\textsuperscript 2
   h j p-a c f\textsuperscript 2 h j p-d f h j p+b f\textsuperscript 2 h q-a c f\textsuperscript 2 h q+b h q-a c h q-d f h q-a i q+a f j
   q)+k (c f\textsuperscript 2 m a\textsuperscript 2+c i\textsuperscript 2 m a\textsuperscript 2+c f\textsuperscript 2 j\textsuperscript 2 m a\textsuperscript 2-e f j\textsuperscript 2 m a\textsuperscript 2+c m a\textsuperscript 2-e f m
   a\textsuperscript 2-h i m a\textsuperscript 2+f h j m a\textsuperscript 2+e i j m a\textsuperscript 2-2 c f i j m a\textsuperscript 2+e\textsuperscript 2 n a\textsuperscript 2+h\textsuperscript 2 n a\textsuperscript 2+e\textsuperscript 2 j\textsuperscript 2 n a\textsuperscript 2-c
   e f j\textsuperscript 2 n a\textsuperscript 2-c e f n a\textsuperscript 2-c h i n a\textsuperscript 2-2 e h j n a\textsuperscript 2+c f h j n a\textsuperscript 2+c e i j n a\textsuperscript 2+n a\textsuperscript 2-f
   h\textsuperscript 2 p a\textsuperscript 2-c e i\textsuperscript 2 p a\textsuperscript 2-c e p a\textsuperscript 2-f p a\textsuperscript 2+e h i p a\textsuperscript 2+c f h i p a\textsuperscript 2-c f\textsuperscript 2 h j p a\textsuperscript 2+e f h
   j p a\textsuperscript 2-e\textsuperscript 2 i j p a\textsuperscript 2+c e f i j p a\textsuperscript 2-c f\textsuperscript 2 h q a\textsuperscript 2-c h q a\textsuperscript 2+e f h q a\textsuperscript 2-e\textsuperscript 2 i q a\textsuperscript 2+c e
   f i q a\textsuperscript 2-i q a\textsuperscript 2+c e j q a\textsuperscript 2+f j q a\textsuperscript 2-c f\textsuperscript 2 k a-c i\textsuperscript 2 k a-c f\textsuperscript 2 j\textsuperscript 2 k a+e f j\textsuperscript 2 k a-c k
   a+e f k a+h i k a-f h j k a-e i j k a+2 c f i j k a-b f\textsuperscript 2 m a-b i\textsuperscript 2 m a-b f\textsuperscript 2 j\textsuperscript 2 m a+d f
   j\textsuperscript 2 m a-b m a+d f m a+g i m a-f g j m a-d i j m a+2 b f i j m a-2 d e j\textsuperscript 2 n a+c d f j\textsuperscript 2 n
   a+b e f j\textsuperscript 2 n a-2 d e n a+c d f n a+b e f n a-2 g h n a+c g i n a+b h i n a+2 e g j n a-c
   f g j n a+2 d h j n a-b f h j n a-c d i j n a-b e i j n a+c d i\textsuperscript 2 p a+b e i\textsuperscript 2 p a+c d p
   a+b e p a+2 f g h p a-e g i p a-c f g i p a-d h i p a-b f h i p a+c f\textsuperscript 2 g j p a-e f g j p
   a+b f\textsuperscript 2 h j p a-d f h j p a+2 d e i j p a-c d f i j p a-b e f i j p a+c f\textsuperscript 2 g q a+c g q
   a-e f g q a+b f\textsuperscript 2 h q a+b h q a-d f h q a+2 d e i q a-c d f i q a-b e f i q a-c d j q a-b
   e j q a+b f\textsuperscript 2 k+b i\textsuperscript 2 k+b f\textsuperscript 2 j\textsuperscript 2 k-d f j\textsuperscript 2 k+b k-d f k-g i k+f g j k+d i j k-2 b f i j
   k+d\textsuperscript 2 n+g\textsuperscript 2 n+d\textsuperscript 2 j\textsuperscript 2 n-b d f j\textsuperscript 2 n-b d f n-b g i n-2 d g j n+b f g j n+b d i j n+n-f g\textsuperscript 2
   p-b d i\textsuperscript 2 p-b d p-f p+d g i p+b f g i p-b f\textsuperscript 2 g j p+d f g j p-d\textsuperscript 2 i j p+b d f i j p-b f\textsuperscript 2
   g q-b g q+d f g q-d\textsuperscript 2 i q+b d f i q-i q+b d j q+f j q)+(-b j\textsuperscript 2 f\textsuperscript 2+a c j\textsuperscript 2
   f\textsuperscript 2-b f\textsuperscript 2+a c f\textsuperscript 2+d j\textsuperscript 2 f-a e j\textsuperscript 2 f+d f-a e f-g j f+a h j f+2 b i j f-2 a c i j f-b i\textsuperscript 2+a
   c i\textsuperscript 2-b+a c+g i-a h i-d i j+a e i j)
   (k\textsuperscript 2+m\textsuperscript 2+n\textsuperscript 2+p\textsuperscript 2+q\textsuperscript 2+1)))+(d k+e m+f n+p) (j q-p) (a
   (b\textsuperscript 2+c\textsuperscript 2+1) (j\textsuperscript 2 f\textsuperscript 2+p\textsuperscript 2 f\textsuperscript 2+j\textsuperscript 2 q\textsuperscript 2 f\textsuperscript 2-2 j p q f\textsuperscript 2+f\textsuperscript 2-2 i j q\textsuperscript 2 f-2 i
   j f-2 n p f+2 j n q f+2 i p q f+i\textsuperscript 2+n\textsuperscript 2+i\textsuperscript 2 q\textsuperscript 2-2 i n q+1) ((g k+h m+i n+j
   p+q) (a i n c\textsuperscript 2-a f j n c\textsuperscript 2-a f i p c\textsuperscript 2+a f\textsuperscript 2 j p c\textsuperscript 2+a f\textsuperscript 2 q c\textsuperscript 2+a q c\textsuperscript 2-a i m c+a
   f j m c+g n c-a h n c-b i n c-d j n c+a e j n c+b f j n c-f g p c+a f h p c+a e i p c+b f
   i p c-b f\textsuperscript 2 j p c+d f j p c-2 a e f j p c-b f\textsuperscript 2 q c-b q c+d f q c-2 a e f q c-g m+a h m+b
   i m+d j m-a e j m-b f j m+e g p-a e h p-b e i p+a e\textsuperscript 2 j p-d e j p+b e f j p+a e\textsuperscript 2 q+a q-d
   e q+b e f q)-(d k+e m+f n+p) (-a f j\textsuperscript 2 n c\textsuperscript 2-a f n c\textsuperscript 2+a i j n c\textsuperscript 2-a i\textsuperscript 2 p
   c\textsuperscript 2-a p c\textsuperscript 2+a f i j p c\textsuperscript 2+a f i q c\textsuperscript 2+a j q c\textsuperscript 2+a f j\textsuperscript 2 m c+a f m c-a i j m c-d j\textsuperscript 2 n c+a
   e j\textsuperscript 2 n c+b f j\textsuperscript 2 n c-d n c+a e n c+b f n c+g j n c-a h j n c-b i j n c+b i\textsuperscript 2 p c+b p c-g
   i p c+2 a h i p c-a f h j p c+d i j p c-a e i j p c-b f i j p c-a f h q c+d i q c-a e i q
   c-b f i q c-b j q c+d j\textsuperscript 2 m-a e j\textsuperscript 2 m-b f j\textsuperscript 2 m+d m-a e m-b f m-g j m+a h j m+b i j m-a
   h\textsuperscript 2 p-a p+g h p-b h i p-d h j p+a e h j p+b f h j p-d h q+a e h q+b f h q+a j q)+(b
   k+c m+n) (a j\textsuperscript 2 n e\textsuperscript 2+a n e\textsuperscript 2-a i j p e\textsuperscript 2-a i q e\textsuperscript 2-a f j\textsuperscript 2 m e-a f m e+a i j m e-d
   j\textsuperscript 2 n e+b f j\textsuperscript 2 n e-a c f j\textsuperscript 2 n e-d n e+b f n e-a c f n e+g j n e-2 a h j n e-b i j n e+a
   c i j n e+b i\textsuperscript 2 p e-a c i\textsuperscript 2 p e+b p e-a c p e-g i p e+a h i p e+a f h j p e+d i j p e-b f
   i j p e+a c f i j p e+a f h q e+d i q e-b f i q e+a c f i q e-b j q e+a c j q e-b f\textsuperscript 2 m+a
   c f\textsuperscript 2 m-b i\textsuperscript 2 m+a c i\textsuperscript 2 m-b f\textsuperscript 2 j\textsuperscript 2 m+a c f\textsuperscript 2 j\textsuperscript 2 m+d f j\textsuperscript 2 m-b m+a c m+d f m+g i m-a h i
   m-f g j m+a f h j m-d i j m+2 b f i j m-2 a c f i j m+a h\textsuperscript 2 n+a n-g h n+b h i n-a c h i
   n+d h j n-b f h j n+a c f h j n-a f h\textsuperscript 2 p-a f p+f g h p-b f h i p+a c f h i p+b f\textsuperscript 2 h j
   p-a c f\textsuperscript 2 h j p-d f h j p+b f\textsuperscript 2 h q-a c f\textsuperscript 2 h q+b h q-a c h q-d f h q-a i q+a f j
   q)-(a f\textsuperscript 2 c\textsuperscript 2+a i\textsuperscript 2 c\textsuperscript 2+a f\textsuperscript 2 j\textsuperscript 2 c\textsuperscript 2+a c\textsuperscript 2-2 a f i j c\textsuperscript 2-b f\textsuperscript 2 c-b i\textsuperscript 2 c-b
   f\textsuperscript 2 j\textsuperscript 2 c+d f j\textsuperscript 2 c-2 a e f j\textsuperscript 2 c-b c+d f c-2 a e f c+g i c-2 a h i c-f g j c+2 a f h j
   c-d i j c+2 a e i j c+2 b f i j c+a e\textsuperscript 2+a h\textsuperscript 2+a e\textsuperscript 2 j\textsuperscript 2-d e j\textsuperscript 2+b e f j\textsuperscript 2+a-d e+b e f-g
   h+b h i+e g j+d h j-2 a e h j-b f h j-b e i j) (k\textsuperscript 2+m\textsuperscript 2+n\textsuperscript 2+p\textsuperscript 2+q\textsuperscript 2+1)-k
   (-(g\textsuperscript 2+h\textsuperscript 2+i\textsuperscript 2+j\textsuperscript 2+1) (-(d\textsuperscript 2+e\textsuperscript 2+f\textsuperscript 2+1) (-m b\textsuperscript 2+c k
   b+a c m b+a n b-a c\textsuperscript 2 k-a k-m+c n)+(b d+c e+f) (b e k-a c e k-a f k-b d m+a c d m-f
   m+a d n+e n)+(-e b\textsuperscript 2+c d b+a c e b+a f b-a c\textsuperscript 2 d-a d-e+c f) (d k+e m+f
   n+p))+(d g+e h+f i+j) (-(d g+e h+f i+j) (-m b\textsuperscript 2+c k b+a c m b+a n b-a c\textsuperscript 2
   k-a k-m+c n)+(b d+c e+f) (b h k-a c h k-a i k-b g m+a c g m-i m+a g n+h n)+(-h
   b\textsuperscript 2+c g b+a c h b+a i b-a c\textsuperscript 2 g-a g-h+c i) (d k+e m+f n+p))-(b g+c h+i)
   (-(d g+e h+f i+j) (b e k-a c e k-a f k-b d m+a c d m-f m+a d n+e
   n)+(d\textsuperscript 2+e\textsuperscript 2+f\textsuperscript 2+1) (b h k-a c h k-a i k-b g m+a c g m-i m+a g n+h n)+(b e g-a
   c e g-a f g-b d h+a c d h-f h+a d i+e i) (d k+e m+f
   n+p))+(-(d\textsuperscript 2+e\textsuperscript 2+f\textsuperscript 2+1) (-h b\textsuperscript 2+c g b+a c h b+a i b-a c\textsuperscript 2 g-a
   g-h+c i)+(b d+c e+f) (b e g-a c e g-a f g-b d h+a c d h-f h+a d i+e i)+(-e
   b\textsuperscript 2+c d b+a c e b+a f b-a c\textsuperscript 2 d-a d-e+c f) (d g+e h+f i+j)) (g k+h m+i n+j
   p+q)))-b (a b+c) (-c j\textsuperscript 2 f\textsuperscript 2-c p\textsuperscript 2 f\textsuperscript 2-c j\textsuperscript 2 q\textsuperscript 2 f\textsuperscript 2-c f\textsuperscript 2+2 c j p q
   f\textsuperscript 2+e j\textsuperscript 2 f+e p\textsuperscript 2 f+e j\textsuperscript 2 q\textsuperscript 2 f-h j q\textsuperscript 2 f+2 c i j q\textsuperscript 2 f+e f-h j f+2 c i j f-m p f+2 c n p
   f+j m q f-2 c j n q f+h p q f-2 c i p q f-2 e j p q f-c i\textsuperscript 2-c n\textsuperscript 2-c i\textsuperscript 2 q\textsuperscript 2+h i q\textsuperscript 2-e i j
   q\textsuperscript 2-c+h i-e i j+m n-e n p-i m q-h n q+2 c i n q+e j n q+e i p q) (-(g k+h m+i
   n+j p+q) (-b j p f\textsuperscript 2+a c j p f\textsuperscript 2-b q f\textsuperscript 2+a c q f\textsuperscript 2+b j n f-a c j n f-g p f+a h p f+b
   i p f-a c i p f+d j p f-a e j p f+d q f-a e q f+g n-a h n-b i n+a c i n-d j n+a e j n-b
   q+a c q)+(d k+e m+f n+p) (b p i\textsuperscript 2-a c p i\textsuperscript 2-b j n i+a c j n i-g p i+a h p i+d
   j p i-a e j p i-b f j p i+a c f j p i+d q i-a e q i-b f q i+a c f q i-d j\textsuperscript 2 n+a e j\textsuperscript 2 n+b
   f j\textsuperscript 2 n-a c f j\textsuperscript 2 n-d n+a e n+b f n-a c f n+g j n-a h j n+b p-a c p-b j q+a c j
   q)-(a k+m) (a j\textsuperscript 2 n e\textsuperscript 2+a n e\textsuperscript 2-a i j p e\textsuperscript 2-a i q e\textsuperscript 2-a f j\textsuperscript 2 m e-a f m e+a i
   j m e-d j\textsuperscript 2 n e+b f j\textsuperscript 2 n e-a c f j\textsuperscript 2 n e-d n e+b f n e-a c f n e+g j n e-2 a h j n e-b i
   j n e+a c i j n e+b i\textsuperscript 2 p e-a c i\textsuperscript 2 p e+b p e-a c p e-g i p e+a h i p e+a f h j p e+d i j
   p e-b f i j p e+a c f i j p e+a f h q e+d i q e-b f i q e+a c f i q e-b j q e+a c j q e-b
   f\textsuperscript 2 m+a c f\textsuperscript 2 m-b i\textsuperscript 2 m+a c i\textsuperscript 2 m-b f\textsuperscript 2 j\textsuperscript 2 m+a c f\textsuperscript 2 j\textsuperscript 2 m+d f j\textsuperscript 2 m-b m+a c m+d f m+g i
   m-a h i m-f g j m+a f h j m-d i j m+2 b f i j m-2 a c f i j m+a h\textsuperscript 2 n+a n-g h n+b h i n-a
   c h i n+d h j n-b f h j n+a c f h j n-a f h\textsuperscript 2 p-a f p+f g h p-b f h i p+a c f h i p+b f\textsuperscript 2
   h j p-a c f\textsuperscript 2 h j p-d f h j p+b f\textsuperscript 2 h q-a c f\textsuperscript 2 h q+b h q-a c h q-d f h q-a i q+a f j
   q)+k (c f\textsuperscript 2 m a\textsuperscript 2+c i\textsuperscript 2 m a\textsuperscript 2+c f\textsuperscript 2 j\textsuperscript 2 m a\textsuperscript 2-e f j\textsuperscript 2 m a\textsuperscript 2+c m a\textsuperscript 2-e f m
   a\textsuperscript 2-h i m a\textsuperscript 2+f h j m a\textsuperscript 2+e i j m a\textsuperscript 2-2 c f i j m a\textsuperscript 2+e\textsuperscript 2 n a\textsuperscript 2+h\textsuperscript 2 n a\textsuperscript 2+e\textsuperscript 2 j\textsuperscript 2 n a\textsuperscript 2-c
   e f j\textsuperscript 2 n a\textsuperscript 2-c e f n a\textsuperscript 2-c h i n a\textsuperscript 2-2 e h j n a\textsuperscript 2+c f h j n a\textsuperscript 2+c e i j n a\textsuperscript 2+n a\textsuperscript 2-f
   h\textsuperscript 2 p a\textsuperscript 2-c e i\textsuperscript 2 p a\textsuperscript 2-c e p a\textsuperscript 2-f p a\textsuperscript 2+e h i p a\textsuperscript 2+c f h i p a\textsuperscript 2-c f\textsuperscript 2 h j p a\textsuperscript 2+e f h
   j p a\textsuperscript 2-e\textsuperscript 2 i j p a\textsuperscript 2+c e f i j p a\textsuperscript 2-c f\textsuperscript 2 h q a\textsuperscript 2-c h q a\textsuperscript 2+e f h q a\textsuperscript 2-e\textsuperscript 2 i q a\textsuperscript 2+c e
   f i q a\textsuperscript 2-i q a\textsuperscript 2+c e j q a\textsuperscript 2+f j q a\textsuperscript 2-c f\textsuperscript 2 k a-c i\textsuperscript 2 k a-c f\textsuperscript 2 j\textsuperscript 2 k a+e f j\textsuperscript 2 k a-c k
   a+e f k a+h i k a-f h j k a-e i j k a+2 c f i j k a-b f\textsuperscript 2 m a-b i\textsuperscript 2 m a-b f\textsuperscript 2 j\textsuperscript 2 m a+d f
   j\textsuperscript 2 m a-b m a+d f m a+g i m a-f g j m a-d i j m a+2 b f i j m a-2 d e j\textsuperscript 2 n a+c d f j\textsuperscript 2 n
   a+b e f j\textsuperscript 2 n a-2 d e n a+c d f n a+b e f n a-2 g h n a+c g i n a+b h i n a+2 e g j n a-c
   f g j n a+2 d h j n a-b f h j n a-c d i j n a-b e i j n a+c d i\textsuperscript 2 p a+b e i\textsuperscript 2 p a+c d p
   a+b e p a+2 f g h p a-e g i p a-c f g i p a-d h i p a-b f h i p a+c f\textsuperscript 2 g j p a-e f g j p
   a+b f\textsuperscript 2 h j p a-d f h j p a+2 d e i j p a-c d f i j p a-b e f i j p a+c f\textsuperscript 2 g q a+c g q
   a-e f g q a+b f\textsuperscript 2 h q a+b h q a-d f h q a+2 d e i q a-c d f i q a-b e f i q a-c d j q a-b
   e j q a+b f\textsuperscript 2 k+b i\textsuperscript 2 k+b f\textsuperscript 2 j\textsuperscript 2 k-d f j\textsuperscript 2 k+b k-d f k-g i k+f g j k+d i j k-2 b f i j
   k+d\textsuperscript 2 n+g\textsuperscript 2 n+d\textsuperscript 2 j\textsuperscript 2 n-b d f j\textsuperscript 2 n-b d f n-b g i n-2 d g j n+b f g j n+b d i j n+n-f g\textsuperscript 2
   p-b d i\textsuperscript 2 p-b d p-f p+d g i p+b f g i p-b f\textsuperscript 2 g j p+d f g j p-d\textsuperscript 2 i j p+b d f i j p-b f\textsuperscript 2
   g q-b g q+d f g q-d\textsuperscript 2 i q+b d f i q-i q+b d j q+f j q)+(-b j\textsuperscript 2 f\textsuperscript 2+a c j\textsuperscript 2
   f\textsuperscript 2-b f\textsuperscript 2+a c f\textsuperscript 2+d j\textsuperscript 2 f-a e j\textsuperscript 2 f+d f-a e f-g j f+a h j f+2 b i j f-2 a c i j f-b i\textsuperscript 2+a
   c i\textsuperscript 2-b+a c+g i-a h i-d i j+a e i j)
   (k\textsuperscript 2+m\textsuperscript 2+n\textsuperscript 2+p\textsuperscript 2+q\textsuperscript 2+1))))-q (g k+h m+i n+j p+q) (d
   (-d q\textsuperscript 2 j\textsuperscript 2+a e q\textsuperscript 2 j\textsuperscript 2+b f q\textsuperscript 2 j\textsuperscript 2-a c f q\textsuperscript 2 j\textsuperscript 2-d j\textsuperscript 2+a e j\textsuperscript 2+b f j\textsuperscript 2-a c f j\textsuperscript 2+g
   q\textsuperscript 2 j-a h q\textsuperscript 2 j-b i q\textsuperscript 2 j+a c i q\textsuperscript 2 j+g j-a h j-b i j+a c i j-k q j+a m q j+b n q j-a c n
   q j+2 d p q j-2 a e p q j-2 b f p q j+2 a c f p q j-d p\textsuperscript 2+a e p\textsuperscript 2+b f p\textsuperscript 2-a c f p\textsuperscript 2-d+a
   e+b f-a c f+k p-a m p-b n p+a c n p-g p q+a h p q+b i p q-a c i p q) ((a b+c)
   (b g+c h+i) (-i q\textsuperscript 2+f j q\textsuperscript 2+n q-f p q-i+f j) (-c j\textsuperscript 2 f\textsuperscript 2-c p\textsuperscript 2 f\textsuperscript 2-c j\textsuperscript 2
   q\textsuperscript 2 f\textsuperscript 2-c f\textsuperscript 2+2 c j p q f\textsuperscript 2+e j\textsuperscript 2 f+e p\textsuperscript 2 f+e j\textsuperscript 2 q\textsuperscript 2 f-h j q\textsuperscript 2 f+2 c i j q\textsuperscript 2 f+e f-h j
   f+2 c i j f-m p f+2 c n p f+j m q f-2 c j n q f+h p q f-2 c i p q f-2 e j p q f-c i\textsuperscript 2-c
   n\textsuperscript 2-c i\textsuperscript 2 q\textsuperscript 2+h i q\textsuperscript 2-e i j q\textsuperscript 2-c+h i-e i j+m n-e n p-i m q-h n q+2 c i n q+e j n q+e i p
   q)-(b\textsuperscript 2+c\textsuperscript 2+1) (a g+h) (-h q\textsuperscript 2+c i q\textsuperscript 2+e j q\textsuperscript 2-c f j q\textsuperscript 2+m q-c n
   q-e p q+c f p q-h+c i+e j-c f j) (j\textsuperscript 2 f\textsuperscript 2+p\textsuperscript 2 f\textsuperscript 2+j\textsuperscript 2 q\textsuperscript 2 f\textsuperscript 2-2 j p q
   f\textsuperscript 2+f\textsuperscript 2-2 i j q\textsuperscript 2 f-2 i j f-2 n p f+2 j n q f+2 i p q f+i\textsuperscript 2+n\textsuperscript 2+i\textsuperscript 2 q\textsuperscript 2-2 i n
   q+1))-(b d+c e+f) (-f q\textsuperscript 2 j\textsuperscript 2-f j\textsuperscript 2+i q\textsuperscript 2 j+i j-n q j+2 f p q j-f
   p\textsuperscript 2-f+n p-i p q) (a (b g+c h+i) (-i q\textsuperscript 2+f j q\textsuperscript 2+n q-f p q-i+f j)
   ((g k+h m+i n+j p+q) (a i n c\textsuperscript 2-a f j n c\textsuperscript 2-a f i p c\textsuperscript 2+a f\textsuperscript 2 j p c\textsuperscript 2+a f\textsuperscript 2 q
   c\textsuperscript 2+a q c\textsuperscript 2-a i m c+a f j m c+g n c-a h n c-b i n c-d j n c+a e j n c+b f j n c-f g p c+a
   f h p c+a e i p c+b f i p c-b f\textsuperscript 2 j p c+d f j p c-2 a e f j p c-b f\textsuperscript 2 q c-b q c+d f q c-2
   a e f q c-g m+a h m+b i m+d j m-a e j m-b f j m+e g p-a e h p-b e i p+a e\textsuperscript 2 j p-d e j p+b
   e f j p+a e\textsuperscript 2 q+a q-d e q+b e f q)-(d k+e m+f n+p) (-a f j\textsuperscript 2 n c\textsuperscript 2-a f n c\textsuperscript 2+a
   i j n c\textsuperscript 2-a i\textsuperscript 2 p c\textsuperscript 2-a p c\textsuperscript 2+a f i j p c\textsuperscript 2+a f i q c\textsuperscript 2+a j q c\textsuperscript 2+a f j\textsuperscript 2 m c+a f m c-a i
   j m c-d j\textsuperscript 2 n c+a e j\textsuperscript 2 n c+b f j\textsuperscript 2 n c-d n c+a e n c+b f n c+g j n c-a h j n c-b i j n
   c+b i\textsuperscript 2 p c+b p c-g i p c+2 a h i p c-a f h j p c+d i j p c-a e i j p c-b f i j p c-a f h
   q c+d i q c-a e i q c-b f i q c-b j q c+d j\textsuperscript 2 m-a e j\textsuperscript 2 m-b f j\textsuperscript 2 m+d m-a e m-b f m-g j
   m+a h j m+b i j m-a h\textsuperscript 2 p-a p+g h p-b h i p-d h j p+a e h j p+b f h j p-d h q+a e h q+b f
   h q+a j q)+(b k+c m+n) (a j\textsuperscript 2 n e\textsuperscript 2+a n e\textsuperscript 2-a i j p e\textsuperscript 2-a i q e\textsuperscript 2-a f j\textsuperscript 2 m
   e-a f m e+a i j m e-d j\textsuperscript 2 n e+b f j\textsuperscript 2 n e-a c f j\textsuperscript 2 n e-d n e+b f n e-a c f n e+g j n e-2
   a h j n e-b i j n e+a c i j n e+b i\textsuperscript 2 p e-a c i\textsuperscript 2 p e+b p e-a c p e-g i p e+a h i p e+a f
   h j p e+d i j p e-b f i j p e+a c f i j p e+a f h q e+d i q e-b f i q e+a c f i q e-b j q
   e+a c j q e-b f\textsuperscript 2 m+a c f\textsuperscript 2 m-b i\textsuperscript 2 m+a c i\textsuperscript 2 m-b f\textsuperscript 2 j\textsuperscript 2 m+a c f\textsuperscript 2 j\textsuperscript 2 m+d f j\textsuperscript 2 m-b m+a
   c m+d f m+g i m-a h i m-f g j m+a f h j m-d i j m+2 b f i j m-2 a c f i j m+a h\textsuperscript 2 n+a n-g
   h n+b h i n-a c h i n+d h j n-b f h j n+a c f h j n-a f h\textsuperscript 2 p-a f p+f g h p-b f h i p+a c
   f h i p+b f\textsuperscript 2 h j p-a c f\textsuperscript 2 h j p-d f h j p+b f\textsuperscript 2 h q-a c f\textsuperscript 2 h q+b h q-a c h q-d f h q-a
   i q+a f j q)-(a f\textsuperscript 2 c\textsuperscript 2+a i\textsuperscript 2 c\textsuperscript 2+a f\textsuperscript 2 j\textsuperscript 2 c\textsuperscript 2+a c\textsuperscript 2-2 a f i j c\textsuperscript 2-b f\textsuperscript 2 c-b
   i\textsuperscript 2 c-b f\textsuperscript 2 j\textsuperscript 2 c+d f j\textsuperscript 2 c-2 a e f j\textsuperscript 2 c-b c+d f c-2 a e f c+g i c-2 a h i c-f g j c+2 a
   f h j c-d i j c+2 a e i j c+2 b f i j c+a e\textsuperscript 2+a h\textsuperscript 2+a e\textsuperscript 2 j\textsuperscript 2-d e j\textsuperscript 2+b e f j\textsuperscript 2+a-d e+b e
   f-g h+b h i+e g j+d h j-2 a e h j-b f h j-b e i j)
   (k\textsuperscript 2+m\textsuperscript 2+n\textsuperscript 2+p\textsuperscript 2+q\textsuperscript 2+1)-k (-(g\textsuperscript 2+h\textsuperscript 2+i\textsuperscript 2+j\textsuperscript 2+1)
   (-(d\textsuperscript 2+e\textsuperscript 2+f\textsuperscript 2+1) (-m b\textsuperscript 2+c k b+a c m b+a n b-a c\textsuperscript 2 k-a k-m+c
   n)+(b d+c e+f) (b e k-a c e k-a f k-b d m+a c d m-f m+a d n+e n)+(-e b\textsuperscript 2+c d
   b+a c e b+a f b-a c\textsuperscript 2 d-a d-e+c f) (d k+e m+f n+p))+(d g+e h+f i+j) (-(d
   g+e h+f i+j) (-m b\textsuperscript 2+c k b+a c m b+a n b-a c\textsuperscript 2 k-a k-m+c n)+(b d+c e+f) (b h
   k-a c h k-a i k-b g m+a c g m-i m+a g n+h n)+(-h b\textsuperscript 2+c g b+a c h b+a i b-a c\textsuperscript 2 g-a
   g-h+c i) (d k+e m+f n+p))-(b g+c h+i) (-(d g+e h+f i+j) (b e k-a c e k-a
   f k-b d m+a c d m-f m+a d n+e n)+(d\textsuperscript 2+e\textsuperscript 2+f\textsuperscript 2+1) (b h k-a c h k-a i k-b g m+a
   c g m-i m+a g n+h n)+(b e g-a c e g-a f g-b d h+a c d h-f h+a d i+e i) (d k+e m+f
   n+p))+(-(d\textsuperscript 2+e\textsuperscript 2+f\textsuperscript 2+1) (-h b\textsuperscript 2+c g b+a c h b+a i b-a c\textsuperscript 2 g-a
   g-h+c i)+(b d+c e+f) (b e g-a c e g-a f g-b d h+a c d h-f h+a d i+e i)+(-e
   b\textsuperscript 2+c d b+a c e b+a f b-a c\textsuperscript 2 d-a d-e+c f) (d g+e h+f i+j)) (g k+h m+i n+j
   p+q)))-b (a g+h) (-h q\textsuperscript 2+c i q\textsuperscript 2+e j q\textsuperscript 2-c f j q\textsuperscript 2+m q-c n q-e p q+c f p
   q-h+c i+e j-c f j) (-(g k+h m+i n+j p+q) (-b j p f\textsuperscript 2+a c j p f\textsuperscript 2-b q
   f\textsuperscript 2+a c q f\textsuperscript 2+b j n f-a c j n f-g p f+a h p f+b i p f-a c i p f+d j p f-a e j p f+d q f-a
   e q f+g n-a h n-b i n+a c i n-d j n+a e j n-b q+a c q)+(d k+e m+f n+p) (b p
   i\textsuperscript 2-a c p i\textsuperscript 2-b j n i+a c j n i-g p i+a h p i+d j p i-a e j p i-b f j p i+a c f j p i+d q
   i-a e q i-b f q i+a c f q i-d j\textsuperscript 2 n+a e j\textsuperscript 2 n+b f j\textsuperscript 2 n-a c f j\textsuperscript 2 n-d n+a e n+b f n-a c f
   n+g j n-a h j n+b p-a c p-b j q+a c j q)-(a k+m) (a j\textsuperscript 2 n e\textsuperscript 2+a n e\textsuperscript 2-a i j p
   e\textsuperscript 2-a i q e\textsuperscript 2-a f j\textsuperscript 2 m e-a f m e+a i j m e-d j\textsuperscript 2 n e+b f j\textsuperscript 2 n e-a c f j\textsuperscript 2 n e-d n e+b f
   n e-a c f n e+g j n e-2 a h j n e-b i j n e+a c i j n e+b i\textsuperscript 2 p e-a c i\textsuperscript 2 p e+b p e-a c p
   e-g i p e+a h i p e+a f h j p e+d i j p e-b f i j p e+a c f i j p e+a f h q e+d i q e-b f
   i q e+a c f i q e-b j q e+a c j q e-b f\textsuperscript 2 m+a c f\textsuperscript 2 m-b i\textsuperscript 2 m+a c i\textsuperscript 2 m-b f\textsuperscript 2 j\textsuperscript 2 m+a c
   f\textsuperscript 2 j\textsuperscript 2 m+d f j\textsuperscript 2 m-b m+a c m+d f m+g i m-a h i m-f g j m+a f h j m-d i j m+2 b f i j m-2
   a c f i j m+a h\textsuperscript 2 n+a n-g h n+b h i n-a c h i n+d h j n-b f h j n+a c f h j n-a f h\textsuperscript 2 p-a
   f p+f g h p-b f h i p+a c f h i p+b f\textsuperscript 2 h j p-a c f\textsuperscript 2 h j p-d f h j p+b f\textsuperscript 2 h q-a c f\textsuperscript 2 h
   q+b h q-a c h q-d f h q-a i q+a f j q)+k (c f\textsuperscript 2 m a\textsuperscript 2+c i\textsuperscript 2 m a\textsuperscript 2+c f\textsuperscript 2 j\textsuperscript 2 m
   a\textsuperscript 2-e f j\textsuperscript 2 m a\textsuperscript 2+c m a\textsuperscript 2-e f m a\textsuperscript 2-h i m a\textsuperscript 2+f h j m a\textsuperscript 2+e i j m a\textsuperscript 2-2 c f i j m a\textsuperscript 2+e\textsuperscript 2
   n a\textsuperscript 2+h\textsuperscript 2 n a\textsuperscript 2+e\textsuperscript 2 j\textsuperscript 2 n a\textsuperscript 2-c e f j\textsuperscript 2 n a\textsuperscript 2-c e f n a\textsuperscript 2-c h i n a\textsuperscript 2-2 e h j n a\textsuperscript 2+c f h
   j n a\textsuperscript 2+c e i j n a\textsuperscript 2+n a\textsuperscript 2-f h\textsuperscript 2 p a\textsuperscript 2-c e i\textsuperscript 2 p a\textsuperscript 2-c e p a\textsuperscript 2-f p a\textsuperscript 2+e h i p a\textsuperscript 2+c f h
   i p a\textsuperscript 2-c f\textsuperscript 2 h j p a\textsuperscript 2+e f h j p a\textsuperscript 2-e\textsuperscript 2 i j p a\textsuperscript 2+c e f i j p a\textsuperscript 2-c f\textsuperscript 2 h q a\textsuperscript 2-c h q
   a\textsuperscript 2+e f h q a\textsuperscript 2-e\textsuperscript 2 i q a\textsuperscript 2+c e f i q a\textsuperscript 2-i q a\textsuperscript 2+c e j q a\textsuperscript 2+f j q a\textsuperscript 2-c f\textsuperscript 2 k a-c i\textsuperscript 2 k
   a-c f\textsuperscript 2 j\textsuperscript 2 k a+e f j\textsuperscript 2 k a-c k a+e f k a+h i k a-f h j k a-e i j k a+2 c f i j k a-b f\textsuperscript 2
   m a-b i\textsuperscript 2 m a-b f\textsuperscript 2 j\textsuperscript 2 m a+d f j\textsuperscript 2 m a-b m a+d f m a+g i m a-f g j m a-d i j m a+2 b f i
   j m a-2 d e j\textsuperscript 2 n a+c d f j\textsuperscript 2 n a+b e f j\textsuperscript 2 n a-2 d e n a+c d f n a+b e f n a-2 g h n a+c
   g i n a+b h i n a+2 e g j n a-c f g j n a+2 d h j n a-b f h j n a-c d i j n a-b e i j n
   a+c d i\textsuperscript 2 p a+b e i\textsuperscript 2 p a+c d p a+b e p a+2 f g h p a-e g i p a-c f g i p a-d h i p a-b f
   h i p a+c f\textsuperscript 2 g j p a-e f g j p a+b f\textsuperscript 2 h j p a-d f h j p a+2 d e i j p a-c d f i j p a-b
   e f i j p a+c f\textsuperscript 2 g q a+c g q a-e f g q a+b f\textsuperscript 2 h q a+b h q a-d f h q a+2 d e i q a-c d f
   i q a-b e f i q a-c d j q a-b e j q a+b f\textsuperscript 2 k+b i\textsuperscript 2 k+b f\textsuperscript 2 j\textsuperscript 2 k-d f j\textsuperscript 2 k+b k-d f k-g i
   k+f g j k+d i j k-2 b f i j k+d\textsuperscript 2 n+g\textsuperscript 2 n+d\textsuperscript 2 j\textsuperscript 2 n-b d f j\textsuperscript 2 n-b d f n-b g i n-2 d g j
   n+b f g j n+b d i j n+n-f g\textsuperscript 2 p-b d i\textsuperscript 2 p-b d p-f p+d g i p+b f g i p-b f\textsuperscript 2 g j p+d f g j
   p-d\textsuperscript 2 i j p+b d f i j p-b f\textsuperscript 2 g q-b g q+d f g q-d\textsuperscript 2 i q+b d f i q-i q+b d j q+f j
   q)+(-b j\textsuperscript 2 f\textsuperscript 2+a c j\textsuperscript 2 f\textsuperscript 2-b f\textsuperscript 2+a c f\textsuperscript 2+d j\textsuperscript 2 f-a e j\textsuperscript 2 f+d f-a e f-g j f+a h
   j f+2 b i j f-2 a c i j f-b i\textsuperscript 2+a c i\textsuperscript 2-b+a c+g i-a h i-d i j+a e i j)
   (k\textsuperscript 2+m\textsuperscript 2+n\textsuperscript 2+p\textsuperscript 2+q\textsuperscript 2+1)))+(d g+e h+f i+j) (-j q\textsuperscript 2+p
   q-j) (a (b\textsuperscript 2+c\textsuperscript 2+1) (j\textsuperscript 2 f\textsuperscript 2+p\textsuperscript 2 f\textsuperscript 2+j\textsuperscript 2 q\textsuperscript 2 f\textsuperscript 2-2 j p q
   f\textsuperscript 2+f\textsuperscript 2-2 i j q\textsuperscript 2 f-2 i j f-2 n p f+2 j n q f+2 i p q f+i\textsuperscript 2+n\textsuperscript 2+i\textsuperscript 2 q\textsuperscript 2-2 i n q+1)
   ((g k+h m+i n+j p+q) (a i n c\textsuperscript 2-a f j n c\textsuperscript 2-a f i p c\textsuperscript 2+a f\textsuperscript 2 j p c\textsuperscript 2+a f\textsuperscript 2 q
   c\textsuperscript 2+a q c\textsuperscript 2-a i m c+a f j m c+g n c-a h n c-b i n c-d j n c+a e j n c+b f j n c-f g p c+a
   f h p c+a e i p c+b f i p c-b f\textsuperscript 2 j p c+d f j p c-2 a e f j p c-b f\textsuperscript 2 q c-b q c+d f q c-2
   a e f q c-g m+a h m+b i m+d j m-a e j m-b f j m+e g p-a e h p-b e i p+a e\textsuperscript 2 j p-d e j p+b
   e f j p+a e\textsuperscript 2 q+a q-d e q+b e f q)-(d k+e m+f n+p) (-a f j\textsuperscript 2 n c\textsuperscript 2-a f n c\textsuperscript 2+a
   i j n c\textsuperscript 2-a i\textsuperscript 2 p c\textsuperscript 2-a p c\textsuperscript 2+a f i j p c\textsuperscript 2+a f i q c\textsuperscript 2+a j q c\textsuperscript 2+a f j\textsuperscript 2 m c+a f m c-a i
   j m c-d j\textsuperscript 2 n c+a e j\textsuperscript 2 n c+b f j\textsuperscript 2 n c-d n c+a e n c+b f n c+g j n c-a h j n c-b i j n
   c+b i\textsuperscript 2 p c+b p c-g i p c+2 a h i p c-a f h j p c+d i j p c-a e i j p c-b f i j p c-a f h
   q c+d i q c-a e i q c-b f i q c-b j q c+d j\textsuperscript 2 m-a e j\textsuperscript 2 m-b f j\textsuperscript 2 m+d m-a e m-b f m-g j
   m+a h j m+b i j m-a h\textsuperscript 2 p-a p+g h p-b h i p-d h j p+a e h j p+b f h j p-d h q+a e h q+b f
   h q+a j q)+(b k+c m+n) (a j\textsuperscript 2 n e\textsuperscript 2+a n e\textsuperscript 2-a i j p e\textsuperscript 2-a i q e\textsuperscript 2-a f j\textsuperscript 2 m
   e-a f m e+a i j m e-d j\textsuperscript 2 n e+b f j\textsuperscript 2 n e-a c f j\textsuperscript 2 n e-d n e+b f n e-a c f n e+g j n e-2
   a h j n e-b i j n e+a c i j n e+b i\textsuperscript 2 p e-a c i\textsuperscript 2 p e+b p e-a c p e-g i p e+a h i p e+a f
   h j p e+d i j p e-b f i j p e+a c f i j p e+a f h q e+d i q e-b f i q e+a c f i q e-b j q
   e+a c j q e-b f\textsuperscript 2 m+a c f\textsuperscript 2 m-b i\textsuperscript 2 m+a c i\textsuperscript 2 m-b f\textsuperscript 2 j\textsuperscript 2 m+a c f\textsuperscript 2 j\textsuperscript 2 m+d f j\textsuperscript 2 m-b m+a
   c m+d f m+g i m-a h i m-f g j m+a f h j m-d i j m+2 b f i j m-2 a c f i j m+a h\textsuperscript 2 n+a n-g
   h n+b h i n-a c h i n+d h j n-b f h j n+a c f h j n-a f h\textsuperscript 2 p-a f p+f g h p-b f h i p+a c
   f h i p+b f\textsuperscript 2 h j p-a c f\textsuperscript 2 h j p-d f h j p+b f\textsuperscript 2 h q-a c f\textsuperscript 2 h q+b h q-a c h q-d f h q-a
   i q+a f j q)-(a f\textsuperscript 2 c\textsuperscript 2+a i\textsuperscript 2 c\textsuperscript 2+a f\textsuperscript 2 j\textsuperscript 2 c\textsuperscript 2+a c\textsuperscript 2-2 a f i j c\textsuperscript 2-b f\textsuperscript 2 c-b
   i\textsuperscript 2 c-b f\textsuperscript 2 j\textsuperscript 2 c+d f j\textsuperscript 2 c-2 a e f j\textsuperscript 2 c-b c+d f c-2 a e f c+g i c-2 a h i c-f g j c+2 a
   f h j c-d i j c+2 a e i j c+2 b f i j c+a e\textsuperscript 2+a h\textsuperscript 2+a e\textsuperscript 2 j\textsuperscript 2-d e j\textsuperscript 2+b e f j\textsuperscript 2+a-d e+b e
   f-g h+b h i+e g j+d h j-2 a e h j-b f h j-b e i j)
   (k\textsuperscript 2+m\textsuperscript 2+n\textsuperscript 2+p\textsuperscript 2+q\textsuperscript 2+1)-k (-(g\textsuperscript 2+h\textsuperscript 2+i\textsuperscript 2+j\textsuperscript 2+1)
   (-(d\textsuperscript 2+e\textsuperscript 2+f\textsuperscript 2+1) (-m b\textsuperscript 2+c k b+a c m b+a n b-a c\textsuperscript 2 k-a k-m+c
   n)+(b d+c e+f) (b e k-a c e k-a f k-b d m+a c d m-f m+a d n+e n)+(-e b\textsuperscript 2+c d
   b+a c e b+a f b-a c\textsuperscript 2 d-a d-e+c f) (d k+e m+f n+p))+(d g+e h+f i+j) (-(d
   g+e h+f i+j) (-m b\textsuperscript 2+c k b+a c m b+a n b-a c\textsuperscript 2 k-a k-m+c n)+(b d+c e+f) (b h
   k-a c h k-a i k-b g m+a c g m-i m+a g n+h n)+(-h b\textsuperscript 2+c g b+a c h b+a i b-a c\textsuperscript 2 g-a
   g-h+c i) (d k+e m+f n+p))-(b g+c h+i) (-(d g+e h+f i+j) (b e k-a c e k-a
   f k-b d m+a c d m-f m+a d n+e n)+(d\textsuperscript 2+e\textsuperscript 2+f\textsuperscript 2+1) (b h k-a c h k-a i k-b g m+a
   c g m-i m+a g n+h n)+(b e g-a c e g-a f g-b d h+a c d h-f h+a d i+e i) (d k+e m+f
   n+p))+(-(d\textsuperscript 2+e\textsuperscript 2+f\textsuperscript 2+1) (-h b\textsuperscript 2+c g b+a c h b+a i b-a c\textsuperscript 2 g-a
   g-h+c i)+(b d+c e+f) (b e g-a c e g-a f g-b d h+a c d h-f h+a d i+e i)+(-e
   b\textsuperscript 2+c d b+a c e b+a f b-a c\textsuperscript 2 d-a d-e+c f) (d g+e h+f i+j)) (g k+h m+i n+j
   p+q)))-b (a b+c) (-c j\textsuperscript 2 f\textsuperscript 2-c p\textsuperscript 2 f\textsuperscript 2-c j\textsuperscript 2 q\textsuperscript 2 f\textsuperscript 2-c f\textsuperscript 2+2 c j p q
   f\textsuperscript 2+e j\textsuperscript 2 f+e p\textsuperscript 2 f+e j\textsuperscript 2 q\textsuperscript 2 f-h j q\textsuperscript 2 f+2 c i j q\textsuperscript 2 f+e f-h j f+2 c i j f-m p f+2 c n p
   f+j m q f-2 c j n q f+h p q f-2 c i p q f-2 e j p q f-c i\textsuperscript 2-c n\textsuperscript 2-c i\textsuperscript 2 q\textsuperscript 2+h i q\textsuperscript 2-e i j
   q\textsuperscript 2-c+h i-e i j+m n-e n p-i m q-h n q+2 c i n q+e j n q+e i p q) (-(g k+h m+i
   n+j p+q) (-b j p f\textsuperscript 2+a c j p f\textsuperscript 2-b q f\textsuperscript 2+a c q f\textsuperscript 2+b j n f-a c j n f-g p f+a h p f+b
   i p f-a c i p f+d j p f-a e j p f+d q f-a e q f+g n-a h n-b i n+a c i n-d j n+a e j n-b
   q+a c q)+(d k+e m+f n+p) (b p i\textsuperscript 2-a c p i\textsuperscript 2-b j n i+a c j n i-g p i+a h p i+d
   j p i-a e j p i-b f j p i+a c f j p i+d q i-a e q i-b f q i+a c f q i-d j\textsuperscript 2 n+a e j\textsuperscript 2 n+b
   f j\textsuperscript 2 n-a c f j\textsuperscript 2 n-d n+a e n+b f n-a c f n+g j n-a h j n+b p-a c p-b j q+a c j
   q)-(a k+m) (a j\textsuperscript 2 n e\textsuperscript 2+a n e\textsuperscript 2-a i j p e\textsuperscript 2-a i q e\textsuperscript 2-a f j\textsuperscript 2 m e-a f m e+a i
   j m e-d j\textsuperscript 2 n e+b f j\textsuperscript 2 n e-a c f j\textsuperscript 2 n e-d n e+b f n e-a c f n e+g j n e-2 a h j n e-b i
   j n e+a c i j n e+b i\textsuperscript 2 p e-a c i\textsuperscript 2 p e+b p e-a c p e-g i p e+a h i p e+a f h j p e+d i j
   p e-b f i j p e+a c f i j p e+a f h q e+d i q e-b f i q e+a c f i q e-b j q e+a c j q e-b
   f\textsuperscript 2 m+a c f\textsuperscript 2 m-b i\textsuperscript 2 m+a c i\textsuperscript 2 m-b f\textsuperscript 2 j\textsuperscript 2 m+a c f\textsuperscript 2 j\textsuperscript 2 m+d f j\textsuperscript 2 m-b m+a c m+d f m+g i
   m-a h i m-f g j m+a f h j m-d i j m+2 b f i j m-2 a c f i j m+a h\textsuperscript 2 n+a n-g h n+b h i n-a
   c h i n+d h j n-b f h j n+a c f h j n-a f h\textsuperscript 2 p-a f p+f g h p-b f h i p+a c f h i p+b f\textsuperscript 2
   h j p-a c f\textsuperscript 2 h j p-d f h j p+b f\textsuperscript 2 h q-a c f\textsuperscript 2 h q+b h q-a c h q-d f h q-a i q+a f j
   q)+k (c f\textsuperscript 2 m a\textsuperscript 2+c i\textsuperscript 2 m a\textsuperscript 2+c f\textsuperscript 2 j\textsuperscript 2 m a\textsuperscript 2-e f j\textsuperscript 2 m a\textsuperscript 2+c m a\textsuperscript 2-e f m
   a\textsuperscript 2-h i m a\textsuperscript 2+f h j m a\textsuperscript 2+e i j m a\textsuperscript 2-2 c f i j m a\textsuperscript 2+e\textsuperscript 2 n a\textsuperscript 2+h\textsuperscript 2 n a\textsuperscript 2+e\textsuperscript 2 j\textsuperscript 2 n a\textsuperscript 2-c
   e f j\textsuperscript 2 n a\textsuperscript 2-c e f n a\textsuperscript 2-c h i n a\textsuperscript 2-2 e h j n a\textsuperscript 2+c f h j n a\textsuperscript 2+c e i j n a\textsuperscript 2+n a\textsuperscript 2-f
   h\textsuperscript 2 p a\textsuperscript 2-c e i\textsuperscript 2 p a\textsuperscript 2-c e p a\textsuperscript 2-f p a\textsuperscript 2+e h i p a\textsuperscript 2+c f h i p a\textsuperscript 2-c f\textsuperscript 2 h j p a\textsuperscript 2+e f h
   j p a\textsuperscript 2-e\textsuperscript 2 i j p a\textsuperscript 2+c e f i j p a\textsuperscript 2-c f\textsuperscript 2 h q a\textsuperscript 2-c h q a\textsuperscript 2+e f h q a\textsuperscript 2-e\textsuperscript 2 i q a\textsuperscript 2+c e
   f i q a\textsuperscript 2-i q a\textsuperscript 2+c e j q a\textsuperscript 2+f j q a\textsuperscript 2-c f\textsuperscript 2 k a-c i\textsuperscript 2 k a-c f\textsuperscript 2 j\textsuperscript 2 k a+e f j\textsuperscript 2 k a-c k
   a+e f k a+h i k a-f h j k a-e i j k a+2 c f i j k a-b f\textsuperscript 2 m a-b i\textsuperscript 2 m a-b f\textsuperscript 2 j\textsuperscript 2 m a+d f
   j\textsuperscript 2 m a-b m a+d f m a+g i m a-f g j m a-d i j m a+2 b f i j m a-2 d e j\textsuperscript 2 n a+c d f j\textsuperscript 2 n
   a+b e f j\textsuperscript 2 n a-2 d e n a+c d f n a+b e f n a-2 g h n a+c g i n a+b h i n a+2 e g j n a-c
   f g j n a+2 d h j n a-b f h j n a-c d i j n a-b e i j n a+c d i\textsuperscript 2 p a+b e i\textsuperscript 2 p a+c d p
   a+b e p a+2 f g h p a-e g i p a-c f g i p a-d h i p a-b f h i p a+c f\textsuperscript 2 g j p a-e f g j p
   a+b f\textsuperscript 2 h j p a-d f h j p a+2 d e i j p a-c d f i j p a-b e f i j p a+c f\textsuperscript 2 g q a+c g q
   a-e f g q a+b f\textsuperscript 2 h q a+b h q a-d f h q a+2 d e i q a-c d f i q a-b e f i q a-c d j q a-b
   e j q a+b f\textsuperscript 2 k+b i\textsuperscript 2 k+b f\textsuperscript 2 j\textsuperscript 2 k-d f j\textsuperscript 2 k+b k-d f k-g i k+f g j k+d i j k-2 b f i j
   k+d\textsuperscript 2 n+g\textsuperscript 2 n+d\textsuperscript 2 j\textsuperscript 2 n-b d f j\textsuperscript 2 n-b d f n-b g i n-2 d g j n+b f g j n+b d i j n+n-f g\textsuperscript 2
   p-b d i\textsuperscript 2 p-b d p-f p+d g i p+b f g i p-b f\textsuperscript 2 g j p+d f g j p-d\textsuperscript 2 i j p+b d f i j p-b f\textsuperscript 2
   g q-b g q+d f g q-d\textsuperscript 2 i q+b d f i q-i q+b d j q+f j q)+(-b j\textsuperscript 2 f\textsuperscript 2+a c j\textsuperscript 2
   f\textsuperscript 2-b f\textsuperscript 2+a c f\textsuperscript 2+d j\textsuperscript 2 f-a e j\textsuperscript 2 f+d f-a e f-g j f+a h j f+2 b i j f-2 a c i j f-b i\textsuperscript 2+a
   c i\textsuperscript 2-b+a c+g i-a h i-d i j+a e i j)
   (k\textsuperscript 2+m\textsuperscript 2+n\textsuperscript 2+p\textsuperscript 2+q\textsuperscript 2+1)))))-q (g k+h m+i n+j p+q)
   (-g (-g q\textsuperscript 2+a h q\textsuperscript 2+b i q\textsuperscript 2-a c i q\textsuperscript 2+d j q\textsuperscript 2-a e j q\textsuperscript 2-b f j q\textsuperscript 2+a c f j q\textsuperscript 2+k
   q-a m q-b n q+a c n q-d p q+a e p q+b f p q-a c f p q-g+a h+b i-a c i+d j-a e j-b f j+a c
   f j) ((b d+c e+f) (-f q\textsuperscript 2 j\textsuperscript 2-f j\textsuperscript 2+i q\textsuperscript 2 j+i j-n q j+2 f p q j-f p\textsuperscript 2-f+n
   p-i p q) ((a d+e) (b k+c m+n) (-n+f p+i q-f j q) (-e q\textsuperscript 2 j\textsuperscript 2+c f q\textsuperscript 2
   j\textsuperscript 2-e j\textsuperscript 2+c f j\textsuperscript 2+h q\textsuperscript 2 j-c i q\textsuperscript 2 j+h j-c i j-m q j+c n q j+2 e p q j-2 c f p q j-e p\textsuperscript 2+c
   f p\textsuperscript 2-e+c f+m p-c n p-h p q+c i p q)-(b d+c e+f) (a k+m) (-m+c n+e p-c f p+h q-c i
   q-e j q+c f j q) (-f q\textsuperscript 2 j\textsuperscript 2-f j\textsuperscript 2+i q\textsuperscript 2 j+i j-n q j+2 f p q j-f p\textsuperscript 2-f+n p-i p
   q))-(d\textsuperscript 2+e\textsuperscript 2+f\textsuperscript 2+1) (q\textsuperscript 2 j\textsuperscript 2+j\textsuperscript 2-2 p q j+p\textsuperscript 2+1)
   ((a b+c) (b k+c m+n) (-n+f p+i q-f j q) (-c j\textsuperscript 2 f\textsuperscript 2-c p\textsuperscript 2 f\textsuperscript 2-c j\textsuperscript 2 q\textsuperscript 2 f\textsuperscript 2-c
   f\textsuperscript 2+2 c j p q f\textsuperscript 2+e j\textsuperscript 2 f+e p\textsuperscript 2 f+e j\textsuperscript 2 q\textsuperscript 2 f-h j q\textsuperscript 2 f+2 c i j q\textsuperscript 2 f+e f-h j f+2 c i j
   f-m p f+2 c n p f+j m q f-2 c j n q f+h p q f-2 c i p q f-2 e j p q f-c i\textsuperscript 2-c n\textsuperscript 2-c i\textsuperscript 2
   q\textsuperscript 2+h i q\textsuperscript 2-e i j q\textsuperscript 2-c+h i-e i j+m n-e n p-i m q-h n q+2 c i n q+e j n q+e i p
   q)-(b\textsuperscript 2+c\textsuperscript 2+1) (a k+m) (-m+c n+e p-c f p+h q-c i q-e j q+c f j q)
   (j\textsuperscript 2 f\textsuperscript 2+p\textsuperscript 2 f\textsuperscript 2+j\textsuperscript 2 q\textsuperscript 2 f\textsuperscript 2-2 j p q f\textsuperscript 2+f\textsuperscript 2-2 i j q\textsuperscript 2 f-2 i j f-2 n p f+2 j n q f+2
   i p q f+i\textsuperscript 2+n\textsuperscript 2+i\textsuperscript 2 q\textsuperscript 2-2 i n q+1))+(d k+e m+f n+p) (j q-p) ((a b+c) (b
   d+c e+f) (-f q\textsuperscript 2 j\textsuperscript 2-f j\textsuperscript 2+i q\textsuperscript 2 j+i j-n q j+2 f p q j-f p\textsuperscript 2-f+n p-i p q)
   (-c j\textsuperscript 2 f\textsuperscript 2-c p\textsuperscript 2 f\textsuperscript 2-c j\textsuperscript 2 q\textsuperscript 2 f\textsuperscript 2-c f\textsuperscript 2+2 c j p q f\textsuperscript 2+e j\textsuperscript 2 f+e p\textsuperscript 2 f+e j\textsuperscript 2 q\textsuperscript 2
   f-h j q\textsuperscript 2 f+2 c i j q\textsuperscript 2 f+e f-h j f+2 c i j f-m p f+2 c n p f+j m q f-2 c j n q f+h p q
   f-2 c i p q f-2 e j p q f-c i\textsuperscript 2-c n\textsuperscript 2-c i\textsuperscript 2 q\textsuperscript 2+h i q\textsuperscript 2-e i j q\textsuperscript 2-c+h i-e i j+m n-e n p-i
   m q-h n q+2 c i n q+e j n q+e i p q)-(b\textsuperscript 2+c\textsuperscript 2+1) (a d+e) (-e q\textsuperscript 2
   j\textsuperscript 2+c f q\textsuperscript 2 j\textsuperscript 2-e j\textsuperscript 2+c f j\textsuperscript 2+h q\textsuperscript 2 j-c i q\textsuperscript 2 j+h j-c i j-m q j+c n q j+2 e p q j-2 c f p
   q j-e p\textsuperscript 2+c f p\textsuperscript 2-e+c f+m p-c n p-h p q+c i p q) (j\textsuperscript 2 f\textsuperscript 2+p\textsuperscript 2 f\textsuperscript 2+j\textsuperscript 2 q\textsuperscript 2
   f\textsuperscript 2-2 j p q f\textsuperscript 2+f\textsuperscript 2-2 i j q\textsuperscript 2 f-2 i j f-2 n p f+2 j n q f+2 i p q f+i\textsuperscript 2+n\textsuperscript 2+i\textsuperscript 2 q\textsuperscript 2-2 i n
   q+1)))+(b g+c h+i) (-i q\textsuperscript 2+f j q\textsuperscript 2+n q-f p q-i+f j) (d
   (-d q\textsuperscript 2 j\textsuperscript 2+a e q\textsuperscript 2 j\textsuperscript 2+b f q\textsuperscript 2 j\textsuperscript 2-a c f q\textsuperscript 2 j\textsuperscript 2-d j\textsuperscript 2+a e j\textsuperscript 2+b f j\textsuperscript 2-a c f j\textsuperscript 2+g
   q\textsuperscript 2 j-a h q\textsuperscript 2 j-b i q\textsuperscript 2 j+a c i q\textsuperscript 2 j+g j-a h j-b i j+a c i j-k q j+a m q j+b n q j-a c n
   q j+2 d p q j-2 a e p q j-2 b f p q j+2 a c f p q j-d p\textsuperscript 2+a e p\textsuperscript 2+b f p\textsuperscript 2-a c f p\textsuperscript 2-d+a
   e+b f-a c f+k p-a m p-b n p+a c n p-g p q+a h p q+b i p q-a c i p q) ((a d+e)
   (b k+c m+n) (-n+f p+i q-f j q) (-e q\textsuperscript 2 j\textsuperscript 2+c f q\textsuperscript 2 j\textsuperscript 2-e j\textsuperscript 2+c f j\textsuperscript 2+h q\textsuperscript 2 j-c i q\textsuperscript 2
   j+h j-c i j-m q j+c n q j+2 e p q j-2 c f p q j-e p\textsuperscript 2+c f p\textsuperscript 2-e+c f+m p-c n p-h p q+c i p
   q)-(b d+c e+f) (a k+m) (-m+c n+e p-c f p+h q-c i q-e j q+c f j q) (-f q\textsuperscript 2
   j\textsuperscript 2-f j\textsuperscript 2+i q\textsuperscript 2 j+i j-n q j+2 f p q j-f p\textsuperscript 2-f+n p-i p
   q))-(d\textsuperscript 2+e\textsuperscript 2+f\textsuperscript 2+1) (q\textsuperscript 2 j\textsuperscript 2+j\textsuperscript 2-2 p q j+p\textsuperscript 2+1) (a
   (b k+c m+n) (-n+f p+i q-f j q) ((g k+h m+i n+j p+q) (a i n c\textsuperscript 2-a f j n c\textsuperscript 2-a f
   i p c\textsuperscript 2+a f\textsuperscript 2 j p c\textsuperscript 2+a f\textsuperscript 2 q c\textsuperscript 2+a q c\textsuperscript 2-a i m c+a f j m c+g n c-a h n c-b i n c-d j n
   c+a e j n c+b f j n c-f g p c+a f h p c+a e i p c+b f i p c-b f\textsuperscript 2 j p c+d f j p c-2 a e f
   j p c-b f\textsuperscript 2 q c-b q c+d f q c-2 a e f q c-g m+a h m+b i m+d j m-a e j m-b f j m+e g p-a e
   h p-b e i p+a e\textsuperscript 2 j p-d e j p+b e f j p+a e\textsuperscript 2 q+a q-d e q+b e f q)-(d k+e m+f n+p)
   (-a f j\textsuperscript 2 n c\textsuperscript 2-a f n c\textsuperscript 2+a i j n c\textsuperscript 2-a i\textsuperscript 2 p c\textsuperscript 2-a p c\textsuperscript 2+a f i j p c\textsuperscript 2+a f i q
   c\textsuperscript 2+a j q c\textsuperscript 2+a f j\textsuperscript 2 m c+a f m c-a i j m c-d j\textsuperscript 2 n c+a e j\textsuperscript 2 n c+b f j\textsuperscript 2 n c-d n c+a e n
   c+b f n c+g j n c-a h j n c-b i j n c+b i\textsuperscript 2 p c+b p c-g i p c+2 a h i p c-a f h j p c+d i
   j p c-a e i j p c-b f i j p c-a f h q c+d i q c-a e i q c-b f i q c-b j q c+d j\textsuperscript 2 m-a e
   j\textsuperscript 2 m-b f j\textsuperscript 2 m+d m-a e m-b f m-g j m+a h j m+b i j m-a h\textsuperscript 2 p-a p+g h p-b h i p-d h j p+a
   e h j p+b f h j p-d h q+a e h q+b f h q+a j q)+(b k+c m+n) (a j\textsuperscript 2 n e\textsuperscript 2+a n
   e\textsuperscript 2-a i j p e\textsuperscript 2-a i q e\textsuperscript 2-a f j\textsuperscript 2 m e-a f m e+a i j m e-d j\textsuperscript 2 n e+b f j\textsuperscript 2 n e-a c f j\textsuperscript 2 n
   e-d n e+b f n e-a c f n e+g j n e-2 a h j n e-b i j n e+a c i j n e+b i\textsuperscript 2 p e-a c i\textsuperscript 2 p
   e+b p e-a c p e-g i p e+a h i p e+a f h j p e+d i j p e-b f i j p e+a c f i j p e+a f h q
   e+d i q e-b f i q e+a c f i q e-b j q e+a c j q e-b f\textsuperscript 2 m+a c f\textsuperscript 2 m-b i\textsuperscript 2 m+a c i\textsuperscript 2 m-b
   f\textsuperscript 2 j\textsuperscript 2 m+a c f\textsuperscript 2 j\textsuperscript 2 m+d f j\textsuperscript 2 m-b m+a c m+d f m+g i m-a h i m-f g j m+a f h j m-d i j
   m+2 b f i j m-2 a c f i j m+a h\textsuperscript 2 n+a n-g h n+b h i n-a c h i n+d h j n-b f h j n+a c f h
   j n-a f h\textsuperscript 2 p-a f p+f g h p-b f h i p+a c f h i p+b f\textsuperscript 2 h j p-a c f\textsuperscript 2 h j p-d f h j p+b
   f\textsuperscript 2 h q-a c f\textsuperscript 2 h q+b h q-a c h q-d f h q-a i q+a f j q)-(a f\textsuperscript 2 c\textsuperscript 2+a i\textsuperscript 2
   c\textsuperscript 2+a f\textsuperscript 2 j\textsuperscript 2 c\textsuperscript 2+a c\textsuperscript 2-2 a f i j c\textsuperscript 2-b f\textsuperscript 2 c-b i\textsuperscript 2 c-b f\textsuperscript 2 j\textsuperscript 2 c+d f j\textsuperscript 2 c-2 a e f j\textsuperscript 2
   c-b c+d f c-2 a e f c+g i c-2 a h i c-f g j c+2 a f h j c-d i j c+2 a e i j c+2 b f i j
   c+a e\textsuperscript 2+a h\textsuperscript 2+a e\textsuperscript 2 j\textsuperscript 2-d e j\textsuperscript 2+b e f j\textsuperscript 2+a-d e+b e f-g h+b h i+e g j+d h j-2 a e h j-b f
   h j-b e i j) (k\textsuperscript 2+m\textsuperscript 2+n\textsuperscript 2+p\textsuperscript 2+q\textsuperscript 2+1)-k
   (-(g\textsuperscript 2+h\textsuperscript 2+i\textsuperscript 2+j\textsuperscript 2+1) (-(d\textsuperscript 2+e\textsuperscript 2+f\textsuperscript 2+1) (-m b\textsuperscript 2+c k
   b+a c m b+a n b-a c\textsuperscript 2 k-a k-m+c n)+(b d+c e+f) (b e k-a c e k-a f k-b d m+a c d m-f
   m+a d n+e n)+(-e b\textsuperscript 2+c d b+a c e b+a f b-a c\textsuperscript 2 d-a d-e+c f) (d k+e m+f
   n+p))+(d g+e h+f i+j) (-(d g+e h+f i+j) (-m b\textsuperscript 2+c k b+a c m b+a n b-a c\textsuperscript 2
   k-a k-m+c n)+(b d+c e+f) (b h k-a c h k-a i k-b g m+a c g m-i m+a g n+h n)+(-h
   b\textsuperscript 2+c g b+a c h b+a i b-a c\textsuperscript 2 g-a g-h+c i) (d k+e m+f n+p))-(b g+c h+i)
   (-(d g+e h+f i+j) (b e k-a c e k-a f k-b d m+a c d m-f m+a d n+e
   n)+(d\textsuperscript 2+e\textsuperscript 2+f\textsuperscript 2+1) (b h k-a c h k-a i k-b g m+a c g m-i m+a g n+h n)+(b e g-a
   c e g-a f g-b d h+a c d h-f h+a d i+e i) (d k+e m+f
   n+p))+(-(d\textsuperscript 2+e\textsuperscript 2+f\textsuperscript 2+1) (-h b\textsuperscript 2+c g b+a c h b+a i b-a c\textsuperscript 2 g-a
   g-h+c i)+(b d+c e+f) (b e g-a c e g-a f g-b d h+a c d h-f h+a d i+e i)+(-e
   b\textsuperscript 2+c d b+a c e b+a f b-a c\textsuperscript 2 d-a d-e+c f) (d g+e h+f i+j)) (g k+h m+i n+j
   p+q)))-b (a k+m) (-m+c n+e p-c f p+h q-c i q-e j q+c f j q) (-(g k+h m+i
   n+j p+q) (-b j p f\textsuperscript 2+a c j p f\textsuperscript 2-b q f\textsuperscript 2+a c q f\textsuperscript 2+b j n f-a c j n f-g p f+a h p f+b
   i p f-a c i p f+d j p f-a e j p f+d q f-a e q f+g n-a h n-b i n+a c i n-d j n+a e j n-b
   q+a c q)+(d k+e m+f n+p) (b p i\textsuperscript 2-a c p i\textsuperscript 2-b j n i+a c j n i-g p i+a h p i+d
   j p i-a e j p i-b f j p i+a c f j p i+d q i-a e q i-b f q i+a c f q i-d j\textsuperscript 2 n+a e j\textsuperscript 2 n+b
   f j\textsuperscript 2 n-a c f j\textsuperscript 2 n-d n+a e n+b f n-a c f n+g j n-a h j n+b p-a c p-b j q+a c j
   q)-(a k+m) (a j\textsuperscript 2 n e\textsuperscript 2+a n e\textsuperscript 2-a i j p e\textsuperscript 2-a i q e\textsuperscript 2-a f j\textsuperscript 2 m e-a f m e+a i
   j m e-d j\textsuperscript 2 n e+b f j\textsuperscript 2 n e-a c f j\textsuperscript 2 n e-d n e+b f n e-a c f n e+g j n e-2 a h j n e-b i
   j n e+a c i j n e+b i\textsuperscript 2 p e-a c i\textsuperscript 2 p e+b p e-a c p e-g i p e+a h i p e+a f h j p e+d i j
   p e-b f i j p e+a c f i j p e+a f h q e+d i q e-b f i q e+a c f i q e-b j q e+a c j q e-b
   f\textsuperscript 2 m+a c f\textsuperscript 2 m-b i\textsuperscript 2 m+a c i\textsuperscript 2 m-b f\textsuperscript 2 j\textsuperscript 2 m+a c f\textsuperscript 2 j\textsuperscript 2 m+d f j\textsuperscript 2 m-b m+a c m+d f m+g i
   m-a h i m-f g j m+a f h j m-d i j m+2 b f i j m-2 a c f i j m+a h\textsuperscript 2 n+a n-g h n+b h i n-a
   c h i n+d h j n-b f h j n+a c f h j n-a f h\textsuperscript 2 p-a f p+f g h p-b f h i p+a c f h i p+b f\textsuperscript 2
   h j p-a c f\textsuperscript 2 h j p-d f h j p+b f\textsuperscript 2 h q-a c f\textsuperscript 2 h q+b h q-a c h q-d f h q-a i q+a f j
   q)+k (c f\textsuperscript 2 m a\textsuperscript 2+c i\textsuperscript 2 m a\textsuperscript 2+c f\textsuperscript 2 j\textsuperscript 2 m a\textsuperscript 2-e f j\textsuperscript 2 m a\textsuperscript 2+c m a\textsuperscript 2-e f m
   a\textsuperscript 2-h i m a\textsuperscript 2+f h j m a\textsuperscript 2+e i j m a\textsuperscript 2-2 c f i j m a\textsuperscript 2+e\textsuperscript 2 n a\textsuperscript 2+h\textsuperscript 2 n a\textsuperscript 2+e\textsuperscript 2 j\textsuperscript 2 n a\textsuperscript 2-c
   e f j\textsuperscript 2 n a\textsuperscript 2-c e f n a\textsuperscript 2-c h i n a\textsuperscript 2-2 e h j n a\textsuperscript 2+c f h j n a\textsuperscript 2+c e i j n a\textsuperscript 2+n a\textsuperscript 2-f
   h\textsuperscript 2 p a\textsuperscript 2-c e i\textsuperscript 2 p a\textsuperscript 2-c e p a\textsuperscript 2-f p a\textsuperscript 2+e h i p a\textsuperscript 2+c f h i p a\textsuperscript 2-c f\textsuperscript 2 h j p a\textsuperscript 2+e f h
   j p a\textsuperscript 2-e\textsuperscript 2 i j p a\textsuperscript 2+c e f i j p a\textsuperscript 2-c f\textsuperscript 2 h q a\textsuperscript 2-c h q a\textsuperscript 2+e f h q a\textsuperscript 2-e\textsuperscript 2 i q a\textsuperscript 2+c e
   f i q a\textsuperscript 2-i q a\textsuperscript 2+c e j q a\textsuperscript 2+f j q a\textsuperscript 2-c f\textsuperscript 2 k a-c i\textsuperscript 2 k a-c f\textsuperscript 2 j\textsuperscript 2 k a+e f j\textsuperscript 2 k a-c k
   a+e f k a+h i k a-f h j k a-e i j k a+2 c f i j k a-b f\textsuperscript 2 m a-b i\textsuperscript 2 m a-b f\textsuperscript 2 j\textsuperscript 2 m a+d f
   j\textsuperscript 2 m a-b m a+d f m a+g i m a-f g j m a-d i j m a+2 b f i j m a-2 d e j\textsuperscript 2 n a+c d f j\textsuperscript 2 n
   a+b e f j\textsuperscript 2 n a-2 d e n a+c d f n a+b e f n a-2 g h n a+c g i n a+b h i n a+2 e g j n a-c
   f g j n a+2 d h j n a-b f h j n a-c d i j n a-b e i j n a+c d i\textsuperscript 2 p a+b e i\textsuperscript 2 p a+c d p
   a+b e p a+2 f g h p a-e g i p a-c f g i p a-d h i p a-b f h i p a+c f\textsuperscript 2 g j p a-e f g j p
   a+b f\textsuperscript 2 h j p a-d f h j p a+2 d e i j p a-c d f i j p a-b e f i j p a+c f\textsuperscript 2 g q a+c g q
   a-e f g q a+b f\textsuperscript 2 h q a+b h q a-d f h q a+2 d e i q a-c d f i q a-b e f i q a-c d j q a-b
   e j q a+b f\textsuperscript 2 k+b i\textsuperscript 2 k+b f\textsuperscript 2 j\textsuperscript 2 k-d f j\textsuperscript 2 k+b k-d f k-g i k+f g j k+d i j k-2 b f i j
   k+d\textsuperscript 2 n+g\textsuperscript 2 n+d\textsuperscript 2 j\textsuperscript 2 n-b d f j\textsuperscript 2 n-b d f n-b g i n-2 d g j n+b f g j n+b d i j n+n-f g\textsuperscript 2
   p-b d i\textsuperscript 2 p-b d p-f p+d g i p+b f g i p-b f\textsuperscript 2 g j p+d f g j p-d\textsuperscript 2 i j p+b d f i j p-b f\textsuperscript 2
   g q-b g q+d f g q-d\textsuperscript 2 i q+b d f i q-i q+b d j q+f j q)+(-b j\textsuperscript 2 f\textsuperscript 2+a c j\textsuperscript 2
   f\textsuperscript 2-b f\textsuperscript 2+a c f\textsuperscript 2+d j\textsuperscript 2 f-a e j\textsuperscript 2 f+d f-a e f-g j f+a h j f+2 b i j f-2 a c i j f-b i\textsuperscript 2+a
   c i\textsuperscript 2-b+a c+g i-a h i-d i j+a e i j)
   (k\textsuperscript 2+m\textsuperscript 2+n\textsuperscript 2+p\textsuperscript 2+q\textsuperscript 2+1)))+(d k+e m+f n+p) (j q-p) (a (b d+c
   e+f) (-f q\textsuperscript 2 j\textsuperscript 2-f j\textsuperscript 2+i q\textsuperscript 2 j+i j-n q j+2 f p q j-f p\textsuperscript 2-f+n p-i p q) ((g
   k+h m+i n+j p+q) (a i n c\textsuperscript 2-a f j n c\textsuperscript 2-a f i p c\textsuperscript 2+a f\textsuperscript 2 j p c\textsuperscript 2+a f\textsuperscript 2 q c\textsuperscript 2+a q
   c\textsuperscript 2-a i m c+a f j m c+g n c-a h n c-b i n c-d j n c+a e j n c+b f j n c-f g p c+a f h p
   c+a e i p c+b f i p c-b f\textsuperscript 2 j p c+d f j p c-2 a e f j p c-b f\textsuperscript 2 q c-b q c+d f q c-2 a e f
   q c-g m+a h m+b i m+d j m-a e j m-b f j m+e g p-a e h p-b e i p+a e\textsuperscript 2 j p-d e j p+b e f j
   p+a e\textsuperscript 2 q+a q-d e q+b e f q)-(d k+e m+f n+p) (-a f j\textsuperscript 2 n c\textsuperscript 2-a f n c\textsuperscript 2+a i j n
   c\textsuperscript 2-a i\textsuperscript 2 p c\textsuperscript 2-a p c\textsuperscript 2+a f i j p c\textsuperscript 2+a f i q c\textsuperscript 2+a j q c\textsuperscript 2+a f j\textsuperscript 2 m c+a f m c-a i j m
   c-d j\textsuperscript 2 n c+a e j\textsuperscript 2 n c+b f j\textsuperscript 2 n c-d n c+a e n c+b f n c+g j n c-a h j n c-b i j n c+b
   i\textsuperscript 2 p c+b p c-g i p c+2 a h i p c-a f h j p c+d i j p c-a e i j p c-b f i j p c-a f h q
   c+d i q c-a e i q c-b f i q c-b j q c+d j\textsuperscript 2 m-a e j\textsuperscript 2 m-b f j\textsuperscript 2 m+d m-a e m-b f m-g j m+a
   h j m+b i j m-a h\textsuperscript 2 p-a p+g h p-b h i p-d h j p+a e h j p+b f h j p-d h q+a e h q+b f h
   q+a j q)+(b k+c m+n) (a j\textsuperscript 2 n e\textsuperscript 2+a n e\textsuperscript 2-a i j p e\textsuperscript 2-a i q e\textsuperscript 2-a f j\textsuperscript 2 m e-a
   f m e+a i j m e-d j\textsuperscript 2 n e+b f j\textsuperscript 2 n e-a c f j\textsuperscript 2 n e-d n e+b f n e-a c f n e+g j n e-2 a h
   j n e-b i j n e+a c i j n e+b i\textsuperscript 2 p e-a c i\textsuperscript 2 p e+b p e-a c p e-g i p e+a h i p e+a f h j
   p e+d i j p e-b f i j p e+a c f i j p e+a f h q e+d i q e-b f i q e+a c f i q e-b j q e+a
   c j q e-b f\textsuperscript 2 m+a c f\textsuperscript 2 m-b i\textsuperscript 2 m+a c i\textsuperscript 2 m-b f\textsuperscript 2 j\textsuperscript 2 m+a c f\textsuperscript 2 j\textsuperscript 2 m+d f j\textsuperscript 2 m-b m+a c
   m+d f m+g i m-a h i m-f g j m+a f h j m-d i j m+2 b f i j m-2 a c f i j m+a h\textsuperscript 2 n+a n-g h
   n+b h i n-a c h i n+d h j n-b f h j n+a c f h j n-a f h\textsuperscript 2 p-a f p+f g h p-b f h i p+a c f
   h i p+b f\textsuperscript 2 h j p-a c f\textsuperscript 2 h j p-d f h j p+b f\textsuperscript 2 h q-a c f\textsuperscript 2 h q+b h q-a c h q-d f h q-a i
   q+a f j q)-(a f\textsuperscript 2 c\textsuperscript 2+a i\textsuperscript 2 c\textsuperscript 2+a f\textsuperscript 2 j\textsuperscript 2 c\textsuperscript 2+a c\textsuperscript 2-2 a f i j c\textsuperscript 2-b f\textsuperscript 2 c-b
   i\textsuperscript 2 c-b f\textsuperscript 2 j\textsuperscript 2 c+d f j\textsuperscript 2 c-2 a e f j\textsuperscript 2 c-b c+d f c-2 a e f c+g i c-2 a h i c-f g j c+2 a
   f h j c-d i j c+2 a e i j c+2 b f i j c+a e\textsuperscript 2+a h\textsuperscript 2+a e\textsuperscript 2 j\textsuperscript 2-d e j\textsuperscript 2+b e f j\textsuperscript 2+a-d e+b e
   f-g h+b h i+e g j+d h j-2 a e h j-b f h j-b e i j)
   (k\textsuperscript 2+m\textsuperscript 2+n\textsuperscript 2+p\textsuperscript 2+q\textsuperscript 2+1)-k (-(g\textsuperscript 2+h\textsuperscript 2+i\textsuperscript 2+j\textsuperscript 2+1)
   (-(d\textsuperscript 2+e\textsuperscript 2+f\textsuperscript 2+1) (-m b\textsuperscript 2+c k b+a c m b+a n b-a c\textsuperscript 2 k-a k-m+c
   n)+(b d+c e+f) (b e k-a c e k-a f k-b d m+a c d m-f m+a d n+e n)+(-e b\textsuperscript 2+c d
   b+a c e b+a f b-a c\textsuperscript 2 d-a d-e+c f) (d k+e m+f n+p))+(d g+e h+f i+j) (-(d
   g+e h+f i+j) (-m b\textsuperscript 2+c k b+a c m b+a n b-a c\textsuperscript 2 k-a k-m+c n)+(b d+c e+f) (b h
   k-a c h k-a i k-b g m+a c g m-i m+a g n+h n)+(-h b\textsuperscript 2+c g b+a c h b+a i b-a c\textsuperscript 2 g-a
   g-h+c i) (d k+e m+f n+p))-(b g+c h+i) (-(d g+e h+f i+j) (b e k-a c e k-a
   f k-b d m+a c d m-f m+a d n+e n)+(d\textsuperscript 2+e\textsuperscript 2+f\textsuperscript 2+1) (b h k-a c h k-a i k-b g m+a
   c g m-i m+a g n+h n)+(b e g-a c e g-a f g-b d h+a c d h-f h+a d i+e i) (d k+e m+f
   n+p))+(-(d\textsuperscript 2+e\textsuperscript 2+f\textsuperscript 2+1) (-h b\textsuperscript 2+c g b+a c h b+a i b-a c\textsuperscript 2 g-a
   g-h+c i)+(b d+c e+f) (b e g-a c e g-a f g-b d h+a c d h-f h+a d i+e i)+(-e
   b\textsuperscript 2+c d b+a c e b+a f b-a c\textsuperscript 2 d-a d-e+c f) (d g+e h+f i+j)) (g k+h m+i n+j
   p+q)))-b (a d+e) (-e q\textsuperscript 2 j\textsuperscript 2+c f q\textsuperscript 2 j\textsuperscript 2-e j\textsuperscript 2+c f j\textsuperscript 2+h q\textsuperscript 2 j-c i q\textsuperscript 2
   j+h j-c i j-m q j+c n q j+2 e p q j-2 c f p q j-e p\textsuperscript 2+c f p\textsuperscript 2-e+c f+m p-c n p-h p q+c i p
   q) (-(g k+h m+i n+j p+q) (-b j p f\textsuperscript 2+a c j p f\textsuperscript 2-b q f\textsuperscript 2+a c q f\textsuperscript 2+b j n
   f-a c j n f-g p f+a h p f+b i p f-a c i p f+d j p f-a e j p f+d q f-a e q f+g n-a h n-b i
   n+a c i n-d j n+a e j n-b q+a c q)+(d k+e m+f n+p) (b p i\textsuperscript 2-a c p i\textsuperscript 2-b j n
   i+a c j n i-g p i+a h p i+d j p i-a e j p i-b f j p i+a c f j p i+d q i-a e q i-b f q i+a
   c f q i-d j\textsuperscript 2 n+a e j\textsuperscript 2 n+b f j\textsuperscript 2 n-a c f j\textsuperscript 2 n-d n+a e n+b f n-a c f n+g j n-a h j n+b
   p-a c p-b j q+a c j q)-(a k+m) (a j\textsuperscript 2 n e\textsuperscript 2+a n e\textsuperscript 2-a i j p e\textsuperscript 2-a i q e\textsuperscript 2-a f
   j\textsuperscript 2 m e-a f m e+a i j m e-d j\textsuperscript 2 n e+b f j\textsuperscript 2 n e-a c f j\textsuperscript 2 n e-d n e+b f n e-a c f n e+g j
   n e-2 a h j n e-b i j n e+a c i j n e+b i\textsuperscript 2 p e-a c i\textsuperscript 2 p e+b p e-a c p e-g i p e+a h i p
   e+a f h j p e+d i j p e-b f i j p e+a c f i j p e+a f h q e+d i q e-b f i q e+a c f i q
   e-b j q e+a c j q e-b f\textsuperscript 2 m+a c f\textsuperscript 2 m-b i\textsuperscript 2 m+a c i\textsuperscript 2 m-b f\textsuperscript 2 j\textsuperscript 2 m+a c f\textsuperscript 2 j\textsuperscript 2 m+d f j\textsuperscript 2
   m-b m+a c m+d f m+g i m-a h i m-f g j m+a f h j m-d i j m+2 b f i j m-2 a c f i j m+a h\textsuperscript 2
   n+a n-g h n+b h i n-a c h i n+d h j n-b f h j n+a c f h j n-a f h\textsuperscript 2 p-a f p+f g h p-b f h
   i p+a c f h i p+b f\textsuperscript 2 h j p-a c f\textsuperscript 2 h j p-d f h j p+b f\textsuperscript 2 h q-a c f\textsuperscript 2 h q+b h q-a c h q-d
   f h q-a i q+a f j q)+k (c f\textsuperscript 2 m a\textsuperscript 2+c i\textsuperscript 2 m a\textsuperscript 2+c f\textsuperscript 2 j\textsuperscript 2 m a\textsuperscript 2-e f j\textsuperscript 2 m
   a\textsuperscript 2+c m a\textsuperscript 2-e f m a\textsuperscript 2-h i m a\textsuperscript 2+f h j m a\textsuperscript 2+e i j m a\textsuperscript 2-2 c f i j m a\textsuperscript 2+e\textsuperscript 2 n a\textsuperscript 2+h\textsuperscript 2 n
   a\textsuperscript 2+e\textsuperscript 2 j\textsuperscript 2 n a\textsuperscript 2-c e f j\textsuperscript 2 n a\textsuperscript 2-c e f n a\textsuperscript 2-c h i n a\textsuperscript 2-2 e h j n a\textsuperscript 2+c f h j n a\textsuperscript 2+c e
   i j n a\textsuperscript 2+n a\textsuperscript 2-f h\textsuperscript 2 p a\textsuperscript 2-c e i\textsuperscript 2 p a\textsuperscript 2-c e p a\textsuperscript 2-f p a\textsuperscript 2+e h i p a\textsuperscript 2+c f h i p a\textsuperscript 2-c
   f\textsuperscript 2 h j p a\textsuperscript 2+e f h j p a\textsuperscript 2-e\textsuperscript 2 i j p a\textsuperscript 2+c e f i j p a\textsuperscript 2-c f\textsuperscript 2 h q a\textsuperscript 2-c h q a\textsuperscript 2+e f h q
   a\textsuperscript 2-e\textsuperscript 2 i q a\textsuperscript 2+c e f i q a\textsuperscript 2-i q a\textsuperscript 2+c e j q a\textsuperscript 2+f j q a\textsuperscript 2-c f\textsuperscript 2 k a-c i\textsuperscript 2 k a-c f\textsuperscript 2 j\textsuperscript 2
   k a+e f j\textsuperscript 2 k a-c k a+e f k a+h i k a-f h j k a-e i j k a+2 c f i j k a-b f\textsuperscript 2 m a-b i\textsuperscript 2 m
   a-b f\textsuperscript 2 j\textsuperscript 2 m a+d f j\textsuperscript 2 m a-b m a+d f m a+g i m a-f g j m a-d i j m a+2 b f i j m a-2 d e
   j\textsuperscript 2 n a+c d f j\textsuperscript 2 n a+b e f j\textsuperscript 2 n a-2 d e n a+c d f n a+b e f n a-2 g h n a+c g i n a+b h
   i n a+2 e g j n a-c f g j n a+2 d h j n a-b f h j n a-c d i j n a-b e i j n a+c d i\textsuperscript 2 p
   a+b e i\textsuperscript 2 p a+c d p a+b e p a+2 f g h p a-e g i p a-c f g i p a-d h i p a-b f h i p a+c
   f\textsuperscript 2 g j p a-e f g j p a+b f\textsuperscript 2 h j p a-d f h j p a+2 d e i j p a-c d f i j p a-b e f i j p
   a+c f\textsuperscript 2 g q a+c g q a-e f g q a+b f\textsuperscript 2 h q a+b h q a-d f h q a+2 d e i q a-c d f i q a-b e
   f i q a-c d j q a-b e j q a+b f\textsuperscript 2 k+b i\textsuperscript 2 k+b f\textsuperscript 2 j\textsuperscript 2 k-d f j\textsuperscript 2 k+b k-d f k-g i k+f g j
   k+d i j k-2 b f i j k+d\textsuperscript 2 n+g\textsuperscript 2 n+d\textsuperscript 2 j\textsuperscript 2 n-b d f j\textsuperscript 2 n-b d f n-b g i n-2 d g j n+b f g j
   n+b d i j n+n-f g\textsuperscript 2 p-b d i\textsuperscript 2 p-b d p-f p+d g i p+b f g i p-b f\textsuperscript 2 g j p+d f g j p-d\textsuperscript 2 i j
   p+b d f i j p-b f\textsuperscript 2 g q-b g q+d f g q-d\textsuperscript 2 i q+b d f i q-i q+b d j q+f j q)+(-b
   j\textsuperscript 2 f\textsuperscript 2+a c j\textsuperscript 2 f\textsuperscript 2-b f\textsuperscript 2+a c f\textsuperscript 2+d j\textsuperscript 2 f-a e j\textsuperscript 2 f+d f-a e f-g j f+a h j f+2 b i j f-2 a
   c i j f-b i\textsuperscript 2+a c i\textsuperscript 2-b+a c+g i-a h i-d i j+a e i j)
   (k\textsuperscript 2+m\textsuperscript 2+n\textsuperscript 2+p\textsuperscript 2+q\textsuperscript 2+1))))-(d g+e h+f i+j) (-j q\textsuperscript 2+p
   q-j) (d (-d q\textsuperscript 2 j\textsuperscript 2+a e q\textsuperscript 2 j\textsuperscript 2+b f q\textsuperscript 2 j\textsuperscript 2-a c f q\textsuperscript 2 j\textsuperscript 2-d j\textsuperscript 2+a e j\textsuperscript 2+b
   f j\textsuperscript 2-a c f j\textsuperscript 2+g q\textsuperscript 2 j-a h q\textsuperscript 2 j-b i q\textsuperscript 2 j+a c i q\textsuperscript 2 j+g j-a h j-b i j+a c i j-k q j+a m
   q j+b n q j-a c n q j+2 d p q j-2 a e p q j-2 b f p q j+2 a c f p q j-d p\textsuperscript 2+a e p\textsuperscript 2+b f
   p\textsuperscript 2-a c f p\textsuperscript 2-d+a e+b f-a c f+k p-a m p-b n p+a c n p-g p q+a h p q+b i p q-a c i p
   q) ((a b+c) (b k+c m+n) (-n+f p+i q-f j q) (-c j\textsuperscript 2 f\textsuperscript 2-c p\textsuperscript 2 f\textsuperscript 2-c j\textsuperscript 2
   q\textsuperscript 2 f\textsuperscript 2-c f\textsuperscript 2+2 c j p q f\textsuperscript 2+e j\textsuperscript 2 f+e p\textsuperscript 2 f+e j\textsuperscript 2 q\textsuperscript 2 f-h j q\textsuperscript 2 f+2 c i j q\textsuperscript 2 f+e f-h j
   f+2 c i j f-m p f+2 c n p f+j m q f-2 c j n q f+h p q f-2 c i p q f-2 e j p q f-c i\textsuperscript 2-c
   n\textsuperscript 2-c i\textsuperscript 2 q\textsuperscript 2+h i q\textsuperscript 2-e i j q\textsuperscript 2-c+h i-e i j+m n-e n p-i m q-h n q+2 c i n q+e j n q+e i p
   q)-(b\textsuperscript 2+c\textsuperscript 2+1) (a k+m) (-m+c n+e p-c f p+h q-c i q-e j q+c f j q)
   (j\textsuperscript 2 f\textsuperscript 2+p\textsuperscript 2 f\textsuperscript 2+j\textsuperscript 2 q\textsuperscript 2 f\textsuperscript 2-2 j p q f\textsuperscript 2+f\textsuperscript 2-2 i j q\textsuperscript 2 f-2 i j f-2 n p f+2 j n q f+2
   i p q f+i\textsuperscript 2+n\textsuperscript 2+i\textsuperscript 2 q\textsuperscript 2-2 i n q+1))-(b d+c e+f) (-f q\textsuperscript 2 j\textsuperscript 2-f j\textsuperscript 2+i q\textsuperscript 2
   j+i j-n q j+2 f p q j-f p\textsuperscript 2-f+n p-i p q) (a (b k+c m+n) (-n+f p+i q-f j q)
   ((g k+h m+i n+j p+q) (a i n c\textsuperscript 2-a f j n c\textsuperscript 2-a f i p c\textsuperscript 2+a f\textsuperscript 2 j p c\textsuperscript 2+a f\textsuperscript 2 q
   c\textsuperscript 2+a q c\textsuperscript 2-a i m c+a f j m c+g n c-a h n c-b i n c-d j n c+a e j n c+b f j n c-f g p c+a
   f h p c+a e i p c+b f i p c-b f\textsuperscript 2 j p c+d f j p c-2 a e f j p c-b f\textsuperscript 2 q c-b q c+d f q c-2
   a e f q c-g m+a h m+b i m+d j m-a e j m-b f j m+e g p-a e h p-b e i p+a e\textsuperscript 2 j p-d e j p+b
   e f j p+a e\textsuperscript 2 q+a q-d e q+b e f q)-(d k+e m+f n+p) (-a f j\textsuperscript 2 n c\textsuperscript 2-a f n c\textsuperscript 2+a
   i j n c\textsuperscript 2-a i\textsuperscript 2 p c\textsuperscript 2-a p c\textsuperscript 2+a f i j p c\textsuperscript 2+a f i q c\textsuperscript 2+a j q c\textsuperscript 2+a f j\textsuperscript 2 m c+a f m c-a i
   j m c-d j\textsuperscript 2 n c+a e j\textsuperscript 2 n c+b f j\textsuperscript 2 n c-d n c+a e n c+b f n c+g j n c-a h j n c-b i j n
   c+b i\textsuperscript 2 p c+b p c-g i p c+2 a h i p c-a f h j p c+d i j p c-a e i j p c-b f i j p c-a f h
   q c+d i q c-a e i q c-b f i q c-b j q c+d j\textsuperscript 2 m-a e j\textsuperscript 2 m-b f j\textsuperscript 2 m+d m-a e m-b f m-g j
   m+a h j m+b i j m-a h\textsuperscript 2 p-a p+g h p-b h i p-d h j p+a e h j p+b f h j p-d h q+a e h q+b f
   h q+a j q)+(b k+c m+n) (a j\textsuperscript 2 n e\textsuperscript 2+a n e\textsuperscript 2-a i j p e\textsuperscript 2-a i q e\textsuperscript 2-a f j\textsuperscript 2 m
   e-a f m e+a i j m e-d j\textsuperscript 2 n e+b f j\textsuperscript 2 n e-a c f j\textsuperscript 2 n e-d n e+b f n e-a c f n e+g j n e-2
   a h j n e-b i j n e+a c i j n e+b i\textsuperscript 2 p e-a c i\textsuperscript 2 p e+b p e-a c p e-g i p e+a h i p e+a f
   h j p e+d i j p e-b f i j p e+a c f i j p e+a f h q e+d i q e-b f i q e+a c f i q e-b j q
   e+a c j q e-b f\textsuperscript 2 m+a c f\textsuperscript 2 m-b i\textsuperscript 2 m+a c i\textsuperscript 2 m-b f\textsuperscript 2 j\textsuperscript 2 m+a c f\textsuperscript 2 j\textsuperscript 2 m+d f j\textsuperscript 2 m-b m+a
   c m+d f m+g i m-a h i m-f g j m+a f h j m-d i j m+2 b f i j m-2 a c f i j m+a h\textsuperscript 2 n+a n-g
   h n+b h i n-a c h i n+d h j n-b f h j n+a c f h j n-a f h\textsuperscript 2 p-a f p+f g h p-b f h i p+a c
   f h i p+b f\textsuperscript 2 h j p-a c f\textsuperscript 2 h j p-d f h j p+b f\textsuperscript 2 h q-a c f\textsuperscript 2 h q+b h q-a c h q-d f h q-a
   i q+a f j q)-(a f\textsuperscript 2 c\textsuperscript 2+a i\textsuperscript 2 c\textsuperscript 2+a f\textsuperscript 2 j\textsuperscript 2 c\textsuperscript 2+a c\textsuperscript 2-2 a f i j c\textsuperscript 2-b f\textsuperscript 2 c-b
   i\textsuperscript 2 c-b f\textsuperscript 2 j\textsuperscript 2 c+d f j\textsuperscript 2 c-2 a e f j\textsuperscript 2 c-b c+d f c-2 a e f c+g i c-2 a h i c-f g j c+2 a
   f h j c-d i j c+2 a e i j c+2 b f i j c+a e\textsuperscript 2+a h\textsuperscript 2+a e\textsuperscript 2 j\textsuperscript 2-d e j\textsuperscript 2+b e f j\textsuperscript 2+a-d e+b e
   f-g h+b h i+e g j+d h j-2 a e h j-b f h j-b e i j)
   (k\textsuperscript 2+m\textsuperscript 2+n\textsuperscript 2+p\textsuperscript 2+q\textsuperscript 2+1)-k (-(g\textsuperscript 2+h\textsuperscript 2+i\textsuperscript 2+j\textsuperscript 2+1)
   (-(d\textsuperscript 2+e\textsuperscript 2+f\textsuperscript 2+1) (-m b\textsuperscript 2+c k b+a c m b+a n b-a c\textsuperscript 2 k-a k-m+c
   n)+(b d+c e+f) (b e k-a c e k-a f k-b d m+a c d m-f m+a d n+e n)+(-e b\textsuperscript 2+c d
   b+a c e b+a f b-a c\textsuperscript 2 d-a d-e+c f) (d k+e m+f n+p))+(d g+e h+f i+j) (-(d
   g+e h+f i+j) (-m b\textsuperscript 2+c k b+a c m b+a n b-a c\textsuperscript 2 k-a k-m+c n)+(b d+c e+f) (b h
   k-a c h k-a i k-b g m+a c g m-i m+a g n+h n)+(-h b\textsuperscript 2+c g b+a c h b+a i b-a c\textsuperscript 2 g-a
   g-h+c i) (d k+e m+f n+p))-(b g+c h+i) (-(d g+e h+f i+j) (b e k-a c e k-a
   f k-b d m+a c d m-f m+a d n+e n)+(d\textsuperscript 2+e\textsuperscript 2+f\textsuperscript 2+1) (b h k-a c h k-a i k-b g m+a
   c g m-i m+a g n+h n)+(b e g-a c e g-a f g-b d h+a c d h-f h+a d i+e i) (d k+e m+f
   n+p))+(-(d\textsuperscript 2+e\textsuperscript 2+f\textsuperscript 2+1) (-h b\textsuperscript 2+c g b+a c h b+a i b-a c\textsuperscript 2 g-a
   g-h+c i)+(b d+c e+f) (b e g-a c e g-a f g-b d h+a c d h-f h+a d i+e i)+(-e
   b\textsuperscript 2+c d b+a c e b+a f b-a c\textsuperscript 2 d-a d-e+c f) (d g+e h+f i+j)) (g k+h m+i n+j
   p+q)))-b (a k+m) (-m+c n+e p-c f p+h q-c i q-e j q+c f j q) (-(g k+h m+i
   n+j p+q) (-b j p f\textsuperscript 2+a c j p f\textsuperscript 2-b q f\textsuperscript 2+a c q f\textsuperscript 2+b j n f-a c j n f-g p f+a h p f+b
   i p f-a c i p f+d j p f-a e j p f+d q f-a e q f+g n-a h n-b i n+a c i n-d j n+a e j n-b
   q+a c q)+(d k+e m+f n+p) (b p i\textsuperscript 2-a c p i\textsuperscript 2-b j n i+a c j n i-g p i+a h p i+d
   j p i-a e j p i-b f j p i+a c f j p i+d q i-a e q i-b f q i+a c f q i-d j\textsuperscript 2 n+a e j\textsuperscript 2 n+b
   f j\textsuperscript 2 n-a c f j\textsuperscript 2 n-d n+a e n+b f n-a c f n+g j n-a h j n+b p-a c p-b j q+a c j
   q)-(a k+m) (a j\textsuperscript 2 n e\textsuperscript 2+a n e\textsuperscript 2-a i j p e\textsuperscript 2-a i q e\textsuperscript 2-a f j\textsuperscript 2 m e-a f m e+a i
   j m e-d j\textsuperscript 2 n e+b f j\textsuperscript 2 n e-a c f j\textsuperscript 2 n e-d n e+b f n e-a c f n e+g j n e-2 a h j n e-b i
   j n e+a c i j n e+b i\textsuperscript 2 p e-a c i\textsuperscript 2 p e+b p e-a c p e-g i p e+a h i p e+a f h j p e+d i j
   p e-b f i j p e+a c f i j p e+a f h q e+d i q e-b f i q e+a c f i q e-b j q e+a c j q e-b
   f\textsuperscript 2 m+a c f\textsuperscript 2 m-b i\textsuperscript 2 m+a c i\textsuperscript 2 m-b f\textsuperscript 2 j\textsuperscript 2 m+a c f\textsuperscript 2 j\textsuperscript 2 m+d f j\textsuperscript 2 m-b m+a c m+d f m+g i
   m-a h i m-f g j m+a f h j m-d i j m+2 b f i j m-2 a c f i j m+a h\textsuperscript 2 n+a n-g h n+b h i n-a
   c h i n+d h j n-b f h j n+a c f h j n-a f h\textsuperscript 2 p-a f p+f g h p-b f h i p+a c f h i p+b f\textsuperscript 2
   h j p-a c f\textsuperscript 2 h j p-d f h j p+b f\textsuperscript 2 h q-a c f\textsuperscript 2 h q+b h q-a c h q-d f h q-a i q+a f j
   q)+k (c f\textsuperscript 2 m a\textsuperscript 2+c i\textsuperscript 2 m a\textsuperscript 2+c f\textsuperscript 2 j\textsuperscript 2 m a\textsuperscript 2-e f j\textsuperscript 2 m a\textsuperscript 2+c m a\textsuperscript 2-e f m
   a\textsuperscript 2-h i m a\textsuperscript 2+f h j m a\textsuperscript 2+e i j m a\textsuperscript 2-2 c f i j m a\textsuperscript 2+e\textsuperscript 2 n a\textsuperscript 2+h\textsuperscript 2 n a\textsuperscript 2+e\textsuperscript 2 j\textsuperscript 2 n a\textsuperscript 2-c
   e f j\textsuperscript 2 n a\textsuperscript 2-c e f n a\textsuperscript 2-c h i n a\textsuperscript 2-2 e h j n a\textsuperscript 2+c f h j n a\textsuperscript 2+c e i j n a\textsuperscript 2+n a\textsuperscript 2-f
   h\textsuperscript 2 p a\textsuperscript 2-c e i\textsuperscript 2 p a\textsuperscript 2-c e p a\textsuperscript 2-f p a\textsuperscript 2+e h i p a\textsuperscript 2+c f h i p a\textsuperscript 2-c f\textsuperscript 2 h j p a\textsuperscript 2+e f h
   j p a\textsuperscript 2-e\textsuperscript 2 i j p a\textsuperscript 2+c e f i j p a\textsuperscript 2-c f\textsuperscript 2 h q a\textsuperscript 2-c h q a\textsuperscript 2+e f h q a\textsuperscript 2-e\textsuperscript 2 i q a\textsuperscript 2+c e
   f i q a\textsuperscript 2-i q a\textsuperscript 2+c e j q a\textsuperscript 2+f j q a\textsuperscript 2-c f\textsuperscript 2 k a-c i\textsuperscript 2 k a-c f\textsuperscript 2 j\textsuperscript 2 k a+e f j\textsuperscript 2 k a-c k
   a+e f k a+h i k a-f h j k a-e i j k a+2 c f i j k a-b f\textsuperscript 2 m a-b i\textsuperscript 2 m a-b f\textsuperscript 2 j\textsuperscript 2 m a+d f
   j\textsuperscript 2 m a-b m a+d f m a+g i m a-f g j m a-d i j m a+2 b f i j m a-2 d e j\textsuperscript 2 n a+c d f j\textsuperscript 2 n
   a+b e f j\textsuperscript 2 n a-2 d e n a+c d f n a+b e f n a-2 g h n a+c g i n a+b h i n a+2 e g j n a-c
   f g j n a+2 d h j n a-b f h j n a-c d i j n a-b e i j n a+c d i\textsuperscript 2 p a+b e i\textsuperscript 2 p a+c d p
   a+b e p a+2 f g h p a-e g i p a-c f g i p a-d h i p a-b f h i p a+c f\textsuperscript 2 g j p a-e f g j p
   a+b f\textsuperscript 2 h j p a-d f h j p a+2 d e i j p a-c d f i j p a-b e f i j p a+c f\textsuperscript 2 g q a+c g q
   a-e f g q a+b f\textsuperscript 2 h q a+b h q a-d f h q a+2 d e i q a-c d f i q a-b e f i q a-c d j q a-b
   e j q a+b f\textsuperscript 2 k+b i\textsuperscript 2 k+b f\textsuperscript 2 j\textsuperscript 2 k-d f j\textsuperscript 2 k+b k-d f k-g i k+f g j k+d i j k-2 b f i j
   k+d\textsuperscript 2 n+g\textsuperscript 2 n+d\textsuperscript 2 j\textsuperscript 2 n-b d f j\textsuperscript 2 n-b d f n-b g i n-2 d g j n+b f g j n+b d i j n+n-f g\textsuperscript 2
   p-b d i\textsuperscript 2 p-b d p-f p+d g i p+b f g i p-b f\textsuperscript 2 g j p+d f g j p-d\textsuperscript 2 i j p+b d f i j p-b f\textsuperscript 2
   g q-b g q+d f g q-d\textsuperscript 2 i q+b d f i q-i q+b d j q+f j q)+(-b j\textsuperscript 2 f\textsuperscript 2+a c j\textsuperscript 2
   f\textsuperscript 2-b f\textsuperscript 2+a c f\textsuperscript 2+d j\textsuperscript 2 f-a e j\textsuperscript 2 f+d f-a e f-g j f+a h j f+2 b i j f-2 a c i j f-b i\textsuperscript 2+a
   c i\textsuperscript 2-b+a c+g i-a h i-d i j+a e i j)
   (k\textsuperscript 2+m\textsuperscript 2+n\textsuperscript 2+p\textsuperscript 2+q\textsuperscript 2+1)))+(d k+e m+f n+p) (j q-p) (a
   (b\textsuperscript 2+c\textsuperscript 2+1) (j\textsuperscript 2 f\textsuperscript 2+p\textsuperscript 2 f\textsuperscript 2+j\textsuperscript 2 q\textsuperscript 2 f\textsuperscript 2-2 j p q f\textsuperscript 2+f\textsuperscript 2-2 i j q\textsuperscript 2 f-2 i
   j f-2 n p f+2 j n q f+2 i p q f+i\textsuperscript 2+n\textsuperscript 2+i\textsuperscript 2 q\textsuperscript 2-2 i n q+1) ((g k+h m+i n+j
   p+q) (a i n c\textsuperscript 2-a f j n c\textsuperscript 2-a f i p c\textsuperscript 2+a f\textsuperscript 2 j p c\textsuperscript 2+a f\textsuperscript 2 q c\textsuperscript 2+a q c\textsuperscript 2-a i m c+a
   f j m c+g n c-a h n c-b i n c-d j n c+a e j n c+b f j n c-f g p c+a f h p c+a e i p c+b f
   i p c-b f\textsuperscript 2 j p c+d f j p c-2 a e f j p c-b f\textsuperscript 2 q c-b q c+d f q c-2 a e f q c-g m+a h m+b
   i m+d j m-a e j m-b f j m+e g p-a e h p-b e i p+a e\textsuperscript 2 j p-d e j p+b e f j p+a e\textsuperscript 2 q+a q-d
   e q+b e f q)-(d k+e m+f n+p) (-a f j\textsuperscript 2 n c\textsuperscript 2-a f n c\textsuperscript 2+a i j n c\textsuperscript 2-a i\textsuperscript 2 p
   c\textsuperscript 2-a p c\textsuperscript 2+a f i j p c\textsuperscript 2+a f i q c\textsuperscript 2+a j q c\textsuperscript 2+a f j\textsuperscript 2 m c+a f m c-a i j m c-d j\textsuperscript 2 n c+a
   e j\textsuperscript 2 n c+b f j\textsuperscript 2 n c-d n c+a e n c+b f n c+g j n c-a h j n c-b i j n c+b i\textsuperscript 2 p c+b p c-g
   i p c+2 a h i p c-a f h j p c+d i j p c-a e i j p c-b f i j p c-a f h q c+d i q c-a e i q
   c-b f i q c-b j q c+d j\textsuperscript 2 m-a e j\textsuperscript 2 m-b f j\textsuperscript 2 m+d m-a e m-b f m-g j m+a h j m+b i j m-a
   h\textsuperscript 2 p-a p+g h p-b h i p-d h j p+a e h j p+b f h j p-d h q+a e h q+b f h q+a j q)+(b
   k+c m+n) (a j\textsuperscript 2 n e\textsuperscript 2+a n e\textsuperscript 2-a i j p e\textsuperscript 2-a i q e\textsuperscript 2-a f j\textsuperscript 2 m e-a f m e+a i j m e-d
   j\textsuperscript 2 n e+b f j\textsuperscript 2 n e-a c f j\textsuperscript 2 n e-d n e+b f n e-a c f n e+g j n e-2 a h j n e-b i j n e+a
   c i j n e+b i\textsuperscript 2 p e-a c i\textsuperscript 2 p e+b p e-a c p e-g i p e+a h i p e+a f h j p e+d i j p e-b f
   i j p e+a c f i j p e+a f h q e+d i q e-b f i q e+a c f i q e-b j q e+a c j q e-b f\textsuperscript 2 m+a
   c f\textsuperscript 2 m-b i\textsuperscript 2 m+a c i\textsuperscript 2 m-b f\textsuperscript 2 j\textsuperscript 2 m+a c f\textsuperscript 2 j\textsuperscript 2 m+d f j\textsuperscript 2 m-b m+a c m+d f m+g i m-a h i
   m-f g j m+a f h j m-d i j m+2 b f i j m-2 a c f i j m+a h\textsuperscript 2 n+a n-g h n+b h i n-a c h i
   n+d h j n-b f h j n+a c f h j n-a f h\textsuperscript 2 p-a f p+f g h p-b f h i p+a c f h i p+b f\textsuperscript 2 h j
   p-a c f\textsuperscript 2 h j p-d f h j p+b f\textsuperscript 2 h q-a c f\textsuperscript 2 h q+b h q-a c h q-d f h q-a i q+a f j
   q)-(a f\textsuperscript 2 c\textsuperscript 2+a i\textsuperscript 2 c\textsuperscript 2+a f\textsuperscript 2 j\textsuperscript 2 c\textsuperscript 2+a c\textsuperscript 2-2 a f i j c\textsuperscript 2-b f\textsuperscript 2 c-b i\textsuperscript 2 c-b
   f\textsuperscript 2 j\textsuperscript 2 c+d f j\textsuperscript 2 c-2 a e f j\textsuperscript 2 c-b c+d f c-2 a e f c+g i c-2 a h i c-f g j c+2 a f h j
   c-d i j c+2 a e i j c+2 b f i j c+a e\textsuperscript 2+a h\textsuperscript 2+a e\textsuperscript 2 j\textsuperscript 2-d e j\textsuperscript 2+b e f j\textsuperscript 2+a-d e+b e f-g
   h+b h i+e g j+d h j-2 a e h j-b f h j-b e i j) (k\textsuperscript 2+m\textsuperscript 2+n\textsuperscript 2+p\textsuperscript 2+q\textsuperscript 2+1)-k
   (-(g\textsuperscript 2+h\textsuperscript 2+i\textsuperscript 2+j\textsuperscript 2+1) (-(d\textsuperscript 2+e\textsuperscript 2+f\textsuperscript 2+1) (-m b\textsuperscript 2+c k
   b+a c m b+a n b-a c\textsuperscript 2 k-a k-m+c n)+(b d+c e+f) (b e k-a c e k-a f k-b d m+a c d m-f
   m+a d n+e n)+(-e b\textsuperscript 2+c d b+a c e b+a f b-a c\textsuperscript 2 d-a d-e+c f) (d k+e m+f
   n+p))+(d g+e h+f i+j) (-(d g+e h+f i+j) (-m b\textsuperscript 2+c k b+a c m b+a n b-a c\textsuperscript 2
   k-a k-m+c n)+(b d+c e+f) (b h k-a c h k-a i k-b g m+a c g m-i m+a g n+h n)+(-h
   b\textsuperscript 2+c g b+a c h b+a i b-a c\textsuperscript 2 g-a g-h+c i) (d k+e m+f n+p))-(b g+c h+i)
   (-(d g+e h+f i+j) (b e k-a c e k-a f k-b d m+a c d m-f m+a d n+e
   n)+(d\textsuperscript 2+e\textsuperscript 2+f\textsuperscript 2+1) (b h k-a c h k-a i k-b g m+a c g m-i m+a g n+h n)+(b e g-a
   c e g-a f g-b d h+a c d h-f h+a d i+e i) (d k+e m+f
   n+p))+(-(d\textsuperscript 2+e\textsuperscript 2+f\textsuperscript 2+1) (-h b\textsuperscript 2+c g b+a c h b+a i b-a c\textsuperscript 2 g-a
   g-h+c i)+(b d+c e+f) (b e g-a c e g-a f g-b d h+a c d h-f h+a d i+e i)+(-e
   b\textsuperscript 2+c d b+a c e b+a f b-a c\textsuperscript 2 d-a d-e+c f) (d g+e h+f i+j)) (g k+h m+i n+j
   p+q)))-b (a b+c) (-c j\textsuperscript 2 f\textsuperscript 2-c p\textsuperscript 2 f\textsuperscript 2-c j\textsuperscript 2 q\textsuperscript 2 f\textsuperscript 2-c f\textsuperscript 2+2 c j p q
   f\textsuperscript 2+e j\textsuperscript 2 f+e p\textsuperscript 2 f+e j\textsuperscript 2 q\textsuperscript 2 f-h j q\textsuperscript 2 f+2 c i j q\textsuperscript 2 f+e f-h j f+2 c i j f-m p f+2 c n p
   f+j m q f-2 c j n q f+h p q f-2 c i p q f-2 e j p q f-c i\textsuperscript 2-c n\textsuperscript 2-c i\textsuperscript 2 q\textsuperscript 2+h i q\textsuperscript 2-e i j
   q\textsuperscript 2-c+h i-e i j+m n-e n p-i m q-h n q+2 c i n q+e j n q+e i p q) (-(g k+h m+i
   n+j p+q) (-b j p f\textsuperscript 2+a c j p f\textsuperscript 2-b q f\textsuperscript 2+a c q f\textsuperscript 2+b j n f-a c j n f-g p f+a h p f+b
   i p f-a c i p f+d j p f-a e j p f+d q f-a e q f+g n-a h n-b i n+a c i n-d j n+a e j n-b
   q+a c q)+(d k+e m+f n+p) (b p i\textsuperscript 2-a c p i\textsuperscript 2-b j n i+a c j n i-g p i+a h p i+d
   j p i-a e j p i-b f j p i+a c f j p i+d q i-a e q i-b f q i+a c f q i-d j\textsuperscript 2 n+a e j\textsuperscript 2 n+b
   f j\textsuperscript 2 n-a c f j\textsuperscript 2 n-d n+a e n+b f n-a c f n+g j n-a h j n+b p-a c p-b j q+a c j
   q)-(a k+m) (a j\textsuperscript 2 n e\textsuperscript 2+a n e\textsuperscript 2-a i j p e\textsuperscript 2-a i q e\textsuperscript 2-a f j\textsuperscript 2 m e-a f m e+a i
   j m e-d j\textsuperscript 2 n e+b f j\textsuperscript 2 n e-a c f j\textsuperscript 2 n e-d n e+b f n e-a c f n e+g j n e-2 a h j n e-b i
   j n e+a c i j n e+b i\textsuperscript 2 p e-a c i\textsuperscript 2 p e+b p e-a c p e-g i p e+a h i p e+a f h j p e+d i j
   p e-b f i j p e+a c f i j p e+a f h q e+d i q e-b f i q e+a c f i q e-b j q e+a c j q e-b
   f\textsuperscript 2 m+a c f\textsuperscript 2 m-b i\textsuperscript 2 m+a c i\textsuperscript 2 m-b f\textsuperscript 2 j\textsuperscript 2 m+a c f\textsuperscript 2 j\textsuperscript 2 m+d f j\textsuperscript 2 m-b m+a c m+d f m+g i
   m-a h i m-f g j m+a f h j m-d i j m+2 b f i j m-2 a c f i j m+a h\textsuperscript 2 n+a n-g h n+b h i n-a
   c h i n+d h j n-b f h j n+a c f h j n-a f h\textsuperscript 2 p-a f p+f g h p-b f h i p+a c f h i p+b f\textsuperscript 2
   h j p-a c f\textsuperscript 2 h j p-d f h j p+b f\textsuperscript 2 h q-a c f\textsuperscript 2 h q+b h q-a c h q-d f h q-a i q+a f j
   q)+k (c f\textsuperscript 2 m a\textsuperscript 2+c i\textsuperscript 2 m a\textsuperscript 2+c f\textsuperscript 2 j\textsuperscript 2 m a\textsuperscript 2-e f j\textsuperscript 2 m a\textsuperscript 2+c m a\textsuperscript 2-e f m
   a\textsuperscript 2-h i m a\textsuperscript 2+f h j m a\textsuperscript 2+e i j m a\textsuperscript 2-2 c f i j m a\textsuperscript 2+e\textsuperscript 2 n a\textsuperscript 2+h\textsuperscript 2 n a\textsuperscript 2+e\textsuperscript 2 j\textsuperscript 2 n a\textsuperscript 2-c
   e f j\textsuperscript 2 n a\textsuperscript 2-c e f n a\textsuperscript 2-c h i n a\textsuperscript 2-2 e h j n a\textsuperscript 2+c f h j n a\textsuperscript 2+c e i j n a\textsuperscript 2+n a\textsuperscript 2-f
   h\textsuperscript 2 p a\textsuperscript 2-c e i\textsuperscript 2 p a\textsuperscript 2-c e p a\textsuperscript 2-f p a\textsuperscript 2+e h i p a\textsuperscript 2+c f h i p a\textsuperscript 2-c f\textsuperscript 2 h j p a\textsuperscript 2+e f h
   j p a\textsuperscript 2-e\textsuperscript 2 i j p a\textsuperscript 2+c e f i j p a\textsuperscript 2-c f\textsuperscript 2 h q a\textsuperscript 2-c h q a\textsuperscript 2+e f h q a\textsuperscript 2-e\textsuperscript 2 i q a\textsuperscript 2+c e
   f i q a\textsuperscript 2-i q a\textsuperscript 2+c e j q a\textsuperscript 2+f j q a\textsuperscript 2-c f\textsuperscript 2 k a-c i\textsuperscript 2 k a-c f\textsuperscript 2 j\textsuperscript 2 k a+e f j\textsuperscript 2 k a-c k
   a+e f k a+h i k a-f h j k a-e i j k a+2 c f i j k a-b f\textsuperscript 2 m a-b i\textsuperscript 2 m a-b f\textsuperscript 2 j\textsuperscript 2 m a+d f
   j\textsuperscript 2 m a-b m a+d f m a+g i m a-f g j m a-d i j m a+2 b f i j m a-2 d e j\textsuperscript 2 n a+c d f j\textsuperscript 2 n
   a+b e f j\textsuperscript 2 n a-2 d e n a+c d f n a+b e f n a-2 g h n a+c g i n a+b h i n a+2 e g j n a-c
   f g j n a+2 d h j n a-b f h j n a-c d i j n a-b e i j n a+c d i\textsuperscript 2 p a+b e i\textsuperscript 2 p a+c d p
   a+b e p a+2 f g h p a-e g i p a-c f g i p a-d h i p a-b f h i p a+c f\textsuperscript 2 g j p a-e f g j p
   a+b f\textsuperscript 2 h j p a-d f h j p a+2 d e i j p a-c d f i j p a-b e f i j p a+c f\textsuperscript 2 g q a+c g q
   a-e f g q a+b f\textsuperscript 2 h q a+b h q a-d f h q a+2 d e i q a-c d f i q a-b e f i q a-c d j q a-b
   e j q a+b f\textsuperscript 2 k+b i\textsuperscript 2 k+b f\textsuperscript 2 j\textsuperscript 2 k-d f j\textsuperscript 2 k+b k-d f k-g i k+f g j k+d i j k-2 b f i j
   k+d\textsuperscript 2 n+g\textsuperscript 2 n+d\textsuperscript 2 j\textsuperscript 2 n-b d f j\textsuperscript 2 n-b d f n-b g i n-2 d g j n+b f g j n+b d i j n+n-f g\textsuperscript 2
   p-b d i\textsuperscript 2 p-b d p-f p+d g i p+b f g i p-b f\textsuperscript 2 g j p+d f g j p-d\textsuperscript 2 i j p+b d f i j p-b f\textsuperscript 2
   g q-b g q+d f g q-d\textsuperscript 2 i q+b d f i q-i q+b d j q+f j q)+(-b j\textsuperscript 2 f\textsuperscript 2+a c j\textsuperscript 2
   f\textsuperscript 2-b f\textsuperscript 2+a c f\textsuperscript 2+d j\textsuperscript 2 f-a e j\textsuperscript 2 f+d f-a e f-g j f+a h j f+2 b i j f-2 a c i j f-b i\textsuperscript 2+a
   c i\textsuperscript 2-b+a c+g i-a h i-d i j+a e i j)
   (k\textsuperscript 2+m\textsuperscript 2+n\textsuperscript 2+p\textsuperscript 2+q\textsuperscript 2+1))))-q (g k+h m+i n+j p+q) (d
   (-d q\textsuperscript 2 j\textsuperscript 2+a e q\textsuperscript 2 j\textsuperscript 2+b f q\textsuperscript 2 j\textsuperscript 2-a c f q\textsuperscript 2 j\textsuperscript 2-d j\textsuperscript 2+a e j\textsuperscript 2+b f j\textsuperscript 2-a c f j\textsuperscript 2+g
   q\textsuperscript 2 j-a h q\textsuperscript 2 j-b i q\textsuperscript 2 j+a c i q\textsuperscript 2 j+g j-a h j-b i j+a c i j-k q j+a m q j+b n q j-a c n
   q j+2 d p q j-2 a e p q j-2 b f p q j+2 a c f p q j-d p\textsuperscript 2+a e p\textsuperscript 2+b f p\textsuperscript 2-a c f p\textsuperscript 2-d+a
   e+b f-a c f+k p-a m p-b n p+a c n p-g p q+a h p q+b i p q-a c i p q) ((a b+c)
   (b d+c e+f) (-f q\textsuperscript 2 j\textsuperscript 2-f j\textsuperscript 2+i q\textsuperscript 2 j+i j-n q j+2 f p q j-f p\textsuperscript 2-f+n p-i p q)
   (-c j\textsuperscript 2 f\textsuperscript 2-c p\textsuperscript 2 f\textsuperscript 2-c j\textsuperscript 2 q\textsuperscript 2 f\textsuperscript 2-c f\textsuperscript 2+2 c j p q f\textsuperscript 2+e j\textsuperscript 2 f+e p\textsuperscript 2 f+e j\textsuperscript 2 q\textsuperscript 2
   f-h j q\textsuperscript 2 f+2 c i j q\textsuperscript 2 f+e f-h j f+2 c i j f-m p f+2 c n p f+j m q f-2 c j n q f+h p q
   f-2 c i p q f-2 e j p q f-c i\textsuperscript 2-c n\textsuperscript 2-c i\textsuperscript 2 q\textsuperscript 2+h i q\textsuperscript 2-e i j q\textsuperscript 2-c+h i-e i j+m n-e n p-i
   m q-h n q+2 c i n q+e j n q+e i p q)-(b\textsuperscript 2+c\textsuperscript 2+1) (a d+e) (-e q\textsuperscript 2
   j\textsuperscript 2+c f q\textsuperscript 2 j\textsuperscript 2-e j\textsuperscript 2+c f j\textsuperscript 2+h q\textsuperscript 2 j-c i q\textsuperscript 2 j+h j-c i j-m q j+c n q j+2 e p q j-2 c f p
   q j-e p\textsuperscript 2+c f p\textsuperscript 2-e+c f+m p-c n p-h p q+c i p q) (j\textsuperscript 2 f\textsuperscript 2+p\textsuperscript 2 f\textsuperscript 2+j\textsuperscript 2 q\textsuperscript 2
   f\textsuperscript 2-2 j p q f\textsuperscript 2+f\textsuperscript 2-2 i j q\textsuperscript 2 f-2 i j f-2 n p f+2 j n q f+2 i p q f+i\textsuperscript 2+n\textsuperscript 2+i\textsuperscript 2 q\textsuperscript 2-2 i n
   q+1))-(b d+c e+f) (-f q\textsuperscript 2 j\textsuperscript 2-f j\textsuperscript 2+i q\textsuperscript 2 j+i j-n q j+2 f p q j-f
   p\textsuperscript 2-f+n p-i p q) (a (b d+c e+f) (-f q\textsuperscript 2 j\textsuperscript 2-f j\textsuperscript 2+i q\textsuperscript 2 j+i j-n q j+2 f p
   q j-f p\textsuperscript 2-f+n p-i p q) ((g k+h m+i n+j p+q) (a i n c\textsuperscript 2-a f j n c\textsuperscript 2-a f i
   p c\textsuperscript 2+a f\textsuperscript 2 j p c\textsuperscript 2+a f\textsuperscript 2 q c\textsuperscript 2+a q c\textsuperscript 2-a i m c+a f j m c+g n c-a h n c-b i n c-d j n c+a
   e j n c+b f j n c-f g p c+a f h p c+a e i p c+b f i p c-b f\textsuperscript 2 j p c+d f j p c-2 a e f j p
   c-b f\textsuperscript 2 q c-b q c+d f q c-2 a e f q c-g m+a h m+b i m+d j m-a e j m-b f j m+e g p-a e h
   p-b e i p+a e\textsuperscript 2 j p-d e j p+b e f j p+a e\textsuperscript 2 q+a q-d e q+b e f q)-(d k+e m+f n+p)
   (-a f j\textsuperscript 2 n c\textsuperscript 2-a f n c\textsuperscript 2+a i j n c\textsuperscript 2-a i\textsuperscript 2 p c\textsuperscript 2-a p c\textsuperscript 2+a f i j p c\textsuperscript 2+a f i q
   c\textsuperscript 2+a j q c\textsuperscript 2+a f j\textsuperscript 2 m c+a f m c-a i j m c-d j\textsuperscript 2 n c+a e j\textsuperscript 2 n c+b f j\textsuperscript 2 n c-d n c+a e n
   c+b f n c+g j n c-a h j n c-b i j n c+b i\textsuperscript 2 p c+b p c-g i p c+2 a h i p c-a f h j p c+d i
   j p c-a e i j p c-b f i j p c-a f h q c+d i q c-a e i q c-b f i q c-b j q c+d j\textsuperscript 2 m-a e
   j\textsuperscript 2 m-b f j\textsuperscript 2 m+d m-a e m-b f m-g j m+a h j m+b i j m-a h\textsuperscript 2 p-a p+g h p-b h i p-d h j p+a
   e h j p+b f h j p-d h q+a e h q+b f h q+a j q)+(b k+c m+n) (a j\textsuperscript 2 n e\textsuperscript 2+a n
   e\textsuperscript 2-a i j p e\textsuperscript 2-a i q e\textsuperscript 2-a f j\textsuperscript 2 m e-a f m e+a i j m e-d j\textsuperscript 2 n e+b f j\textsuperscript 2 n e-a c f j\textsuperscript 2 n
   e-d n e+b f n e-a c f n e+g j n e-2 a h j n e-b i j n e+a c i j n e+b i\textsuperscript 2 p e-a c i\textsuperscript 2 p
   e+b p e-a c p e-g i p e+a h i p e+a f h j p e+d i j p e-b f i j p e+a c f i j p e+a f h q
   e+d i q e-b f i q e+a c f i q e-b j q e+a c j q e-b f\textsuperscript 2 m+a c f\textsuperscript 2 m-b i\textsuperscript 2 m+a c i\textsuperscript 2 m-b
   f\textsuperscript 2 j\textsuperscript 2 m+a c f\textsuperscript 2 j\textsuperscript 2 m+d f j\textsuperscript 2 m-b m+a c m+d f m+g i m-a h i m-f g j m+a f h j m-d i j
   m+2 b f i j m-2 a c f i j m+a h\textsuperscript 2 n+a n-g h n+b h i n-a c h i n+d h j n-b f h j n+a c f h
   j n-a f h\textsuperscript 2 p-a f p+f g h p-b f h i p+a c f h i p+b f\textsuperscript 2 h j p-a c f\textsuperscript 2 h j p-d f h j p+b
   f\textsuperscript 2 h q-a c f\textsuperscript 2 h q+b h q-a c h q-d f h q-a i q+a f j q)-(a f\textsuperscript 2 c\textsuperscript 2+a i\textsuperscript 2
   c\textsuperscript 2+a f\textsuperscript 2 j\textsuperscript 2 c\textsuperscript 2+a c\textsuperscript 2-2 a f i j c\textsuperscript 2-b f\textsuperscript 2 c-b i\textsuperscript 2 c-b f\textsuperscript 2 j\textsuperscript 2 c+d f j\textsuperscript 2 c-2 a e f j\textsuperscript 2
   c-b c+d f c-2 a e f c+g i c-2 a h i c-f g j c+2 a f h j c-d i j c+2 a e i j c+2 b f i j
   c+a e\textsuperscript 2+a h\textsuperscript 2+a e\textsuperscript 2 j\textsuperscript 2-d e j\textsuperscript 2+b e f j\textsuperscript 2+a-d e+b e f-g h+b h i+e g j+d h j-2 a e h j-b f
   h j-b e i j) (k\textsuperscript 2+m\textsuperscript 2+n\textsuperscript 2+p\textsuperscript 2+q\textsuperscript 2+1)-k
   (-(g\textsuperscript 2+h\textsuperscript 2+i\textsuperscript 2+j\textsuperscript 2+1) (-(d\textsuperscript 2+e\textsuperscript 2+f\textsuperscript 2+1) (-m b\textsuperscript 2+c k
   b+a c m b+a n b-a c\textsuperscript 2 k-a k-m+c n)+(b d+c e+f) (b e k-a c e k-a f k-b d m+a c d m-f
   m+a d n+e n)+(-e b\textsuperscript 2+c d b+a c e b+a f b-a c\textsuperscript 2 d-a d-e+c f) (d k+e m+f
   n+p))+(d g+e h+f i+j) (-(d g+e h+f i+j) (-m b\textsuperscript 2+c k b+a c m b+a n b-a c\textsuperscript 2
   k-a k-m+c n)+(b d+c e+f) (b h k-a c h k-a i k-b g m+a c g m-i m+a g n+h n)+(-h
   b\textsuperscript 2+c g b+a c h b+a i b-a c\textsuperscript 2 g-a g-h+c i) (d k+e m+f n+p))-(b g+c h+i)
   (-(d g+e h+f i+j) (b e k-a c e k-a f k-b d m+a c d m-f m+a d n+e
   n)+(d\textsuperscript 2+e\textsuperscript 2+f\textsuperscript 2+1) (b h k-a c h k-a i k-b g m+a c g m-i m+a g n+h n)+(b e g-a
   c e g-a f g-b d h+a c d h-f h+a d i+e i) (d k+e m+f
   n+p))+(-(d\textsuperscript 2+e\textsuperscript 2+f\textsuperscript 2+1) (-h b\textsuperscript 2+c g b+a c h b+a i b-a c\textsuperscript 2 g-a
   g-h+c i)+(b d+c e+f) (b e g-a c e g-a f g-b d h+a c d h-f h+a d i+e i)+(-e
   b\textsuperscript 2+c d b+a c e b+a f b-a c\textsuperscript 2 d-a d-e+c f) (d g+e h+f i+j)) (g k+h m+i n+j
   p+q)))-b (a d+e) (-e q\textsuperscript 2 j\textsuperscript 2+c f q\textsuperscript 2 j\textsuperscript 2-e j\textsuperscript 2+c f j\textsuperscript 2+h q\textsuperscript 2 j-c i q\textsuperscript 2
   j+h j-c i j-m q j+c n q j+2 e p q j-2 c f p q j-e p\textsuperscript 2+c f p\textsuperscript 2-e+c f+m p-c n p-h p q+c i p
   q) (-(g k+h m+i n+j p+q) (-b j p f\textsuperscript 2+a c j p f\textsuperscript 2-b q f\textsuperscript 2+a c q f\textsuperscript 2+b j n
   f-a c j n f-g p f+a h p f+b i p f-a c i p f+d j p f-a e j p f+d q f-a e q f+g n-a h n-b i
   n+a c i n-d j n+a e j n-b q+a c q)+(d k+e m+f n+p) (b p i\textsuperscript 2-a c p i\textsuperscript 2-b j n
   i+a c j n i-g p i+a h p i+d j p i-a e j p i-b f j p i+a c f j p i+d q i-a e q i-b f q i+a
   c f q i-d j\textsuperscript 2 n+a e j\textsuperscript 2 n+b f j\textsuperscript 2 n-a c f j\textsuperscript 2 n-d n+a e n+b f n-a c f n+g j n-a h j n+b
   p-a c p-b j q+a c j q)-(a k+m) (a j\textsuperscript 2 n e\textsuperscript 2+a n e\textsuperscript 2-a i j p e\textsuperscript 2-a i q e\textsuperscript 2-a f
   j\textsuperscript 2 m e-a f m e+a i j m e-d j\textsuperscript 2 n e+b f j\textsuperscript 2 n e-a c f j\textsuperscript 2 n e-d n e+b f n e-a c f n e+g j
   n e-2 a h j n e-b i j n e+a c i j n e+b i\textsuperscript 2 p e-a c i\textsuperscript 2 p e+b p e-a c p e-g i p e+a h i p
   e+a f h j p e+d i j p e-b f i j p e+a c f i j p e+a f h q e+d i q e-b f i q e+a c f i q
   e-b j q e+a c j q e-b f\textsuperscript 2 m+a c f\textsuperscript 2 m-b i\textsuperscript 2 m+a c i\textsuperscript 2 m-b f\textsuperscript 2 j\textsuperscript 2 m+a c f\textsuperscript 2 j\textsuperscript 2 m+d f j\textsuperscript 2
   m-b m+a c m+d f m+g i m-a h i m-f g j m+a f h j m-d i j m+2 b f i j m-2 a c f i j m+a h\textsuperscript 2
   n+a n-g h n+b h i n-a c h i n+d h j n-b f h j n+a c f h j n-a f h\textsuperscript 2 p-a f p+f g h p-b f h
   i p+a c f h i p+b f\textsuperscript 2 h j p-a c f\textsuperscript 2 h j p-d f h j p+b f\textsuperscript 2 h q-a c f\textsuperscript 2 h q+b h q-a c h q-d
   f h q-a i q+a f j q)+k (c f\textsuperscript 2 m a\textsuperscript 2+c i\textsuperscript 2 m a\textsuperscript 2+c f\textsuperscript 2 j\textsuperscript 2 m a\textsuperscript 2-e f j\textsuperscript 2 m
   a\textsuperscript 2+c m a\textsuperscript 2-e f m a\textsuperscript 2-h i m a\textsuperscript 2+f h j m a\textsuperscript 2+e i j m a\textsuperscript 2-2 c f i j m a\textsuperscript 2+e\textsuperscript 2 n a\textsuperscript 2+h\textsuperscript 2 n
   a\textsuperscript 2+e\textsuperscript 2 j\textsuperscript 2 n a\textsuperscript 2-c e f j\textsuperscript 2 n a\textsuperscript 2-c e f n a\textsuperscript 2-c h i n a\textsuperscript 2-2 e h j n a\textsuperscript 2+c f h j n a\textsuperscript 2+c e
   i j n a\textsuperscript 2+n a\textsuperscript 2-f h\textsuperscript 2 p a\textsuperscript 2-c e i\textsuperscript 2 p a\textsuperscript 2-c e p a\textsuperscript 2-f p a\textsuperscript 2+e h i p a\textsuperscript 2+c f h i p a\textsuperscript 2-c
   f\textsuperscript 2 h j p a\textsuperscript 2+e f h j p a\textsuperscript 2-e\textsuperscript 2 i j p a\textsuperscript 2+c e f i j p a\textsuperscript 2-c f\textsuperscript 2 h q a\textsuperscript 2-c h q a\textsuperscript 2+e f h q
   a\textsuperscript 2-e\textsuperscript 2 i q a\textsuperscript 2+c e f i q a\textsuperscript 2-i q a\textsuperscript 2+c e j q a\textsuperscript 2+f j q a\textsuperscript 2-c f\textsuperscript 2 k a-c i\textsuperscript 2 k a-c f\textsuperscript 2 j\textsuperscript 2
   k a+e f j\textsuperscript 2 k a-c k a+e f k a+h i k a-f h j k a-e i j k a+2 c f i j k a-b f\textsuperscript 2 m a-b i\textsuperscript 2 m
   a-b f\textsuperscript 2 j\textsuperscript 2 m a+d f j\textsuperscript 2 m a-b m a+d f m a+g i m a-f g j m a-d i j m a+2 b f i j m a-2 d e
   j\textsuperscript 2 n a+c d f j\textsuperscript 2 n a+b e f j\textsuperscript 2 n a-2 d e n a+c d f n a+b e f n a-2 g h n a+c g i n a+b h
   i n a+2 e g j n a-c f g j n a+2 d h j n a-b f h j n a-c d i j n a-b e i j n a+c d i\textsuperscript 2 p
   a+b e i\textsuperscript 2 p a+c d p a+b e p a+2 f g h p a-e g i p a-c f g i p a-d h i p a-b f h i p a+c
   f\textsuperscript 2 g j p a-e f g j p a+b f\textsuperscript 2 h j p a-d f h j p a+2 d e i j p a-c d f i j p a-b e f i j p
   a+c f\textsuperscript 2 g q a+c g q a-e f g q a+b f\textsuperscript 2 h q a+b h q a-d f h q a+2 d e i q a-c d f i q a-b e
   f i q a-c d j q a-b e j q a+b f\textsuperscript 2 k+b i\textsuperscript 2 k+b f\textsuperscript 2 j\textsuperscript 2 k-d f j\textsuperscript 2 k+b k-d f k-g i k+f g j
   k+d i j k-2 b f i j k+d\textsuperscript 2 n+g\textsuperscript 2 n+d\textsuperscript 2 j\textsuperscript 2 n-b d f j\textsuperscript 2 n-b d f n-b g i n-2 d g j n+b f g j
   n+b d i j n+n-f g\textsuperscript 2 p-b d i\textsuperscript 2 p-b d p-f p+d g i p+b f g i p-b f\textsuperscript 2 g j p+d f g j p-d\textsuperscript 2 i j
   p+b d f i j p-b f\textsuperscript 2 g q-b g q+d f g q-d\textsuperscript 2 i q+b d f i q-i q+b d j q+f j q)+(-b
   j\textsuperscript 2 f\textsuperscript 2+a c j\textsuperscript 2 f\textsuperscript 2-b f\textsuperscript 2+a c f\textsuperscript 2+d j\textsuperscript 2 f-a e j\textsuperscript 2 f+d f-a e f-g j f+a h j f+2 b i j f-2 a
   c i j f-b i\textsuperscript 2+a c i\textsuperscript 2-b+a c+g i-a h i-d i j+a e i j)
   (k\textsuperscript 2+m\textsuperscript 2+n\textsuperscript 2+p\textsuperscript 2+q\textsuperscript 2+1)))+(d\textsuperscript 2+e\textsuperscript 2+f\textsuperscript 2+1) (q\textsuperscript 2
   j\textsuperscript 2+j\textsuperscript 2-2 p q j+p\textsuperscript 2+1) (a (b\textsuperscript 2+c\textsuperscript 2+1) (j\textsuperscript 2 f\textsuperscript 2+p\textsuperscript 2 f\textsuperscript 2+j\textsuperscript 2 q\textsuperscript 2
   f\textsuperscript 2-2 j p q f\textsuperscript 2+f\textsuperscript 2-2 i j q\textsuperscript 2 f-2 i j f-2 n p f+2 j n q f+2 i p q f+i\textsuperscript 2+n\textsuperscript 2+i\textsuperscript 2 q\textsuperscript 2-2 i n
   q+1) ((g k+h m+i n+j p+q) (a i n c\textsuperscript 2-a f j n c\textsuperscript 2-a f i p c\textsuperscript 2+a f\textsuperscript 2 j p
   c\textsuperscript 2+a f\textsuperscript 2 q c\textsuperscript 2+a q c\textsuperscript 2-a i m c+a f j m c+g n c-a h n c-b i n c-d j n c+a e j n c+b f j n
   c-f g p c+a f h p c+a e i p c+b f i p c-b f\textsuperscript 2 j p c+d f j p c-2 a e f j p c-b f\textsuperscript 2 q c-b q
   c+d f q c-2 a e f q c-g m+a h m+b i m+d j m-a e j m-b f j m+e g p-a e h p-b e i p+a e\textsuperscript 2 j
   p-d e j p+b e f j p+a e\textsuperscript 2 q+a q-d e q+b e f q)-(d k+e m+f n+p) (-a f j\textsuperscript 2 n
   c\textsuperscript 2-a f n c\textsuperscript 2+a i j n c\textsuperscript 2-a i\textsuperscript 2 p c\textsuperscript 2-a p c\textsuperscript 2+a f i j p c\textsuperscript 2+a f i q c\textsuperscript 2+a j q c\textsuperscript 2+a f j\textsuperscript 2
   m c+a f m c-a i j m c-d j\textsuperscript 2 n c+a e j\textsuperscript 2 n c+b f j\textsuperscript 2 n c-d n c+a e n c+b f n c+g j n c-a h
   j n c-b i j n c+b i\textsuperscript 2 p c+b p c-g i p c+2 a h i p c-a f h j p c+d i j p c-a e i j p c-b f
   i j p c-a f h q c+d i q c-a e i q c-b f i q c-b j q c+d j\textsuperscript 2 m-a e j\textsuperscript 2 m-b f j\textsuperscript 2 m+d m-a e
   m-b f m-g j m+a h j m+b i j m-a h\textsuperscript 2 p-a p+g h p-b h i p-d h j p+a e h j p+b f h j p-d h
   q+a e h q+b f h q+a j q)+(b k+c m+n) (a j\textsuperscript 2 n e\textsuperscript 2+a n e\textsuperscript 2-a i j p e\textsuperscript 2-a i q
   e\textsuperscript 2-a f j\textsuperscript 2 m e-a f m e+a i j m e-d j\textsuperscript 2 n e+b f j\textsuperscript 2 n e-a c f j\textsuperscript 2 n e-d n e+b f n e-a c f
   n e+g j n e-2 a h j n e-b i j n e+a c i j n e+b i\textsuperscript 2 p e-a c i\textsuperscript 2 p e+b p e-a c p e-g i p
   e+a h i p e+a f h j p e+d i j p e-b f i j p e+a c f i j p e+a f h q e+d i q e-b f i q e+a
   c f i q e-b j q e+a c j q e-b f\textsuperscript 2 m+a c f\textsuperscript 2 m-b i\textsuperscript 2 m+a c i\textsuperscript 2 m-b f\textsuperscript 2 j\textsuperscript 2 m+a c f\textsuperscript 2 j\textsuperscript 2
   m+d f j\textsuperscript 2 m-b m+a c m+d f m+g i m-a h i m-f g j m+a f h j m-d i j m+2 b f i j m-2 a c f i
   j m+a h\textsuperscript 2 n+a n-g h n+b h i n-a c h i n+d h j n-b f h j n+a c f h j n-a f h\textsuperscript 2 p-a f p+f g
   h p-b f h i p+a c f h i p+b f\textsuperscript 2 h j p-a c f\textsuperscript 2 h j p-d f h j p+b f\textsuperscript 2 h q-a c f\textsuperscript 2 h q+b h
   q-a c h q-d f h q-a i q+a f j q)-(a f\textsuperscript 2 c\textsuperscript 2+a i\textsuperscript 2 c\textsuperscript 2+a f\textsuperscript 2 j\textsuperscript 2 c\textsuperscript 2+a c\textsuperscript 2-2 a
   f i j c\textsuperscript 2-b f\textsuperscript 2 c-b i\textsuperscript 2 c-b f\textsuperscript 2 j\textsuperscript 2 c+d f j\textsuperscript 2 c-2 a e f j\textsuperscript 2 c-b c+d f c-2 a e f c+g i c-2
   a h i c-f g j c+2 a f h j c-d i j c+2 a e i j c+2 b f i j c+a e\textsuperscript 2+a h\textsuperscript 2+a e\textsuperscript 2 j\textsuperscript 2-d e
   j\textsuperscript 2+b e f j\textsuperscript 2+a-d e+b e f-g h+b h i+e g j+d h j-2 a e h j-b f h j-b e i j)
   (k\textsuperscript 2+m\textsuperscript 2+n\textsuperscript 2+p\textsuperscript 2+q\textsuperscript 2+1)-k (-(g\textsuperscript 2+h\textsuperscript 2+i\textsuperscript 2+j\textsuperscript 2+1)
   (-(d\textsuperscript 2+e\textsuperscript 2+f\textsuperscript 2+1) (-m b\textsuperscript 2+c k b+a c m b+a n b-a c\textsuperscript 2 k-a k-m+c
   n)+(b d+c e+f) (b e k-a c e k-a f k-b d m+a c d m-f m+a d n+e n)+(-e b\textsuperscript 2+c d
   b+a c e b+a f b-a c\textsuperscript 2 d-a d-e+c f) (d k+e m+f n+p))+(d g+e h+f i+j) (-(d
   g+e h+f i+j) (-m b\textsuperscript 2+c k b+a c m b+a n b-a c\textsuperscript 2 k-a k-m+c n)+(b d+c e+f) (b h
   k-a c h k-a i k-b g m+a c g m-i m+a g n+h n)+(-h b\textsuperscript 2+c g b+a c h b+a i b-a c\textsuperscript 2 g-a
   g-h+c i) (d k+e m+f n+p))-(b g+c h+i) (-(d g+e h+f i+j) (b e k-a c e k-a
   f k-b d m+a c d m-f m+a d n+e n)+(d\textsuperscript 2+e\textsuperscript 2+f\textsuperscript 2+1) (b h k-a c h k-a i k-b g m+a
   c g m-i m+a g n+h n)+(b e g-a c e g-a f g-b d h+a c d h-f h+a d i+e i) (d k+e m+f
   n+p))+(-(d\textsuperscript 2+e\textsuperscript 2+f\textsuperscript 2+1) (-h b\textsuperscript 2+c g b+a c h b+a i b-a c\textsuperscript 2 g-a
   g-h+c i)+(b d+c e+f) (b e g-a c e g-a f g-b d h+a c d h-f h+a d i+e i)+(-e
   b\textsuperscript 2+c d b+a c e b+a f b-a c\textsuperscript 2 d-a d-e+c f) (d g+e h+f i+j)) (g k+h m+i n+j
   p+q)))-b (a b+c) (-c j\textsuperscript 2 f\textsuperscript 2-c p\textsuperscript 2 f\textsuperscript 2-c j\textsuperscript 2 q\textsuperscript 2 f\textsuperscript 2-c f\textsuperscript 2+2 c j p q
   f\textsuperscript 2+e j\textsuperscript 2 f+e p\textsuperscript 2 f+e j\textsuperscript 2 q\textsuperscript 2 f-h j q\textsuperscript 2 f+2 c i j q\textsuperscript 2 f+e f-h j f+2 c i j f-m p f+2 c n p
   f+j m q f-2 c j n q f+h p q f-2 c i p q f-2 e j p q f-c i\textsuperscript 2-c n\textsuperscript 2-c i\textsuperscript 2 q\textsuperscript 2+h i q\textsuperscript 2-e i j
   q\textsuperscript 2-c+h i-e i j+m n-e n p-i m q-h n q+2 c i n q+e j n q+e i p q) (-(g k+h m+i
   n+j p+q) (-b j p f\textsuperscript 2+a c j p f\textsuperscript 2-b q f\textsuperscript 2+a c q f\textsuperscript 2+b j n f-a c j n f-g p f+a h p f+b
   i p f-a c i p f+d j p f-a e j p f+d q f-a e q f+g n-a h n-b i n+a c i n-d j n+a e j n-b
   q+a c q)+(d k+e m+f n+p) (b p i\textsuperscript 2-a c p i\textsuperscript 2-b j n i+a c j n i-g p i+a h p i+d
   j p i-a e j p i-b f j p i+a c f j p i+d q i-a e q i-b f q i+a c f q i-d j\textsuperscript 2 n+a e j\textsuperscript 2 n+b
   f j\textsuperscript 2 n-a c f j\textsuperscript 2 n-d n+a e n+b f n-a c f n+g j n-a h j n+b p-a c p-b j q+a c j
   q)-(a k+m) (a j\textsuperscript 2 n e\textsuperscript 2+a n e\textsuperscript 2-a i j p e\textsuperscript 2-a i q e\textsuperscript 2-a f j\textsuperscript 2 m e-a f m e+a i
   j m e-d j\textsuperscript 2 n e+b f j\textsuperscript 2 n e-a c f j\textsuperscript 2 n e-d n e+b f n e-a c f n e+g j n e-2 a h j n e-b i
   j n e+a c i j n e+b i\textsuperscript 2 p e-a c i\textsuperscript 2 p e+b p e-a c p e-g i p e+a h i p e+a f h j p e+d i j
   p e-b f i j p e+a c f i j p e+a f h q e+d i q e-b f i q e+a c f i q e-b j q e+a c j q e-b
   f\textsuperscript 2 m+a c f\textsuperscript 2 m-b i\textsuperscript 2 m+a c i\textsuperscript 2 m-b f\textsuperscript 2 j\textsuperscript 2 m+a c f\textsuperscript 2 j\textsuperscript 2 m+d f j\textsuperscript 2 m-b m+a c m+d f m+g i
   m-a h i m-f g j m+a f h j m-d i j m+2 b f i j m-2 a c f i j m+a h\textsuperscript 2 n+a n-g h n+b h i n-a
   c h i n+d h j n-b f h j n+a c f h j n-a f h\textsuperscript 2 p-a f p+f g h p-b f h i p+a c f h i p+b f\textsuperscript 2
   h j p-a c f\textsuperscript 2 h j p-d f h j p+b f\textsuperscript 2 h q-a c f\textsuperscript 2 h q+b h q-a c h q-d f h q-a i q+a f j
   q)+k (c f\textsuperscript 2 m a\textsuperscript 2+c i\textsuperscript 2 m a\textsuperscript 2+c f\textsuperscript 2 j\textsuperscript 2 m a\textsuperscript 2-e f j\textsuperscript 2 m a\textsuperscript 2+c m a\textsuperscript 2-e f m
   a\textsuperscript 2-h i m a\textsuperscript 2+f h j m a\textsuperscript 2+e i j m a\textsuperscript 2-2 c f i j m a\textsuperscript 2+e\textsuperscript 2 n a\textsuperscript 2+h\textsuperscript 2 n a\textsuperscript 2+e\textsuperscript 2 j\textsuperscript 2 n a\textsuperscript 2-c
   e f j\textsuperscript 2 n a\textsuperscript 2-c e f n a\textsuperscript 2-c h i n a\textsuperscript 2-2 e h j n a\textsuperscript 2+c f h j n a\textsuperscript 2+c e i j n a\textsuperscript 2+n a\textsuperscript 2-f
   h\textsuperscript 2 p a\textsuperscript 2-c e i\textsuperscript 2 p a\textsuperscript 2-c e p a\textsuperscript 2-f p a\textsuperscript 2+e h i p a\textsuperscript 2+c f h i p a\textsuperscript 2-c f\textsuperscript 2 h j p a\textsuperscript 2+e f h
   j p a\textsuperscript 2-e\textsuperscript 2 i j p a\textsuperscript 2+c e f i j p a\textsuperscript 2-c f\textsuperscript 2 h q a\textsuperscript 2-c h q a\textsuperscript 2+e f h q a\textsuperscript 2-e\textsuperscript 2 i q a\textsuperscript 2+c e
   f i q a\textsuperscript 2-i q a\textsuperscript 2+c e j q a\textsuperscript 2+f j q a\textsuperscript 2-c f\textsuperscript 2 k a-c i\textsuperscript 2 k a-c f\textsuperscript 2 j\textsuperscript 2 k a+e f j\textsuperscript 2 k a-c k
   a+e f k a+h i k a-f h j k a-e i j k a+2 c f i j k a-b f\textsuperscript 2 m a-b i\textsuperscript 2 m a-b f\textsuperscript 2 j\textsuperscript 2 m a+d f
   j\textsuperscript 2 m a-b m a+d f m a+g i m a-f g j m a-d i j m a+2 b f i j m a-2 d e j\textsuperscript 2 n a+c d f j\textsuperscript 2 n
   a+b e f j\textsuperscript 2 n a-2 d e n a+c d f n a+b e f n a-2 g h n a+c g i n a+b h i n a+2 e g j n a-c
   f g j n a+2 d h j n a-b f h j n a-c d i j n a-b e i j n a+c d i\textsuperscript 2 p a+b e i\textsuperscript 2 p a+c d p
   a+b e p a+2 f g h p a-e g i p a-c f g i p a-d h i p a-b f h i p a+c f\textsuperscript 2 g j p a-e f g j p
   a+b f\textsuperscript 2 h j p a-d f h j p a+2 d e i j p a-c d f i j p a-b e f i j p a+c f\textsuperscript 2 g q a+c g q
   a-e f g q a+b f\textsuperscript 2 h q a+b h q a-d f h q a+2 d e i q a-c d f i q a-b e f i q a-c d j q a-b
   e j q a+b f\textsuperscript 2 k+b i\textsuperscript 2 k+b f\textsuperscript 2 j\textsuperscript 2 k-d f j\textsuperscript 2 k+b k-d f k-g i k+f g j k+d i j k-2 b f i j
   k+d\textsuperscript 2 n+g\textsuperscript 2 n+d\textsuperscript 2 j\textsuperscript 2 n-b d f j\textsuperscript 2 n-b d f n-b g i n-2 d g j n+b f g j n+b d i j n+n-f g\textsuperscript 2
   p-b d i\textsuperscript 2 p-b d p-f p+d g i p+b f g i p-b f\textsuperscript 2 g j p+d f g j p-d\textsuperscript 2 i j p+b d f i j p-b f\textsuperscript 2
   g q-b g q+d f g q-d\textsuperscript 2 i q+b d f i q-i q+b d j q+f j q)+(-b j\textsuperscript 2 f\textsuperscript 2+a c j\textsuperscript 2
   f\textsuperscript 2-b f\textsuperscript 2+a c f\textsuperscript 2+d j\textsuperscript 2 f-a e j\textsuperscript 2 f+d f-a e f-g j f+a h j f+2 b i j f-2 a c i j f-b i\textsuperscript 2+a
   c i\textsuperscript 2-b+a c+g i-a h i-d i j+a e i j)
   (k\textsuperscript 2+m\textsuperscript 2+n\textsuperscript 2+p\textsuperscript 2+q\textsuperscript 2+1)))))-(k\textsuperscript 2+m\textsuperscript 2+n\textsuperscript 2+p\textsuperscript 2+q\textsuperscript 2+
   1) (-g (-g q\textsuperscript 2+a h q\textsuperscript 2+b i q\textsuperscript 2-a c i q\textsuperscript 2+d j q\textsuperscript 2-a e j q\textsuperscript 2-b f j q\textsuperscript 2+a c
   f j q\textsuperscript 2+k q-a m q-b n q+a c n q-d p q+a e p q+b f p q-a c f p q-g+a h+b i-a c i+d j-a e
   j-b f j+a c f j) ((b d+c e+f) (-f q\textsuperscript 2 j\textsuperscript 2-f j\textsuperscript 2+i q\textsuperscript 2 j+i j-n q j+2 f p q
   j-f p\textsuperscript 2-f+n p-i p q) ((a d+e) (b g+c h+i) (-i q\textsuperscript 2+f j q\textsuperscript 2+n q-f p q-i+f
   j) (-e q\textsuperscript 2 j\textsuperscript 2+c f q\textsuperscript 2 j\textsuperscript 2-e j\textsuperscript 2+c f j\textsuperscript 2+h q\textsuperscript 2 j-c i q\textsuperscript 2 j+h j-c i j-m q j+c n
   q j+2 e p q j-2 c f p q j-e p\textsuperscript 2+c f p\textsuperscript 2-e+c f+m p-c n p-h p q+c i p q)-(b d+c e+f)
   (a g+h) (-h q\textsuperscript 2+c i q\textsuperscript 2+e j q\textsuperscript 2-c f j q\textsuperscript 2+m q-c n q-e p q+c f p q-h+c i+e j-c f
   j) (-f q\textsuperscript 2 j\textsuperscript 2-f j\textsuperscript 2+i q\textsuperscript 2 j+i j-n q j+2 f p q j-f p\textsuperscript 2-f+n p-i p
   q))-(d\textsuperscript 2+e\textsuperscript 2+f\textsuperscript 2+1) (q\textsuperscript 2 j\textsuperscript 2+j\textsuperscript 2-2 p q j+p\textsuperscript 2+1)
   ((a b+c) (b g+c h+i) (-i q\textsuperscript 2+f j q\textsuperscript 2+n q-f p q-i+f j) (-c j\textsuperscript 2 f\textsuperscript 2-c
   p\textsuperscript 2 f\textsuperscript 2-c j\textsuperscript 2 q\textsuperscript 2 f\textsuperscript 2-c f\textsuperscript 2+2 c j p q f\textsuperscript 2+e j\textsuperscript 2 f+e p\textsuperscript 2 f+e j\textsuperscript 2 q\textsuperscript 2 f-h j q\textsuperscript 2 f+2 c i j
   q\textsuperscript 2 f+e f-h j f+2 c i j f-m p f+2 c n p f+j m q f-2 c j n q f+h p q f-2 c i p q f-2 e j p
   q f-c i\textsuperscript 2-c n\textsuperscript 2-c i\textsuperscript 2 q\textsuperscript 2+h i q\textsuperscript 2-e i j q\textsuperscript 2-c+h i-e i j+m n-e n p-i m q-h n q+2 c i n q+e
   j n q+e i p q)-(b\textsuperscript 2+c\textsuperscript 2+1) (a g+h) (-h q\textsuperscript 2+c i q\textsuperscript 2+e j q\textsuperscript 2-c f j
   q\textsuperscript 2+m q-c n q-e p q+c f p q-h+c i+e j-c f j) (j\textsuperscript 2 f\textsuperscript 2+p\textsuperscript 2 f\textsuperscript 2+j\textsuperscript 2 q\textsuperscript 2 f\textsuperscript 2-2 j
   p q f\textsuperscript 2+f\textsuperscript 2-2 i j q\textsuperscript 2 f-2 i j f-2 n p f+2 j n q f+2 i p q f+i\textsuperscript 2+n\textsuperscript 2+i\textsuperscript 2 q\textsuperscript 2-2 i n
   q+1))+(d g+e h+f i+j) (-j q\textsuperscript 2+p q-j) ((a b+c) (b d+c e+f)
   (-f q\textsuperscript 2 j\textsuperscript 2-f j\textsuperscript 2+i q\textsuperscript 2 j+i j-n q j+2 f p q j-f p\textsuperscript 2-f+n p-i p q) (-c j\textsuperscript 2
   f\textsuperscript 2-c p\textsuperscript 2 f\textsuperscript 2-c j\textsuperscript 2 q\textsuperscript 2 f\textsuperscript 2-c f\textsuperscript 2+2 c j p q f\textsuperscript 2+e j\textsuperscript 2 f+e p\textsuperscript 2 f+e j\textsuperscript 2 q\textsuperscript 2 f-h j q\textsuperscript 2 f+2 c
   i j q\textsuperscript 2 f+e f-h j f+2 c i j f-m p f+2 c n p f+j m q f-2 c j n q f+h p q f-2 c i p q f-2 e
   j p q f-c i\textsuperscript 2-c n\textsuperscript 2-c i\textsuperscript 2 q\textsuperscript 2+h i q\textsuperscript 2-e i j q\textsuperscript 2-c+h i-e i j+m n-e n p-i m q-h n q+2 c i n
   q+e j n q+e i p q)-(b\textsuperscript 2+c\textsuperscript 2+1) (a d+e) (-e q\textsuperscript 2 j\textsuperscript 2+c f q\textsuperscript 2 j\textsuperscript 2-e
   j\textsuperscript 2+c f j\textsuperscript 2+h q\textsuperscript 2 j-c i q\textsuperscript 2 j+h j-c i j-m q j+c n q j+2 e p q j-2 c f p q j-e p\textsuperscript 2+c f
   p\textsuperscript 2-e+c f+m p-c n p-h p q+c i p q) (j\textsuperscript 2 f\textsuperscript 2+p\textsuperscript 2 f\textsuperscript 2+j\textsuperscript 2 q\textsuperscript 2 f\textsuperscript 2-2 j p q
   f\textsuperscript 2+f\textsuperscript 2-2 i j q\textsuperscript 2 f-2 i j f-2 n p f+2 j n q f+2 i p q f+i\textsuperscript 2+n\textsuperscript 2+i\textsuperscript 2 q\textsuperscript 2-2 i n
   q+1)))+(b g+c h+i) (-i q\textsuperscript 2+f j q\textsuperscript 2+n q-f p q-i+f j) (d
   (-d q\textsuperscript 2 j\textsuperscript 2+a e q\textsuperscript 2 j\textsuperscript 2+b f q\textsuperscript 2 j\textsuperscript 2-a c f q\textsuperscript 2 j\textsuperscript 2-d j\textsuperscript 2+a e j\textsuperscript 2+b f j\textsuperscript 2-a c f j\textsuperscript 2+g
   q\textsuperscript 2 j-a h q\textsuperscript 2 j-b i q\textsuperscript 2 j+a c i q\textsuperscript 2 j+g j-a h j-b i j+a c i j-k q j+a m q j+b n q j-a c n
   q j+2 d p q j-2 a e p q j-2 b f p q j+2 a c f p q j-d p\textsuperscript 2+a e p\textsuperscript 2+b f p\textsuperscript 2-a c f p\textsuperscript 2-d+a
   e+b f-a c f+k p-a m p-b n p+a c n p-g p q+a h p q+b i p q-a c i p q) ((a d+e)
   (b g+c h+i) (-i q\textsuperscript 2+f j q\textsuperscript 2+n q-f p q-i+f j) (-e q\textsuperscript 2 j\textsuperscript 2+c f q\textsuperscript 2 j\textsuperscript 2-e
   j\textsuperscript 2+c f j\textsuperscript 2+h q\textsuperscript 2 j-c i q\textsuperscript 2 j+h j-c i j-m q j+c n q j+2 e p q j-2 c f p q j-e p\textsuperscript 2+c f
   p\textsuperscript 2-e+c f+m p-c n p-h p q+c i p q)-(b d+c e+f) (a g+h) (-h q\textsuperscript 2+c i q\textsuperscript 2+e j
   q\textsuperscript 2-c f j q\textsuperscript 2+m q-c n q-e p q+c f p q-h+c i+e j-c f j) (-f q\textsuperscript 2 j\textsuperscript 2-f j\textsuperscript 2+i q\textsuperscript 2
   j+i j-n q j+2 f p q j-f p\textsuperscript 2-f+n p-i p q))-(d\textsuperscript 2+e\textsuperscript 2+f\textsuperscript 2+1)
   (q\textsuperscript 2 j\textsuperscript 2+j\textsuperscript 2-2 p q j+p\textsuperscript 2+1) (a (b g+c h+i) (-i q\textsuperscript 2+f j q\textsuperscript 2+n q-f p
   q-i+f j) ((g k+h m+i n+j p+q) (a i n c\textsuperscript 2-a f j n c\textsuperscript 2-a f i p c\textsuperscript 2+a f\textsuperscript 2 j
   p c\textsuperscript 2+a f\textsuperscript 2 q c\textsuperscript 2+a q c\textsuperscript 2-a i m c+a f j m c+g n c-a h n c-b i n c-d j n c+a e j n c+b f j
   n c-f g p c+a f h p c+a e i p c+b f i p c-b f\textsuperscript 2 j p c+d f j p c-2 a e f j p c-b f\textsuperscript 2 q c-b
   q c+d f q c-2 a e f q c-g m+a h m+b i m+d j m-a e j m-b f j m+e g p-a e h p-b e i p+a e\textsuperscript 2
   j p-d e j p+b e f j p+a e\textsuperscript 2 q+a q-d e q+b e f q)-(d k+e m+f n+p) (-a f j\textsuperscript 2 n
   c\textsuperscript 2-a f n c\textsuperscript 2+a i j n c\textsuperscript 2-a i\textsuperscript 2 p c\textsuperscript 2-a p c\textsuperscript 2+a f i j p c\textsuperscript 2+a f i q c\textsuperscript 2+a j q c\textsuperscript 2+a f j\textsuperscript 2
   m c+a f m c-a i j m c-d j\textsuperscript 2 n c+a e j\textsuperscript 2 n c+b f j\textsuperscript 2 n c-d n c+a e n c+b f n c+g j n c-a h
   j n c-b i j n c+b i\textsuperscript 2 p c+b p c-g i p c+2 a h i p c-a f h j p c+d i j p c-a e i j p c-b f
   i j p c-a f h q c+d i q c-a e i q c-b f i q c-b j q c+d j\textsuperscript 2 m-a e j\textsuperscript 2 m-b f j\textsuperscript 2 m+d m-a e
   m-b f m-g j m+a h j m+b i j m-a h\textsuperscript 2 p-a p+g h p-b h i p-d h j p+a e h j p+b f h j p-d h
   q+a e h q+b f h q+a j q)+(b k+c m+n) (a j\textsuperscript 2 n e\textsuperscript 2+a n e\textsuperscript 2-a i j p e\textsuperscript 2-a i q
   e\textsuperscript 2-a f j\textsuperscript 2 m e-a f m e+a i j m e-d j\textsuperscript 2 n e+b f j\textsuperscript 2 n e-a c f j\textsuperscript 2 n e-d n e+b f n e-a c f
   n e+g j n e-2 a h j n e-b i j n e+a c i j n e+b i\textsuperscript 2 p e-a c i\textsuperscript 2 p e+b p e-a c p e-g i p
   e+a h i p e+a f h j p e+d i j p e-b f i j p e+a c f i j p e+a f h q e+d i q e-b f i q e+a
   c f i q e-b j q e+a c j q e-b f\textsuperscript 2 m+a c f\textsuperscript 2 m-b i\textsuperscript 2 m+a c i\textsuperscript 2 m-b f\textsuperscript 2 j\textsuperscript 2 m+a c f\textsuperscript 2 j\textsuperscript 2
   m+d f j\textsuperscript 2 m-b m+a c m+d f m+g i m-a h i m-f g j m+a f h j m-d i j m+2 b f i j m-2 a c f i
   j m+a h\textsuperscript 2 n+a n-g h n+b h i n-a c h i n+d h j n-b f h j n+a c f h j n-a f h\textsuperscript 2 p-a f p+f g
   h p-b f h i p+a c f h i p+b f\textsuperscript 2 h j p-a c f\textsuperscript 2 h j p-d f h j p+b f\textsuperscript 2 h q-a c f\textsuperscript 2 h q+b h
   q-a c h q-d f h q-a i q+a f j q)-(a f\textsuperscript 2 c\textsuperscript 2+a i\textsuperscript 2 c\textsuperscript 2+a f\textsuperscript 2 j\textsuperscript 2 c\textsuperscript 2+a c\textsuperscript 2-2 a
   f i j c\textsuperscript 2-b f\textsuperscript 2 c-b i\textsuperscript 2 c-b f\textsuperscript 2 j\textsuperscript 2 c+d f j\textsuperscript 2 c-2 a e f j\textsuperscript 2 c-b c+d f c-2 a e f c+g i c-2
   a h i c-f g j c+2 a f h j c-d i j c+2 a e i j c+2 b f i j c+a e\textsuperscript 2+a h\textsuperscript 2+a e\textsuperscript 2 j\textsuperscript 2-d e
   j\textsuperscript 2+b e f j\textsuperscript 2+a-d e+b e f-g h+b h i+e g j+d h j-2 a e h j-b f h j-b e i j)
   (k\textsuperscript 2+m\textsuperscript 2+n\textsuperscript 2+p\textsuperscript 2+q\textsuperscript 2+1)-k (-(g\textsuperscript 2+h\textsuperscript 2+i\textsuperscript 2+j\textsuperscript 2+1)
   (-(d\textsuperscript 2+e\textsuperscript 2+f\textsuperscript 2+1) (-m b\textsuperscript 2+c k b+a c m b+a n b-a c\textsuperscript 2 k-a k-m+c
   n)+(b d+c e+f) (b e k-a c e k-a f k-b d m+a c d m-f m+a d n+e n)+(-e b\textsuperscript 2+c d
   b+a c e b+a f b-a c\textsuperscript 2 d-a d-e+c f) (d k+e m+f n+p))+(d g+e h+f i+j) (-(d
   g+e h+f i+j) (-m b\textsuperscript 2+c k b+a c m b+a n b-a c\textsuperscript 2 k-a k-m+c n)+(b d+c e+f) (b h
   k-a c h k-a i k-b g m+a c g m-i m+a g n+h n)+(-h b\textsuperscript 2+c g b+a c h b+a i b-a c\textsuperscript 2 g-a
   g-h+c i) (d k+e m+f n+p))-(b g+c h+i) (-(d g+e h+f i+j) (b e k-a c e k-a
   f k-b d m+a c d m-f m+a d n+e n)+(d\textsuperscript 2+e\textsuperscript 2+f\textsuperscript 2+1) (b h k-a c h k-a i k-b g m+a
   c g m-i m+a g n+h n)+(b e g-a c e g-a f g-b d h+a c d h-f h+a d i+e i) (d k+e m+f
   n+p))+(-(d\textsuperscript 2+e\textsuperscript 2+f\textsuperscript 2+1) (-h b\textsuperscript 2+c g b+a c h b+a i b-a c\textsuperscript 2 g-a
   g-h+c i)+(b d+c e+f) (b e g-a c e g-a f g-b d h+a c d h-f h+a d i+e i)+(-e
   b\textsuperscript 2+c d b+a c e b+a f b-a c\textsuperscript 2 d-a d-e+c f) (d g+e h+f i+j)) (g k+h m+i n+j
   p+q)))-b (a g+h) (-h q\textsuperscript 2+c i q\textsuperscript 2+e j q\textsuperscript 2-c f j q\textsuperscript 2+m q-c n q-e p q+c f p
   q-h+c i+e j-c f j) (-(g k+h m+i n+j p+q) (-b j p f\textsuperscript 2+a c j p f\textsuperscript 2-b q
   f\textsuperscript 2+a c q f\textsuperscript 2+b j n f-a c j n f-g p f+a h p f+b i p f-a c i p f+d j p f-a e j p f+d q f-a
   e q f+g n-a h n-b i n+a c i n-d j n+a e j n-b q+a c q)+(d k+e m+f n+p) (b p
   i\textsuperscript 2-a c p i\textsuperscript 2-b j n i+a c j n i-g p i+a h p i+d j p i-a e j p i-b f j p i+a c f j p i+d q
   i-a e q i-b f q i+a c f q i-d j\textsuperscript 2 n+a e j\textsuperscript 2 n+b f j\textsuperscript 2 n-a c f j\textsuperscript 2 n-d n+a e n+b f n-a c f
   n+g j n-a h j n+b p-a c p-b j q+a c j q)-(a k+m) (a j\textsuperscript 2 n e\textsuperscript 2+a n e\textsuperscript 2-a i j p
   e\textsuperscript 2-a i q e\textsuperscript 2-a f j\textsuperscript 2 m e-a f m e+a i j m e-d j\textsuperscript 2 n e+b f j\textsuperscript 2 n e-a c f j\textsuperscript 2 n e-d n e+b f
   n e-a c f n e+g j n e-2 a h j n e-b i j n e+a c i j n e+b i\textsuperscript 2 p e-a c i\textsuperscript 2 p e+b p e-a c p
   e-g i p e+a h i p e+a f h j p e+d i j p e-b f i j p e+a c f i j p e+a f h q e+d i q e-b f
   i q e+a c f i q e-b j q e+a c j q e-b f\textsuperscript 2 m+a c f\textsuperscript 2 m-b i\textsuperscript 2 m+a c i\textsuperscript 2 m-b f\textsuperscript 2 j\textsuperscript 2 m+a c
   f\textsuperscript 2 j\textsuperscript 2 m+d f j\textsuperscript 2 m-b m+a c m+d f m+g i m-a h i m-f g j m+a f h j m-d i j m+2 b f i j m-2
   a c f i j m+a h\textsuperscript 2 n+a n-g h n+b h i n-a c h i n+d h j n-b f h j n+a c f h j n-a f h\textsuperscript 2 p-a
   f p+f g h p-b f h i p+a c f h i p+b f\textsuperscript 2 h j p-a c f\textsuperscript 2 h j p-d f h j p+b f\textsuperscript 2 h q-a c f\textsuperscript 2 h
   q+b h q-a c h q-d f h q-a i q+a f j q)+k (c f\textsuperscript 2 m a\textsuperscript 2+c i\textsuperscript 2 m a\textsuperscript 2+c f\textsuperscript 2 j\textsuperscript 2 m
   a\textsuperscript 2-e f j\textsuperscript 2 m a\textsuperscript 2+c m a\textsuperscript 2-e f m a\textsuperscript 2-h i m a\textsuperscript 2+f h j m a\textsuperscript 2+e i j m a\textsuperscript 2-2 c f i j m a\textsuperscript 2+e\textsuperscript 2
   n a\textsuperscript 2+h\textsuperscript 2 n a\textsuperscript 2+e\textsuperscript 2 j\textsuperscript 2 n a\textsuperscript 2-c e f j\textsuperscript 2 n a\textsuperscript 2-c e f n a\textsuperscript 2-c h i n a\textsuperscript 2-2 e h j n a\textsuperscript 2+c f h
   j n a\textsuperscript 2+c e i j n a\textsuperscript 2+n a\textsuperscript 2-f h\textsuperscript 2 p a\textsuperscript 2-c e i\textsuperscript 2 p a\textsuperscript 2-c e p a\textsuperscript 2-f p a\textsuperscript 2+e h i p a\textsuperscript 2+c f h
   i p a\textsuperscript 2-c f\textsuperscript 2 h j p a\textsuperscript 2+e f h j p a\textsuperscript 2-e\textsuperscript 2 i j p a\textsuperscript 2+c e f i j p a\textsuperscript 2-c f\textsuperscript 2 h q a\textsuperscript 2-c h q
   a\textsuperscript 2+e f h q a\textsuperscript 2-e\textsuperscript 2 i q a\textsuperscript 2+c e f i q a\textsuperscript 2-i q a\textsuperscript 2+c e j q a\textsuperscript 2+f j q a\textsuperscript 2-c f\textsuperscript 2 k a-c i\textsuperscript 2 k
   a-c f\textsuperscript 2 j\textsuperscript 2 k a+e f j\textsuperscript 2 k a-c k a+e f k a+h i k a-f h j k a-e i j k a+2 c f i j k a-b f\textsuperscript 2
   m a-b i\textsuperscript 2 m a-b f\textsuperscript 2 j\textsuperscript 2 m a+d f j\textsuperscript 2 m a-b m a+d f m a+g i m a-f g j m a-d i j m a+2 b f i
   j m a-2 d e j\textsuperscript 2 n a+c d f j\textsuperscript 2 n a+b e f j\textsuperscript 2 n a-2 d e n a+c d f n a+b e f n a-2 g h n a+c
   g i n a+b h i n a+2 e g j n a-c f g j n a+2 d h j n a-b f h j n a-c d i j n a-b e i j n
   a+c d i\textsuperscript 2 p a+b e i\textsuperscript 2 p a+c d p a+b e p a+2 f g h p a-e g i p a-c f g i p a-d h i p a-b f
   h i p a+c f\textsuperscript 2 g j p a-e f g j p a+b f\textsuperscript 2 h j p a-d f h j p a+2 d e i j p a-c d f i j p a-b
   e f i j p a+c f\textsuperscript 2 g q a+c g q a-e f g q a+b f\textsuperscript 2 h q a+b h q a-d f h q a+2 d e i q a-c d f
   i q a-b e f i q a-c d j q a-b e j q a+b f\textsuperscript 2 k+b i\textsuperscript 2 k+b f\textsuperscript 2 j\textsuperscript 2 k-d f j\textsuperscript 2 k+b k-d f k-g i
   k+f g j k+d i j k-2 b f i j k+d\textsuperscript 2 n+g\textsuperscript 2 n+d\textsuperscript 2 j\textsuperscript 2 n-b d f j\textsuperscript 2 n-b d f n-b g i n-2 d g j
   n+b f g j n+b d i j n+n-f g\textsuperscript 2 p-b d i\textsuperscript 2 p-b d p-f p+d g i p+b f g i p-b f\textsuperscript 2 g j p+d f g j
   p-d\textsuperscript 2 i j p+b d f i j p-b f\textsuperscript 2 g q-b g q+d f g q-d\textsuperscript 2 i q+b d f i q-i q+b d j q+f j
   q)+(-b j\textsuperscript 2 f\textsuperscript 2+a c j\textsuperscript 2 f\textsuperscript 2-b f\textsuperscript 2+a c f\textsuperscript 2+d j\textsuperscript 2 f-a e j\textsuperscript 2 f+d f-a e f-g j f+a h
   j f+2 b i j f-2 a c i j f-b i\textsuperscript 2+a c i\textsuperscript 2-b+a c+g i-a h i-d i j+a e i j)
   (k\textsuperscript 2+m\textsuperscript 2+n\textsuperscript 2+p\textsuperscript 2+q\textsuperscript 2+1)))+(d g+e h+f i+j) (-j q\textsuperscript 2+p
   q-j) (a (b d+c e+f) (-f q\textsuperscript 2 j\textsuperscript 2-f j\textsuperscript 2+i q\textsuperscript 2 j+i j-n q j+2 f p q j-f
   p\textsuperscript 2-f+n p-i p q) ((g k+h m+i n+j p+q) (a i n c\textsuperscript 2-a f j n c\textsuperscript 2-a f i p
   c\textsuperscript 2+a f\textsuperscript 2 j p c\textsuperscript 2+a f\textsuperscript 2 q c\textsuperscript 2+a q c\textsuperscript 2-a i m c+a f j m c+g n c-a h n c-b i n c-d j n c+a e
   j n c+b f j n c-f g p c+a f h p c+a e i p c+b f i p c-b f\textsuperscript 2 j p c+d f j p c-2 a e f j p
   c-b f\textsuperscript 2 q c-b q c+d f q c-2 a e f q c-g m+a h m+b i m+d j m-a e j m-b f j m+e g p-a e h
   p-b e i p+a e\textsuperscript 2 j p-d e j p+b e f j p+a e\textsuperscript 2 q+a q-d e q+b e f q)-(d k+e m+f n+p)
   (-a f j\textsuperscript 2 n c\textsuperscript 2-a f n c\textsuperscript 2+a i j n c\textsuperscript 2-a i\textsuperscript 2 p c\textsuperscript 2-a p c\textsuperscript 2+a f i j p c\textsuperscript 2+a f i q
   c\textsuperscript 2+a j q c\textsuperscript 2+a f j\textsuperscript 2 m c+a f m c-a i j m c-d j\textsuperscript 2 n c+a e j\textsuperscript 2 n c+b f j\textsuperscript 2 n c-d n c+a e n
   c+b f n c+g j n c-a h j n c-b i j n c+b i\textsuperscript 2 p c+b p c-g i p c+2 a h i p c-a f h j p c+d i
   j p c-a e i j p c-b f i j p c-a f h q c+d i q c-a e i q c-b f i q c-b j q c+d j\textsuperscript 2 m-a e
   j\textsuperscript 2 m-b f j\textsuperscript 2 m+d m-a e m-b f m-g j m+a h j m+b i j m-a h\textsuperscript 2 p-a p+g h p-b h i p-d h j p+a
   e h j p+b f h j p-d h q+a e h q+b f h q+a j q)+(b k+c m+n) (a j\textsuperscript 2 n e\textsuperscript 2+a n
   e\textsuperscript 2-a i j p e\textsuperscript 2-a i q e\textsuperscript 2-a f j\textsuperscript 2 m e-a f m e+a i j m e-d j\textsuperscript 2 n e+b f j\textsuperscript 2 n e-a c f j\textsuperscript 2 n
   e-d n e+b f n e-a c f n e+g j n e-2 a h j n e-b i j n e+a c i j n e+b i\textsuperscript 2 p e-a c i\textsuperscript 2 p
   e+b p e-a c p e-g i p e+a h i p e+a f h j p e+d i j p e-b f i j p e+a c f i j p e+a f h q
   e+d i q e-b f i q e+a c f i q e-b j q e+a c j q e-b f\textsuperscript 2 m+a c f\textsuperscript 2 m-b i\textsuperscript 2 m+a c i\textsuperscript 2 m-b
   f\textsuperscript 2 j\textsuperscript 2 m+a c f\textsuperscript 2 j\textsuperscript 2 m+d f j\textsuperscript 2 m-b m+a c m+d f m+g i m-a h i m-f g j m+a f h j m-d i j
   m+2 b f i j m-2 a c f i j m+a h\textsuperscript 2 n+a n-g h n+b h i n-a c h i n+d h j n-b f h j n+a c f h
   j n-a f h\textsuperscript 2 p-a f p+f g h p-b f h i p+a c f h i p+b f\textsuperscript 2 h j p-a c f\textsuperscript 2 h j p-d f h j p+b
   f\textsuperscript 2 h q-a c f\textsuperscript 2 h q+b h q-a c h q-d f h q-a i q+a f j q)-(a f\textsuperscript 2 c\textsuperscript 2+a i\textsuperscript 2
   c\textsuperscript 2+a f\textsuperscript 2 j\textsuperscript 2 c\textsuperscript 2+a c\textsuperscript 2-2 a f i j c\textsuperscript 2-b f\textsuperscript 2 c-b i\textsuperscript 2 c-b f\textsuperscript 2 j\textsuperscript 2 c+d f j\textsuperscript 2 c-2 a e f j\textsuperscript 2
   c-b c+d f c-2 a e f c+g i c-2 a h i c-f g j c+2 a f h j c-d i j c+2 a e i j c+2 b f i j
   c+a e\textsuperscript 2+a h\textsuperscript 2+a e\textsuperscript 2 j\textsuperscript 2-d e j\textsuperscript 2+b e f j\textsuperscript 2+a-d e+b e f-g h+b h i+e g j+d h j-2 a e h j-b f
   h j-b e i j) (k\textsuperscript 2+m\textsuperscript 2+n\textsuperscript 2+p\textsuperscript 2+q\textsuperscript 2+1)-k
   (-(g\textsuperscript 2+h\textsuperscript 2+i\textsuperscript 2+j\textsuperscript 2+1) (-(d\textsuperscript 2+e\textsuperscript 2+f\textsuperscript 2+1) (-m b\textsuperscript 2+c k
   b+a c m b+a n b-a c\textsuperscript 2 k-a k-m+c n)+(b d+c e+f) (b e k-a c e k-a f k-b d m+a c d m-f
   m+a d n+e n)+(-e b\textsuperscript 2+c d b+a c e b+a f b-a c\textsuperscript 2 d-a d-e+c f) (d k+e m+f
   n+p))+(d g+e h+f i+j) (-(d g+e h+f i+j) (-m b\textsuperscript 2+c k b+a c m b+a n b-a c\textsuperscript 2
   k-a k-m+c n)+(b d+c e+f) (b h k-a c h k-a i k-b g m+a c g m-i m+a g n+h n)+(-h
   b\textsuperscript 2+c g b+a c h b+a i b-a c\textsuperscript 2 g-a g-h+c i) (d k+e m+f n+p))-(b g+c h+i)
   (-(d g+e h+f i+j) (b e k-a c e k-a f k-b d m+a c d m-f m+a d n+e
   n)+(d\textsuperscript 2+e\textsuperscript 2+f\textsuperscript 2+1) (b h k-a c h k-a i k-b g m+a c g m-i m+a g n+h n)+(b e g-a
   c e g-a f g-b d h+a c d h-f h+a d i+e i) (d k+e m+f
   n+p))+(-(d\textsuperscript 2+e\textsuperscript 2+f\textsuperscript 2+1) (-h b\textsuperscript 2+c g b+a c h b+a i b-a c\textsuperscript 2 g-a
   g-h+c i)+(b d+c e+f) (b e g-a c e g-a f g-b d h+a c d h-f h+a d i+e i)+(-e
   b\textsuperscript 2+c d b+a c e b+a f b-a c\textsuperscript 2 d-a d-e+c f) (d g+e h+f i+j)) (g k+h m+i n+j
   p+q)))-b (a d+e) (-e q\textsuperscript 2 j\textsuperscript 2+c f q\textsuperscript 2 j\textsuperscript 2-e j\textsuperscript 2+c f j\textsuperscript 2+h q\textsuperscript 2 j-c i q\textsuperscript 2
   j+h j-c i j-m q j+c n q j+2 e p q j-2 c f p q j-e p\textsuperscript 2+c f p\textsuperscript 2-e+c f+m p-c n p-h p q+c i p
   q) (-(g k+h m+i n+j p+q) (-b j p f\textsuperscript 2+a c j p f\textsuperscript 2-b q f\textsuperscript 2+a c q f\textsuperscript 2+b j n
   f-a c j n f-g p f+a h p f+b i p f-a c i p f+d j p f-a e j p f+d q f-a e q f+g n-a h n-b i
   n+a c i n-d j n+a e j n-b q+a c q)+(d k+e m+f n+p) (b p i\textsuperscript 2-a c p i\textsuperscript 2-b j n
   i+a c j n i-g p i+a h p i+d j p i-a e j p i-b f j p i+a c f j p i+d q i-a e q i-b f q i+a
   c f q i-d j\textsuperscript 2 n+a e j\textsuperscript 2 n+b f j\textsuperscript 2 n-a c f j\textsuperscript 2 n-d n+a e n+b f n-a c f n+g j n-a h j n+b
   p-a c p-b j q+a c j q)-(a k+m) (a j\textsuperscript 2 n e\textsuperscript 2+a n e\textsuperscript 2-a i j p e\textsuperscript 2-a i q e\textsuperscript 2-a f
   j\textsuperscript 2 m e-a f m e+a i j m e-d j\textsuperscript 2 n e+b f j\textsuperscript 2 n e-a c f j\textsuperscript 2 n e-d n e+b f n e-a c f n e+g j
   n e-2 a h j n e-b i j n e+a c i j n e+b i\textsuperscript 2 p e-a c i\textsuperscript 2 p e+b p e-a c p e-g i p e+a h i p
   e+a f h j p e+d i j p e-b f i j p e+a c f i j p e+a f h q e+d i q e-b f i q e+a c f i q
   e-b j q e+a c j q e-b f\textsuperscript 2 m+a c f\textsuperscript 2 m-b i\textsuperscript 2 m+a c i\textsuperscript 2 m-b f\textsuperscript 2 j\textsuperscript 2 m+a c f\textsuperscript 2 j\textsuperscript 2 m+d f j\textsuperscript 2
   m-b m+a c m+d f m+g i m-a h i m-f g j m+a f h j m-d i j m+2 b f i j m-2 a c f i j m+a h\textsuperscript 2
   n+a n-g h n+b h i n-a c h i n+d h j n-b f h j n+a c f h j n-a f h\textsuperscript 2 p-a f p+f g h p-b f h
   i p+a c f h i p+b f\textsuperscript 2 h j p-a c f\textsuperscript 2 h j p-d f h j p+b f\textsuperscript 2 h q-a c f\textsuperscript 2 h q+b h q-a c h q-d
   f h q-a i q+a f j q)+k (c f\textsuperscript 2 m a\textsuperscript 2+c i\textsuperscript 2 m a\textsuperscript 2+c f\textsuperscript 2 j\textsuperscript 2 m a\textsuperscript 2-e f j\textsuperscript 2 m
   a\textsuperscript 2+c m a\textsuperscript 2-e f m a\textsuperscript 2-h i m a\textsuperscript 2+f h j m a\textsuperscript 2+e i j m a\textsuperscript 2-2 c f i j m a\textsuperscript 2+e\textsuperscript 2 n a\textsuperscript 2+h\textsuperscript 2 n
   a\textsuperscript 2+e\textsuperscript 2 j\textsuperscript 2 n a\textsuperscript 2-c e f j\textsuperscript 2 n a\textsuperscript 2-c e f n a\textsuperscript 2-c h i n a\textsuperscript 2-2 e h j n a\textsuperscript 2+c f h j n a\textsuperscript 2+c e
   i j n a\textsuperscript 2+n a\textsuperscript 2-f h\textsuperscript 2 p a\textsuperscript 2-c e i\textsuperscript 2 p a\textsuperscript 2-c e p a\textsuperscript 2-f p a\textsuperscript 2+e h i p a\textsuperscript 2+c f h i p a\textsuperscript 2-c
   f\textsuperscript 2 h j p a\textsuperscript 2+e f h j p a\textsuperscript 2-e\textsuperscript 2 i j p a\textsuperscript 2+c e f i j p a\textsuperscript 2-c f\textsuperscript 2 h q a\textsuperscript 2-c h q a\textsuperscript 2+e f h q
   a\textsuperscript 2-e\textsuperscript 2 i q a\textsuperscript 2+c e f i q a\textsuperscript 2-i q a\textsuperscript 2+c e j q a\textsuperscript 2+f j q a\textsuperscript 2-c f\textsuperscript 2 k a-c i\textsuperscript 2 k a-c f\textsuperscript 2 j\textsuperscript 2
   k a+e f j\textsuperscript 2 k a-c k a+e f k a+h i k a-f h j k a-e i j k a+2 c f i j k a-b f\textsuperscript 2 m a-b i\textsuperscript 2 m
   a-b f\textsuperscript 2 j\textsuperscript 2 m a+d f j\textsuperscript 2 m a-b m a+d f m a+g i m a-f g j m a-d i j m a+2 b f i j m a-2 d e
   j\textsuperscript 2 n a+c d f j\textsuperscript 2 n a+b e f j\textsuperscript 2 n a-2 d e n a+c d f n a+b e f n a-2 g h n a+c g i n a+b h
   i n a+2 e g j n a-c f g j n a+2 d h j n a-b f h j n a-c d i j n a-b e i j n a+c d i\textsuperscript 2 p
   a+b e i\textsuperscript 2 p a+c d p a+b e p a+2 f g h p a-e g i p a-c f g i p a-d h i p a-b f h i p a+c
   f\textsuperscript 2 g j p a-e f g j p a+b f\textsuperscript 2 h j p a-d f h j p a+2 d e i j p a-c d f i j p a-b e f i j p
   a+c f\textsuperscript 2 g q a+c g q a-e f g q a+b f\textsuperscript 2 h q a+b h q a-d f h q a+2 d e i q a-c d f i q a-b e
   f i q a-c d j q a-b e j q a+b f\textsuperscript 2 k+b i\textsuperscript 2 k+b f\textsuperscript 2 j\textsuperscript 2 k-d f j\textsuperscript 2 k+b k-d f k-g i k+f g j
   k+d i j k-2 b f i j k+d\textsuperscript 2 n+g\textsuperscript 2 n+d\textsuperscript 2 j\textsuperscript 2 n-b d f j\textsuperscript 2 n-b d f n-b g i n-2 d g j n+b f g j
   n+b d i j n+n-f g\textsuperscript 2 p-b d i\textsuperscript 2 p-b d p-f p+d g i p+b f g i p-b f\textsuperscript 2 g j p+d f g j p-d\textsuperscript 2 i j
   p+b d f i j p-b f\textsuperscript 2 g q-b g q+d f g q-d\textsuperscript 2 i q+b d f i q-i q+b d j q+f j q)+(-b
   j\textsuperscript 2 f\textsuperscript 2+a c j\textsuperscript 2 f\textsuperscript 2-b f\textsuperscript 2+a c f\textsuperscript 2+d j\textsuperscript 2 f-a e j\textsuperscript 2 f+d f-a e f-g j f+a h j f+2 b i j f-2 a
   c i j f-b i\textsuperscript 2+a c i\textsuperscript 2-b+a c+g i-a h i-d i j+a e i j)
   (k\textsuperscript 2+m\textsuperscript 2+n\textsuperscript 2+p\textsuperscript 2+q\textsuperscript 2+1))))-(d g+e h+f i+j) (-j q\textsuperscript 2+p
   q-j) (d (-d q\textsuperscript 2 j\textsuperscript 2+a e q\textsuperscript 2 j\textsuperscript 2+b f q\textsuperscript 2 j\textsuperscript 2-a c f q\textsuperscript 2 j\textsuperscript 2-d j\textsuperscript 2+a e j\textsuperscript 2+b
   f j\textsuperscript 2-a c f j\textsuperscript 2+g q\textsuperscript 2 j-a h q\textsuperscript 2 j-b i q\textsuperscript 2 j+a c i q\textsuperscript 2 j+g j-a h j-b i j+a c i j-k q j+a m
   q j+b n q j-a c n q j+2 d p q j-2 a e p q j-2 b f p q j+2 a c f p q j-d p\textsuperscript 2+a e p\textsuperscript 2+b f
   p\textsuperscript 2-a c f p\textsuperscript 2-d+a e+b f-a c f+k p-a m p-b n p+a c n p-g p q+a h p q+b i p q-a c i p
   q) ((a b+c) (b g+c h+i) (-i q\textsuperscript 2+f j q\textsuperscript 2+n q-f p q-i+f j) (-c
   j\textsuperscript 2 f\textsuperscript 2-c p\textsuperscript 2 f\textsuperscript 2-c j\textsuperscript 2 q\textsuperscript 2 f\textsuperscript 2-c f\textsuperscript 2+2 c j p q f\textsuperscript 2+e j\textsuperscript 2 f+e p\textsuperscript 2 f+e j\textsuperscript 2 q\textsuperscript 2 f-h j q\textsuperscript 2
   f+2 c i j q\textsuperscript 2 f+e f-h j f+2 c i j f-m p f+2 c n p f+j m q f-2 c j n q f+h p q f-2 c i p q
   f-2 e j p q f-c i\textsuperscript 2-c n\textsuperscript 2-c i\textsuperscript 2 q\textsuperscript 2+h i q\textsuperscript 2-e i j q\textsuperscript 2-c+h i-e i j+m n-e n p-i m q-h n q+2
   c i n q+e j n q+e i p q)-(b\textsuperscript 2+c\textsuperscript 2+1) (a g+h) (-h q\textsuperscript 2+c i q\textsuperscript 2+e j
   q\textsuperscript 2-c f j q\textsuperscript 2+m q-c n q-e p q+c f p q-h+c i+e j-c f j) (j\textsuperscript 2 f\textsuperscript 2+p\textsuperscript 2 f\textsuperscript 2+j\textsuperscript 2
   q\textsuperscript 2 f\textsuperscript 2-2 j p q f\textsuperscript 2+f\textsuperscript 2-2 i j q\textsuperscript 2 f-2 i j f-2 n p f+2 j n q f+2 i p q f+i\textsuperscript 2+n\textsuperscript 2+i\textsuperscript 2 q\textsuperscript 2-2
   i n q+1))-(b d+c e+f) (-f q\textsuperscript 2 j\textsuperscript 2-f j\textsuperscript 2+i q\textsuperscript 2 j+i j-n q j+2 f p q j-f
   p\textsuperscript 2-f+n p-i p q) (a (b g+c h+i) (-i q\textsuperscript 2+f j q\textsuperscript 2+n q-f p q-i+f j)
   ((g k+h m+i n+j p+q) (a i n c\textsuperscript 2-a f j n c\textsuperscript 2-a f i p c\textsuperscript 2+a f\textsuperscript 2 j p c\textsuperscript 2+a f\textsuperscript 2 q
   c\textsuperscript 2+a q c\textsuperscript 2-a i m c+a f j m c+g n c-a h n c-b i n c-d j n c+a e j n c+b f j n c-f g p c+a
   f h p c+a e i p c+b f i p c-b f\textsuperscript 2 j p c+d f j p c-2 a e f j p c-b f\textsuperscript 2 q c-b q c+d f q c-2
   a e f q c-g m+a h m+b i m+d j m-a e j m-b f j m+e g p-a e h p-b e i p+a e\textsuperscript 2 j p-d e j p+b
   e f j p+a e\textsuperscript 2 q+a q-d e q+b e f q)-(d k+e m+f n+p) (-a f j\textsuperscript 2 n c\textsuperscript 2-a f n c\textsuperscript 2+a
   i j n c\textsuperscript 2-a i\textsuperscript 2 p c\textsuperscript 2-a p c\textsuperscript 2+a f i j p c\textsuperscript 2+a f i q c\textsuperscript 2+a j q c\textsuperscript 2+a f j\textsuperscript 2 m c+a f m c-a i
   j m c-d j\textsuperscript 2 n c+a e j\textsuperscript 2 n c+b f j\textsuperscript 2 n c-d n c+a e n c+b f n c+g j n c-a h j n c-b i j n
   c+b i\textsuperscript 2 p c+b p c-g i p c+2 a h i p c-a f h j p c+d i j p c-a e i j p c-b f i j p c-a f h
   q c+d i q c-a e i q c-b f i q c-b j q c+d j\textsuperscript 2 m-a e j\textsuperscript 2 m-b f j\textsuperscript 2 m+d m-a e m-b f m-g j
   m+a h j m+b i j m-a h\textsuperscript 2 p-a p+g h p-b h i p-d h j p+a e h j p+b f h j p-d h q+a e h q+b f
   h q+a j q)+(b k+c m+n) (a j\textsuperscript 2 n e\textsuperscript 2+a n e\textsuperscript 2-a i j p e\textsuperscript 2-a i q e\textsuperscript 2-a f j\textsuperscript 2 m
   e-a f m e+a i j m e-d j\textsuperscript 2 n e+b f j\textsuperscript 2 n e-a c f j\textsuperscript 2 n e-d n e+b f n e-a c f n e+g j n e-2
   a h j n e-b i j n e+a c i j n e+b i\textsuperscript 2 p e-a c i\textsuperscript 2 p e+b p e-a c p e-g i p e+a h i p e+a f
   h j p e+d i j p e-b f i j p e+a c f i j p e+a f h q e+d i q e-b f i q e+a c f i q e-b j q
   e+a c j q e-b f\textsuperscript 2 m+a c f\textsuperscript 2 m-b i\textsuperscript 2 m+a c i\textsuperscript 2 m-b f\textsuperscript 2 j\textsuperscript 2 m+a c f\textsuperscript 2 j\textsuperscript 2 m+d f j\textsuperscript 2 m-b m+a
   c m+d f m+g i m-a h i m-f g j m+a f h j m-d i j m+2 b f i j m-2 a c f i j m+a h\textsuperscript 2 n+a n-g
   h n+b h i n-a c h i n+d h j n-b f h j n+a c f h j n-a f h\textsuperscript 2 p-a f p+f g h p-b f h i p+a c
   f h i p+b f\textsuperscript 2 h j p-a c f\textsuperscript 2 h j p-d f h j p+b f\textsuperscript 2 h q-a c f\textsuperscript 2 h q+b h q-a c h q-d f h q-a
   i q+a f j q)-(a f\textsuperscript 2 c\textsuperscript 2+a i\textsuperscript 2 c\textsuperscript 2+a f\textsuperscript 2 j\textsuperscript 2 c\textsuperscript 2+a c\textsuperscript 2-2 a f i j c\textsuperscript 2-b f\textsuperscript 2 c-b
   i\textsuperscript 2 c-b f\textsuperscript 2 j\textsuperscript 2 c+d f j\textsuperscript 2 c-2 a e f j\textsuperscript 2 c-b c+d f c-2 a e f c+g i c-2 a h i c-f g j c+2 a
   f h j c-d i j c+2 a e i j c+2 b f i j c+a e\textsuperscript 2+a h\textsuperscript 2+a e\textsuperscript 2 j\textsuperscript 2-d e j\textsuperscript 2+b e f j\textsuperscript 2+a-d e+b e
   f-g h+b h i+e g j+d h j-2 a e h j-b f h j-b e i j)
   (k\textsuperscript 2+m\textsuperscript 2+n\textsuperscript 2+p\textsuperscript 2+q\textsuperscript 2+1)-k (-(g\textsuperscript 2+h\textsuperscript 2+i\textsuperscript 2+j\textsuperscript 2+1)
   (-(d\textsuperscript 2+e\textsuperscript 2+f\textsuperscript 2+1) (-m b\textsuperscript 2+c k b+a c m b+a n b-a c\textsuperscript 2 k-a k-m+c
   n)+(b d+c e+f) (b e k-a c e k-a f k-b d m+a c d m-f m+a d n+e n)+(-e b\textsuperscript 2+c d
   b+a c e b+a f b-a c\textsuperscript 2 d-a d-e+c f) (d k+e m+f n+p))+(d g+e h+f i+j) (-(d
   g+e h+f i+j) (-m b\textsuperscript 2+c k b+a c m b+a n b-a c\textsuperscript 2 k-a k-m+c n)+(b d+c e+f) (b h
   k-a c h k-a i k-b g m+a c g m-i m+a g n+h n)+(-h b\textsuperscript 2+c g b+a c h b+a i b-a c\textsuperscript 2 g-a
   g-h+c i) (d k+e m+f n+p))-(b g+c h+i) (-(d g+e h+f i+j) (b e k-a c e k-a
   f k-b d m+a c d m-f m+a d n+e n)+(d\textsuperscript 2+e\textsuperscript 2+f\textsuperscript 2+1) (b h k-a c h k-a i k-b g m+a
   c g m-i m+a g n+h n)+(b e g-a c e g-a f g-b d h+a c d h-f h+a d i+e i) (d k+e m+f
   n+p))+(-(d\textsuperscript 2+e\textsuperscript 2+f\textsuperscript 2+1) (-h b\textsuperscript 2+c g b+a c h b+a i b-a c\textsuperscript 2 g-a
   g-h+c i)+(b d+c e+f) (b e g-a c e g-a f g-b d h+a c d h-f h+a d i+e i)+(-e
   b\textsuperscript 2+c d b+a c e b+a f b-a c\textsuperscript 2 d-a d-e+c f) (d g+e h+f i+j)) (g k+h m+i n+j
   p+q)))-b (a g+h) (-h q\textsuperscript 2+c i q\textsuperscript 2+e j q\textsuperscript 2-c f j q\textsuperscript 2+m q-c n q-e p q+c f p
   q-h+c i+e j-c f j) (-(g k+h m+i n+j p+q) (-b j p f\textsuperscript 2+a c j p f\textsuperscript 2-b q
   f\textsuperscript 2+a c q f\textsuperscript 2+b j n f-a c j n f-g p f+a h p f+b i p f-a c i p f+d j p f-a e j p f+d q f-a
   e q f+g n-a h n-b i n+a c i n-d j n+a e j n-b q+a c q)+(d k+e m+f n+p) (b p
   i\textsuperscript 2-a c p i\textsuperscript 2-b j n i+a c j n i-g p i+a h p i+d j p i-a e j p i-b f j p i+a c f j p i+d q
   i-a e q i-b f q i+a c f q i-d j\textsuperscript 2 n+a e j\textsuperscript 2 n+b f j\textsuperscript 2 n-a c f j\textsuperscript 2 n-d n+a e n+b f n-a c f
   n+g j n-a h j n+b p-a c p-b j q+a c j q)-(a k+m) (a j\textsuperscript 2 n e\textsuperscript 2+a n e\textsuperscript 2-a i j p
   e\textsuperscript 2-a i q e\textsuperscript 2-a f j\textsuperscript 2 m e-a f m e+a i j m e-d j\textsuperscript 2 n e+b f j\textsuperscript 2 n e-a c f j\textsuperscript 2 n e-d n e+b f
   n e-a c f n e+g j n e-2 a h j n e-b i j n e+a c i j n e+b i\textsuperscript 2 p e-a c i\textsuperscript 2 p e+b p e-a c p
   e-g i p e+a h i p e+a f h j p e+d i j p e-b f i j p e+a c f i j p e+a f h q e+d i q e-b f
   i q e+a c f i q e-b j q e+a c j q e-b f\textsuperscript 2 m+a c f\textsuperscript 2 m-b i\textsuperscript 2 m+a c i\textsuperscript 2 m-b f\textsuperscript 2 j\textsuperscript 2 m+a c
   f\textsuperscript 2 j\textsuperscript 2 m+d f j\textsuperscript 2 m-b m+a c m+d f m+g i m-a h i m-f g j m+a f h j m-d i j m+2 b f i j m-2
   a c f i j m+a h\textsuperscript 2 n+a n-g h n+b h i n-a c h i n+d h j n-b f h j n+a c f h j n-a f h\textsuperscript 2 p-a
   f p+f g h p-b f h i p+a c f h i p+b f\textsuperscript 2 h j p-a c f\textsuperscript 2 h j p-d f h j p+b f\textsuperscript 2 h q-a c f\textsuperscript 2 h
   q+b h q-a c h q-d f h q-a i q+a f j q)+k (c f\textsuperscript 2 m a\textsuperscript 2+c i\textsuperscript 2 m a\textsuperscript 2+c f\textsuperscript 2 j\textsuperscript 2 m
   a\textsuperscript 2-e f j\textsuperscript 2 m a\textsuperscript 2+c m a\textsuperscript 2-e f m a\textsuperscript 2-h i m a\textsuperscript 2+f h j m a\textsuperscript 2+e i j m a\textsuperscript 2-2 c f i j m a\textsuperscript 2+e\textsuperscript 2
   n a\textsuperscript 2+h\textsuperscript 2 n a\textsuperscript 2+e\textsuperscript 2 j\textsuperscript 2 n a\textsuperscript 2-c e f j\textsuperscript 2 n a\textsuperscript 2-c e f n a\textsuperscript 2-c h i n a\textsuperscript 2-2 e h j n a\textsuperscript 2+c f h
   j n a\textsuperscript 2+c e i j n a\textsuperscript 2+n a\textsuperscript 2-f h\textsuperscript 2 p a\textsuperscript 2-c e i\textsuperscript 2 p a\textsuperscript 2-c e p a\textsuperscript 2-f p a\textsuperscript 2+e h i p a\textsuperscript 2+c f h
   i p a\textsuperscript 2-c f\textsuperscript 2 h j p a\textsuperscript 2+e f h j p a\textsuperscript 2-e\textsuperscript 2 i j p a\textsuperscript 2+c e f i j p a\textsuperscript 2-c f\textsuperscript 2 h q a\textsuperscript 2-c h q
   a\textsuperscript 2+e f h q a\textsuperscript 2-e\textsuperscript 2 i q a\textsuperscript 2+c e f i q a\textsuperscript 2-i q a\textsuperscript 2+c e j q a\textsuperscript 2+f j q a\textsuperscript 2-c f\textsuperscript 2 k a-c i\textsuperscript 2 k
   a-c f\textsuperscript 2 j\textsuperscript 2 k a+e f j\textsuperscript 2 k a-c k a+e f k a+h i k a-f h j k a-e i j k a+2 c f i j k a-b f\textsuperscript 2
   m a-b i\textsuperscript 2 m a-b f\textsuperscript 2 j\textsuperscript 2 m a+d f j\textsuperscript 2 m a-b m a+d f m a+g i m a-f g j m a-d i j m a+2 b f i
   j m a-2 d e j\textsuperscript 2 n a+c d f j\textsuperscript 2 n a+b e f j\textsuperscript 2 n a-2 d e n a+c d f n a+b e f n a-2 g h n a+c
   g i n a+b h i n a+2 e g j n a-c f g j n a+2 d h j n a-b f h j n a-c d i j n a-b e i j n
   a+c d i\textsuperscript 2 p a+b e i\textsuperscript 2 p a+c d p a+b e p a+2 f g h p a-e g i p a-c f g i p a-d h i p a-b f
   h i p a+c f\textsuperscript 2 g j p a-e f g j p a+b f\textsuperscript 2 h j p a-d f h j p a+2 d e i j p a-c d f i j p a-b
   e f i j p a+c f\textsuperscript 2 g q a+c g q a-e f g q a+b f\textsuperscript 2 h q a+b h q a-d f h q a+2 d e i q a-c d f
   i q a-b e f i q a-c d j q a-b e j q a+b f\textsuperscript 2 k+b i\textsuperscript 2 k+b f\textsuperscript 2 j\textsuperscript 2 k-d f j\textsuperscript 2 k+b k-d f k-g i
   k+f g j k+d i j k-2 b f i j k+d\textsuperscript 2 n+g\textsuperscript 2 n+d\textsuperscript 2 j\textsuperscript 2 n-b d f j\textsuperscript 2 n-b d f n-b g i n-2 d g j
   n+b f g j n+b d i j n+n-f g\textsuperscript 2 p-b d i\textsuperscript 2 p-b d p-f p+d g i p+b f g i p-b f\textsuperscript 2 g j p+d f g j
   p-d\textsuperscript 2 i j p+b d f i j p-b f\textsuperscript 2 g q-b g q+d f g q-d\textsuperscript 2 i q+b d f i q-i q+b d j q+f j
   q)+(-b j\textsuperscript 2 f\textsuperscript 2+a c j\textsuperscript 2 f\textsuperscript 2-b f\textsuperscript 2+a c f\textsuperscript 2+d j\textsuperscript 2 f-a e j\textsuperscript 2 f+d f-a e f-g j f+a h
   j f+2 b i j f-2 a c i j f-b i\textsuperscript 2+a c i\textsuperscript 2-b+a c+g i-a h i-d i j+a e i j)
   (k\textsuperscript 2+m\textsuperscript 2+n\textsuperscript 2+p\textsuperscript 2+q\textsuperscript 2+1)))+(d g+e h+f i+j) (-j q\textsuperscript 2+p
   q-j) (a (b\textsuperscript 2+c\textsuperscript 2+1) (j\textsuperscript 2 f\textsuperscript 2+p\textsuperscript 2 f\textsuperscript 2+j\textsuperscript 2 q\textsuperscript 2 f\textsuperscript 2-2 j p q
   f\textsuperscript 2+f\textsuperscript 2-2 i j q\textsuperscript 2 f-2 i j f-2 n p f+2 j n q f+2 i p q f+i\textsuperscript 2+n\textsuperscript 2+i\textsuperscript 2 q\textsuperscript 2-2 i n q+1)
   ((g k+h m+i n+j p+q) (a i n c\textsuperscript 2-a f j n c\textsuperscript 2-a f i p c\textsuperscript 2+a f\textsuperscript 2 j p c\textsuperscript 2+a f\textsuperscript 2 q
   c\textsuperscript 2+a q c\textsuperscript 2-a i m c+a f j m c+g n c-a h n c-b i n c-d j n c+a e j n c+b f j n c-f g p c+a
   f h p c+a e i p c+b f i p c-b f\textsuperscript 2 j p c+d f j p c-2 a e f j p c-b f\textsuperscript 2 q c-b q c+d f q c-2
   a e f q c-g m+a h m+b i m+d j m-a e j m-b f j m+e g p-a e h p-b e i p+a e\textsuperscript 2 j p-d e j p+b
   e f j p+a e\textsuperscript 2 q+a q-d e q+b e f q)-(d k+e m+f n+p) (-a f j\textsuperscript 2 n c\textsuperscript 2-a f n c\textsuperscript 2+a
   i j n c\textsuperscript 2-a i\textsuperscript 2 p c\textsuperscript 2-a p c\textsuperscript 2+a f i j p c\textsuperscript 2+a f i q c\textsuperscript 2+a j q c\textsuperscript 2+a f j\textsuperscript 2 m c+a f m c-a i
   j m c-d j\textsuperscript 2 n c+a e j\textsuperscript 2 n c+b f j\textsuperscript 2 n c-d n c+a e n c+b f n c+g j n c-a h j n c-b i j n
   c+b i\textsuperscript 2 p c+b p c-g i p c+2 a h i p c-a f h j p c+d i j p c-a e i j p c-b f i j p c-a f h
   q c+d i q c-a e i q c-b f i q c-b j q c+d j\textsuperscript 2 m-a e j\textsuperscript 2 m-b f j\textsuperscript 2 m+d m-a e m-b f m-g j
   m+a h j m+b i j m-a h\textsuperscript 2 p-a p+g h p-b h i p-d h j p+a e h j p+b f h j p-d h q+a e h q+b f
   h q+a j q)+(b k+c m+n) (a j\textsuperscript 2 n e\textsuperscript 2+a n e\textsuperscript 2-a i j p e\textsuperscript 2-a i q e\textsuperscript 2-a f j\textsuperscript 2 m
   e-a f m e+a i j m e-d j\textsuperscript 2 n e+b f j\textsuperscript 2 n e-a c f j\textsuperscript 2 n e-d n e+b f n e-a c f n e+g j n e-2
   a h j n e-b i j n e+a c i j n e+b i\textsuperscript 2 p e-a c i\textsuperscript 2 p e+b p e-a c p e-g i p e+a h i p e+a f
   h j p e+d i j p e-b f i j p e+a c f i j p e+a f h q e+d i q e-b f i q e+a c f i q e-b j q
   e+a c j q e-b f\textsuperscript 2 m+a c f\textsuperscript 2 m-b i\textsuperscript 2 m+a c i\textsuperscript 2 m-b f\textsuperscript 2 j\textsuperscript 2 m+a c f\textsuperscript 2 j\textsuperscript 2 m+d f j\textsuperscript 2 m-b m+a
   c m+d f m+g i m-a h i m-f g j m+a f h j m-d i j m+2 b f i j m-2 a c f i j m+a h\textsuperscript 2 n+a n-g
   h n+b h i n-a c h i n+d h j n-b f h j n+a c f h j n-a f h\textsuperscript 2 p-a f p+f g h p-b f h i p+a c
   f h i p+b f\textsuperscript 2 h j p-a c f\textsuperscript 2 h j p-d f h j p+b f\textsuperscript 2 h q-a c f\textsuperscript 2 h q+b h q-a c h q-d f h q-a
   i q+a f j q)-(a f\textsuperscript 2 c\textsuperscript 2+a i\textsuperscript 2 c\textsuperscript 2+a f\textsuperscript 2 j\textsuperscript 2 c\textsuperscript 2+a c\textsuperscript 2-2 a f i j c\textsuperscript 2-b f\textsuperscript 2 c-b
   i\textsuperscript 2 c-b f\textsuperscript 2 j\textsuperscript 2 c+d f j\textsuperscript 2 c-2 a e f j\textsuperscript 2 c-b c+d f c-2 a e f c+g i c-2 a h i c-f g j c+2 a
   f h j c-d i j c+2 a e i j c+2 b f i j c+a e\textsuperscript 2+a h\textsuperscript 2+a e\textsuperscript 2 j\textsuperscript 2-d e j\textsuperscript 2+b e f j\textsuperscript 2+a-d e+b e
   f-g h+b h i+e g j+d h j-2 a e h j-b f h j-b e i j)
   (k\textsuperscript 2+m\textsuperscript 2+n\textsuperscript 2+p\textsuperscript 2+q\textsuperscript 2+1)-k (-(g\textsuperscript 2+h\textsuperscript 2+i\textsuperscript 2+j\textsuperscript 2+1)
   (-(d\textsuperscript 2+e\textsuperscript 2+f\textsuperscript 2+1) (-m b\textsuperscript 2+c k b+a c m b+a n b-a c\textsuperscript 2 k-a k-m+c
   n)+(b d+c e+f) (b e k-a c e k-a f k-b d m+a c d m-f m+a d n+e n)+(-e b\textsuperscript 2+c d
   b+a c e b+a f b-a c\textsuperscript 2 d-a d-e+c f) (d k+e m+f n+p))+(d g+e h+f i+j) (-(d
   g+e h+f i+j) (-m b\textsuperscript 2+c k b+a c m b+a n b-a c\textsuperscript 2 k-a k-m+c n)+(b d+c e+f) (b h
   k-a c h k-a i k-b g m+a c g m-i m+a g n+h n)+(-h b\textsuperscript 2+c g b+a c h b+a i b-a c\textsuperscript 2 g-a
   g-h+c i) (d k+e m+f n+p))-(b g+c h+i) (-(d g+e h+f i+j) (b e k-a c e k-a
   f k-b d m+a c d m-f m+a d n+e n)+(d\textsuperscript 2+e\textsuperscript 2+f\textsuperscript 2+1) (b h k-a c h k-a i k-b g m+a
   c g m-i m+a g n+h n)+(b e g-a c e g-a f g-b d h+a c d h-f h+a d i+e i) (d k+e m+f
   n+p))+(-(d\textsuperscript 2+e\textsuperscript 2+f\textsuperscript 2+1) (-h b\textsuperscript 2+c g b+a c h b+a i b-a c\textsuperscript 2 g-a
   g-h+c i)+(b d+c e+f) (b e g-a c e g-a f g-b d h+a c d h-f h+a d i+e i)+(-e
   b\textsuperscript 2+c d b+a c e b+a f b-a c\textsuperscript 2 d-a d-e+c f) (d g+e h+f i+j)) (g k+h m+i n+j
   p+q)))-b (a b+c) (-c j\textsuperscript 2 f\textsuperscript 2-c p\textsuperscript 2 f\textsuperscript 2-c j\textsuperscript 2 q\textsuperscript 2 f\textsuperscript 2-c f\textsuperscript 2+2 c j p q
   f\textsuperscript 2+e j\textsuperscript 2 f+e p\textsuperscript 2 f+e j\textsuperscript 2 q\textsuperscript 2 f-h j q\textsuperscript 2 f+2 c i j q\textsuperscript 2 f+e f-h j f+2 c i j f-m p f+2 c n p
   f+j m q f-2 c j n q f+h p q f-2 c i p q f-2 e j p q f-c i\textsuperscript 2-c n\textsuperscript 2-c i\textsuperscript 2 q\textsuperscript 2+h i q\textsuperscript 2-e i j
   q\textsuperscript 2-c+h i-e i j+m n-e n p-i m q-h n q+2 c i n q+e j n q+e i p q) (-(g k+h m+i
   n+j p+q) (-b j p f\textsuperscript 2+a c j p f\textsuperscript 2-b q f\textsuperscript 2+a c q f\textsuperscript 2+b j n f-a c j n f-g p f+a h p f+b
   i p f-a c i p f+d j p f-a e j p f+d q f-a e q f+g n-a h n-b i n+a c i n-d j n+a e j n-b
   q+a c q)+(d k+e m+f n+p) (b p i\textsuperscript 2-a c p i\textsuperscript 2-b j n i+a c j n i-g p i+a h p i+d
   j p i-a e j p i-b f j p i+a c f j p i+d q i-a e q i-b f q i+a c f q i-d j\textsuperscript 2 n+a e j\textsuperscript 2 n+b
   f j\textsuperscript 2 n-a c f j\textsuperscript 2 n-d n+a e n+b f n-a c f n+g j n-a h j n+b p-a c p-b j q+a c j
   q)-(a k+m) (a j\textsuperscript 2 n e\textsuperscript 2+a n e\textsuperscript 2-a i j p e\textsuperscript 2-a i q e\textsuperscript 2-a f j\textsuperscript 2 m e-a f m e+a i
   j m e-d j\textsuperscript 2 n e+b f j\textsuperscript 2 n e-a c f j\textsuperscript 2 n e-d n e+b f n e-a c f n e+g j n e-2 a h j n e-b i
   j n e+a c i j n e+b i\textsuperscript 2 p e-a c i\textsuperscript 2 p e+b p e-a c p e-g i p e+a h i p e+a f h j p e+d i j
   p e-b f i j p e+a c f i j p e+a f h q e+d i q e-b f i q e+a c f i q e-b j q e+a c j q e-b
   f\textsuperscript 2 m+a c f\textsuperscript 2 m-b i\textsuperscript 2 m+a c i\textsuperscript 2 m-b f\textsuperscript 2 j\textsuperscript 2 m+a c f\textsuperscript 2 j\textsuperscript 2 m+d f j\textsuperscript 2 m-b m+a c m+d f m+g i
   m-a h i m-f g j m+a f h j m-d i j m+2 b f i j m-2 a c f i j m+a h\textsuperscript 2 n+a n-g h n+b h i n-a
   c h i n+d h j n-b f h j n+a c f h j n-a f h\textsuperscript 2 p-a f p+f g h p-b f h i p+a c f h i p+b f\textsuperscript 2
   h j p-a c f\textsuperscript 2 h j p-d f h j p+b f\textsuperscript 2 h q-a c f\textsuperscript 2 h q+b h q-a c h q-d f h q-a i q+a f j
   q)+k (c f\textsuperscript 2 m a\textsuperscript 2+c i\textsuperscript 2 m a\textsuperscript 2+c f\textsuperscript 2 j\textsuperscript 2 m a\textsuperscript 2-e f j\textsuperscript 2 m a\textsuperscript 2+c m a\textsuperscript 2-e f m
   a\textsuperscript 2-h i m a\textsuperscript 2+f h j m a\textsuperscript 2+e i j m a\textsuperscript 2-2 c f i j m a\textsuperscript 2+e\textsuperscript 2 n a\textsuperscript 2+h\textsuperscript 2 n a\textsuperscript 2+e\textsuperscript 2 j\textsuperscript 2 n a\textsuperscript 2-c
   e f j\textsuperscript 2 n a\textsuperscript 2-c e f n a\textsuperscript 2-c h i n a\textsuperscript 2-2 e h j n a\textsuperscript 2+c f h j n a\textsuperscript 2+c e i j n a\textsuperscript 2+n a\textsuperscript 2-f
   h\textsuperscript 2 p a\textsuperscript 2-c e i\textsuperscript 2 p a\textsuperscript 2-c e p a\textsuperscript 2-f p a\textsuperscript 2+e h i p a\textsuperscript 2+c f h i p a\textsuperscript 2-c f\textsuperscript 2 h j p a\textsuperscript 2+e f h
   j p a\textsuperscript 2-e\textsuperscript 2 i j p a\textsuperscript 2+c e f i j p a\textsuperscript 2-c f\textsuperscript 2 h q a\textsuperscript 2-c h q a\textsuperscript 2+e f h q a\textsuperscript 2-e\textsuperscript 2 i q a\textsuperscript 2+c e
   f i q a\textsuperscript 2-i q a\textsuperscript 2+c e j q a\textsuperscript 2+f j q a\textsuperscript 2-c f\textsuperscript 2 k a-c i\textsuperscript 2 k a-c f\textsuperscript 2 j\textsuperscript 2 k a+e f j\textsuperscript 2 k a-c k
   a+e f k a+h i k a-f h j k a-e i j k a+2 c f i j k a-b f\textsuperscript 2 m a-b i\textsuperscript 2 m a-b f\textsuperscript 2 j\textsuperscript 2 m a+d f
   j\textsuperscript 2 m a-b m a+d f m a+g i m a-f g j m a-d i j m a+2 b f i j m a-2 d e j\textsuperscript 2 n a+c d f j\textsuperscript 2 n
   a+b e f j\textsuperscript 2 n a-2 d e n a+c d f n a+b e f n a-2 g h n a+c g i n a+b h i n a+2 e g j n a-c
   f g j n a+2 d h j n a-b f h j n a-c d i j n a-b e i j n a+c d i\textsuperscript 2 p a+b e i\textsuperscript 2 p a+c d p
   a+b e p a+2 f g h p a-e g i p a-c f g i p a-d h i p a-b f h i p a+c f\textsuperscript 2 g j p a-e f g j p
   a+b f\textsuperscript 2 h j p a-d f h j p a+2 d e i j p a-c d f i j p a-b e f i j p a+c f\textsuperscript 2 g q a+c g q
   a-e f g q a+b f\textsuperscript 2 h q a+b h q a-d f h q a+2 d e i q a-c d f i q a-b e f i q a-c d j q a-b
   e j q a+b f\textsuperscript 2 k+b i\textsuperscript 2 k+b f\textsuperscript 2 j\textsuperscript 2 k-d f j\textsuperscript 2 k+b k-d f k-g i k+f g j k+d i j k-2 b f i j
   k+d\textsuperscript 2 n+g\textsuperscript 2 n+d\textsuperscript 2 j\textsuperscript 2 n-b d f j\textsuperscript 2 n-b d f n-b g i n-2 d g j n+b f g j n+b d i j n+n-f g\textsuperscript 2
   p-b d i\textsuperscript 2 p-b d p-f p+d g i p+b f g i p-b f\textsuperscript 2 g j p+d f g j p-d\textsuperscript 2 i j p+b d f i j p-b f\textsuperscript 2
   g q-b g q+d f g q-d\textsuperscript 2 i q+b d f i q-i q+b d j q+f j q)+(-b j\textsuperscript 2 f\textsuperscript 2+a c j\textsuperscript 2
   f\textsuperscript 2-b f\textsuperscript 2+a c f\textsuperscript 2+d j\textsuperscript 2 f-a e j\textsuperscript 2 f+d f-a e f-g j f+a h j f+2 b i j f-2 a c i j f-b i\textsuperscript 2+a
   c i\textsuperscript 2-b+a c+g i-a h i-d i j+a e i j)
   (k\textsuperscript 2+m\textsuperscript 2+n\textsuperscript 2+p\textsuperscript 2+q\textsuperscript 2+1))))+(g\textsuperscript 2+h\textsuperscript 2+i\textsuperscript 2+j\textsuperscript 2+1)
   (q\textsuperscript 2+1) (d (-d q\textsuperscript 2 j\textsuperscript 2+a e q\textsuperscript 2 j\textsuperscript 2+b f q\textsuperscript 2 j\textsuperscript 2-a c f q\textsuperscript 2 j\textsuperscript 2-d j\textsuperscript 2+a
   e j\textsuperscript 2+b f j\textsuperscript 2-a c f j\textsuperscript 2+g q\textsuperscript 2 j-a h q\textsuperscript 2 j-b i q\textsuperscript 2 j+a c i q\textsuperscript 2 j+g j-a h j-b i j+a c i j-k
   q j+a m q j+b n q j-a c n q j+2 d p q j-2 a e p q j-2 b f p q j+2 a c f p q j-d p\textsuperscript 2+a e
   p\textsuperscript 2+b f p\textsuperscript 2-a c f p\textsuperscript 2-d+a e+b f-a c f+k p-a m p-b n p+a c n p-g p q+a h p q+b i p q-a c i
   p q) ((a b+c) (b d+c e+f) (-f q\textsuperscript 2 j\textsuperscript 2-f j\textsuperscript 2+i q\textsuperscript 2 j+i j-n q j+2 f p q j-f
   p\textsuperscript 2-f+n p-i p q) (-c j\textsuperscript 2 f\textsuperscript 2-c p\textsuperscript 2 f\textsuperscript 2-c j\textsuperscript 2 q\textsuperscript 2 f\textsuperscript 2-c f\textsuperscript 2+2 c j p q f\textsuperscript 2+e j\textsuperscript 2
   f+e p\textsuperscript 2 f+e j\textsuperscript 2 q\textsuperscript 2 f-h j q\textsuperscript 2 f+2 c i j q\textsuperscript 2 f+e f-h j f+2 c i j f-m p f+2 c n p f+j m q
   f-2 c j n q f+h p q f-2 c i p q f-2 e j p q f-c i\textsuperscript 2-c n\textsuperscript 2-c i\textsuperscript 2 q\textsuperscript 2+h i q\textsuperscript 2-e i j q\textsuperscript 2-c+h
   i-e i j+m n-e n p-i m q-h n q+2 c i n q+e j n q+e i p q)-(b\textsuperscript 2+c\textsuperscript 2+1) (a
   d+e) (-e q\textsuperscript 2 j\textsuperscript 2+c f q\textsuperscript 2 j\textsuperscript 2-e j\textsuperscript 2+c f j\textsuperscript 2+h q\textsuperscript 2 j-c i q\textsuperscript 2 j+h j-c i j-m q j+c n q
   j+2 e p q j-2 c f p q j-e p\textsuperscript 2+c f p\textsuperscript 2-e+c f+m p-c n p-h p q+c i p q) (j\textsuperscript 2
   f\textsuperscript 2+p\textsuperscript 2 f\textsuperscript 2+j\textsuperscript 2 q\textsuperscript 2 f\textsuperscript 2-2 j p q f\textsuperscript 2+f\textsuperscript 2-2 i j q\textsuperscript 2 f-2 i j f-2 n p f+2 j n q f+2 i p q
   f+i\textsuperscript 2+n\textsuperscript 2+i\textsuperscript 2 q\textsuperscript 2-2 i n q+1))-(b d+c e+f) (-f q\textsuperscript 2 j\textsuperscript 2-f j\textsuperscript 2+i q\textsuperscript 2 j+i
   j-n q j+2 f p q j-f p\textsuperscript 2-f+n p-i p q) (a (b d+c e+f) (-f q\textsuperscript 2 j\textsuperscript 2-f j\textsuperscript 2+i
   q\textsuperscript 2 j+i j-n q j+2 f p q j-f p\textsuperscript 2-f+n p-i p q) ((g k+h m+i n+j p+q) (a i n
   c\textsuperscript 2-a f j n c\textsuperscript 2-a f i p c\textsuperscript 2+a f\textsuperscript 2 j p c\textsuperscript 2+a f\textsuperscript 2 q c\textsuperscript 2+a q c\textsuperscript 2-a i m c+a f j m c+g n c-a h
   n c-b i n c-d j n c+a e j n c+b f j n c-f g p c+a f h p c+a e i p c+b f i p c-b f\textsuperscript 2 j p
   c+d f j p c-2 a e f j p c-b f\textsuperscript 2 q c-b q c+d f q c-2 a e f q c-g m+a h m+b i m+d j m-a e j
   m-b f j m+e g p-a e h p-b e i p+a e\textsuperscript 2 j p-d e j p+b e f j p+a e\textsuperscript 2 q+a q-d e q+b e f
   q)-(d k+e m+f n+p) (-a f j\textsuperscript 2 n c\textsuperscript 2-a f n c\textsuperscript 2+a i j n c\textsuperscript 2-a i\textsuperscript 2 p c\textsuperscript 2-a p c\textsuperscript 2+a
   f i j p c\textsuperscript 2+a f i q c\textsuperscript 2+a j q c\textsuperscript 2+a f j\textsuperscript 2 m c+a f m c-a i j m c-d j\textsuperscript 2 n c+a e j\textsuperscript 2 n c+b f
   j\textsuperscript 2 n c-d n c+a e n c+b f n c+g j n c-a h j n c-b i j n c+b i\textsuperscript 2 p c+b p c-g i p c+2 a h i
   p c-a f h j p c+d i j p c-a e i j p c-b f i j p c-a f h q c+d i q c-a e i q c-b f i q c-b
   j q c+d j\textsuperscript 2 m-a e j\textsuperscript 2 m-b f j\textsuperscript 2 m+d m-a e m-b f m-g j m+a h j m+b i j m-a h\textsuperscript 2 p-a p+g h
   p-b h i p-d h j p+a e h j p+b f h j p-d h q+a e h q+b f h q+a j q)+(b k+c m+n)
   (a j\textsuperscript 2 n e\textsuperscript 2+a n e\textsuperscript 2-a i j p e\textsuperscript 2-a i q e\textsuperscript 2-a f j\textsuperscript 2 m e-a f m e+a i j m e-d j\textsuperscript 2 n e+b
   f j\textsuperscript 2 n e-a c f j\textsuperscript 2 n e-d n e+b f n e-a c f n e+g j n e-2 a h j n e-b i j n e+a c i j n
   e+b i\textsuperscript 2 p e-a c i\textsuperscript 2 p e+b p e-a c p e-g i p e+a h i p e+a f h j p e+d i j p e-b f i j p
   e+a c f i j p e+a f h q e+d i q e-b f i q e+a c f i q e-b j q e+a c j q e-b f\textsuperscript 2 m+a c f\textsuperscript 2
   m-b i\textsuperscript 2 m+a c i\textsuperscript 2 m-b f\textsuperscript 2 j\textsuperscript 2 m+a c f\textsuperscript 2 j\textsuperscript 2 m+d f j\textsuperscript 2 m-b m+a c m+d f m+g i m-a h i m-f g
   j m+a f h j m-d i j m+2 b f i j m-2 a c f i j m+a h\textsuperscript 2 n+a n-g h n+b h i n-a c h i n+d h j
   n-b f h j n+a c f h j n-a f h\textsuperscript 2 p-a f p+f g h p-b f h i p+a c f h i p+b f\textsuperscript 2 h j p-a c f\textsuperscript 2
   h j p-d f h j p+b f\textsuperscript 2 h q-a c f\textsuperscript 2 h q+b h q-a c h q-d f h q-a i q+a f j q)-(a
   f\textsuperscript 2 c\textsuperscript 2+a i\textsuperscript 2 c\textsuperscript 2+a f\textsuperscript 2 j\textsuperscript 2 c\textsuperscript 2+a c\textsuperscript 2-2 a f i j c\textsuperscript 2-b f\textsuperscript 2 c-b i\textsuperscript 2 c-b f\textsuperscript 2 j\textsuperscript 2 c+d f j\textsuperscript 2
   c-2 a e f j\textsuperscript 2 c-b c+d f c-2 a e f c+g i c-2 a h i c-f g j c+2 a f h j c-d i j c+2 a e i j
   c+2 b f i j c+a e\textsuperscript 2+a h\textsuperscript 2+a e\textsuperscript 2 j\textsuperscript 2-d e j\textsuperscript 2+b e f j\textsuperscript 2+a-d e+b e f-g h+b h i+e g j+d h j-2
   a e h j-b f h j-b e i j) (k\textsuperscript 2+m\textsuperscript 2+n\textsuperscript 2+p\textsuperscript 2+q\textsuperscript 2+1)-k
   (-(g\textsuperscript 2+h\textsuperscript 2+i\textsuperscript 2+j\textsuperscript 2+1) (-(d\textsuperscript 2+e\textsuperscript 2+f\textsuperscript 2+1) (-m b\textsuperscript 2+c k
   b+a c m b+a n b-a c\textsuperscript 2 k-a k-m+c n)+(b d+c e+f) (b e k-a c e k-a f k-b d m+a c d m-f
   m+a d n+e n)+(-e b\textsuperscript 2+c d b+a c e b+a f b-a c\textsuperscript 2 d-a d-e+c f) (d k+e m+f
   n+p))+(d g+e h+f i+j) (-(d g+e h+f i+j) (-m b\textsuperscript 2+c k b+a c m b+a n b-a c\textsuperscript 2
   k-a k-m+c n)+(b d+c e+f) (b h k-a c h k-a i k-b g m+a c g m-i m+a g n+h n)+(-h
   b\textsuperscript 2+c g b+a c h b+a i b-a c\textsuperscript 2 g-a g-h+c i) (d k+e m+f n+p))-(b g+c h+i)
   (-(d g+e h+f i+j) (b e k-a c e k-a f k-b d m+a c d m-f m+a d n+e
   n)+(d\textsuperscript 2+e\textsuperscript 2+f\textsuperscript 2+1) (b h k-a c h k-a i k-b g m+a c g m-i m+a g n+h n)+(b e g-a
   c e g-a f g-b d h+a c d h-f h+a d i+e i) (d k+e m+f
   n+p))+(-(d\textsuperscript 2+e\textsuperscript 2+f\textsuperscript 2+1) (-h b\textsuperscript 2+c g b+a c h b+a i b-a c\textsuperscript 2 g-a
   g-h+c i)+(b d+c e+f) (b e g-a c e g-a f g-b d h+a c d h-f h+a d i+e i)+(-e
   b\textsuperscript 2+c d b+a c e b+a f b-a c\textsuperscript 2 d-a d-e+c f) (d g+e h+f i+j)) (g k+h m+i n+j
   p+q)))-b (a d+e) (-e q\textsuperscript 2 j\textsuperscript 2+c f q\textsuperscript 2 j\textsuperscript 2-e j\textsuperscript 2+c f j\textsuperscript 2+h q\textsuperscript 2 j-c i q\textsuperscript 2
   j+h j-c i j-m q j+c n q j+2 e p q j-2 c f p q j-e p\textsuperscript 2+c f p\textsuperscript 2-e+c f+m p-c n p-h p q+c i p
   q) (-(g k+h m+i n+j p+q) (-b j p f\textsuperscript 2+a c j p f\textsuperscript 2-b q f\textsuperscript 2+a c q f\textsuperscript 2+b j n
   f-a c j n f-g p f+a h p f+b i p f-a c i p f+d j p f-a e j p f+d q f-a e q f+g n-a h n-b i
   n+a c i n-d j n+a e j n-b q+a c q)+(d k+e m+f n+p) (b p i\textsuperscript 2-a c p i\textsuperscript 2-b j n
   i+a c j n i-g p i+a h p i+d j p i-a e j p i-b f j p i+a c f j p i+d q i-a e q i-b f q i+a
   c f q i-d j\textsuperscript 2 n+a e j\textsuperscript 2 n+b f j\textsuperscript 2 n-a c f j\textsuperscript 2 n-d n+a e n+b f n-a c f n+g j n-a h j n+b
   p-a c p-b j q+a c j q)-(a k+m) (a j\textsuperscript 2 n e\textsuperscript 2+a n e\textsuperscript 2-a i j p e\textsuperscript 2-a i q e\textsuperscript 2-a f
   j\textsuperscript 2 m e-a f m e+a i j m e-d j\textsuperscript 2 n e+b f j\textsuperscript 2 n e-a c f j\textsuperscript 2 n e-d n e+b f n e-a c f n e+g j
   n e-2 a h j n e-b i j n e+a c i j n e+b i\textsuperscript 2 p e-a c i\textsuperscript 2 p e+b p e-a c p e-g i p e+a h i p
   e+a f h j p e+d i j p e-b f i j p e+a c f i j p e+a f h q e+d i q e-b f i q e+a c f i q
   e-b j q e+a c j q e-b f\textsuperscript 2 m+a c f\textsuperscript 2 m-b i\textsuperscript 2 m+a c i\textsuperscript 2 m-b f\textsuperscript 2 j\textsuperscript 2 m+a c f\textsuperscript 2 j\textsuperscript 2 m+d f j\textsuperscript 2
   m-b m+a c m+d f m+g i m-a h i m-f g j m+a f h j m-d i j m+2 b f i j m-2 a c f i j m+a h\textsuperscript 2
   n+a n-g h n+b h i n-a c h i n+d h j n-b f h j n+a c f h j n-a f h\textsuperscript 2 p-a f p+f g h p-b f h
   i p+a c f h i p+b f\textsuperscript 2 h j p-a c f\textsuperscript 2 h j p-d f h j p+b f\textsuperscript 2 h q-a c f\textsuperscript 2 h q+b h q-a c h q-d
   f h q-a i q+a f j q)+k (c f\textsuperscript 2 m a\textsuperscript 2+c i\textsuperscript 2 m a\textsuperscript 2+c f\textsuperscript 2 j\textsuperscript 2 m a\textsuperscript 2-e f j\textsuperscript 2 m
   a\textsuperscript 2+c m a\textsuperscript 2-e f m a\textsuperscript 2-h i m a\textsuperscript 2+f h j m a\textsuperscript 2+e i j m a\textsuperscript 2-2 c f i j m a\textsuperscript 2+e\textsuperscript 2 n a\textsuperscript 2+h\textsuperscript 2 n
   a\textsuperscript 2+e\textsuperscript 2 j\textsuperscript 2 n a\textsuperscript 2-c e f j\textsuperscript 2 n a\textsuperscript 2-c e f n a\textsuperscript 2-c h i n a\textsuperscript 2-2 e h j n a\textsuperscript 2+c f h j n a\textsuperscript 2+c e
   i j n a\textsuperscript 2+n a\textsuperscript 2-f h\textsuperscript 2 p a\textsuperscript 2-c e i\textsuperscript 2 p a\textsuperscript 2-c e p a\textsuperscript 2-f p a\textsuperscript 2+e h i p a\textsuperscript 2+c f h i p a\textsuperscript 2-c
   f\textsuperscript 2 h j p a\textsuperscript 2+e f h j p a\textsuperscript 2-e\textsuperscript 2 i j p a\textsuperscript 2+c e f i j p a\textsuperscript 2-c f\textsuperscript 2 h q a\textsuperscript 2-c h q a\textsuperscript 2+e f h q
   a\textsuperscript 2-e\textsuperscript 2 i q a\textsuperscript 2+c e f i q a\textsuperscript 2-i q a\textsuperscript 2+c e j q a\textsuperscript 2+f j q a\textsuperscript 2-c f\textsuperscript 2 k a-c i\textsuperscript 2 k a-c f\textsuperscript 2 j\textsuperscript 2
   k a+e f j\textsuperscript 2 k a-c k a+e f k a+h i k a-f h j k a-e i j k a+2 c f i j k a-b f\textsuperscript 2 m a-b i\textsuperscript 2 m
   a-b f\textsuperscript 2 j\textsuperscript 2 m a+d f j\textsuperscript 2 m a-b m a+d f m a+g i m a-f g j m a-d i j m a+2 b f i j m a-2 d e
   j\textsuperscript 2 n a+c d f j\textsuperscript 2 n a+b e f j\textsuperscript 2 n a-2 d e n a+c d f n a+b e f n a-2 g h n a+c g i n a+b h
   i n a+2 e g j n a-c f g j n a+2 d h j n a-b f h j n a-c d i j n a-b e i j n a+c d i\textsuperscript 2 p
   a+b e i\textsuperscript 2 p a+c d p a+b e p a+2 f g h p a-e g i p a-c f g i p a-d h i p a-b f h i p a+c
   f\textsuperscript 2 g j p a-e f g j p a+b f\textsuperscript 2 h j p a-d f h j p a+2 d e i j p a-c d f i j p a-b e f i j p
   a+c f\textsuperscript 2 g q a+c g q a-e f g q a+b f\textsuperscript 2 h q a+b h q a-d f h q a+2 d e i q a-c d f i q a-b e
   f i q a-c d j q a-b e j q a+b f\textsuperscript 2 k+b i\textsuperscript 2 k+b f\textsuperscript 2 j\textsuperscript 2 k-d f j\textsuperscript 2 k+b k-d f k-g i k+f g j
   k+d i j k-2 b f i j k+d\textsuperscript 2 n+g\textsuperscript 2 n+d\textsuperscript 2 j\textsuperscript 2 n-b d f j\textsuperscript 2 n-b d f n-b g i n-2 d g j n+b f g j
   n+b d i j n+n-f g\textsuperscript 2 p-b d i\textsuperscript 2 p-b d p-f p+d g i p+b f g i p-b f\textsuperscript 2 g j p+d f g j p-d\textsuperscript 2 i j
   p+b d f i j p-b f\textsuperscript 2 g q-b g q+d f g q-d\textsuperscript 2 i q+b d f i q-i q+b d j q+f j q)+(-b
   j\textsuperscript 2 f\textsuperscript 2+a c j\textsuperscript 2 f\textsuperscript 2-b f\textsuperscript 2+a c f\textsuperscript 2+d j\textsuperscript 2 f-a e j\textsuperscript 2 f+d f-a e f-g j f+a h j f+2 b i j f-2 a
   c i j f-b i\textsuperscript 2+a c i\textsuperscript 2-b+a c+g i-a h i-d i j+a e i j)
   (k\textsuperscript 2+m\textsuperscript 2+n\textsuperscript 2+p\textsuperscript 2+q\textsuperscript 2+1)))+(d\textsuperscript 2+e\textsuperscript 2+f\textsuperscript 2+1) (q\textsuperscript 2
   j\textsuperscript 2+j\textsuperscript 2-2 p q j+p\textsuperscript 2+1) (a (b\textsuperscript 2+c\textsuperscript 2+1) (j\textsuperscript 2 f\textsuperscript 2+p\textsuperscript 2 f\textsuperscript 2+j\textsuperscript 2 q\textsuperscript 2
   f\textsuperscript 2-2 j p q f\textsuperscript 2+f\textsuperscript 2-2 i j q\textsuperscript 2 f-2 i j f-2 n p f+2 j n q f+2 i p q f+i\textsuperscript 2+n\textsuperscript 2+i\textsuperscript 2 q\textsuperscript 2-2 i n
   q+1) ((g k+h m+i n+j p+q) (a i n c\textsuperscript 2-a f j n c\textsuperscript 2-a f i p c\textsuperscript 2+a f\textsuperscript 2 j p
   c\textsuperscript 2+a f\textsuperscript 2 q c\textsuperscript 2+a q c\textsuperscript 2-a i m c+a f j m c+g n c-a h n c-b i n c-d j n c+a e j n c+b f j n
   c-f g p c+a f h p c+a e i p c+b f i p c-b f\textsuperscript 2 j p c+d f j p c-2 a e f j p c-b f\textsuperscript 2 q c-b q
   c+d f q c-2 a e f q c-g m+a h m+b i m+d j m-a e j m-b f j m+e g p-a e h p-b e i p+a e\textsuperscript 2 j
   p-d e j p+b e f j p+a e\textsuperscript 2 q+a q-d e q+b e f q)-(d k+e m+f n+p) (-a f j\textsuperscript 2 n
   c\textsuperscript 2-a f n c\textsuperscript 2+a i j n c\textsuperscript 2-a i\textsuperscript 2 p c\textsuperscript 2-a p c\textsuperscript 2+a f i j p c\textsuperscript 2+a f i q c\textsuperscript 2+a j q c\textsuperscript 2+a f j\textsuperscript 2
   m c+a f m c-a i j m c-d j\textsuperscript 2 n c+a e j\textsuperscript 2 n c+b f j\textsuperscript 2 n c-d n c+a e n c+b f n c+g j n c-a h
   j n c-b i j n c+b i\textsuperscript 2 p c+b p c-g i p c+2 a h i p c-a f h j p c+d i j p c-a e i j p c-b f
   i j p c-a f h q c+d i q c-a e i q c-b f i q c-b j q c+d j\textsuperscript 2 m-a e j\textsuperscript 2 m-b f j\textsuperscript 2 m+d m-a e
   m-b f m-g j m+a h j m+b i j m-a h\textsuperscript 2 p-a p+g h p-b h i p-d h j p+a e h j p+b f h j p-d h
   q+a e h q+b f h q+a j q)+(b k+c m+n) (a j\textsuperscript 2 n e\textsuperscript 2+a n e\textsuperscript 2-a i j p e\textsuperscript 2-a i q
   e\textsuperscript 2-a f j\textsuperscript 2 m e-a f m e+a i j m e-d j\textsuperscript 2 n e+b f j\textsuperscript 2 n e-a c f j\textsuperscript 2 n e-d n e+b f n e-a c f
   n e+g j n e-2 a h j n e-b i j n e+a c i j n e+b i\textsuperscript 2 p e-a c i\textsuperscript 2 p e+b p e-a c p e-g i p
   e+a h i p e+a f h j p e+d i j p e-b f i j p e+a c f i j p e+a f h q e+d i q e-b f i q e+a
   c f i q e-b j q e+a c j q e-b f\textsuperscript 2 m+a c f\textsuperscript 2 m-b i\textsuperscript 2 m+a c i\textsuperscript 2 m-b f\textsuperscript 2 j\textsuperscript 2 m+a c f\textsuperscript 2 j\textsuperscript 2
   m+d f j\textsuperscript 2 m-b m+a c m+d f m+g i m-a h i m-f g j m+a f h j m-d i j m+2 b f i j m-2 a c f i
   j m+a h\textsuperscript 2 n+a n-g h n+b h i n-a c h i n+d h j n-b f h j n+a c f h j n-a f h\textsuperscript 2 p-a f p+f g
   h p-b f h i p+a c f h i p+b f\textsuperscript 2 h j p-a c f\textsuperscript 2 h j p-d f h j p+b f\textsuperscript 2 h q-a c f\textsuperscript 2 h q+b h
   q-a c h q-d f h q-a i q+a f j q)-(a f\textsuperscript 2 c\textsuperscript 2+a i\textsuperscript 2 c\textsuperscript 2+a f\textsuperscript 2 j\textsuperscript 2 c\textsuperscript 2+a c\textsuperscript 2-2 a
   f i j c\textsuperscript 2-b f\textsuperscript 2 c-b i\textsuperscript 2 c-b f\textsuperscript 2 j\textsuperscript 2 c+d f j\textsuperscript 2 c-2 a e f j\textsuperscript 2 c-b c+d f c-2 a e f c+g i c-2
   a h i c-f g j c+2 a f h j c-d i j c+2 a e i j c+2 b f i j c+a e\textsuperscript 2+a h\textsuperscript 2+a e\textsuperscript 2 j\textsuperscript 2-d e
   j\textsuperscript 2+b e f j\textsuperscript 2+a-d e+b e f-g h+b h i+e g j+d h j-2 a e h j-b f h j-b e i j)
   (k\textsuperscript 2+m\textsuperscript 2+n\textsuperscript 2+p\textsuperscript 2+q\textsuperscript 2+1)-k (-(g\textsuperscript 2+h\textsuperscript 2+i\textsuperscript 2+j\textsuperscript 2+1)
   (-(d\textsuperscript 2+e\textsuperscript 2+f\textsuperscript 2+1) (-m b\textsuperscript 2+c k b+a c m b+a n b-a c\textsuperscript 2 k-a k-m+c
   n)+(b d+c e+f) (b e k-a c e k-a f k-b d m+a c d m-f m+a d n+e n)+(-e b\textsuperscript 2+c d
   b+a c e b+a f b-a c\textsuperscript 2 d-a d-e+c f) (d k+e m+f n+p))+(d g+e h+f i+j) (-(d
   g+e h+f i+j) (-m b\textsuperscript 2+c k b+a c m b+a n b-a c\textsuperscript 2 k-a k-m+c n)+(b d+c e+f) (b h
   k-a c h k-a i k-b g m+a c g m-i m+a g n+h n)+(-h b\textsuperscript 2+c g b+a c h b+a i b-a c\textsuperscript 2 g-a
   g-h+c i) (d k+e m+f n+p))-(b g+c h+i) (-(d g+e h+f i+j) (b e k-a c e k-a
   f k-b d m+a c d m-f m+a d n+e n)+(d\textsuperscript 2+e\textsuperscript 2+f\textsuperscript 2+1) (b h k-a c h k-a i k-b g m+a
   c g m-i m+a g n+h n)+(b e g-a c e g-a f g-b d h+a c d h-f h+a d i+e i) (d k+e m+f
   n+p))+(-(d\textsuperscript 2+e\textsuperscript 2+f\textsuperscript 2+1) (-h b\textsuperscript 2+c g b+a c h b+a i b-a c\textsuperscript 2 g-a
   g-h+c i)+(b d+c e+f) (b e g-a c e g-a f g-b d h+a c d h-f h+a d i+e i)+(-e
   b\textsuperscript 2+c d b+a c e b+a f b-a c\textsuperscript 2 d-a d-e+c f) (d g+e h+f i+j)) (g k+h m+i n+j
   p+q)))-b (a b+c) (-c j\textsuperscript 2 f\textsuperscript 2-c p\textsuperscript 2 f\textsuperscript 2-c j\textsuperscript 2 q\textsuperscript 2 f\textsuperscript 2-c f\textsuperscript 2+2 c j p q
   f\textsuperscript 2+e j\textsuperscript 2 f+e p\textsuperscript 2 f+e j\textsuperscript 2 q\textsuperscript 2 f-h j q\textsuperscript 2 f+2 c i j q\textsuperscript 2 f+e f-h j f+2 c i j f-m p f+2 c n p
   f+j m q f-2 c j n q f+h p q f-2 c i p q f-2 e j p q f-c i\textsuperscript 2-c n\textsuperscript 2-c i\textsuperscript 2 q\textsuperscript 2+h i q\textsuperscript 2-e i j
   q\textsuperscript 2-c+h i-e i j+m n-e n p-i m q-h n q+2 c i n q+e j n q+e i p q) (-(g k+h m+i
   n+j p+q) (-b j p f\textsuperscript 2+a c j p f\textsuperscript 2-b q f\textsuperscript 2+a c q f\textsuperscript 2+b j n f-a c j n f-g p f+a h p f+b
   i p f-a c i p f+d j p f-a e j p f+d q f-a e q f+g n-a h n-b i n+a c i n-d j n+a e j n-b
   q+a c q)+(d k+e m+f n+p) (b p i\textsuperscript 2-a c p i\textsuperscript 2-b j n i+a c j n i-g p i+a h p i+d
   j p i-a e j p i-b f j p i+a c f j p i+d q i-a e q i-b f q i+a c f q i-d j\textsuperscript 2 n+a e j\textsuperscript 2 n+b
   f j\textsuperscript 2 n-a c f j\textsuperscript 2 n-d n+a e n+b f n-a c f n+g j n-a h j n+b p-a c p-b j q+a c j
   q)-(a k+m) (a j\textsuperscript 2 n e\textsuperscript 2+a n e\textsuperscript 2-a i j p e\textsuperscript 2-a i q e\textsuperscript 2-a f j\textsuperscript 2 m e-a f m e+a i
   j m e-d j\textsuperscript 2 n e+b f j\textsuperscript 2 n e-a c f j\textsuperscript 2 n e-d n e+b f n e-a c f n e+g j n e-2 a h j n e-b i
   j n e+a c i j n e+b i\textsuperscript 2 p e-a c i\textsuperscript 2 p e+b p e-a c p e-g i p e+a h i p e+a f h j p e+d i j
   p e-b f i j p e+a c f i j p e+a f h q e+d i q e-b f i q e+a c f i q e-b j q e+a c j q e-b
   f\textsuperscript 2 m+a c f\textsuperscript 2 m-b i\textsuperscript 2 m+a c i\textsuperscript 2 m-b f\textsuperscript 2 j\textsuperscript 2 m+a c f\textsuperscript 2 j\textsuperscript 2 m+d f j\textsuperscript 2 m-b m+a c m+d f m+g i
   m-a h i m-f g j m+a f h j m-d i j m+2 b f i j m-2 a c f i j m+a h\textsuperscript 2 n+a n-g h n+b h i n-a
   c h i n+d h j n-b f h j n+a c f h j n-a f h\textsuperscript 2 p-a f p+f g h p-b f h i p+a c f h i p+b f\textsuperscript 2
   h j p-a c f\textsuperscript 2 h j p-d f h j p+b f\textsuperscript 2 h q-a c f\textsuperscript 2 h q+b h q-a c h q-d f h q-a i q+a f j
   q)+k (c f\textsuperscript 2 m a\textsuperscript 2+c i\textsuperscript 2 m a\textsuperscript 2+c f\textsuperscript 2 j\textsuperscript 2 m a\textsuperscript 2-e f j\textsuperscript 2 m a\textsuperscript 2+c m a\textsuperscript 2-e f m
   a\textsuperscript 2-h i m a\textsuperscript 2+f h j m a\textsuperscript 2+e i j m a\textsuperscript 2-2 c f i j m a\textsuperscript 2+e\textsuperscript 2 n a\textsuperscript 2+h\textsuperscript 2 n a\textsuperscript 2+e\textsuperscript 2 j\textsuperscript 2 n a\textsuperscript 2-c
   e f j\textsuperscript 2 n a\textsuperscript 2-c e f n a\textsuperscript 2-c h i n a\textsuperscript 2-2 e h j n a\textsuperscript 2+c f h j n a\textsuperscript 2+c e i j n a\textsuperscript 2+n a\textsuperscript 2-f
   h\textsuperscript 2 p a\textsuperscript 2-c e i\textsuperscript 2 p a\textsuperscript 2-c e p a\textsuperscript 2-f p a\textsuperscript 2+e h i p a\textsuperscript 2+c f h i p a\textsuperscript 2-c f\textsuperscript 2 h j p a\textsuperscript 2+e f h
   j p a\textsuperscript 2-e\textsuperscript 2 i j p a\textsuperscript 2+c e f i j p a\textsuperscript 2-c f\textsuperscript 2 h q a\textsuperscript 2-c h q a\textsuperscript 2+e f h q a\textsuperscript 2-e\textsuperscript 2 i q a\textsuperscript 2+c e
   f i q a\textsuperscript 2-i q a\textsuperscript 2+c e j q a\textsuperscript 2+f j q a\textsuperscript 2-c f\textsuperscript 2 k a-c i\textsuperscript 2 k a-c f\textsuperscript 2 j\textsuperscript 2 k a+e f j\textsuperscript 2 k a-c k
   a+e f k a+h i k a-f h j k a-e i j k a+2 c f i j k a-b f\textsuperscript 2 m a-b i\textsuperscript 2 m a-b f\textsuperscript 2 j\textsuperscript 2 m a+d f
   j\textsuperscript 2 m a-b m a+d f m a+g i m a-f g j m a-d i j m a+2 b f i j m a-2 d e j\textsuperscript 2 n a+c d f j\textsuperscript 2 n
   a+b e f j\textsuperscript 2 n a-2 d e n a+c d f n a+b e f n a-2 g h n a+c g i n a+b h i n a+2 e g j n a-c
   f g j n a+2 d h j n a-b f h j n a-c d i j n a-b e i j n a+c d i\textsuperscript 2 p a+b e i\textsuperscript 2 p a+c d p
   a+b e p a+2 f g h p a-e g i p a-c f g i p a-d h i p a-b f h i p a+c f\textsuperscript 2 g j p a-e f g j p
   a+b f\textsuperscript 2 h j p a-d f h j p a+2 d e i j p a-c d f i j p a-b e f i j p a+c f\textsuperscript 2 g q a+c g q
   a-e f g q a+b f\textsuperscript 2 h q a+b h q a-d f h q a+2 d e i q a-c d f i q a-b e f i q a-c d j q a-b
   e j q a+b f\textsuperscript 2 k+b i\textsuperscript 2 k+b f\textsuperscript 2 j\textsuperscript 2 k-d f j\textsuperscript 2 k+b k-d f k-g i k+f g j k+d i j k-2 b f i j
   k+d\textsuperscript 2 n+g\textsuperscript 2 n+d\textsuperscript 2 j\textsuperscript 2 n-b d f j\textsuperscript 2 n-b d f n-b g i n-2 d g j n+b f g j n+b d i j n+n-f g\textsuperscript 2
   p-b d i\textsuperscript 2 p-b d p-f p+d g i p+b f g i p-b f\textsuperscript 2 g j p+d f g j p-d\textsuperscript 2 i j p+b d f i j p-b f\textsuperscript 2
   g q-b g q+d f g q-d\textsuperscript 2 i q+b d f i q-i q+b d j q+f j q)+(-b j\textsuperscript 2 f\textsuperscript 2+a c j\textsuperscript 2
   f\textsuperscript 2-b f\textsuperscript 2+a c f\textsuperscript 2+d j\textsuperscript 2 f-a e j\textsuperscript 2 f+d f-a e f-g j f+a h j f+2 b i j f-2 a c i j f-b i\textsuperscript 2+a
   c i\textsuperscript 2-b+a c+g i-a h i-d i j+a e i j)
   (k\textsuperscript 2+m\textsuperscript 2+n\textsuperscript 2+p\textsuperscript 2+q\textsuperscript 2+1)))))

\end{document}